\documentclass[a4paper,11pt]{article}
\usepackage{jheppubMod}
\usepackage{amsmath,amsfonts,amssymb,array}
\usepackage{bm,bbm,bbm,bbold}
\usepackage{float}
\usepackage[T1]{fontenc}
\usepackage{dsfont}
\usepackage{enumerate}
\usepackage{float}
\usepackage{graphicx}
\usepackage{lineno,lipsum}
\usepackage{multirow,multicol}
\usepackage{subfigure,slashed}
\usepackage{tabu}
\usepackage{datetime}
\usepackage{ulem}
\hypersetup{bookmarks=true,unicode=true,pdftoolbar=true,pdfmenubar=true,
  pdffitwindow=false,pdfstartview={FitH},
  pdftitle={Heavy Neutrinos with Dynamic Jet Vetoes},pdfauthor={Pascoli, Ruiz, Weiland},
  pdfsubject={Subject},pdfcreator={Creator},pdfproducer={Producer},
  pdfkeywords={Heavy Neutrinos}{Charged Lepton Flavor Violation}{Hadron Colliders}{Jet Vetoes},
  pdfnewwindow=true,colorlinks=true
}

\providecommand{\tabularnewline}{\\}

\newcolumntype{x}[1]{
{\centering\hspace{0pt}}p{#1}}

\newcommand{\MeV}{{\rm ~MeV}}
\newcommand{\GeV}{{\rm ~GeV}}
\newcommand{\TeV}{{\rm ~TeV}}
\newcommand{\pb}{{\rm ~pb}}
\newcommand{\fb}{{\rm ~fb}}

\newcommand{\invfb}{{\rm ~fb^{-1}}}
\newcommand{\invab}{{\rm ~ab^{-1}}}

\newcommand{\vareps}{\varepsilon}

\newcommand{\I}{\rm 1\kern-.24em l}

\newcommand{\pTVeto}{p_T^{\rm Veto}}
\newcommand{\nkll}[1]{{\rm N}^{#1}{\rm LL}}

\newcommand{\met}{{\rm MET}}

\newcommand{\mgamc}{\texttt{mg5amc}}

\newcommand{\mgpyma}{\texttt{MG5aMC@NLO+PY8+MA5}}
\newcommand{\confirm}[1]{{\color{black}#1}}
\title{Heavy Neutrinos with Dynamic Jet Vetoes:\\ 
Multilepton Searches at $\sqrt{s} = 14,~27,$ and $100$ TeV}

\author[a]{Silvia Pascoli,}
\author[b,a]{Richard Ruiz,}
\author[c,a]{and Cedric Weiland}

\affiliation[a]{Institute for Particle Physics Phenomenology {\rm(IPPP)},\\
Department of Physics, Durham University, South Road, Durham, DH1 3LE, UK}
\affiliation[b]{Centre for Cosmology, Particle Physics and Phenomenology {\rm (CP3)},\\
Universit\'e Catholique de Louvain, Chemin du Cyclotron, Louvain la Neuve, B-1348, Belgium,}
\affiliation[c]{PITTsburgh Particle Physics, Astrophysics, and Cosmology Center (PITT PACC),\\ 
Department of Physics and Astronomy, University of Pittsburgh, 3941 O'Hara Street,\\
 Pittsburgh, PA, 15260, USA}

\emailAdd{silvia.pascoli@durham.ac.uk}
\emailAdd{richard.ruiz@uclouvain.be}
\emailAdd{cew64@pitt.edu}

\abstract{Heavy neutrinos $(N)$ 
remain one of most promising explanations for the origin of neutrinos' tiny masses and large mixing angles.
In light of broad advances in understanding and modeling of hadron collisions at large momentum transfers,
we revisit the long-standard search strategy for heavy $N$  decaying to multiple charged leptons $(\ell)$, $pp \to N\ell X \to 3\ell \nu X$.
For electroweak and TeV-scale $N$, 
we propose a qualitatively new collider analysis premised on a dynamic jet veto
and discriminating,  on an event-by-event basis, according to the relative amount of hadronic and leptonic activity.
We report that the sensitivity to heavy neutrinos at the CERN Large Hadron Collider (LHC) can be improved by roughly an order of magnitude at both $\mathcal{L}=300\invfb$ and $3\invab$.
At $\sqrt{s}=14$ TeV with $\mathcal{L}=3~{\rm ab}^{-1}$, 
we find active-sterile mixing as small as $\vert V_{\ell N}\vert^2 = 10^{-2} ~(10^{-3})~[5\times10^{-4}]$ can be probed at 95\%~CL
for heavy Dirac neutrinos masses $m_N \lesssim 1200~(300)~[200]$ GeV,
well beyond the present $\vert V_{\ell N}\vert^2 \lesssim 10^{-3} -10^{-1}$ constraints for such heavy states set by indirect searches and precision measurements.
The improvement holds also for Majorana $N$, and is largely independent of whether charged lepton flavor is conserved or violated.
The analysis, built almost entirely from inclusive, transverse observables, is designed to be robust across increasing collider energies, 
and hence serves as a basis for searches at future colliders:
With $\mathcal{L}=15~{\rm ab}^{-1}$ at $\sqrt{s}=27$ TeV,
one  can probe mixing below $\vert V_{\ell N}\vert^2 = 10^{-2} ~(10^{-3})~[2\times10^{-4}]$ for $m_N \lesssim 3500~(700)~[200]$ GeV.
At a hypothetical 100 TeV $pp$ collider with $\mathcal{L}=30~{\rm ab}^{-1}$, one can probe mixing 
down to $9\times10^{-5}$ for $m_N \lesssim 200$ GeV,
below $10^{-3}$ for $m_N \lesssim 4$ TeV, and 
below $10^{-2}$ for $m_N \lesssim 15$ TeV.
We anticipate these results can be further improved with detector-specific tuning and application of multi-variant / machines learning techniques.
To facilitate such investigations, we make publicly available Monte Carlo libraries needed for the precision computations/simulations used in our study.
\setcounter{page}{1}
}

\preprint{CP3-18-77, ~IPPP/18/111, ~PITT-PACC-1821, ~VBSCAN-PUB-10-18}
\arxivnumber{1812.08750}

\begin{document}
\maketitle
\flushbottom
\setcounter{page}{2}

\section{Introduction}\label{sec:intro}

The origin of tiny neutrino masses and clarity on neutrinos' Majorana nature remain some of the most pressing mysteries in particle physics today.
To this end, the hypothesized existence of $j\geq2$ right-handed (RH) neutrinos 
$(\nu_R^j)$~\cite{Minkowski:1977sc, Yanagida:1979as, GellMann:1980vs, Glashow:1979nm,Mohapatra:1979ia, Shrock:1980ct, Schechter:1980gr}, 
represents one of the best, though far from only~\cite{Konetschny:1977bn, Cheng:1980qt,Lazarides:1980nt,Schechter:1980gr, Mohapatra:1980yp,
Foot:1988aq, Zee:1980ai, Hall:1983id, Zee:1985id, Babu:1988ki,Krauss:2002px}, solution to these questions.
With RH neutrinos, electroweak (EW)-scale Dirac masses can be generated spontaneously during EW symmetry breaking (EWSB),
so long as there exists Yukawa couplings to the left-handed (LH) leptons and the Standard Model (SM) Higgs field, $\mathcal{L}_Y = y^{ij}_\nu \overline{L^i} \tilde{\Phi}_{\rm SM} \nu_R^j + {\rm H.c.}$
Barring a new fundamental symmetry that ensures lepton number conservation (LNC) below the EW scale, 
RH neutrinos invariably possess RH Majorana masses $\mu_R^j$, but possibly also additional Dirac masses $(m_R)$ depending on the precise field content, 
that can suppress neutrino masses in two distinct limits~\cite{Pilaftsis:1991ug,Kersten:2007vk,Moffat:2017feq}:
For $\mu_R^j$ much larger than the EW scale $v_{\rm EW} = \sqrt{2}\langle\Phi_{\rm SM}\rangle\approx246\GeV$, the canonical, high-scale Type I Seesaw 
mechanism~\cite{Minkowski:1977sc, Yanagida:1979as, GellMann:1980vs, Glashow:1979nm,Mohapatra:1979ia, Shrock:1980ct, Schechter:1980gr} 
is triggered, leading to light neutrino mass eigenvalues inversely proportional to $\mu_R^j$,
\begin{equation}
\tilde{M}^{\rm light}_\nu \approx - \tilde{m}_D ~ \tilde{\mu}_R^{-1} ~ \tilde{m}_D^T.
\end{equation}
Here, the tilde $(\sim)$ denotes matrix-valued objects, 
with $\tilde{M}^{\rm light}$ representing the diagonal, light neutrino mass matrix and $\tilde{m}_D = \tilde{y}_\nu  \langle \Phi_{\rm SM}  \rangle$ the neutrino Dirac mass matrix.
For  $\mu_R^i$ that are of the EW scale or much smaller, one realizes the low-scale Type I Seesaw mechanisms, among which are the 
Inverse~\cite{Mohapatra:1986aw,Mohapatra:1986bd,Bernabeu:1987gr,Gavela:2009cd}
or Linear~\cite{Akhmedov:1995ip,Akhmedov:1995vm} Seesaws,
and finds light neutrino mass eigenstates with masses  proportional to small lepton number violating masses or couplings. For example, in the Inverse Seesaw one has
\begin{equation}
 \tilde{M}^{\rm light}_\nu  \approx  \tilde{m}_D ~\tilde{m}_R^{T-1} ~\tilde{\mu}_R ~\tilde{m}_R^{-1} ~\tilde{m}_D^T.
\end{equation}
In all cases, one predicts the existence of heavy neutrino mass eigenstates $(N_m)$  that can couple to SM particles through mass mixing, perhaps even appreciably.
While $\mu_R^j$ is the scale at which lepton number violation (LNV) occurs, without additional inputs, their values in relation to the EW scale are \textit{a~priori} arbitrary.
(Though in the vanishing $\mu_R,~m_D$ limits, accidental global symmetries are recovered.)
Hence, $N_m$ may be kinematically accessible at any number of laboratory-based experiments,
including collider facilities like the CERN Large Hadron Collider (LHC) 
and its potential successors~\cite{Avetisyan:2013onh,Blondel:2014bra,Arkani-Hamed:2015vfh,Mangano:2016jyj,Golling:2016gvc,CEPCStudyGroup:2018ghi,CidVidal:2018eel}.

\begin{figure}[!t]
\begin{center}
\includegraphics[width=.85\textwidth]{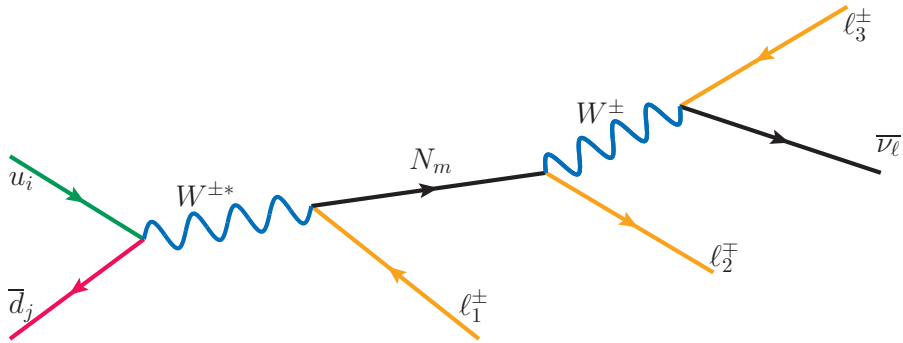}
\end{center}
\caption{Born diagram for charged current Drell-Yan production of heavy $N$ 
with subsequent decay to the three charged lepton (trilepton) final state. 
}
\label{fig:feynman_DYX_Trilep}
\end{figure}

As such, the discovery potential at colliders of particles responsible for neutrino masses, even if only partially accessible, 
is tremendous~\cite{delAguila:2008cj,Atre:2009rg,Deppisch:2015qwa,Antusch:2016ejd,Cai:2017jrq,Cai:2017mow,Dev:2018kpa}.
More specifically, in the absence of new force carriers and additional sources of spontaneous symmetry breaking, 
hadron colliders searches for heavy neutrinos cover an impressive range of particle masses, from the \MeV~up to the \TeV,  and a multitude of active-sterile mixing angles.
(With new gauge bosons and scalars, this can be pushed further.)
The limitations of this sensitivity, however, are subtle but substantial:

(a) In hadron collisions, 
projections~\cite{Gronau:1984ct,
Petcov:1984nf,Willenbrock:1985tj,Dicus:1991wj,Datta:1991mf,Datta:1993nm,
Cvetic:2002jy,
Han:2006ip,
delAguila:2007qnc,
delAguila:2008pw,delAguila:2008cj,delAguila:2008hw,Si:2008jd,
Atre:2009rg,Chao:2009ef,
Cvetic:2010rw,
BhupalDev:2012zg,Das:2012ze,
Dev:2013wba,
Arganda:2014dta,Wang:2014lda,Hessler:2014ssa,Alva:2014gxa,
Deppisch:2015qwa,Arganda:2015ija,Ruiz:2015gsa,
Degrande:2016aje,Dib:2016wge,DeRomeri:2016gum,Antusch:2016ejd,
Dib:2017iva,Ruiz:2017yyf,Caputo:2017pit,Marcano:2017ucg,Cai:2017mow,
Pascoli:2018rsg,Antusch:2018bgr,
Abada:2018sfh,Li:2018pag}
and 
searches~\cite{Aad:2015xaa,Sirunyan:2018mtv,Sirunyan:2018xiv,Aaboud:2018spl} 
closely follow the seminal works of Refs.~\cite{Keung:1983uu,Han:2006ip,delAguila:2008cj,Atre:2009rg}.
Of these, nearly all are based on the resonant production of a single heavy neutrino $N_m$ and a charged lepton $(\ell)$ 
through the charged current (CC) Drell-Yan (DY) process, i.e.,  $q\overline{q'}\to W^{(*)}\to N \ell$, as proposed by Ref.~\cite{Keung:1983uu}. 
At the Born level, the process is shown diagrammatically in Fig.~\ref{fig:feynman_DYX_Trilep}(a).
However, the CC DY process is no longer seen as the only viable means for producing heavy neutrinos at colliders.
It is now known that the $W\gamma \to N \ell^\pm$ vector boson fusion (VBF) process~\cite{Datta:1993nm,Dev:2013wba,Alva:2014gxa,Degrande:2016aje}, 
shown in Fig.~\ref{fig:feynman_DYX_Trilep}(c), 
is the dominant production mechanism for $N_m$ with TeV-scale masses at $\sqrt{s}=14\TeV$~\cite{Alva:2014gxa,Degrande:2016aje} 
and sizably enhances the inclusive production rate for lighter $N_m$.
At higher collider energies, 
neutral current (NC) processes~\cite{Gronau:1984ct,Willenbrock:1985tj,Dicus:1991wj,Datta:1991mf,Datta:1993nm}, 
including the gluon fusion (GF) channel $gg \to Z^*/h^* \to N\nu_\ell$~\cite{Willenbrock:1985tj,Dicus:1991wj,Hessler:2014ssa},
as shown in Fig.~\ref{fig:feynman_DYX_Trilep}(b), can surpass the CC DY cross section for masses below 1 TeV~\cite{Hessler:2014ssa,Ruiz:2017yyf}.
For much lighter sterile neutrino masses, 
the importance of NC and non-resonant processes has likewise been stressed elsewhere~\cite{BhupalDev:2012zg,Antusch:2016ejd,Caputo:2017pit,Cai:2017mow,Abada:2018sfh}.

\begin{figure}[!t]
\begin{center}
\includegraphics[width=.98\textwidth]{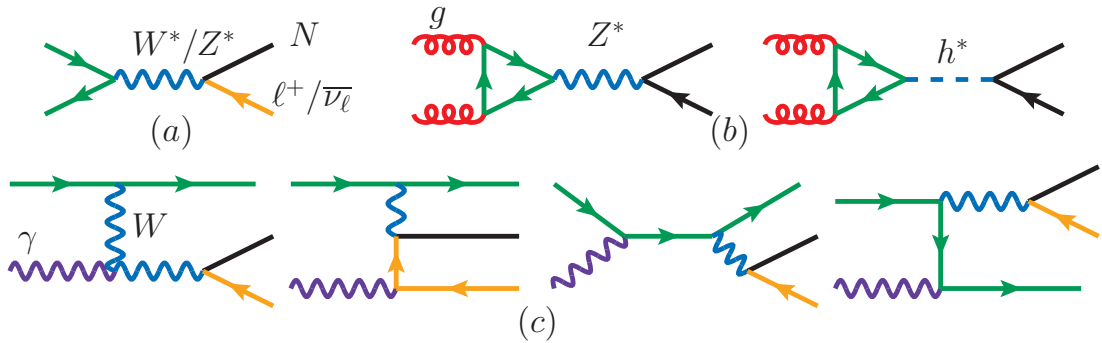}
\end{center}
\caption{Born diagrams for leading heavy neutrino production mechanisms in hadron collisions.
Diagrams throughout this text are drawn using JaxoDraw~\cite{Binosi:2003yf}.
}
\label{fig:feynmanMulti}
\end{figure}

(b) The search strategies prescribed by Refs.~\cite{Keung:1983uu,Han:2006ip,delAguila:2008cj,Atre:2009rg} 
are highly effective in discriminating against leading SM background processes, but only when reconstructed particles are correctly identified.
Misidentified and mis-reconstructed objects, often neglected in theoretical studies, have a substantial impact on sensitivity 
at hadron colliders~\cite{Abachi:1995yi,delAguila:2007qnc,Alva:2014gxa,Aad:2015xaa,Sirunyan:2018mtv,Sirunyan:2018xiv,Aaboud:2018spl,Pascoli:2018rsg}.
A prime example of this is in multilepton searches for leptonic decays of  $N_m$.
For production through the CC DY mode, the full process chain is given by
\begin{equation}
 q \overline{q'} \to W^{\pm~(*)} 
 \to N_m \ell_1^\pm 
 \to \ell_1^\pm \ell_2^\mp W^{\pm~(*)} 
 \to \ell_1^\pm \ell_2^\mp \ell_3^\pm \overset{(-)}{\nu_\ell},
 \label{eq:HeavyN_DYX_TrilepX}
\end{equation}
and shown diagrammatically in Fig.~\ref{fig:feynman_DYX_Trilep}.
Collider searches for Eq.~\ref{eq:HeavyN_DYX_TrilepX} 
typically require three ``analysis-quality'' leptons and veto events with additional central, energetic charged leptons.
After standard selection cuts, the remaining backgrounds consist of~\cite{delAguila:2008cj,Sirunyan:2018mtv}:
EW processes with many charged leptons, e.g., $pp\to nW \to n\ell+X$, wherein one or more charged leptons are too soft or too forward to be readily identified;
top quark processes wherein one fails to successfully tag bottom jets;
light jets misidentified as electrons or hadronically decaying $\tau$ leptons $(\tau_h)$;
and electrons/positrons whose charges are mis-measured.
Undoubtedly, reduction of these so-called irreducible and ``fake" backgrounds would have substantial positive impact on 
the discovery potential of heavy neutrinos.

(c) Standard multilepton search strategies for heavy neutrinos are premised almost solely on the existence of high-$p_T$ charged leptons originating from heavy $N_m$ decays.
(This is similarly the case searches of other colorless Seesaw particles~\cite{Keung:1983uu,delAguila:2008cj,Cai:2017mow}.) 
Aside from vetoing central jets that have additionally been $b$-tagged, 
essentially no information about jet activity in exploited in searches for high-mass $N_m$.
This is despite the CC DY and $W\gamma$ fusion mechanisms having qualitatively different QCD radiation patterns than their leading backgrounds.
(It is softer, more forward than in leading backgrounds~\cite{Alva:2014gxa,Ruiz:2015gsa,Ruiz:2015zca,Pascoli:2018rsg}.)
This is also despite the incredible improvement~\cite{Frixione:2002ik,Gleisberg:2008ta,Buckley:2011ms,Hamilton:2012np,Hoeche:2012yf,Alwall:2014hca,Bellm:2015jjp} 
in efficiently modeling QCD in hadron collisions for both SM and beyond the SM (BSM) processes
since the publication of Refs.~\cite{Keung:1983uu,Han:2006ip,delAguila:2008cj,Atre:2009rg}.
Notably, such improvements include better control over perturbative corrections and uncertainties associated 
with selection cuts based on the presence of energetic hadronic activity, i.e., jet vetoes~\cite{Barger:1990py,Barger:1991ar,Bjorken:1992er,Fletcher:1993ij,Barger:1994zq}, 
in both SM measurements~\cite{
Stewart:2010pd,Berger:2010xi,Becher:2012qa,Banfi:2012jm,Becher:2013xia,Stewart:2013faa,Meade:2014fca,Jaiswal:2014yba,Becher:2014aya,
Monni:2014zra,Gangal:2014qda,Gangal:2016kuo,Michel:2018hui}
and new physics searches~\cite{Tackmann:2016jyb,Ebert:2016idf,Fuks:2017vtl,Pascoli:2018rsg,Fuks:2019iaj}.
Moreover, while never intended for such circumstances,
with the incredible increase of top quark and multi-jet cross sections at higher $\sqrt{s}$,
it is doubtful that current search strategies will remain sufficient for future $pp$ colliders~\cite{Avetisyan:2013onh,Arkani-Hamed:2015vfh,Mangano:2016jyj,Golling:2016gvc,CidVidal:2018eel}.

In light of these observations, we have revisited the long-standard, hadron collider search strategy for heavy neutrinos that decay to fully leptonic final states,
as given in Eq.~\ref{eq:HeavyN_DYX_TrilepX}.
We report the construction of a qualitatively new collider analysis that is built almost entirely from inclusive, transverse observables whose shapes remain unchanged with varying collider energies.
Unlike past trilepton studies~\cite{delAguila:2008cj,BhupalDev:2012zg,Das:2012ze,Deppisch:2015qwa,Dib:2016wge,Antusch:2016ejd,Dib:2017iva,Sirunyan:2018mtv,Abada:2018sfh}, 
we premise the analysis on the near absence of central, energetic hadronic activity in the CC DY and $W\gamma$ fusion processes on an event-by-event basis.
Using state-of-the-art Monte Carlo (MC) tools, which enable fully differential event generation up to next-to-leading (NLO) in QCD with parton shower (PS) matching,
this discrimination criterion is implemented by employing a dynamic jet veto, 
one where the jet veto $p_T$ threshold $(\pTVeto)$ for each event is set to the $p_T$ of the leading charged lepton in the event~\cite{Pascoli:2018rsg}.
In doing so, we are able to build an analysis that ultimately selects for the relative amounts of hadronic and leptonic activities.
Importantly, we find irreducible and ``fake'' background suppression is improved.

For simplicity, we have restricted ourselves to a single, heavy Dirac-like / pseudo-Dirac neutrino as one would find in low-scale Type I Seesaws.
For completeness we have also considered a single, heavy Majorana neutrino one would find in baseline phenomenological models.
Using as benchmark an analysis inspired by the 2018 CMS collaboration's trilepton search for heavy neutrinos~\cite{Sirunyan:2018mtv}, 
we report that the sensitivity to active-sterile mixing $(\vert V_{\ell N}\vert^2)$ at the 14 TeV LHC can be improved by an order of magnitude at both $\mathcal{L}=300\invfb$ and $3\invab$.
At $\sqrt{s}=14$ TeV with $\mathcal{L}=3~{\rm ab}^{-1}$, 
we find active-sterile mixing as small as $\vert V_{\ell N}\vert^2 = 10^{-2} ~(10^{-3})~[5\times10^{-4}]$ can be probed  at 95\%~CL
for heavy Dirac neutrinos masses $m_N \lesssim 1200~(300)~[200]$ GeV,
well  beyond the present $\vert V_{\ell N}\vert^2 \lesssim 10^{-3} -10^{-1}$ constraints for such heavy states set by indirect searches and precision measurements.
By design, the analysis exhibits an unusual robustness against increasing collider energies, and therefore offers to serve as a baseline analysis for future collider searches.
At $\sqrt{s}=27~(100)$ TeV, we find that fixing down to $\vert V_{\ell N}\vert^2 \lesssim 10^{-5}$ for $m_N \lesssim 200\GeV$ 
and masses as large as  $m_N\sim 3.5~(15)\TeV$ for $\vert V_{\ell N}\vert^2 \lesssim 10^{-2}$ can be probed.
We anticipate the results we report can be further improved with detector-specific tuning and application of multi-variant / machines learning techniques~\cite{Guest:2018yhq}.
To facilitate such investigations, we make publicly available the
MC libraries\footnote[2]{\href{https://feynrules.irmp.ucl.ac.be/wiki/HeavyN}{feynrules.irmp.ucl.ac.be/wiki/HeavyN}} used in our computations.

The remainder of this report is organized the following order:
In Sec.~\ref{sec:Theory} we summarize the key components of the benchmark heavy neutrino models considered in this investigation, commenting also on present constraints.
In Sec.~\ref{sec:Setup} is a dedication discussion of our MC modeling, benchmark numerical inputs, and public accessibility of our MC libraries.
Inclusive, parton- and particle-level properties of heavy neutrinos are described in Sec.~\ref{sec:Colliders}.
Jet properties and the theoretical impact of dynamic jet vetoes for both signal and background processes are discussed extensively in Sec.~\ref{sec:jetVeto}.
We present the quantitative impact of dynamic jet vetoes in searches for heavy neutrinos at the LHC in Sec.~\ref{sec:Observability}.
In Sec.~\ref{sec:Summary}, we summarize and discuss our findings; we conclude in Sec.~\ref{sec:Conclusions}.

\section{Neutrino Mass Models}\label{sec:Theory}
If sterile neutrinos participate in the origin of neutrino masses, and if their mixing with active neutrinos is not too small, 
then it may be possible to produce them at laboratory-based experiments, like the LHC.
To model such interactions with SM particles, we consider two benchmark scenarios:
The first is the Inverse Seesaw (ISS) model~\cite{Mohapatra:1986aw,Mohapatra:1986bd,Bernabeu:1987gr},
which is a realistic, low-scale Type I neutrino mass model that permits naturally large Yukawa couplings and heavy neutrinos with masses accessible by colliders experiments.
The second is the Phenomenological Type I Seesaw, as investigated by Refs.~\cite{delAguila:2008cj,Atre:2009rg}.
The qualitative distinction between the two is that the former features pseudo-Dirac / Dirac-like heavy mass eigenstates whereas the latter features heavy Majorana mass eigenstates.

Despite well-established decoupling of EW- and TeV-scale Majorana neutrinos from collider experiments 
in scenarios where the SM is augmented by only singlet fermions~\cite{Pilaftsis:1991ug,Kersten:2007vk,Moffat:2017feq}, the latter is considered for two reasons:
(i) The discovery of LNV mediated by heavy neutrinos would unambiguously imply a particle field content that is richer than that hypothesized~\cite{Moffat:2017feq}.
(This is notable as the canonical collider signatures of the Phenomenological Type I Seesaw can be mimicked by non-minimal 
scenarios~\cite{Aparici:2009fh,Bhattacharya:2015vja,Kobach:2016ami,Liao:2016qyd,Elgaard-Clausen:2017xkq,Ruiz:2017nip,Duarte:2018kiv}). 
(ii) More practically, in the absence of new gauge bosons, 
experimental searches are commonly interpreted in the context of the Phenomenological Type I Seesaw~\cite{Aad:2015xaa,Sirunyan:2018mtv,Sirunyan:2018xiv,Aaboud:2018spl}.

The remainder of this section is outlined in the following manner:
In Sec.~\ref{sec:theoryDirac}, we discuss the ISS model.
In Sec.~\ref{sec:theoryMajor}, we briefly highlight the relevant distinctions between the Phenomenological Type I Seesaw and the ISS.
Constraints on heavy sterile neutrinos are summarized in Sec.~\ref{sec:theoryConstraints}.

\subsection{Inverse Seesaw}\label{sec:theoryDirac}

The ISS mechanism~\cite{Mohapatra:1986aw,Mohapatra:1986bd,Bernabeu:1987gr} is built on the consequences of 
the SM being extended by many more SM gauge singlet fermions than the two strictly~\cite{Wyler:1982dd} needed to reproduce oscillation data.
The result is a relationship between the scale of LNV and light neutrino masses that is qualitatively different from the traditional, high-scale Type I Seesaw.
Such a situation can be realized naturally in ultraviolet completions of the SM, 
including scenarios with warped extra dimensions~\cite{Agashe:2015izu,Agashe:2016ttz,Agashe:2017ann} and extended gauge 
symmetries~\cite{Dev:2009aw,Parida:2010wq,Nemevsek:2012iq,Anamiati:2016uxp,Helo:2018rll}.

For our purposes, we consider a model wherein the SM field content is extended by three pairs of fermionic singlets, one for each generation of active neutrinos, 
that possess opposite lepton number, $\nu_R^j$ ($L=+1$) and $X_R^k$ ($L=-1$).
A benefit of this field content is that it explicitly displays a nearly conserved lepton number symmetry that 
is necessarily present~\cite{Moffat:2017feq} in low-scale Seesaw models featuring only singlet fermions.
The Lagrangian for the ISS is given by
\begin{equation}
\mathcal{L}_{\rm ISS} = \mathcal{L}_{\rm SM} + \mathcal{L}_{\rm Kin.} + \mathcal{L}_{\nu},
  \label{eq:LagISS}
\end{equation} 
where $\mathcal{L}_{\rm SM}$ is the usual SM Lagrangian, $\mathcal{L}_{\rm Kin.}$ are kinetic operators for the $\nu_R^j$ and $X_R^k$ fields, and the so-called ``neutrino portal'' couplings
and mass terms are given by 
\begin{equation}
  \mathcal{L}_\nu = - y^{ij}_\nu \overline{L^i} \widetilde{\Phi}_{\rm SM} \nu_{R}^j - m_R^{jk} \overline{\nu_{R}^{jC}} X_{R}^k 
  -  \frac{1}{2} \mu_X^{kk'} \overline{X_R^{kC}} X_R^{k'} 
  - \frac{1}{2} \mu_R^{jj'} \overline{\nu_R^{jC}} \nu_R^{j'}
  + {\rm H.c.}
   \label{eq:LagISSnuPortal}
\end{equation}
In Eq.~\ref{eq:LagISSnuPortal}, $\Phi_{\rm SM}$ is the SM Higgs field, and $\widetilde \Phi= i \sigma_2\Phi^*$ is its SU$(2)_L$ rotation.
Both $\tilde{y}_\nu$ and $\tilde{m}_R$ are complex matrices while $\tilde{\mu}_{X}$ is a complex symmetric matrix whose norm is small compared to the EW scale,
in agreement with the near conservation of lepton number.
In the ISS, the Majorana masses for $\nu_R^j$, i.e., $\tilde{\mu}_R$, contribute to light neutrino masses only at a subleading level.
They likewise have negligible impact on collider phenomenology.
Hence, for simplicity, we take $\tilde{\mu}_R$ to be identically zero for the remainder of this study.

After EWSB,  the $9\times9$ neutrino mass matrix, in the  $(\nu_L^C\,,\;\nu_R\,,\;X_R)$ basis, is given by
\begin{equation}
  \label{ISSmatrix}
  \tilde{\mathcal{M}}_{\mathrm{ISS}} =
  \left(
    \begin{array}{c c c} 0 & \tilde{m}_D & 0 
      \\ \tilde{m}_D^T  &  0 &  \tilde{m}_R 
      \\ 0  & \tilde{m}_R^T  &  \tilde{\mu}_X
    \end{array}\right)\,,
\end{equation}
with $\tilde{m}_D=\tilde{y}_\nu \langle \Phi\rangle$ being the neutrino Dirac mass matrix. 
Since the mass matrix $\tilde{\mathcal{M}}_{\mathrm{ISS}}$ is complex and symmetric, it can be put in a diagonal form using Takagi factorization:
\begin{equation}
U^T_\nu \tilde{\mathcal{M}}_{\mathrm{ISS}} U_\nu \equiv  \tilde{\mathcal{M}}_\nu^{\rm diag} = \text{diag}(m_{\nu_1},\dots,m_{\nu_3},m_{N_1},\dots,m_{N_6}).
\label{eq:TakagiUnu}
\end{equation}
Here $m_{\nu_1},\dots,m_{\nu_3},$ are three light neutrino mass eigenvalues,
 $m_{N_1},\dots,m_{N_6}$ are six heavy neutrino mass eigenvalues,
and  $U_\nu$ is a $9\times9$ unitary matrix.
Due to the near conservation of lepton number symmetry, the diagonal neutrino mass matrix takes a specific form,
with three light neutrino mass eigenstates of Majorana character and six heavy Majorana neutrino mass eigenstates that form three pseudo-Dirac pairs. 

For a single generation, i.e., one $(\nu_R,~X_R)$ pair, this can be seen explicitly.
In the ISS limit, where $\mu_X \ll m_D, m_R$, the neutrino mass spectrum takes the form,
\begin{equation}
 m_\nu 		\simeq \mu_X\frac{m_{D}^2}{m_{D}^2+m_{R}^2},
 \quad\text{and}\quad
 m_{N_1,N_2} 	\simeq \sqrt{m_{R}^2+m_{D}^2} \mp \frac{\mu_X}{2}\frac{m_{R}^2}{(m_{D}^2+m_{R}^2)}.
 \label{eq:ISSMassOneGen}
\end{equation}
Importantly, one sees the presence of a light mass eigenvalue $(m_\nu)$ that is directly proportional to the lepton number violating parameter $\mu_X$, 
which is the inverse of the usual high-scale Type I Seesaw mechanism.
One sees also that this LNV scale also controls the mass splitting of the heavier pair of mass eigenstates.
Thus, in the limit $\mu_X \rightarrow 0$, i.e., in the limit that lepton number is identically conserved, 
the light neutrino is massless while the two heavy neutrinos become exactly degenerate, forming a Dirac fermion. 
 Thus, in the context of the ISS, the net contribution of heavy neutrinos $N_m$ to lepton number violating-processes 
 that occur at a momentum transfer scale $Q$ will be suppressed by the smallness of $(\mu_X/Q)^2$, as required to accommodate light neutrino masses.
 Moreover, this strongly motivates complementary searches strategies at the LHC (and future colliders) 
 for lepton number conserving processes,  such as the trilepton process in Fig.~\ref{fig:feynman_DYX_Trilep} that we consider.

With three generations, the neutrino mass matrix $\tilde{\mathcal{M}}_{\mathrm{ISS}}$ can be diagonalised by block~\cite{GonzalezGarcia:1988rw},
allowing one to write the $3\times3$ light neutrino mass matrix as,
\begin{equation} 
 \tilde{M}_\nu^{\rm light} \simeq \tilde{m}_D ~\tilde{m}_R^{T-1} ~\tilde{\mu}_X ~\tilde{m}_R^{-1} ~\tilde{m}_D^T.
 \label{eq:MlightLO}
\end{equation}
This can be further expressed in diagonal form using the Pontecorvo-Maki-Nakagawa-Sakata (PMNS) rotation matrix, $U_{\rm PMNS}$~\cite{Pontecorvo:1957cp,Maki:1962mu}.
Doing this gives 
\begin{equation} \label{mnulight}
 U_{\rm PMNS}^T   ~\tilde{M}_\nu^{\rm light}  ~U_{\rm PMNS} \equiv \tilde{M}_\nu^{\rm light~diag} = \mathrm{diag}(m_{\nu_1}\,, m_{\nu_2}\,, m_{\nu_3}).
\end{equation}
Importantly, we note that while the PMNS matrix, as historically defined~\cite{Pontecorvo:1957cp,Maki:1962mu}, is strictly unitary.
Here, the $3\times3$ block measured by neutrino oscillation experiments is not guaranteed to be separately unitary within the full $9\times9$, unitary mixing matrix and does not strictly correspond
to the matrix $U_{\rm PMNS}$ defined above. 
(Though indeed neutrino oscillation experiments measure the sub-block $V$ as defined in Eq.~\ref{eq:EffectiveMixing} below.)
With these relations, reproduction of oscillation data can be ensured by judicious parametrization of high energy inputs.
Specifically, we use the $\mu_X$-parameterization ~\cite{Arganda:2014dta}, giving
\begin{equation}
\tilde{\mu}_X = \tilde{m}_R^T ~\tilde{m}_D^{-1}~ U_{\rm PMNS}^* ~\tilde{M}_\nu^{\rm light~diag}~  U_{\rm PMNS}^\dagger~ \tilde{m}_D^{T-1} ~\tilde{m}_R.
\end{equation}
The above is valid only in the regime that individual elements of $\tilde{m}_D$ are much smaller than individual elements of $\tilde{m}_R$. 
For $(m_D^{ij}/m_R^{kl})\gtrsim0.1$, higher order terms in the Seesaw expansion parameter $(\tilde{m}_D \tilde{m}_R^{-1})$ must be included, 
and can be found in Ref.~\cite{Baglio:2016bop}.

In realizing the ISS at collider experiments, we observe first that for the momentum transfer scales $Q$ typically probed in collider experiments, one is insensitive to tiny neutrino masses.
In other words, to a very good approximation, $(m_{\nu_m}/Q)^2\sim0$.
This means, however, that on collider scales, the effective Lagrangian one wants  should consistently and parametrically follow from the $(m_{\nu_m}/Q)^2 \sim (\mu_X^{kk'}/Q)^2 \to 0$ limit.
For the ISS, this means to the zeroth order in $(\mu_X^{kk'}/Q)^2$ for all $k,k'$.
Hence, the relevant heavy neutrino degrees of freedom at collider experiments are the 
three Dirac mass eigenstates that are recovered precisely in the LNC limit, and henceforth denoted by $N_{m'}$ for $m=1,\dots,3$.

Now, working in a basis where the charged lepton mass matrix is diagonal, 
the effective $3+3$ neutrino mixing matrix $U_\nu$ in Eq.~\ref{eq:TakagiUnu} can be generically parameterized by~\cite{Atre:2009rg}
\begin{equation}
 U_\nu^*=\left(\begin{array}{c c}
                U & V \\
                X & Y 
               \end{array} \right).
               \label{eq:EffectiveMixing}
\end{equation}
Here, $U_{\ell \nu_m}$ is the $3\times3$ matrix describing light neutrinos $(\nu_m)$ coupling to SM weak currents (and measured by oscillation experiments),
and $V_{\ell N_{m'}}$ is the $3\times 3$ matrix describing heavy neutrinos $(N_{m'})$ coupling to SM weak currents.
Subsequently, neutrino flavor states $\nu_\ell$ are related to the mass eigenstates by the decomposition~\cite{Atre:2009rg}
\begin{equation}
\nu_\ell = \sum_{m=1}^3 U_{\ell m} \nu_m ~+~ \sum_{m'=4}^6 V_{\ell m'} N_{m'},
\label{eq:nuDecomposition}
\end{equation}
which should be understood as applying to the LH components of Dirac spinors.
Eq.~\ref{eq:nuDecomposition} is the master relationship that allows us to build up SM couplings to  heavy Dirac neutrinos $N_{m'}$ at high-$Q^2$ experiments.
In particular, after EWSB and to leading order in $V_{\ell N_{m'}}$, the relevant interactions with EW bosons in the Feynman-'t Hooft gauge are 
\begin{equation}
 \mathcal{L}^{\rm Int.}_{\rm ISS} =  \mathcal{L}^{W^\pm}+ \mathcal{L}^{Z}  +  \mathcal{L}^{H} + \mathcal{L}^{G^\pm} + \mathcal{L}^{G^0} ,
 \label{eq:DiracLag}
\end{equation}
where the charged current and charged Goldstone couplings to $N_{m'}$ are
\begin{equation}
 \mathcal{L}^{W^\pm} + \mathcal{L}^{G^\pm} 
 = -\frac{g}{\sqrt{2}} \sum_{\ell=e}^\tau \sum_{m'=4}^6 \overline{N_{m'}} V^*_{\ell m'}\left( W_\mu^+ \gamma^\mu + i\frac{m_{N_{m'}}}{M_W}G^+\right)P_L  \ell^- + {\rm H.c.},
\end{equation}
the analogous neutral current and neutral Goldstone couplings are 
\begin{equation}
 \mathcal{L}^{Z} + \mathcal{L}^{G^0} 
 = -\frac{g}{2\cos\theta_W} \sum_{\ell=e}^\tau \sum_{m'=4}^6 \overline{N_{m'}} V^*_{\ell m'}\left( Z_\mu \gamma^\mu + i\cos\theta_W\frac{m_{N_{m'}}}{M_W}G^0\right)P_L  \nu_\ell + {\rm H.c.},
\end{equation}
and the coupling to the SM Higgs boson is
\begin{equation}
\mathcal{L}^{H} = -\frac{g}{2M_W} h \sum_{\ell=e}^\tau \sum_{m'=4}^6 \overline{N_{m'}} V^*_{\ell m'} m_{N_{m'}}  P_L  \nu_\ell + {\rm H.c.}
\end{equation}
Here, $g\approx0.65$ is the SU$(2)$ coupling constant, $\theta_W$ is the weak mixing angle, 
and $P_{L/R}=(1\mp\gamma^5)/2$ are the usual LH/RH chiral projection operators for Dirac fermions.

With Eq.~\ref{eq:DiracLag}, one can determine many properties of heavy Dirac neutrinos, including their partial widths for decays to on-shell EW bosons, 
which we list here for completeness:
\begin{eqnarray}
\Gamma\left(N_{m'} \to W^+ \ell^-\right) & \equiv \Gamma_{N_{m'}}^W		 = & \cfrac{g^2}{64\pi M_W^2}\vert V_{\ell N}\vert^2 m_{N_{m'}}^3(1-r_W)^2\left(1+2r_W\right), \\
\Gamma\left(N_{m'} \to Z \nu_\ell \right) & \equiv \Gamma_{N_{m'}}^Z		 = & \cfrac{g^2}{128\pi M_W^2}\vert V_{\ell N}\vert^2 m_{N_{m'}}^3(1-r_Z)^2\left(1+2r_Z\right), \\
\Gamma\left(N_{m'} \to h \nu_\ell \right) & \equiv \Gamma_{N_{m'}}^h		 = & \cfrac{g^2}{128\pi M_W^2}\vert V_{\ell N}\vert^2 m_{N_{m'}}^3(1-r_H)^2, \quad  r_X \equiv \frac{M_X^2}{m_{N_{m'}}^2}.
\label{eq:DiracWidths}
\end{eqnarray}
The total decay width of $N_{m'}$ is then given by the sum over all $N_{m'} \to X$ partial widths:
\begin{equation}
\Gamma_{N_{m'}}^{\rm Tot.} 	\equiv \sum_X  \Gamma\left(N_{m'} \to X\right) \approx \Gamma_{N_{m'}}^W + \Gamma_{N_{m'}}^Z + \Gamma_{N_{m'}}^h.
\end{equation}
Likewise the $N_{m'} \to X$ branching rate (BR), or branching fraction, is defined as the ratio
\begin{equation}
{\rm BR}\left(N_{m'} \to X\right) = \cfrac{\Gamma\left(N_{m'} \to X\right)}{\sum_Y \Gamma\left(N_{m'} \to Y\right)}.
\end{equation}
For fixed $\vert V_{\ell N_{m'}}\vert$, 
in the high-energy / infinite $m_{N_{m'}}$ limit, the Goldstone Equivalence Theorem becomes manifest, and one obtains the branching relationship~\cite{Atre:2009rg}
\begin{equation}
{\rm BR}\left(N_{m'} \to W^+ \ell^-\right) \approx 2\times {\rm BR}\left(N_{m'} \to Z \nu_\ell\right) \approx 2\times  {\rm BR}\left(N_{m'} \to h \nu_\ell\right).
\end{equation}
Contrary to common belief, the $2:1:1$ ratio holds even for Dirac neutrinos due to the fact that the $N_{m'}-W$ charged current coupling is a factor of $\sqrt{2}$ larger 
than the neutral current and Higgs couplings at the Lagrangian level, as seen above.
This itself is a manifestation of the fact that the $W_\mu$ boson is a linear combination of the $W_\mu^1$ and $W_\mu^2$  vector fields.

\subsection{Phenomenological Type I Seesaw}\label{sec:theoryMajor}

The formalism of the Phenomenological Type I Seesaw is well-documented~\cite{Atre:2009rg} and will not be repeated here.
For three heavy Majorana neutrinos, the relevant interaction Lagrangian strongly resembles Eq.~\ref{eq:DiracLag}.
The crucial difference is the Majorana condition, which states that in the mass basis the field $N_{m'}$ and its conjugate are related by $N_{m'}^c = N_{m'}$.
This of course means that, if sufficiently heavy, both the $N_{m'} \to W^+ \ell^- ,~ Z \nu_\ell, ~h \nu_\ell$ 
and  $N_{m'} \to W^- \ell^+ ,~ Z \overline{\nu_\ell}, ~h \overline{\nu_\ell}$ decay channels are open.
(Note that to leading order in $V_{\ell N_{m'}}$, the light Majorana mass eigenstates $\nu_{m}$ are approximated by the flavor eigenstates $\nu_\ell$.)
By CP-invariance, the partial widths for $N_{m'} \to W^- \ell^+$ and $N_{m'} \to W^+ \ell^-$ are equal, as are the analogous $Z$ and $h$ partial widths.
Hence, the total width for a heavy Majorana neutrino and a heavy Dirac neutrino of the same mass are related by
\begin{equation}
\Gamma_{N_{m'}}^{\rm Tot.~-~Majorana}  = 2\times \Gamma_{N_{m'}}^{\rm Tot.~-~Dirac}.
\end{equation} 
This preserves the well-known branching relationship,
\begin{eqnarray}
{\rm BR}\left(N_{m'} \to W^+ \ell^-  + W^-\ell^-\right) 
&\approx& 2\times {\rm BR}\left(N_{m'} \to Z \nu_\ell + Z\overline{\nu_\ell}\right) \nonumber\\
&\approx& 2\times  {\rm BR}\left(N_{m'} \to h \nu_\ell + h\overline{\nu_\ell}\right).
\end{eqnarray}

\subsection{Constraints on Heavy Sterile Neutrinos}\label{sec:theoryConstraints}
Heavy neutrinos can manifest themselves in a variety of terrestrial- and space-base experiments. 
Hence their existence is constrained from a number of sources.
For reviews on sterile neutrino constraints, see 
Ref.~\cite{Antusch:2006vwa,delAguila:2008pw,Antusch:2014woa,Drewes:2015iva,Fernandez-Martinez:2016lgt,Drewes:2016jae,Cai:2017mow}.
We now summarize such constraints:

\begin{itemize}

\item \textbf{Global Constraints on Non-Unitarity of PMNS Matrix:}
Deviation from unitarity of the light neutrino mixing matrix $U$ induced by active-sterile mixing can be expressed generically by
the Hermitian matrix $\eta_{\ell\ell'}$ and the relationship
\begin{equation}
 U^{*}=(I-\eta)\,U_\mathrm{PMNS}.
\end{equation}
In a general Seesaw scenario, constraints from a global fit~\cite{Fernandez-Martinez:2016lgt} 
to EW precision data, tests of CKM unitarity, and tests of lepton universality limit $\eta_{\ell\ell'}$ to be:
\begin{align}
 \sqrt{2|\eta_{ee}|}&<0.050\,, 		 & \sqrt{2|\eta_{e\mu}|}<0.026\,, \nonumber \\
 \sqrt{2|\eta_{\mu\mu}|}&<0.021\,, 	 & \sqrt{2|\eta_{e\tau}|}<0.052\,, \nonumber \\
 \sqrt{2|\eta_{\tau\tau}|}&<0.075\,,	 & \sqrt{2|\eta_{\mu\tau}|}<0.035,
\label{eq:EWPOconstraints}
\end{align}
at 95\% CL.
In a general Seesaw scenario, 
$\eta_{\ell\ell'}$ is related to mixing matrices $V_{\ell N_{m'}}$ by the relationship $\sqrt{2\vert \eta_{\ell\ell'}\vert } = \sum_{m'=4}^{6}\sqrt{V_{\ell m'} V_{\ell'm'}^*}$.
Under the benchmark assumption that $\ell\ell'$ mixing is dominated/saturated by the lightest heavy mass eigenstate $N_{m'=4}$, as we invoke, 
the relationship simplifies to $2\vert \eta_{\ell\ell'}\vert  \approx \vert V_{\ell 4} V_{\ell' 4}^{*} \vert $.

\item \textbf{Searches for Charged Lepton Flavor Violation:}
Heavy neutrinos with a mass close the EW scale and large off-diagonal Yukawa couplings can induce cLFV in  decays of charged leptons.
Searches for such processes have set the following $90\%$~C.L.~upper limit on decay branching rates:
\begin{table}[!h]
\begin{center}
 \begin{tabular}{c c}
    BR$(\mu^+\rightarrow e^+ \gamma)$ 			& $<4.2\times 10^{-13}$~\cite{TheMEG:2016wtm}\\
    BR$(\tau^\pm \rightarrow e^\pm \gamma)$ 		& $<3.3 \times 10^{-8}$~\cite{Aubert:2009ag}\\
    BR$(\tau^\pm \rightarrow \mu^\pm \gamma)$ 	& $<4.4 \times 10^{-8}$~\cite{Aubert:2009ag}\\
    BR$(\mu^+\rightarrow e^+e^+e^-)$ 			& $<1.0 \times 10^{-12}$~\cite{Bellgardt:1987du}\\
    BR$(\tau^-\rightarrow \mu^- \mu^+ \mu^-)$ 		& $<2.1\times10^{-8}$~\cite{Hayasaka:2010np}\\
    BR$(\tau^- \rightarrow e^- \mu^+ \mu^-)$ 		& $<2.7\times10^{-8}$~\cite{Hayasaka:2010np}\\
    BR$(\tau^- \rightarrow \mu^- e^+ e^-)$ 			& $<1.8\times10^{-8}$~\cite{Hayasaka:2010np}\\
    BR$(\tau^-\rightarrow e^- e^+ e^- )$ 			& $<2.7\times10^{-8}$~\cite{Hayasaka:2010np}\\
    BR$(\tau^-\rightarrow e^+ \mu^- \mu^-)$ 		& $<1.7\times10^{-8}$~\cite{Hayasaka:2010np}\\
    BR$(\tau^-\rightarrow \mu^+ e^- e^-)$ 			& $<1.5\times10^{-8}$~\cite{Hayasaka:2010np}\\
 \end{tabular}
\end{center}
\end{table}

\item \textbf{Cosmological Constraints on Light Neutrino Masses:} 
Measurements of large scale structure in the universe by the Planck satellite, in conjunction with WMAP + highL + BAO data, 
 yields the upper limit on the sum of all light neutrino masses~\cite{Aghanim:2018eyx}:
\begin{equation}
 \sum_{m} m_{\nu_m}  < 0.12\,\mathrm{eV}, \quad \text{at} ~95\%~\text{CL}.
\label{eq:Planck}
\end{equation}

\item \textbf{}

\item \textbf{Direct Searches at Collider Experiments:}
Sterile neutrinos with masses above $M_W$ decaying into the fully leptonic final state, as shown in Fig.~\ref{fig:feynman_DYX_Trilep},
have been constrained directly for the first time at colliders by the CMS experiment~\cite{Sirunyan:2018mtv}.
\begin{equation}
\text{For}\quad m_{N_4} = 200~(500)\GeV, \qquad \vert V_{e4}\vert^2 < 2\times10^{-2}~(1\times10^{-1}) \quad\text{at}~95\%~\text{CL},
\end{equation}
with $\mathcal{L}\approx 36\invfb$ at $\sqrt{s}=13\TeV$.
For $\vert V_{\mu4}\vert$, limits are more constraining but nonetheless comparable.
Searches for LNV using the same-sign lepton + jets channel 
by both ATLAS~\cite{Aaboud:2018spl} and CMS~\cite{Sirunyan:2018xiv} with the same amount of data 
yield limits on active-sterile mixing that is still slightly more constraining, but again comparable.

\item \textbf{Perturbativity Bounds on Total Width:}
Large Yukawa couplings, in the form of large active-sterile mixing, can lead to too large heavy neutrino total decay widths.
In this analysis, we will constrain the total width of $N_{m'}$ to be:
\begin{equation}
 \Gamma_{N_{m'}}^{\rm Tot.} ~<~ 5\% ~\times~  m_{N_{m'}}.
\label{eq:NuWidth}
\end{equation}

\end{itemize}

\section{Computational Setup}\label{sec:Setup}
We now describe, in detail, our computational setup.
In Sec.~\ref{sec:mcTools}, we describe the general configuration of our Monte Carlo (MC) event generator tool chain as well as the public accessibility of our MC libraries.
In Secs.~\ref{sec:mcSigGen} and ~\ref{sec:mcBkgGen}, we elaborate on our modeling of individual signal and background processes.
Our SM input choices are given in Sec.~\ref{sec:mcSMinputs},
and in Sec.~\ref{sec:mcHeavyNinputs} we give our benchmark heavy neutrino inputs.
Unless noted, our computational setup remains constant for $\sqrt{s}=14,~27$, and 100 TeV.

\subsection{Monte Carlo Event Generation and Public Accessibility}\label{sec:mcTools}
In order to compute hadronic scattering rates involving heavy Dirac neutrinos, 
we use the simplified Lagrangian of Eq.~\ref{eq:DiracLag} to construct a 
Dirac neutrino-variant of the NLO-accurate \texttt{HeavyN}~\cite{Alva:2014gxa,Degrande:2016aje} FeynRules (FR)~\confirm{v2.3.32}~\cite{Alloul:2013bka,Christensen:2008py} model file.
This is then imported into the general-purpose event generator \texttt{MadGraph5\_aMC@NLO} (\mgamc)~\confirm{v2.6.2}~\cite{Alwall:2014hca}.
Similar to the original \texttt{HeavyN} libraries, which assumes heavy Majorana neutrinos, 
the \texttt{HeavyN\_Dirac} model contains three heavy Dirac mass eigenstates that can couple independently to each lepton family in the SM proportional to a mixing factor of $\vert V_{\ell N_m}\vert$.
While such mixing is heavily constrained in realistic neutrino mass models, these extra degrees of freedom provide necessary flexibility in computing rates and distributions in a flavor model-independent manner.
Most all the functionality and labeling of the \texttt{HeavyN} files are retained.\footnote[2]{
A technical exception is a change in the heavy neutrino MC particle identification (PID) codes.
These differ to avoid conflict with standard HEP PID convention established in 2002 for Majorana neutrinos ~\cite{Hagiwara:2002fs}.}
Using \texttt{NLOCT}~\cite{Degrande:2014vpa} and \texttt{FeynArts}~\cite{Hahn:2000kx},
we include QCD renormalization and $R_2$ rational counter terms up to one loop in $\alpha_S$.
Feynman rules are collected into a Universal FR Object (UFO) file~\cite{Degrande:2011ua}
that is publicly available from the FeynRules Model Database, at the URL: \href{http://feynrules.irmp.ucl.ac.be/wiki/HeavyN}{feynrules.irmp.ucl.ac.be/wiki/HeavyN}.
Subsequently, tree-level processes can be computed up to NLO in QCD and QCD loop-induced processes can be computed at LO accuracy 
using the state-of-the-art generators \texttt{HERWIG++}~\cite{Bellm:2015jjp},~\mgamc,~ or \texttt{SHERPA}~\cite{Gleisberg:2008ta}.

For scatterings involving heavy Majorana neutrinos, we use the aforementioned~\texttt{HeavyN}~libraries.
For additional model  details, see Ref.~\cite{Alva:2014gxa,Degrande:2016aje} and references 
therein\footnote[3]{For both Dirac and Majorana libraries, model variants that include six massive quarks, three massive charged leptons, and/or non-diagonal quark mixing are also available.}.

Parton-level event generation at LO and NLO in QCD is handled by \mgamc.
For decays of heavy neutrinos and SM particles $W,Z,t$, we invoke the narrow width approximation (NWA)
and pass partonic events to \texttt{MadSpin}~\cite{Artoisenet:2012st} and \texttt{MadWidth}~\cite{Alwall:2014bza}. 
Spin correlation in decays of intermediate resonances is preserved through the implementation of the spin-correlated NWA in \texttt{MadSpin}.
Parton events are  inputed to \texttt{Pythia}~\confirm{v8.230} (PY8)~\cite{Sjostrand:2014zea} for parton showering and hadronization. 
Decays of $\mathcal{B}$ hadrons and $\tau$ leptons to lighter hadrons and leptons are channeled through PY8.
Following the CMS analysis of Ref.~\cite{Khachatryan:2016jww}, we use the PY8 tune CUETP8M1 ``Monash*''~(\texttt{Tune:pp = 18})~\cite{Skands:2014pea}. 
For a realistic shower simulation, we switch \texttt{on} recoil/primordial momentum and QED shower flags, including backwards evolution for photon-initiated processes.
The impact of underlying event is neglected and left to follow-up investigations.
Hadron-level events are given to \texttt{MadAnalysis5} \confirm{v1.6.33} \cite{Conte:2012fm,Conte:2018vmg} for particle-level clustering via \texttt{FastJet}~\confirm{v3.2.1}~\cite{Cacciari:2011ma}.
Following the jet veto study of Ref.~\cite{Fuks:2017vtl}, we employ the anti-$k_T$ algorithm~\cite{Cacciari:2008gp} with radius parameter $R=1$.
During jet clustering, tagging efficiency for $b$-jets and hadronic decays of $\tau$ leptons $(\tau_h)$ is taken to be 100\% with 0\% misidentification rates of light jets.
(Realistic tagging efficiencies and misidentification rates are applied at the analysis-level; see Sec.~\ref{sec:DetModeling} for details.)
Reconstructed / clustered, particle-level events are ultimately written to file in the Les Houches Event (LHE) format~\cite{Alwall:2006yp}.
Only jets with $p_T>1\GeV$ are recorded to file.
Particle-level events are analyzed using an in-house, \texttt{ROOT}-based analysis framework.

In several instances, we compute cross section normalizations that are beyond the accuracy of NLO in QCD with PS matching to leading logarithmic (LL) precision. 
For such cases, we heavily exploit the factorization properties within the formalism of Soft-Collinear Effective Field Theory (SCET)~\cite{Bauer:2000yr,Bauer:2001yt,Beneke:2002ph},
particularly when working in  momentum space~\cite{Becher:2006nr}.
Approximate next-to-next-to-next-leading logarithmic (N$^3$LL) threshold corrections to the gluon fusion process in Fig.~\ref{fig:feynmanMulti}(b),
which captures the dominant contributions up to next-to-next-to-leading order (N$^2$LO) in $\alpha_s$~\cite{Bonvini:2014qga}, 
are obtained following Ref.~\cite{Becher:2006nr,Becher:2006mr,Ahrens:2008nc,Ruiz:2017yyf}.
In addition, we obtained cross sections with a static jet veto up to NLO in QCD with jet veto resummation matching at N$^2$LL,
using the automated \texttt{MadGraph5\_aMC@NLO+SCET} libraries developed by Ref.~\cite{Alwall:2014hca,Becher:2014aya}.

\subsection{Heavy Neutrino Signal Event Generation}\label{sec:mcSigGen}

We now describe our signal process event generation using \mgamc.
Here,  we also provide instructions for using the \texttt{HeavyN\_Dirac} and \texttt{HeavyN} libraries.

In this study we consider heavy Dirac and Majorana neutrinos  produced through the DY and $W\gamma$ fusion mechanisms, as shown in Fig.~\ref{fig:feynmanMulti}(a) and (c),
and decayed to fully leptonic final states. At the Born level, this corresponds to the production and decay chains~\cite{Degrande:2016aje},
\begin{eqnarray}
 \text{DY}  &:& q\overline{q'} \to \ell_1 N, 	 \quad\text{with} \quad N \to \ell_2 W \quad\text{and}\quad W \to \ell_3 \nu,     \label{eq:dyxSetup}\\
 \text{VBF} &:& q\gamma \to q'  \ell_1 N,	 \quad\text{with} \quad N \to \ell_2 W \quad\text{and}\quad W \to \ell_3 \nu.     \label{eq:vbfSetup}
\end{eqnarray}

For Dirac neutrinos, the syntax of the \texttt{HeavyN\_Dirac} model file is designed to closely match the original \texttt{HeavyN} model file for Majorana neutrinos.
Hence, much of the syntax presented in Ref.~\cite{Degrande:2016aje} to simulate Eqs.~\ref{eq:dyxSetup}-\ref{eq:vbfSetup} for Majorana neutrinos is unchanged.
For the inclusive CC DY process, the appropriate \mgamc~syntax is
\begin{verbatim}
 import model SM_HeavyN_Dirac_NLO
 set gauge Feynman
 define p = 21 1 2 3 4 -1 -2 -3 -4
 define ww = w+ w-
 define ell = e+ mu+ ta+ e- mu- ta-
 define nn = n1 n1~
 generate p p > nn ell QCD=0 [QCD]
 output HeavyNDirac_DYX_Nl_NLO ; launch -f
\end{verbatim}
Here, \texttt{p} here denotes an active parton $p\in\{q,\overline{q},g\}$, with  $q\in\{u,c,d,s\}$, and \texttt{ell} is any one of the three charged leptons, $\ell \in \{e, \mu, \tau\}$.
In this analysis, we aim to consistently model decays of $\tau$ leptons and $\mathcal{B}$ hadrons, and therefore assume all charged generation-III fermions to be massive.
Hence, neither $b$ or $t$ are considered active partons.
The \texttt{[QCD]} flag activates the MC@NLO~\cite{Frixione:2002ik} and \texttt{MadLoop}~\cite{Hirschi:2011pa,Hirschi:2015iia} formalisms in~\mgamc, 
allowing one to compute processes up to one loop in $\alpha_s$ and parton shower (PS) matching to LL accuracy~\cite{Alwall:2014hca}.
For the VBF channel, the corresponding commands are
\begin{verbatim}
 generate    p a > nn ell p QED=3 QCD=0 [QCD]
 add process a p > nn ell p QED=3 QCD=0 [QCD]
\end{verbatim}
Explicitly, we consider collinear, initial-state photons from both hadronic and partonic sources, 
as advised by Refs.~\cite{Martin:2014nqa,Alva:2014gxa,Degrande:2016aje,Harland-Lang:2016kog} and as implemented in Refs.~\cite{Degrande:2016aje}.
This is handled automatically by our use of the LUXqed formalism~\cite{Manohar:2016nzj,Manohar:2017eqh} (see Sec.~\ref{sec:mcSMinputs}), 
and keeping the usual PDFs activated at all times.
During event generation, this means keeping the beam ID codes in the \mgamc~input file \texttt{Cards/run\_card.dat}~to their default values:
\begin{verbatim}
 1     = lpp1 ! beam 1 type (0=No PDF, 1=proton, 2=photon from proton)
 1     = lpp2 ! beam 2 type (0=No PDF, 1=proton, 2=photon from proton)
\end{verbatim}
Contrary to what has frequently been reported in the literature,
setting \texttt{lppX=2} will activate what is known\footnote{
This should not be confused with the ``Effective Photon Approximation,''
which is also known as the ``Weizs\"acker-Williams Approximation'' as well as the ``color-striped Altarelli-Parisi Splitting Functions,''
 see e.g., Ref.~\cite{Peskin:1995ev}, nor should it be confused with the ``Improved Weizs\"acker-Williams Approximation''~\cite{Frixione:1993yw}.
 } 
 as the ``Equivalent Photon Approximation''~\cite{Budnev:1974de}, or ``elastic photon PDF'' in the language of Ref.~\cite{Alva:2014gxa}.
This is a phenomenological model for $pX$ scattering that describes the ultra low-$Q^2$, i.e.,~$\vert Q^2\vert \lesssim m_{\rm proton}^2$, 
emission of collinear photons from an on-shell proton that transitions into a (meta-stable) nucleon, i.e., $p \to \gamma N^{*}$.
The corresponding photon can then be used to model initial-state photons in $\gamma X$ scattering. 
 The Equivalent Photon Approximation formalism has since been superseded by the LUXqed formalism~\cite{Manohar:2016nzj,Manohar:2017eqh}.
 Regardless of formalism, the initial-state photon flux at  $Q^2 \sim  m_{\rm proton}^2$ 
 set the (non-perturbative) boundary condition for the ``inelastic photon PDF''~\cite{Martin:2014nqa,Alva:2014gxa},
 which describes initial-state photons as constituent partons of the proton in high-$Q^2$ scattering~\cite{Martin:2004dh}.
 Using the mixed QCD-QED DGLAP evolution equations~\cite{Gribov:1972ri,Dokshitzer:1977sg,Altarelli:1977zs,Martin:2004dh},
 the (inelastic) photon PDF can be evolved to arbitrarily high $Q^2 \gg  m_{\rm proton}^2$.
 In conjunction with QED parton showering, which we account for explicitly with PY8, 
 initial-state photons are backwards-evolved and matched to collinear $q\to q\gamma$ splittings~(see Ref.~\cite{Sjostrand:2014zea} for precise details).
A consequence is the sizable likelihood of resolving the associated forward jet.

For the production of heavy Majorana neutrinos, we use the original \texttt{HeavyN} libraries.
Two notable (though hopefully obvious) syntax modifications to those above are needed:
(i) One needs to import the \texttt{HeavyN} model file instead of the \texttt{HeavyN\_Dirac} libraries.
(ii) One must omit the charge conjugate of $N$ since they are the same particle.
Following Ref.~\cite{Degrande:2016aje}, the relevant  \mgamc~syntax for the CC DY channel is 
\begin{verbatim}
 import model SM_HeavyN_NLO
 ...
 generate p p > n1 ell QCD=0 [QCD]
\end{verbatim}
Analogously, for the VBF channel, the commands are
\begin{verbatim}
 generate    p a > n1 ell p QED=3 QCD=0 [QCD]
 add process a p > n1 ell p QED=3 QCD=0 [QCD]
\end{verbatim}

For leptonic decays of a Dirac $N$ as given in Eqs.~(\ref{eq:dyxSetup})-(\ref{eq:vbfSetup}), the \texttt{MadSpin} syntax is
\begin{verbatim}
define nn = n1 n1~
define ell = e+ e- mu+ mu- ta+ ta-
define vv = ve ve~ vm vm~ vt vt~
define ww = w+ w-
decay  nn > ww ell, ww > ell vv
\end{verbatim}
For Majorana $N$, the relevant substitutions are:
\begin{verbatim}
...
decay  n1 > ww ell, ww > ell vv
\end{verbatim}

The exception to this procedure is for the kinematic distributions presented in Secs.~\ref{sec:ColliderPartonKin} and ~\ref{sec:ColliderParticleKin}.
There we drop the NWA applied to $N$ in Eq.~\ref{eq:dyxSetup} and consider the fuller process
\begin{eqnarray}
 \text{DY}  &:& q\overline{q'} \to \ell_1 N^{(*)} \to \ell_1 ~\ell_2 ~W, 	 \quad\text{with} \quad W \to \ell_3 \nu.     \label{eq:dyxSetupNoNWA}
\end{eqnarray}
As the signal process proceeds through an $s$-channel heavy neutrino, which may be off-shell, we use the commands
\begin{verbatim}
generate p p > n1 || n1~ > ell ell ww QCD=0 [QCD]
\end{verbatim}
The operator syntax \texttt{> X >} selects only for diagrams/amplitudes with particle(s) \texttt{X} appearing in the $s$-channel; \texttt{||} acts as the Boolean inclusive \texttt{OR} operator.
The decay of the $W$ boson is treated with the NWA.

In both model files, the mass eigenstates $N_m$ couple to all SM lepton flavors $(\ell)$ with a mixing strength of $\vert V_{\ell N_m} \vert$. 
Hence, the above production-level commands will generate matrix elements (ignoring charge multiplicity) for all three $(N\ell_1)$-flavor permutations.
Likewise, the decay-level syntax will generate matrix elements for all $(\ell_1-\ell_2-\ell_3)$-flavor permutations.
To help ensure consistent event generation across the various flavor permutations considered, 
including to correctly account for leptonic decays of $\tau$ leptons,
we choose to control the relative abundance of various charged lepton flavors by tuning the parameters 
$\vert V_{\ell_1 N_1}\vert, ~ \vert V_{\ell_2 N_1}\vert, ~ \vert V_{\ell_3 N_1}\vert $ at runtime.

For fully inclusive DY signal samples, no generator-level cuts are needed nor are applied.
For VBF samples, $t$-channel QED poles involving both quarks and charged leptons are present~\cite{Alva:2014gxa}.
These divergences are regulated with the loose generator-level cuts
\begin{equation}
 p_T^{j,~\text{gen.}} > 20\GeV,    \quad ~p_T^{\ell,~\text{gen.}}>10\GeV, 
 \quad   ~\vert \eta^{j,~\text{gen.}} \vert < 5,    \quad ~\vert \eta^{\ell,~\text{gen.}} \vert < 3.
 \label{cut:vbfGenerator}
\end{equation}
As we are considering $pp$ collisions up to 100 TeV, a brief comment is due:
For the heavy $N$ mass scales under consideration $(m_N<20\TeV)$, the $p_T^{\rm gen.}$ selected are sufficient to regulate QED Sudakov logarithms, 
i.e., the cuts are such that $\alpha(p_T^{\rm gen}) \log^2(m_N^2/p_T^{\rm gen~2})<1$.
For $m_N\gtrsim20\TeV$, more stringent cuts are necessary~\cite{Degrande:2016aje}.
For such scales, one may need to consider factorizing collinear $q\to q'W^*$ splittings in $W\gamma$ fusion into a $W$ boson PDF
as was done for the photon in Refs.~\cite{Alva:2014gxa,Degrande:2016aje}.
For related discussions see Refs.~\cite{Arkani-Hamed:2015vfh,Mangano:2016jyj,Chen:2016wkt,Bauer:2017isx,Bauer:2017bnh,Manohar:2018kfx}.

A final technical comment: \mgamc~is less-than-optimized for phase space integration of VBF topologies, particularly at NLO in QCD.
This is partially due to 
integrating over $t$-channel momentum transfers as well as 
a virtual pentagon diagram that, while nonzero in its own right, 
does not contribute due to color conservation and the lack of color flow at the Born level~\cite{Han:1992hr}.
To aid computation efficiency, the quantity \texttt{\#IRPoleCheckThreshold}
in the file \mgamc~runtime card \texttt{Cards/FKS\_params.dat} is set to \texttt{-1.0d0} from its original value of \texttt{1.0d-5} for all VBF signal computations.
For the DY signal process at $\sqrt{s}=100$ TeV, a value of \texttt{1.0d-3} or \texttt{1.0d-2} is used for $m_N = 150-300\GeV$ to ease numerical stability requirements for so low Bjorken-$x$.

\subsection{Standard Model Background Event Generation}~\label{sec:mcBkgGen}
We now elaborate on the details of our SM background event simulation.

To study the trilepton process $pp\to 3\ell X$ in conjunction with a jet veto, we consider three- and four-charged lepton processes from a variety of SM sources.
In order to better ensure a reliable description of the leading jet at all $p_T$ scales, all processes are considered at NLO+PS, with one exception.
We separate the SM processes into four categories:
(i) top quarks, 
(ii) EW diboson continuum,
(iii) resonant EW multi-boson production, and
(iv) fake charged leptons.
Throughout the text we refer to ``continuum'' processes as those that possess resonant and non-resonant components.
For such samples, full interference between resonant and non-resonant regions of phase space is considered.

\subsubsection*{Top Quarks}
Due to their large rates and inherent mass scales, top quarks produced in association with EW bosons, e.g., $t\overline{t}W$, 
are leading backgrounds to traditional trilepton searches that rely on vetoing only central, $b$-tagged jets~\cite{delAguila:2008cj,Sirunyan:2018mtv}.
Invariably, the rates of these many-lepton processes are able to compensate for the small (but not entirely rare) likelihood 
that $b$-jets fail to be identified while simultaneously one or more charged leptons fail to meet ``analysis quality object'' criteria.
To account also for these potentially soft (low-$p_T$), forward (high-$\vert \eta\vert$) charged leptons, we consider the continuum processes
\begin{eqnarray}
  pp \to t\overline{t}\ell\ell,                             	&\quad\text{with}\quad&  t \to Wb \to \ell\nu_\ell b,\\
  pp \to t\overline{t}\ell \nu,                             	&\quad\text{with}\quad&  t \to Wb \to \ell\nu_\ell b, \quad\text{and} \\
  \overset{(-)}{b}q \to \overset{(-)}{t}q'\ell\ell,        &\quad\text{with}\quad&  t \to Wb \to \ell\nu_\ell b.
\end{eqnarray}
As for the signal process above, $p$ here denotes a parton  $p\in\{q,\overline{q},g\}$, with  $q\in\{u,c,d,s\}$, and $\ell \in \{e, \mu, \tau\}$.
We consider only pure gauge interactions and neglect the contribution of a SM-like Higgs, e.g., $pp\to t\overline{t}h,~h\to\tau^+\tau^-$.
Inclusion of such processes would incredibly complicate our matrix element computations at NLO while contribute only marginally to our background modeling.
Owing to the fact that the $t\overline{t}\ell\ell\to2\ell X$ and $t\overline{t}\ell\ell\to3\ell X$ signals are efficiently cut out by minimal selection cuts,
for MC efficiency, only the fully leptonic $t\overline{t}\ell\ell\to 4\ell X$ decay mode is considered.
For the first two top quark processes, which we model at NLO+PS, the  \mgamc~syntax is:
\begin{verbatim}
 generate p p > t t~ ell ell / h QCD=2 QED=2 [QCD]
 generate p p > t t~ ell vv / h QCD=2 QED=2 [QCD]
\end{verbatim}
For both processes, \texttt{MadSpin} syntax is:
\begin{verbatim}
 decay tt > ww bb, ww > ell vv
\end{verbatim}

The single top process $bq \to tq'\ell\ell$ is a somewhat novel channel as it was only recently observed at Run II of the LHC~\cite{Sirunyan:2017nbr,CMS-PAS-TOP-18-008}.
Modeling this process is nuanced as we are formally working a four-flavor scheme with a massive $b$ quark (in order to preserve modeling of $\mathcal{B}$ meson decays).
Hence, for $tq\ell\ell$, we take \confirm{$m_b = 10^{-10}\GeV$} at the matrix element level in order to use the $b$ quark PDF 
and circumvent internal \mgamc~checks for massive, initial-state partons.
(These checks are necessary to ensure consistency of the Collinear Factorization Theorem~\cite{Collins:1984kg,Collins:1985ue,Stewart:2009yx,Collins:2011zzd,Becher:2012qa,Diehl:2015bca} as implemented in \mgamc.)
On the other hand, implementation of the aMC@NLO formalism~\cite{Frixione:2002ik} 
in \mgamc~requires that the initial-state parton $b$ be identically massless, not approximately massless.
Subsequently, this is the only event simulation that we model at LO+PS (all others are simulated at NLO+PS).
Subsequently, the Born-level $tq\ell\ell$ process is modeled with the following:
\begin{verbatim}
 define qq = u c d s u~ c~ d~ s~
 define bb = b b~
 define tt = t t~
 generate    qq bb > ell ell qq tt / h QCD=0 QED=4
 add process bb qq > ell ell qq tt / h QCD=0 QED=4
\end{verbatim}
The leptonic decay of $t$ is handled in the same manner as the $p p \to t \overline{t}\ell X$ processes.

With massive top quarks, the $p p \to t \overline{t} \ell \nu$ matrix element possesses no IR poles at the Born level.
Hence, no regulating cuts are needed nor are any applied during event generation.
The $p p \to t \overline{t} \ell \ell$ and $b q \to t q' \ell \ell$ matrix elements, on the other hand, do possess IR poles as $m_{\ell\ell}\to0$. 
Hence, the following generator-level cuts are applied:
\begin{eqnarray}
 m(\ell_1,~\text{gen.}, \ell_2,~\text{gen.}) > 10\GeV \quad\text{and}\quad \vert \eta^{\ell,~\text{gen.}} \vert < 3.
 \label{cut:topGenerator}
\end{eqnarray}
The invariant mass boundary is set to 10 GeV in order to enrich the contribution of soft-leptons but high enough to remove contribution of low-mass DY resonances.

\subsubsection*{Electroweak (Diboson) Continuum}
Electroweak production of three- and four-charged lepton events in hadron collisions,
\begin{eqnarray}
 pp \to \ell\ell\ell\nu \quad\text{and}\quad  pp \to \ell\ell\ell\ell,
\end{eqnarray}
which originate from
double-resonant ($VV'$ for $V\in\{W,Z\}$),
single-resonant $(VV^{'*}/V\gamma^*)$, 
and zero-resonant $(V^*V^{'*}/V^*\gamma^*)$
diboson production,
are the largest, pure-EW background to trilepton searches for heavy neutrinos with $m_N$ at or above the EW scale.
In the presence of a jet veto, the $3\ell\nu$ process indeed surpasses top quarks  as the principle irreducible background.
We model the full continuum processes at NLO+PS using:
\begin{verbatim}
 define nX  = n1 n2 n3
 generate p p > ell ell ell ell / h nX QCD=0 QED=4 [QCD]
 generate p p > ell ell ell vv / h nX QCD=0 QED=4 [QCD]
\end{verbatim}
We apply the same generator-level cuts as given in Eq.~\ref{cut:topGenerator}.

We note that as events are generated using the \texttt{HeavyN} model file, it is necessary to also remove their diagrammatic contribution.
In practice, this is done with the flag \texttt{/ nX}, with $nX\in\{N_1,~N_2,~N_3\}$.
Alternatively, one can leave diagrams in place and either reduce active-sterile mixing elements to an incredibly small number, e.g., $10^{-10}$, 
or heavy neutrino masses to an incredibly large value, e.g., $10^{10}\GeV$, to decouple the fields.

\subsubsection*{Electroweak (Multi-Boson) Production}
Beyond the EW diboson continuum, significant but subleading backgrounds are the multi-boson processes, i.e, the production of three or more EW bosons.
As representative backgrounds, we consider specifically the fully resonant and mixed-resonant channels,
\begin{eqnarray}
 pp \to WWW \quad\text{and}\quad  pp \to WW\ell\ell,
\end{eqnarray}
at NLO+PS. The generator-level commands are:
\begin{verbatim}
 generate p p > ww ww ww / h QCD=0 QED=3 [QCD]
 generate p p > ell ell w+ w- / h QCD=0 QED=4 [QCD]
\end{verbatim}
We decay $W$ bosons to leptons using the NWA in the same manner done for top quarks.
Higher vector boson multiplicities, e.g., $WWZZ$, suffer from coupling and phase space suppression, and are ignored.
Similarly, alternative weak boson permutations, e.g., $WZZ$, suffer from branching fraction and coupling suppression, and hence are subleading to the processes above.
For the $WW\ell\ell$ process, we apply the generator-level cuts of Eq.~(\ref{cut:topGenerator}).

\subsubsection*{Fake Leptons}
As discussed in Sec.~\ref{sec:intro}, a substantial limitation to current multilepton searches for heavy neutrinos 
are misidentified and mis-reconstructed/mis-measured objects~\cite{delAguila:2008cj,Sirunyan:2018mtv}, and are often neglected in theoretical studies.
Though unlikely, about 1 in $10^4$ QCD jets with relatively low $p_T$ are misidentified as electrons and positrons~\cite{Aad:2016tuk,Alvarez:2016nrz}.
Similarly, it is possible for QCD jets to be misidentified as hadronically decaying $\tau$ leptons, 
with rates spanning about 1 in $10^{2}-10^{4}$ as a function of jet $p_T$~\cite{CMS-DP-2017-036}.
Hence, assuming that a jet is mis-tagged as a charged lepton, SM processes with genuine hard lepton pairs, such as 
\begin{eqnarray}
 &pp \to t\overline{t}, &\quad\text{with}\quad  \overset{(-)}{t}\to W\overset{(-)}{b} \quad\text{and}\quad W \to \ell \nu,
 \\
 &pp \to WWj,           &\quad\text{with}\quad  W \to \ell \nu, 
\end{eqnarray}
can mimic the $pp\to3\ell X$ signal.
We emphasize that not only can jets from the Born process contribute to the ``fake'' lepton background but also 
softer, wide-angle, initial-state (ISR) and final-state radiation (FSR) that are pulled into the matrix element (event file)
through $\mathcal{O}(\alpha_S)$ fixed order corrections and resummed corrections in the parton shower.
Hence, we believe modeling the progenitors of fake leptons, where possible, to at least NLO+PS is critically necessary.
To model the above processes, we use the~syntax
\begin{verbatim}
 generate p p > t t~ QCD=2 QED=0 [QCD]
 generate p p > w+ w- j QED=2 QCD=1 [QCD]
\end{verbatim}
To regulate the $WWj$ process, we impose the generator-level cuts,
\begin{equation}
 p_T^{j,~\text{gen.}} > 30\GeV \quad\text{and}\quad \vert \eta^{j~\text{gen.}} \vert  < 3.
\end{equation}
Top quarks and $W$ bosons are treated in the NWA and decayed to leptonic final states as above.
We defer details of how fake leptons are modeled at the analysis level to Sec.~\ref{sec:Observability}.

\subsection{Standard Model Inputs}\label{sec:mcSMinputs}
Assuming $n_f=4$ massless quarks and a \confirm{diagonal Cabbibo-Kobayashi-Maskawa (CKM) matrix with unit entries},
we take SM inputs from the 2014 Particle Data Group~\cite{Agashe:2014kda}:
\begin{eqnarray}
 \alpha^{\rm \overline{MS}}(M_{Z})= 1/127.940, 	&\quad& M_{Z}=91.1876\GeV, \quad \sin^{2}_{\rm \overline{MS}}(\theta_{W}) 	= 0.23126, 
 \\
 m_t(m_t) = 173.3\GeV, 				&\quad&	m_b(m_b) = 4.7\GeV, \quad  m_\tau = 1.777\GeV.
 \label{eq:smInputs}
\end{eqnarray}
Despite $m_b\neq0$, we employ PDFs with variable $n_f$.
This technical inconsistency allows us to have a more realistic PDF normalization for multi-TeV $m_N$ across different colliders.

PDF and $\alpha_s(\mu_r)$ evolution are extracted using the \texttt{LHAPDF} \confirm{v6.1.6} libraries~\cite{Buckley:2014ana}.
Throughout this study, we use several PDFs to best match the formal accuracy of our various collider calculations:
For LO and NLO in QCD calculations with and without PS-matching, 
we use the NNPDF 3.1 NLO+LUXqed PDF set~\cite{Bertone:2017bme} (\texttt{lhaid=324900}).
The PDF set features an inelastic photon PDF, i.e., photons as a parton in a hadron, 
matched to an elastic photon PDF, i.e., photons from hadrons, at low momentum transfer $(Q^2)$
 using the LUXqed formalism~\cite{Manohar:2016nzj,Manohar:2017eqh}.
For resummed jet veto calculations at NLO+N$^2$LL(veto), we use the NNPDF 3.1 N$^2$LO+LUXqed PDF set (\texttt{lhaid=325100}) 
to avoid double counting of $\mathcal{O}(\alpha_s^2)$ emissions.
For resummed threshold calculations at N$^3$LL(threshold), we use the NNPDF N$^2$LO+N$^2$LL(threshold) PDF set~\cite{Bonvini:2015ira}.
For discussions of the various PDFs and their impact on BSM collider phenomenology, including neutrino mass models,
see Refs.~\cite{Alva:2014gxa,Beenakker:2015rna,Mitra:2016kov,Fiaschi:2018xdm,Fiaschi:2018hgm} and references therein.
Similarly, while the PY8 tune we use was constructed using the NNPDF 2.3 LO PDF set, 
we expect the error from this mismatch to be sub-leading with respect other sources of uncertainty.

For signal and background processes, we follow the recommendation of the heavy neutrino study of Ref.~\cite{Degrande:2016aje} and 
set the collinear factorization $(\mu_f)$ and QCD renormalization $(\mu_r)$ scales to (half) the sum of all visible final-state particles' transverse 
energy (\texttt{dynamic\_scale\_choice=3} in \mgamc),
\begin{equation}
 \mu_f,\mu_r = \mu_0 = \frac{1}{2}\sum_{k=N,\ell,\text{jets}}E_{T,k} = \frac{1}{2}\sum_k\sqrt{m_k^2 + p_{T,k}^2}.
 \label{eq:scale}
\end{equation}
The impact of this scale choice on TeV-scale leptons produced via Drell-Yan is that the relative size of $\mathcal{O}(\alpha_s)$ corrections 
to the total rate and differential cross sections are largely constant as a function of both the heavy lepton mass and the values of inclusive, leptonic observables.
For further discussions, see Refs.~\cite{Ruiz:2015zca,Degrande:2016aje,Fuks:2017vtl} and classic references therein.
Where appropriate, we estimate the impact of missing perturbative corrections to our various calculations, i.e.,
scale uncertainties, by varying $\mu_f,\mu_r$ between $0.5 < \mu_{r,f}/\mu_0 < 2$.
For PDF uncertainties, we report standard deviations across replica PDFs as prescribed by Ref.~\cite{Buckley:2014ana}.
The exception to this procedure are for color-singlet processes initiated by gluon fusion, 
where we instead follow Ref.~\cite{Ruiz:2017yyf} and references therein due to the threshold resummation applied.
We do not account for the perturbative uncertainties associated with the PS shower matching scale $\mu_s$.
While likely important, we note that they can be brought well into control with the inclusion of additional matrix element matching~\cite{Jones:2017giv,Chakraborty:2018kqn}.

\subsection{Benchmark Heavy Neutrino Inputs}\label{sec:mcHeavyNinputs}
Thought this analysis, we restrict ourselves to searches for the lightest, heavy neutrino mass eigenstate, $N_{m'=4}$, which we denote henceforth simply as $N$.
For discovery purposes, we also take active-sterile neutrino mixing $V_{\ell N}$ to be independent of neutrino masses.
This action requires care to ensure that $N$ of arbitrarily high mass are consistently modeled in MC event generators.
In accordance with the Goldstone Equivalence Theorem, for fixed active-sterile mixing in Type I-like Seesaws models,
heavy neutrinos with masses much larger than the EW scale possess total decay widths that scale as $\Gamma_N^{\rm Tot.} \sim G_F m_N^3 \sum _\ell\vert V_{\ell N}\vert^2$.
Ultimately, this is due to implicitly increasing the magnitude of the Dirac neutrino Yukawa in $V_{\ell N}$ to ensure light neutrino masses are small,
and leads to $N$ being strongly coupled, like the top quark, to the SM Higgs and the EW Goldstone bosons.
 
For a na\"ive benchmark mixing of $\vert V_{\ell N}\vert \sim 1$, this cubic mass dependence is very steep:  
for $m_N \lesssim 1\TeV$, one obtains $\Gamma_N^{\rm Tot.} / m_N \sim 1$, indicating a breakdown of the particle description of $N$.
Even for more modest benchmark values of $m_N$ or $\vert V_{\ell N}\vert$, care is still needed.
For $\vert V_{\ell N}\vert \sim 0.1$, one still finds that $\Gamma_N^{\rm Tot.}\sim \mathcal{O}(10)\GeV$ for $m_N\sim 1\TeV$.
While the NWA and Breit-Wigner approximations are justified for $\Gamma_N^{\rm Tot.} / m_N  \lesssim 1\%$,
a total width of $\Gamma_N^{\rm Tot.}\sim 10\GeV$ means that the virtuality of a propagating, nearly on-shell $N$ can shift naturally by $\delta \sqrt{p_N^2}\sim 20-30\GeV$ on an event-by-event basis.
This is known to broaden the kinematics of a heavy neutrino's decay products in MC simulations at $\sqrt{s}=14\TeV$~\cite{Alva:2014gxa}
to a level that  experimentally  resolvable.
As a result, one obtains an expected experimental sensitivity to EW- and TeV-scale $N$ that deviates greatly from the true sensitivity.

On the other hand, in practice, one usually obeys the relation, $\Gamma_N^{\rm Tot.}/m_N \ll 1$, for realistic values of $\vert V_{\ell N}\vert$~\cite{Fernandez-Martinez:2016lgt}.
Hence, to avoid these ``large width'' artifacts in our MC simulations, 
we develop our selection cuts and estimate our signal-significance assuming a small generator-level mixing. 
Results are reinterpreted for smaller mixing accordingly.
For charged lepton flavor conserving scenarios, we use as our generator-level mixing,
\begin{equation}
\confirm{
\vert V_{\ell_1 N} \vert = \frac{1}{\sqrt{2}}\times10^{-2} \approx 7.07 \times 10^{-3}  \quad\text{and}\quad  \vert V_{\ell_2 N} \vert , \vert V_{\ell_3 N} \vert = 0,
}
\label{eq:mcMixingLFC}
\end{equation}
for $\ell_1 \in\{ e,~\mu~,\tau\}$. For charged lepton flavor-violating event samples, we use,
\begin{equation}
\confirm{
\vert V_{\ell_1 N} \vert = \vert V_{\ell_2 N} \vert = \frac{1}{\sqrt{2}}\times10^{-2}   \quad\text{and}\quad   \vert V_{\ell_3 N} \vert = 0.
}
\label{eq:mcMixingLFV}
\end{equation}
In total, all six benchmark permutations of $\ell_1,\ell_2,\ell_3 \in \{ e,\mu,\tau\}$ are considered.
We anticipate our main conclusions to hold for the democratic mixing scenario, where $\vert V_{\ell_1 N} \vert = \vert V_{\ell_2 N} \vert = \vert V_{\ell_3 N} \vert$,
and therefore neglect this common benchmark scenario.
For more details on flavor-mixing hypotheses, see Sec.~\ref{sec:obsSigDef}.

\section{Heavy Neutrinos at Hadron Colliders}\label{sec:Colliders}

While production- and decay-level kinematics of heavy neutrinos are well-reported throughout the literature for collider energies spanning $\sqrt{s}=2-100\TeV$,
rarely (if at all) has it been done for the express purpose of building a tailored collider analysis that remains robust against varying collider energies, let alone in the context of jet vetoes.
This is a highly nontrivial task and requires one to be simultaneously aware and ignorant (inclusive) with respect to the growing transverse, hadronic activity at increasing $\sqrt{s}$.

Therefore, in this section, we present total and differential cross sections for the production of a single heavy neutrino $N$ and its subsequent decay to the trilepton final state, 
as given in Eq.~\ref{eq:HeavyN_DYX_TrilepX} and shown in Fig.~\ref{fig:feynman_DYX_Trilep}, at the collider energies $\sqrt{s}=14,~27,$ and $100\TeV$.
We begin in Sec.~\ref{sec:ColliderProduction}, where we briefly summarize the collider production formalism that enters into our state-of-the-art predictions.
In a language suitable for broader audiences, we make explicit the connections between 
the formal accuracy of our jet modeling and the terms that appear in the 
Collinear Factorization Theorem~\cite{Collins:1984kg,Collins:1985ue,Stewart:2009yx,Collins:2011zzd,Becher:2012qa,Diehl:2015bca}, 
i.e., the master equation for collider physics.
In Sec.~\ref{sec:ColliderXSec}, we compare the leading heavy neutrino production channels shown in Fig.~\ref{fig:feynmanMulti} 
and reveal a qualitative picture for heavy neutrino production that is different for the LHC and its potential successors.
In Sec.~\ref{sec:ColliderPartonKin}, we present parton-level distributions of elementary leptonic observables that form the basis for several arguments used to build our  analysis.
Remaining robust against increasing hadron collider center-of-mass energies requires one to have a realistic description of QCD activity, even for the signal process.
This is why show in Sec.~\ref{sec:ColliderParticleKin} particle-level\footnote[2]{
Throughout this analysis, ``particle-level'' events, also labeled as generator-level, reconstructed events, 
refer to MC events that have been decayed, parton showered, hadronized, and passed through jet clustering.
}
distributions of both elementary and more complex kinematic observables for leptons at NLO+PS.
Properties related to hadronic activity (jets) are reported in Sec.~\ref{sec:jetVeto}.

We note that due to the final-state light neutrino that appears in our $pp\to N\ell \to 3\ell\nu$ process, events inherently possess some degree of transverse momentum imbalance.
To disentangle this from the presence of light neutrinos from $\tau$ lepton decays,
 we temporarily limit ourselves to the $e-\mu$ mixing / no $\tau$-mixing scenario, with $V_{\ell N}$ set by Eq.~\ref{eq:mcMixingLFV}.
 We also further restrict decays of the intermediate $W$ boson to only $e$- and $\mu$-flavored leptons.
 These restrictions are not imposed in the final sensitivity estimates reported in Sec.~\ref{sec:Observability}.
 Finally, we briefly stress that for the mass range of $N$ considered  $(m_N > m_h)$ and the scales involved with the production mechanisms studied,  
 helicity suppression of lepton number violating currents~\cite{Kayser:1982br,Han:2012vk,Balantekin:2018ukw}, 
 a consequence of the so-called Dirac-Majorana Confusion Theorem~\cite{Kayser:1982br}, is not present. 
 Therefore, what follows here and in Sec.~\ref{sec:jetVeto} holds for both heavy Dirac and Majorana neutrinos produced and decayed via SM currents.

\subsection{Heavy Neutrino Production Formalism}\label{sec:ColliderProduction}

For the inclusive production and decay of a heavy neutrino $N$ in association with $X$ (e.g., $X=\ell$, $\nu$, or $\ell j$), 
we calculate total $(\sigma)$ and differential $(d\sigma)$ scattering rates
in accordance with the Collinear Factorization Theorem~\cite{Collins:1984kg,Collins:1985ue,Stewart:2009yx,Collins:2011zzd,Becher:2012qa}, given by
\begin{eqnarray}
d\sigma(pp \to NX+{\rm Anything}) &=& f \otimes f \otimes \Delta \otimes  d\hat{\sigma} + \mathcal{O}\left(\frac{\Lambda_{\rm NP}^n}{Q^{n+2}}\right)\nonumber\\
\end{eqnarray}
\begin{eqnarray}
		  &=&
\frac{1}{1+\delta_{ik}} \sum_{\substack{i,k=q,g,\gamma\\ \beta,\beta'=S,F,\overline{F},A}} 
\int_{\tau_0}^1 d\tau ~\int_{\tau}^1 d\frac{\xi_1}{\xi_1} ~\int_{\tau/\xi_1}^1 \frac{dz}{z}
\nonumber\\
& & \left[ f_{i/p}(\xi_1,\mu_f)f_{k/p}(\xi_2,\mu_f)+ (1\leftrightarrow2)\right] 
~
\Delta^{\beta\beta'}_{ik}(z)
~
d\hat{\sigma}^{\beta\beta'}(ik\to NX)
~+~  \mathcal{O}\left(\frac{\Lambda_{\rm NP}^n}{Q^{n+2}}\right).
\label{eq:factThm}
\end{eqnarray}
The above states that in $pp$ collisions,
hadron-level observables $(d\sigma)$ can be expressed as the product of conditional probabilities, i.e., a convolution $(\otimes)$,
of process-independent parton distribution functions (PDFs) $f$,
a quasi-universal Sudakov factor $\Delta$,
and a process-specific, parton-level $ik\to NX$ hard scattering observable $d\hat{\sigma}$
that occurs at the hard scattering scale $Q = m_{NX} \sim m_N$,
all up to suppressed corrections that scale as powers of the hadronic, or non-perturbative (NP), scale $\Lambda_{\rm NP}\sim1-2\GeV$.

More precisely, $f_{i/p}(\xi,\mu_f)$ represents the likelihood of observing a massless parton $i\in\{q,\overline{q},g,\gamma\}$,
where $q \in \{u,c,d,s\}$, carrying 
a transverse momentum $p_T^i$ below a collinear factorization scale $\mu_f$ and 
a longitudinal momentum $p_z^i = \xi P_z$ away from a proton $p$ with momentum $P\approx(\sqrt{s}/2,0,0,\pm \sqrt{s}/2)$.
Via the DGLAP renormalization group evolution equations~\cite{Gribov:1972ri, Dokshitzer:1977sg,Altarelli:1977zs},
$f_{i/p}$ accounts for (resums) all collinear $j \to i+j' $ QCD and QED emissions leading to parton $i$ such that $p_T^{i,~j'} < \mu_f$.
That is, $f_{i/p}$ accounts for all  initial-state gluons, photons, and quarks with $p_T < \mu_f$ emitted in association with the $(NX)$ system.
Radiations with $p_T^{j'} > \mu_f$ are included through appropriate $\mathcal{O}(\alpha_s^m \alpha^n)$ perturbative corrections to the hard matrix element $d\hat{\sigma}$.
Inclusion of QED DGLAP evolution~\cite{Martin:2004dh,Martin:2014nqa,Manohar:2016nzj,Manohar:2017eqh}  that is matched 
at $\mu_f \to m_{p}\sim1\GeV$ to a proton's elastic photon PDF implies~\cite{Martin:2014nqa} the photon PDF $f_{\gamma/p}$ is valid for all photon virtualities.
This is needed to correctly model inclusive $N\ell^\pm$ production from $W\gamma$ fusion~\cite{Alva:2014gxa,Degrande:2016aje}.
Together, partons $i$ and $k$ from PDFs $f_{i/p}$ and $f_{k/p}$ generate the parton scattering scale $\sqrt{\hat{s}} = \sqrt{\xi_1 \xi_2 s} \geq Q$.

Intuitively, the Sudakov factor $\Delta(z) = \delta(1-z) + \mathcal{O}(\alpha_s) + \mathcal{O}(\alpha)$, 
through exponentiation and resummation~\cite{Contopanagos:1996nh,Laenen:2000ij,Dasgupta:2018nvj},
acts to ``dress''  (in a quantum field theoretic sense) bare partons with collinear and/or soft, i.e., unresolved, QCD and QED radiation, rendering them more inline with physical quantities.
(In practice and reality, arbitrarily infrared emissions are regulated by hadronization.)
More technically, the Sudakov factor accounts for gluons (and photons) carrying a momentum fraction $(1-z)$ emitted prior to the $ik\to NX$ hard process $(\hat{\sigma}^{\beta\beta'})$. 
Here, $z=Q^2/\hat{s}$ ~ is a dynamic measure of how much of the total partonic energy is carried by the hard $(ik)$ process.
The remainder is carried away by gluons and photons with momenta that are small (soft) compared to $Q$.
$\Delta_{ik}^{\beta\beta'}$ is quasi-process-independent in the sense that it is sensitive to the color and Lorentz structure of the incoming partons, e.g., 
$\beta,\beta'=F\overline{F}$ for $(q\overline{q'})$- or $AA$ for $(gg)$-annihilation into a color-singlet final state, but is insensitive to the hard process itself.
The precise organization and exponentiation
of $\mathcal{O}(\alpha_s)$ (and higher order) terms in $\Delta^{\beta\beta'}_{ik}(z)$ 
leads to the all orders resummation of logarithms associated with collinear and/or soft ISR and FSR.
This includes terms associated with jet vetoes that are present in differential distributions but cancel for the inclusive cross 
section~\cite{Dasgupta:2001sh,Appleby:2002ke, Forshaw:2006fk,
Stewart:2009yx,Stewart:2010pd,Berger:2010xi,
Becher:2012qa,Becher:2013xia,Gaunt:2014ska}.
To recover fixed order (FO) predictions from Eq.~\ref{eq:factThm}, one simply neglects all terms in $\Delta(z)$ beyond the $\delta(1-z)$ kernel.
Notably, collinear ISR and FSR are well-modeled by Sudakov form factors as implemented in modern parton showers (PS) since they formally resum corresponding terms 
to at least leading logarithmic (LL) accuracy, when expanding the color algebra in powers of $1/N_c$~\cite{Dasgupta:2014yra}.

For the $n$-body $ik\to NX$ hard process, the parton-level observables $\hat{\sigma}$ and $d\hat{\sigma}$, for unpolarized hadron collisions, are obtained from the expressions,
\begin{eqnarray}
& & \hat{\sigma} ~=~ \int dPS~\frac{d\hat{\sigma}}{dPS_n},
 \quad\text{where} \\
& & \frac{d\hat{\sigma}}{dPS_n} = \frac{1}{2Q^2}\frac{1}{(2s_i+1)(2s_k+1)N_c^i N_c^k} \sum_{\rm dof} \vert \mathcal{M}(ik\to NX)\vert^2, 
 \quad\text{with} \\
& & dPS_n(p_i+p_k;p_{f=1},\dots,p_{f=n}) = (2\pi)^4 \delta^4\left(p_i + p_j - \sum_{f=1}^n p_f\right)\prod^{n}_{f} \cfrac{d^3 p_f}{(2\pi)^3 2E_f}.
\end{eqnarray}
Here, $s_j = 1/2$ are the helicity degrees of freedom (dof) for massless parton $j=i,k$; $N_c^j$ are similarly the SU$(3)_c$ color dof;
$\mathcal{M}$ is the Lorentz-invariant matrix element calculable via perturbative methods;
and $dPS_n$ is the separately Lorentz-invariant $n$-body phase space.
For NLO in QCD corrections, we employ the MC@NLO~\cite{Frixione:2002ik} formalism as implemented in \mgamc~in order to insure real radiation common to 
the PDFs $f$, Sudakov factor $\Delta$, and $d\hat{\sigma}$ beyond LO are appropriately subtracted, up to our claimed NLO precision.

Beyond perturbative corrections to matrix elements in $d\hat{\sigma}$, 
to all orders, the impact of collinear, initial-state QCD and QED emissions on the normalization of the total, hadronic cross section $(\sigma)$ are, 
by construction, already included in the DGLAP evolution of PDFs\footnote[2]{
For clarity, the normalization increase of the trilepton process in Fig.~\ref{fig:feynman_DYX_Trilep}
at NLO in QCD is almost entirely driven by the virtual correction and results from a combination of QCD color factors, a loop phase space suppression factor, 
and a $\pi^2$ term due DY to being a space-like process~\cite{Altarelli:1979ub,Hamberg:1990np,Anastasiou:2003yy}.
The typical correction of $+20$ to $+40\%$ at NLO does not originate from  ISR; see, e.g., Ref.~\cite{Ruiz:2015zca} and references therein.}.
This is the reason for subtraction of $\mathcal{O}(\alpha_s)$ terms from the PDF when $d\hat{\sigma}$ is known at NLO in QCD.
By unitarity, collinear FSR does not readily impact the normalization.
Qualitatively, however, both ISR and FSR can significantly alter the shape of hadron-level differential distributions $(d\sigma)$ and should not be neglected.
Independent of jet vetoes, neglecting the parton shower entirely and considering searches only at the (unphysical) 
partonic level has a substantial impact on selection cut efficiencies for Seesaw particles~\cite{delAguila:2007qnc,delAguila:2008cj,delAguila:2008hw}.

The impact of soft/threshold logarithms~\cite{Sterman:1986aj,Catani:1989ne,Catani:1990rp} contained in $\Delta$, 
which originate from considering additional partonic phase space contributions~\cite{Forte:2002ni}, 
can sizably increase the total cross section normalization with respect to Born-level predictions for certain 
kinematic regimes~\cite{Appell:1988ie} and initial-state color configurations~\cite{Catani:1996yz}.
(Though these resummed contributions are not entirely independent of FO contributions; see, for example, Ref.~\cite{Ahrens:2008qu}.) 
For new physics processes, this is still the case after removing potential double counting of soft-collinear contributions in 
PDFs~\cite{Bonvini:2015ira,Beenakker:2015rna,Mitra:2016kov,Ruiz:2017yyf,Fiaschi:2018xdm,Fiaschi:2018hgm}.
For the GF production process in Fig.~\ref{fig:feynmanMulti}(b),
and exploiting the renormalization group (RG) evolution properties of the Factorization Theorem in momentum space~\cite{Becher:2006nr,Becher:2006mr,Ahrens:2008nc},
the N$^3$LL threshold corrections can be obtained in a straightforward manner~\cite{Ruiz:2017yyf}.
The procedure ostensibly requires replacing the FO expression for the Sudakov factor $\Delta^{\rm FO} \approx \delta(1-z)$ 
with the resummed / RG-improved expression at N$^3$LL(threshold), $\Delta^{\rm Res.}$ ,
while adding and subtracting the necessary terms to coax numerical stability.
For NLO+N$^2$LL jet veto corrections, the procedure is more involved but nonetheless straightforward~\cite{Becher:2014aya}.
As for pure fixed order corrections, it requires subtraction of collinear splittings from the PDF $f$.
However, instead of simply augmenting $\alpha_s$-subtracted PDFs with $\alpha_s$-corrected matrix elements, 
one introduces  corrections that imbues incoming / initial-state partons with transverse momentum.
These corrections act as an intermediate term connecting (a) the PDFs evolved up to the collinear factorization $\mu_f$ to (b) the $p_T$ of the leading QCD emission in the matrix element.
 (In some sense, the result is akin working with transverse momentum dependent (TMD) PDFs that are matched to the $\alpha_s$-corrected matrix element.)

Aside from hadronization, power-suppressed, non-perturbative terms of the order $\mathcal{O}\left(\Lambda_{\rm NP}^n / Q^{n+2}\right)$, i.e., 
higher twister terms in the operator product expansion~\cite{Collins:1984kg,Collins:1985ue,Sterman:1995fz,Collins:2011zzd}, 
 are neglected in this study.
For jet vetoes, this is  a potential source of sizable theoretical uncertainty.
More precisely, the momenta of reconstructed jets in hadron collisions, and hence jet vetoes cross sections,
receive corrections from Glauber exchanges, e.g., double parton scattering/multiple parton interactions/underlying event (UE).
Such corrections are beyond the theorem's formal accuracy nor are presently known to fully factorize;
for further details, see Refs.~\cite{Dasgupta:2007wa,Stewart:2009yx,Collins:2011zzd,Becher:2012qa,Gaunt:2014ska,Zeng:2015iba,Diehl:2015bca}.
While we neglect their impact, we do believe it is possible to reliably parametrize UE, particularly at larger $\sqrt{s}$, as demonstrated in Ref.~\cite{Brooks:2018tgf}.
A dedicated investigation into the impact of UE on jet vetoes applied to new physics searches is deferred to future studies.
For related details, see Sec.~\ref{sec:jetVeto}.

\subsubsection*{Factorization of Heavy Neutrino Mixing from Collider Observables}

For the resonant production of a single heavy neutrino in $pp$ collisions, the active-sterile mixing element $V_{\ell N}$
enters the scattering matrix elements as an overall multiplicative factor.
Hence, the (squared) norm of the active-sterile mixing, $\vert V_{\ell N}\vert^2$, factors out at the squared matrix element,
and therefore factors out of at the hadronic level.
This gives rise to the so-called ``bare cross section'' $d\sigma_0$, defined, for example for $pp\to N\ell$, as~\cite{Han:2006ip}
\begin{equation}
 d\sigma_0(pp\to N\ell) \equiv d\sigma(pp\to N\ell) / \vert V_{\ell  N}\vert^2  = f\otimes f\otimes\Delta\otimes d\hat{\sigma} / \vert V_{\ell N}\vert^2.
 \label{eq:bareProd}
\end{equation}
Subsequently, semi-flavor-model independent prediction for both Dirac and Majorana neutrinos can be constructed at the $N$ production level.
$\vert V_{\ell N}\vert^2$ also enters into the partial decay widths of $N$, and hence its total decay width.
As this in turn appears in the Breit-Wigner propagator for $N$, the factorization of neutrino mixing elements from decay-level heavy neutrino observables is slight more subtle. 
The analogous ``bare cross section'' for resonant production and decay of $N$, in for example $pp\to N\ell_1 \to \ell_1 \ell_2 V$, is~\cite{Han:2006ip}
\begin{equation}
  d\sigma_0(pp\to N\ell_1 \to \ell_1 \ell_2 V) \equiv d\sigma(pp\to N\ell_1 \to \ell_1 \ell_2 V) / S_{\ell_1\ell_2}, 
 \quad S_{\ell_1\ell_2} = \cfrac{\vert V_{\ell_1 N}\vert^2 \vert V_{\ell_2 N}\vert^2 }{\sum_{\ell'=e}^\tau \vert V_{\ell' N}\vert^2 }.
  \label{eq:bareDecay}
\end{equation}
Notably, both factorizations hold identically to all orders in QCD~\cite{Ruiz:2015zca,Degrande:2016aje}, 
so long as the individual decay products of $N$ remain color neutral; for hadronic decays of $N$, the factorization holds at least to NLO.
Trivially, $d\sigma_0$ can be obtained numerically by setting $\vert V_{\ell N}\vert=1$ in $d\hat{\sigma}$ or by a rescaling for nonzero $\vert V_{\ell N}\vert$.

For multiple interfering heavy neutrino mass eigenstates, the above relations cannot be generically applied.
This is particularly the case for pseudo-Dirac neutrino pairs, 
where relative CP phases in the $V_{\ell N}$ lead to significant destructive interference for $L$-violating processes.
However, as discussed in Sec.~\ref{sec:theoryDirac}, the mass splitting for pseudo-Dirac neutrino pairs in a realistic ISS should be vanishingly small on the collider scale.
Hence, pseudo-Dirac pairs can be justifiably modeled as a single Dirac neutrino in collider environments.

\subsection{Heavy Neutrino Production at the LHC and Beyond}\label{sec:ColliderXSec}

\begin{figure*}[!t]
\begin{center}
\subfigure[]{\includegraphics[width=.48\textwidth]{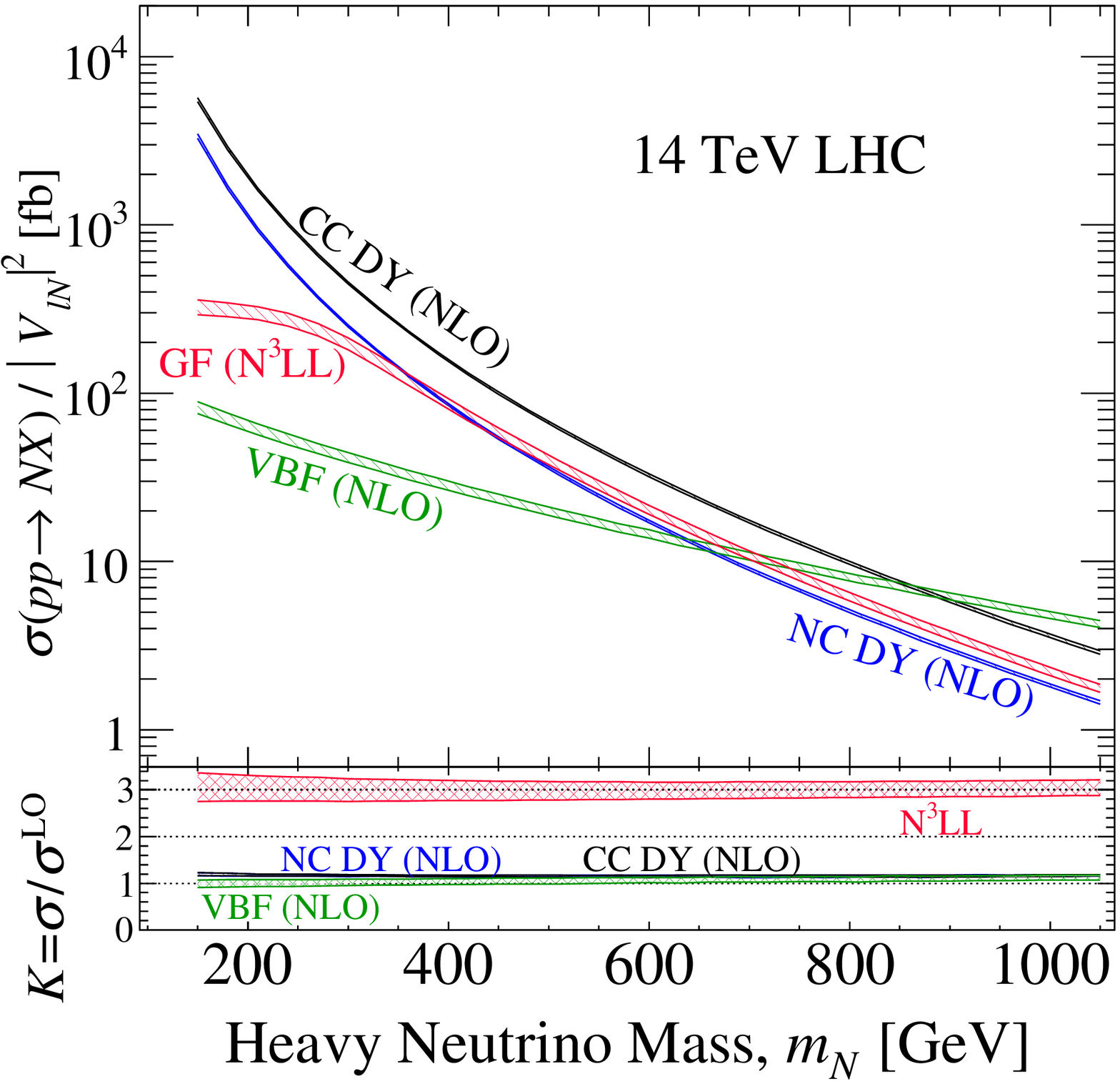}	\label{fig:xsecLHCX14}		}
\subfigure[]{\includegraphics[width=.48\textwidth]{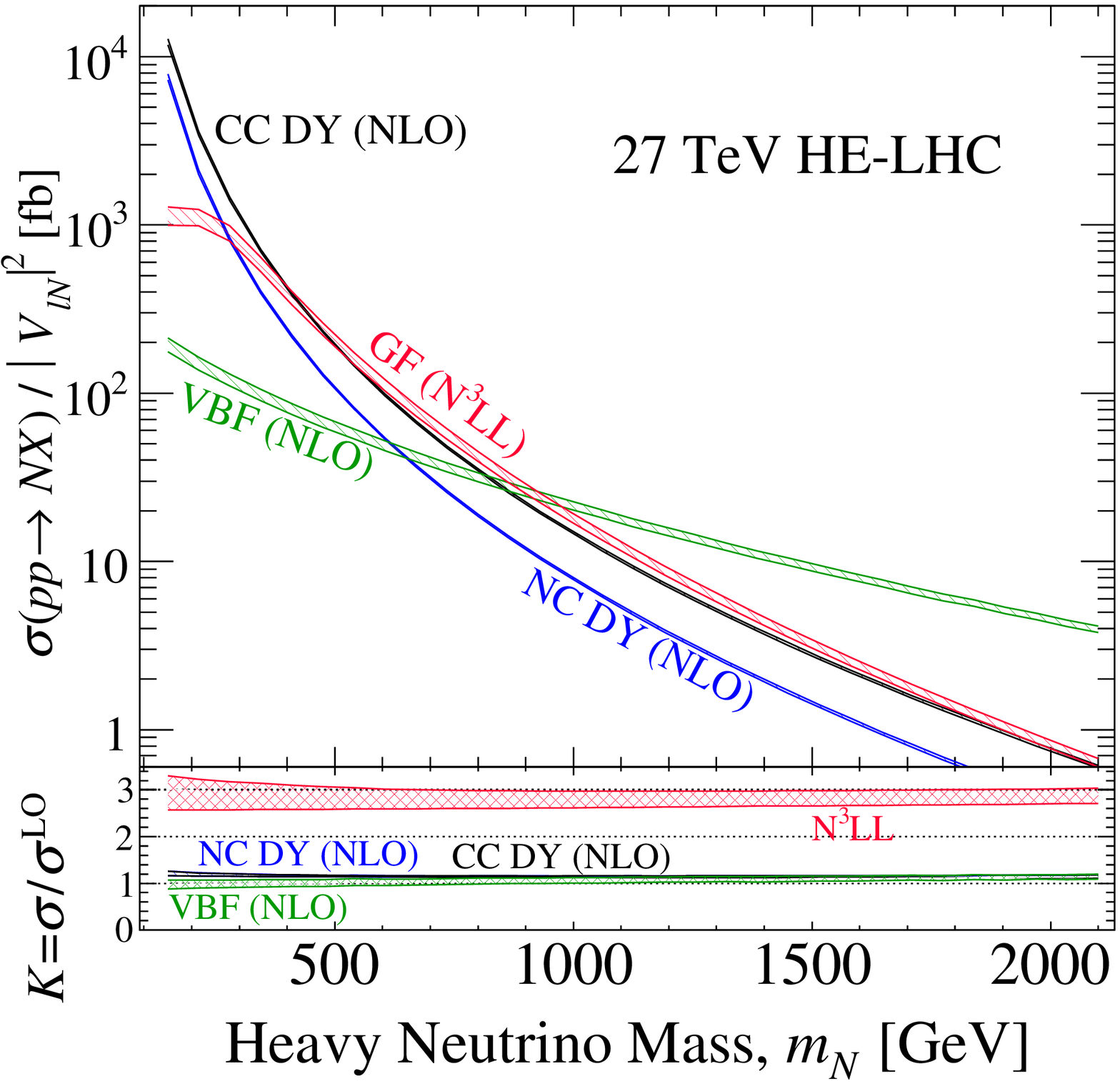}	\label{fig:xsecLHCX27}		}
\\
\subfigure[]{\includegraphics[width=.48\textwidth]{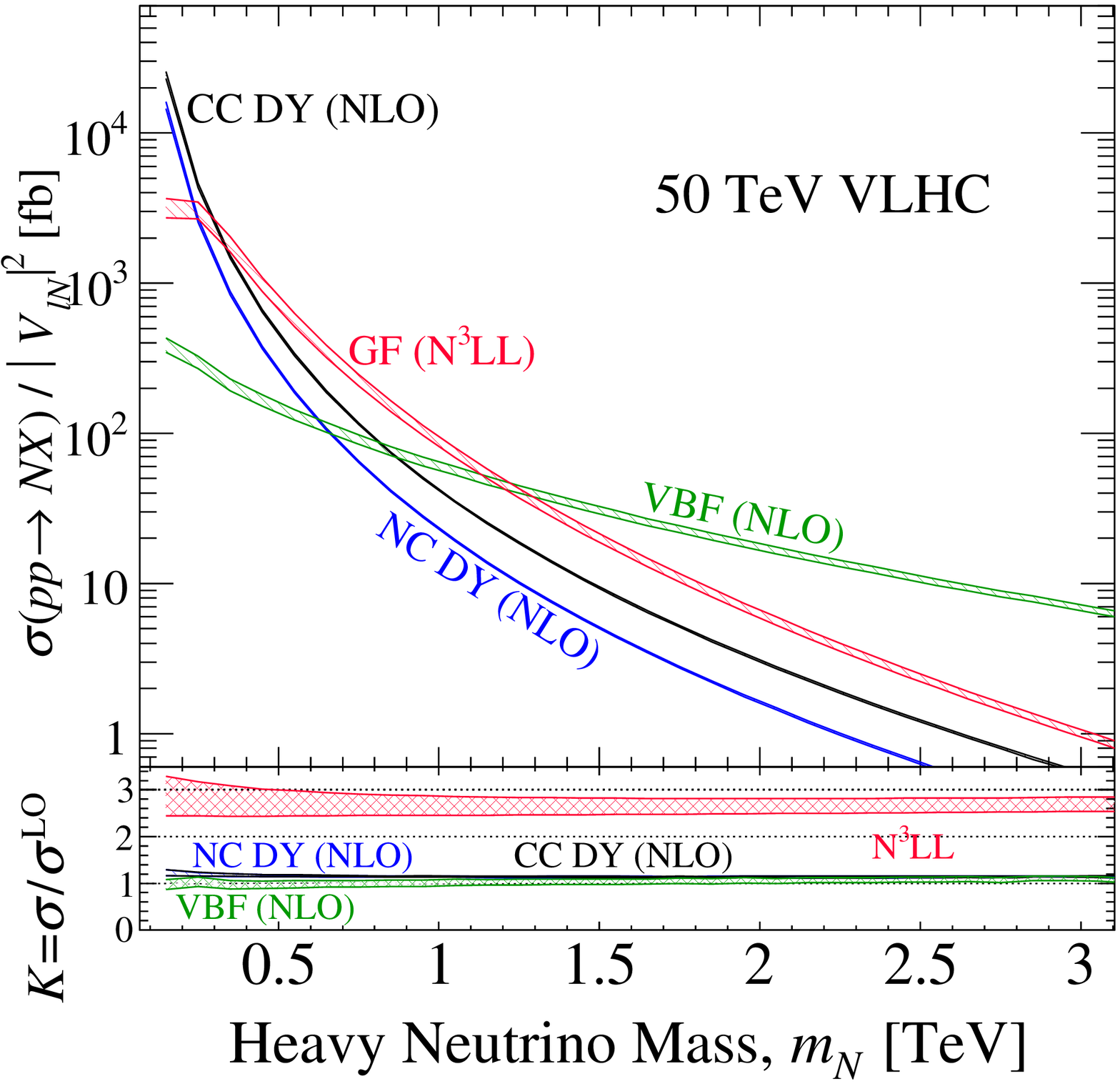}	\label{fig:xsecLHCX50}		}
\subfigure[]{\includegraphics[width=.48\textwidth]{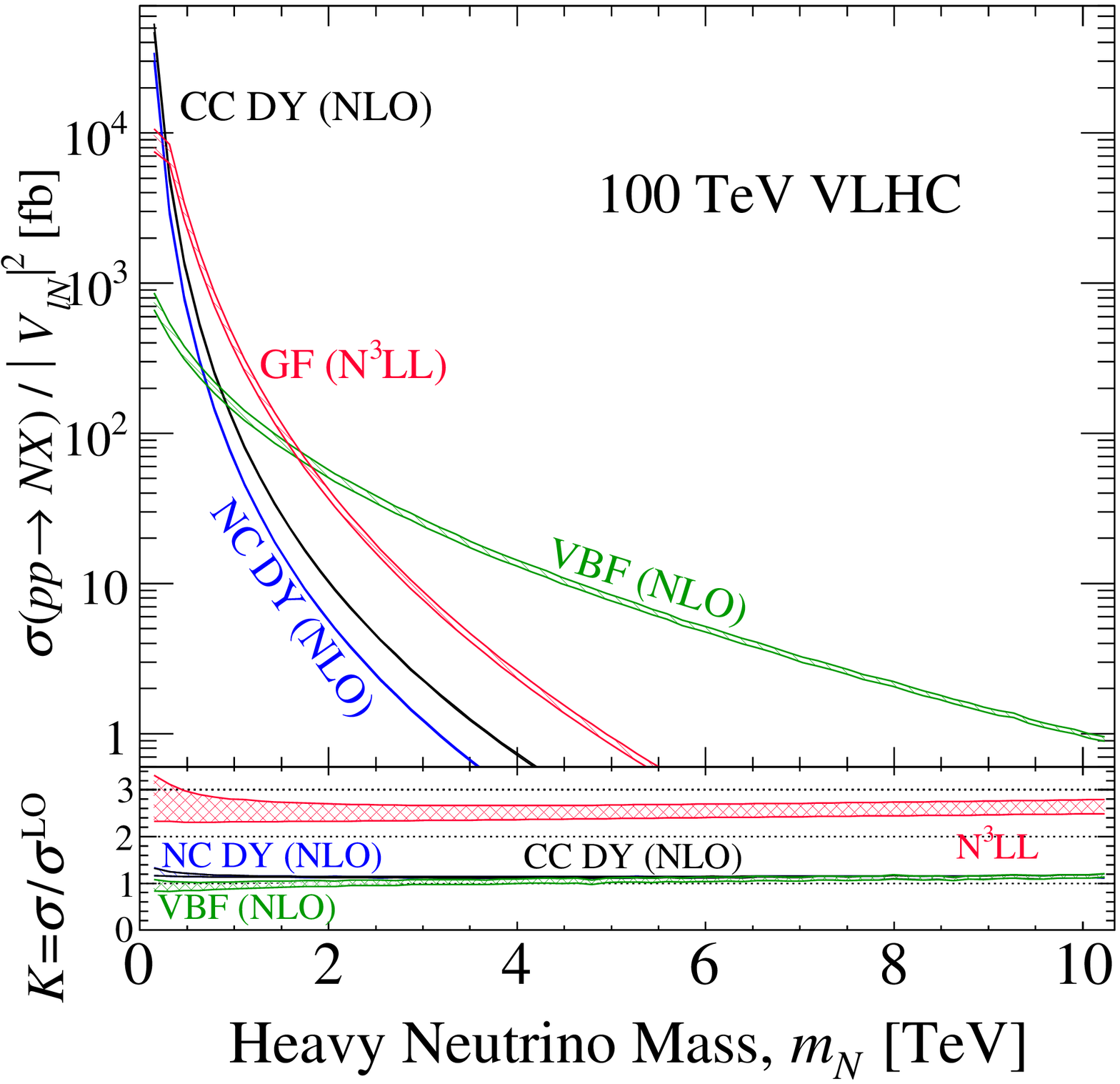}	\label{fig:xsecLHC100}		}
\end{center}
\caption{
Heavy neutrino production cross section, divided by active-sterile mixing $\vert V_{\ell N}\vert^2$, 
via charged (CC) and neutral (NC) current Drell-Yan (DY), $W\gamma$ fusion (VBF), and gluon fusion (GF)
at various accuracies in $pp$ collisions at (a) $\sqrt{s}=14$ TeV, (b) 27 TeV, (c) 50, and (d) 100 TeV.
}
\label{fig:xsecVsMass}
\end{figure*}

To begin quantifying the discovery potential of trilepton searches for heavy neutrinos at $pp$ colliders, 
we compare the leading cross sections for a single resonant $N$ produced through SM currents.
Using Eq.~\ref{eq:factThm} and the methodology outlined in Sec.~\ref{sec:Setup}, we plot in Fig.~\ref{fig:xsecVsMass} 
the bare production cross section $\sigma_0$ (as defined in Eq.~\ref{eq:bareProd}),
for $N$ produced though the CC DY and NC DY channels at NLO in QCD, as shown in Fig.~\ref{fig:feynmanMulti}(a) and given by,
\begin{equation}
 q \overline{q'} \to W^{\pm*} \to N ~\ell^\pm \quad\text{and}\quad q \overline{q} \to Z^* \to N ~\overset{(-)}{\nu_\ell},
\end{equation}
at the Born level. These processes, especially the CC DY, are to serve as baselines for subsequent discussions.
We also plot the N$^3$LL GF mechanism as shown in Fig.~\ref{fig:feynmanMulti}(b),  
\begin{equation}
  g g \to h^* / Z^* \to  N ~\overset{(-)}{\nu_\ell},
\end{equation}
as well as the $W\gamma$ VBF channel at NLO in QCD, shown in Fig.~\ref{fig:feynmanMulti}(c), and given by
\begin{equation}
 q \gamma \to N ~\ell^\pm ~q'.
\end{equation}
We show the bare cross section as a function of heavy neutrino mass $(m_N)$ and at (a) $\sqrt{s} = $ 14, (b) 27, (c) 50, and (d) 100 TeV.
The band of each curve corresponds to the residual perturbative scale dependence.
As a measure of the impact of perturbative corrections on the cross section, in the lower panel of each plot is the QCD N$^j$LO+N$^k$LL $K$-factor,
\begin{equation}
 K^{\rm N^jLO+N^kLL} = \sigma^{\rm N^jLO+N^kLL} / \sigma^{\rm LO},
\end{equation}
where $\sigma^{\rm N^jLO+N^kLL}~(\sigma^{\rm LO})$ is the N$^j$LO+N$^k$LL in QCD ~(LO/Born) cross section.

For the 14 TeV LHC, the nontrivial interplay between matrix elements and PDFs for the above processes has been reported in detail
elsewhere~\cite{Willenbrock:1985tj,Alva:2014gxa,Hessler:2014ssa,Degrande:2016aje,Antusch:2016ejd,Ruiz:2017yyf,Cai:2017mow} and will not be discussed further.
We note simply that for the mass range $m_N = 150-1000\GeV$, the bare cross sections and their relative uncertainties span roughly
\begin{eqnarray}
 {\rm CC~DY~(NLO)}	&:&	\confirm{6\pb-~3\fb 	\quad\pm 2-3\%}, \\
 {\rm NC~DY~(NLO)}	&:&	\confirm{3\pb-1.5\fb	\quad\pm 1-3\%}, \\
 {\rm GF~(N^{3}LL)}		&:&	\confirm{300\fb-2\fb	\quad\pm 3-10\%}, \\
 {\rm VBF~(NLO)}		&:&	\confirm{~80\fb-4\fb	\quad\pm 4-8\%}.
\end{eqnarray}
More interesting is investigating this interplay at potential successors of the LHC,
including the HE-LHC at $\sqrt{s}=27$ TeV, a $\sqrt{s}=50$ TeV variant of the VLHC, and the $\sqrt{s}=100$ TeV VLHC.
At the HE-LHC and for $m_N = 150\GeV-2\TeV$, the rates and uncertainties span
\begin{eqnarray}
 {\rm CC~DY~(NLO)}	&:&	\confirm{10\pb-0.6\fb 	\quad\pm 1-5\%}, \\
 {\rm NC~DY~(NLO)}	&:&	\confirm{~8\pb-0.3\fb		\quad\pm 1-5\%}, \\
 {\rm GF~(N^{3}LL)}		&:&	\confirm{~1\pb-0.7\fb		\quad\pm 3-15\%}, \\
 {\rm VBF~(NLO)}		&:&	\confirm{200\fb-4\fb		\quad\pm 4-10\%}.
\end{eqnarray}
At a 50 TeV VLHC and for $m_N = 150\GeV-3\TeV$, one sees 
\begin{eqnarray}
 {\rm CC~DY~(NLO)}	&:&	\confirm{25\pb-0.5\fb 	\quad\pm 1-6.5\%}, \\
 {\rm NC~DY~(NLO)}	&:&	\confirm{15\pb-0.3\fb		\quad\pm 1-7\%}, \\
 {\rm GF~(N^{3}LL)}		&:&	\confirm{~3\pb-~1\fb~		\quad\pm 3-15\%}, \\
 {\rm VBF~(NLO)}		&:&	\confirm{400\fb-~6\fb		\quad\pm 4-10\%},
\end{eqnarray}
and at 100 TeV with $m_N = 150\GeV-5\TeV$ the rates and uncertainties are
\begin{eqnarray}
 {\rm CC~DY~(NLO)}	&:&	\confirm{50\pb-0.2\fb 	\quad\pm 1-8\%}, \\
 {\rm NC~DY~(NLO)}	&:&	\confirm{30\pb-0.1\fb		\quad\pm 1-9\%}, \\
 {\rm GF~(N^{3}LL)}		&:&	\confirm{10\pb-1\fb~		\quad\pm 4-20\%}.
\end{eqnarray}
Exceptionally, at 100 TeV with $m_N = 150\GeV-10\TeV$, the VBF rate spans about
\begin{eqnarray}
 {\rm VBF~(NLO)}		&:&	\confirm{750\fb-1\fb	\quad\pm 3-10\%}.
\end{eqnarray}
A detailed summary of heavy $N$ production rates, their residual scale dependencies, and their QCD $K$-factors,
for representative $m_N$ and $\sqrt{s}$ is given in Tab.~\ref{tb:xSec}.
In all cases, the largest uncertainties are associated with the lowest mass points and quickly moderate to smaller levels.
This falloff is particularly notable for the DY channels.
The sensitivity at the lowest $(m_N/\sqrt{s})$ is attributed to the opening of $qg$- and $\gamma g$-initiated processes, and hence sensitivity to the gluon PDF, 
which slopes most significantly at low Bjorken-$x$.
Moreover, at LO, the $W\gamma$ fusion process only possesses $\mu_f$ dependence (through the PDF);  $\mu_r$ dependence arises through $\alpha_s$ first at NLO.
The GF channel, despite its high accuracy, possesses a large factorization scale uncertainty due to missing real radiation terms 
that would otherwise stabilize the prediction~\cite{Bonvini:2014qga,Ruiz:2017yyf}.

\begin{table}[!t]
\begin{center}
\resizebox{\columnwidth}{!}{
\begin{tabular}{ c || c | c | c | c | c | c }
\hline
\hline
$m_N$ 	& \multicolumn{2}{c|}{150 GeV} 	
		& \multicolumn{2}{c|}{450 GeV}	
		& \multicolumn{2}{c }{600 GeV} \tabularnewline\hline
Process	& $\sigma/\vert V_{\ell N}\vert^2$ [fb]	& $K$ 	& $\sigma/\vert V_{\ell N}\vert^2$ [fb]	& $K$	& $\sigma/\vert V_{\ell N}\vert^2$ [fb]	& $K$ \tabularnewline\hline		
	 \multicolumn{7}{c}{$\sqrt{s}=14$ TeV} \tabularnewline\hline		
CC DY (NLO) 	&	$5.57\times10^{3}~^{+3\%}_{-3\%}$	& 1.20	& $1.00\times10^{2}~^{+2\%}_{-1\%}$	& 1.15	& $3.25\times10^{1}~^{+2\%}_{-2\%}$	& 1.15	\tabularnewline
NC DY (NLO) 	&	$3.40\times10^{3}~^{+2\%}_{-3\%}$	& 1.20	& $5.41\times10^{1}~^{+2\%}_{-1\%}$	& 1.15	& $1.72\times10^{1}~^{+2\%}_{-2\%}$	& 1.15	\tabularnewline
VBF (NLO)   	&	$8.51\times10^{1}~^{+8\%}_{-8\%}$	& 1.03	& $2.35\times10^{1}~^{+6\%}_{-6\%}$	& 1.04	& $1.44\times10^{1}~^{+6\%}_{-5\%}$	& 1.06	\tabularnewline
GF $(\nkll{3})$	
	    &   $3.36\times10^{2}~^{+7\%}_{-13\%}$   & 3.15 & $6.01\times10^{1}~^{+4\%}_{-10\%}$   & 3.07 & $2.09\times10^{1}~^{+3\%}_{ -9\%}$   & 3.07 \tabularnewline\hline
\multicolumn{7}{c}{$\sqrt{s}=27$ TeV} \tabularnewline\hline
CC DY (NLO) 	&	$1.23\times10^{4}~^{+3\%}_{-5\%}$		& 1.23	& $2.81\times10^{2}~^{+1\%}_{-1\%}$	& 1.16	& $1.00\times10^{2}~^{+1\%}_{-1\%}$	& 1.15	\tabularnewline 
NC DY (NLO) 	&	$7.64\times10^{3}~^{+3\%}_{-5\%}$		& 1.22	& $1.56\times10^{2}~^{+2\%}_{-1\%}$	& 1.16	& $5.54\times10^{1}~^{+1\%}_{-1\%}$	& 1.15	\tabularnewline 
VBF (NLO)   	&	$1.95\times10^{2}~^{+10\%}_{-9\%}$	& 0.98	& $7.20\times10^{1}~^{+7\%}_{-7\%}$	& 1.00	& $4.91\times10^{1}~^{+7\%}_{-6\%}$	& 1.02	\tabularnewline 
GF $(\nkll{3})$	
	    & $1.165\times10^{3}~^{+10\%}_{-15\%}$ & 3.01 & $2.892\times10^{2}~^{+ 6\%}_{-11\%}$ & 2.90 & $1.179\times10^{2}~^{+ 5\%}_{-10\%}$ & 2.88 \tabularnewline\hline
\multicolumn{7}{c}{$\sqrt{s}=50$ TeV} \tabularnewline\hline
CC DY (NLO) 	&	$2.46\times10^{4}~^{+4\%}_{-6\%}$		& 1.24	& $6.55\times10^{2}~^{+2\%}_{-2\%}$	& 1.17	& $2.48\times10^{2}~^{+1\%}_{-2\%}$	& 1.16	\tabularnewline
NC DY (NLO) 	&	$1.55\times10^{4}~^{+4\%}_{-7\%}$		& 1.25	& $3.74\times10^{2}~^{+2\%}_{-2\%}$	& 1.17	& $1.40\times10^{2}~^{+2\%}_{-2\%}$	& 1.16	\tabularnewline
VBF (NLO)   	&	$3.90\times10^{2}~^{+11\%}_{-11\%}$	& 0.97	& $1.66\times10^{2}~^{+8\%}_{-8\%}$	& 0.97	& $1.21\times10^{2}~^{+8\%}_{-8\%}$	& 0.98	\tabularnewline
GF $(\nkll{3})$	
	    &   $3.26\times10^{3}~^{+13\%}_{-16\%}$ & 2.93 & $1.01\times10^{3}~^{+ 8\%}_{-13\%}$ & 2.79 & $4.54\times10^{2}~^{+ 7\%}_{-11\%}$ & 2.75  \tabularnewline\hline
\multicolumn{7}{c}{$\sqrt{s}=100$ TeV} \tabularnewline\hline
CC DY (NLO) 	&	$5.17\times10^{4}~^{+5\%}_{-9\%}$		& 1.28	& $1.57\times10^{3}~^{+2\%}_{-4\%}$	& 1.19	& $6.24\times10^{2}~^{+2\%}_{-3\%}$	& 1.17	\tabularnewline
NC DY (NLO) 	&	$3.28\times10^{4}~^{+5\%}_{-9\%}$		& 1.28	& $9.16\times10^{2}~^{+2\%}_{-4\%}$	& 1.19	& $3.61\times10^{2}~^{+2\%}_{-3\%}$	& 1.17	\tabularnewline
VBF (NLO)   	&	$7.59\times10^{2}~^{+13\%}_{-12\%}$	& 0.95	& $3.60\times10^{2}~^{+10\%}_{-10\%}$	& 0.93	& $2.76\times10^{2}~^{+9\%}_{-9\%}$	& 0.94	\tabularnewline
GF $(\nkll{3})$		
	    & $9.22\times10^{3}~^{+15\%}_{ -19\%}$ & 2.87  & $3.42\times10^{3}~^{+11\%}_{ -14\%}$ & 2.70  & $1.68\times10^{3}~^{+10\%}_{ -13\%}$ & 2.66 \tabularnewline\hline
	    \hline
\end{tabular}
} 
\caption{
Leading heavy $N$ production cross sections [fb] at various accuracies,
divided by mixing $\vert V_{\ell N}\vert^2$, 
with residual scale dependence $(\%)$ and QCD $K$-factor, for representative $m_N$ and $\sqrt{s}$.
}
\label{tb:xSec}
\end{center}
\end{table}

Qualitatively, one sees a number of important changes in the relative importance of heavy neutrino production mechanisms
 when going from $\sqrt{s}=14$ TeV to 27, 50, or 100 TeV.
While in all cases the VBF process overtakes the CC DY process at a largely static $m_N\sim900\GeV$, 
the dominance of GF varies due to the surge in the $gg$ luminosity for diminishing $(m_N / \sqrt{s})$.
The constant crossover for the CC DY and VBF channel is likely tied to the fact that both are driven by valence quark-sea parton scattering,
and therefore receive the same growth in parton luminosity as collider energy increases.
At 14 TeV, GF is sub-leading for all $m_N$ under consideration but comparable to the NC DY process.
Technically, the NC DY and GF should be summed coherently
 as the GF  channel is formally a separately gauge invariant $\mathcal{O}(\alpha_s^2)$ correction to the NC DY mode~\cite{Willenbrock:1985tj}.
(Similarly, the $W\gamma$ channel is a separately gauge invariant EW correction to the inclusive CC DY process~\cite{Alva:2014gxa}.)
 Doing so renders it much closer to the (CC DY+VBF) process at 14 TeV~\cite{Ruiz:2017yyf}.
At 27~(50) TeV, GF emerges as the leading channel for \confirm{$m_N\approx450-900~(300-1200)\GeV$}, beyond which VBF is dominant.
For 100 TeV, the situation is more exaggerated with GF dominating for \confirm{$m_N\approx250-1750\GeV$},
which alone encompasses the entire high-mass range under consideration for $\sqrt{s}=14-27\TeV$.
Despite the prominence of the inclusive (NC DY+GF) $pp\to N\nu X$ process, few (if any) dedicated collider studies exits for $m_N > m_h$.
However, this regretful absence is partly due to the process' large SM backgrounds.

The importance of the VBF channel cannot be overemphasized: 
across the four colliders, for $m_N=1~(2)~[3]~\{10\}$ TeV, the bare cross sections reach about \confirm{$\sigma_0\sim 1-6\fb$}.
In light of the proposed $\mathcal{O}(3-30)\invab$ datasets being earnestly discussed by the community~\cite{Arkani-Hamed:2015vfh,Mangano:2016jyj,Golling:2016gvc,CidVidal:2018eel},
one expects that active-sterile neutrino mixing on the order of $\vert V_{\ell N}\vert^2 \sim 10^{-4}-10^{-2}$ for this mass range can be probed by the VBF channel alone.
In the context of Type I-like Seesaw models, these rates also suggest the first direct sensitivity to $\mathcal{O}(10)$ TeV heavy neutrino masses as colliders,
and compels us to focus on the discovery prospect of the inclusive (CC DY+VBF) $pp\to N\ell X$ signature for the remainder of this study.

\subsection{Parton-Level Kinematics and Observables at LO}\label{sec:ColliderPartonKin}

In this section, we present at $\sqrt{s} = 14,~27,$ and 100 TeV, the parton-level kinematics of the
trilepton process for a single heavy $N$ via the CC DY mechanism, at LO in QCD,
\begin{equation}
 p ~p ~\to~  ~\ell_N ~N^{(*)} ~\to~ ~\ell_N ~\ell_W~ ~W ~\to~ ~\ell_N ~\ell_W ~\ell_\nu ~\nu.
 \label{eq:partonDY}
\end{equation}
The purpose of this section is to establish baseline kinematical features inherent to the $q\overline{q'}\to N\ell \to 3\ell\nu$ hard process 
and explore the dependence, or lack thereof, on increasing collider energies.
Observations here ultimately form the basis for several conclusions we make at the particle-level in Sec.~\ref{sec:ColliderParticleKin},
 as well as for justifications of analysis selection criteria.

As we are working momentarily at the partonic level, 
we have access to the standard HEP MC particle identification codes (PID) of final-state particles and their resonant parent particle in the outputted \mgamc~LHE files.
Therefore, for each charged lepton in the following figures, the subscript label $X\in\{N,W,\nu\}$ denotes the particle with which the charged lepton was produced:
$\ell_N$ (solid) denotes the primary charged lepton produced in association with $N$;
$\ell_W$ (dash) represents the charged lepton from the $N$ decay and produced with an on-shell $W$ boson; and
$\ell_\nu$ (dot) is the charged lepton from the leptonic $W$ decay.
Notably, the recording of the PID and 4-momentum is true even for the intermediate $N$, which is not modeled in this section with the NWA (see Sec.~\ref{sec:mcSigGen}).
In MC event generation with \mgamc, intermediate particles that go on-shell are recorded in the event file independent of the NWA being applied.
As the widths of heavy neutrinos used in MC generation are engineered to be very small due to small active-sterile mixing (see Sec.~\ref{sec:mcHeavyNinputs}), 
in accordance with realistic neutrino mass models,
we find nearly all \confirm{$(\gtrsim98\%)$} events generated contain an approximately on-shell $N$ with its momentum recorded to file.

Briefly, we note that we omit kinematics associated with the $W\gamma$ fusion channel as they have been reported systematically for various $\sqrt{s}$ elsewhere~\cite{Alva:2014gxa}.
In addition, the benchmark (pseudo)rapidities\footnote[2]{For massive 4-momentum $p^\mu$, rapidity is defined as $y = 0.5\log[(p^0+p^3)/(p^0-p^3)]$. 
In the massless limit, $y$ simplifies to pseudorapidity, $y \overset{p^2 \to 0}{=} \eta \equiv -\log[\tan(\theta/2)]$,
where $\theta$ is the polar angle with respect to the beam/$\hat{z}$-axis.} 
for nominal tracker coverage of charged leptons in future collider studies are set to $y,~\eta = \pm2.5$, 
following the Snowmass 2013 recommendations~\cite{Avetisyan:2013onh}.
Hence, throughout this section, guide lines at these values  are inserted in all rapidity plots.

\begin{figure*}[!t]
\begin{center}
\subfigure[]{\includegraphics[width=.48\textwidth]{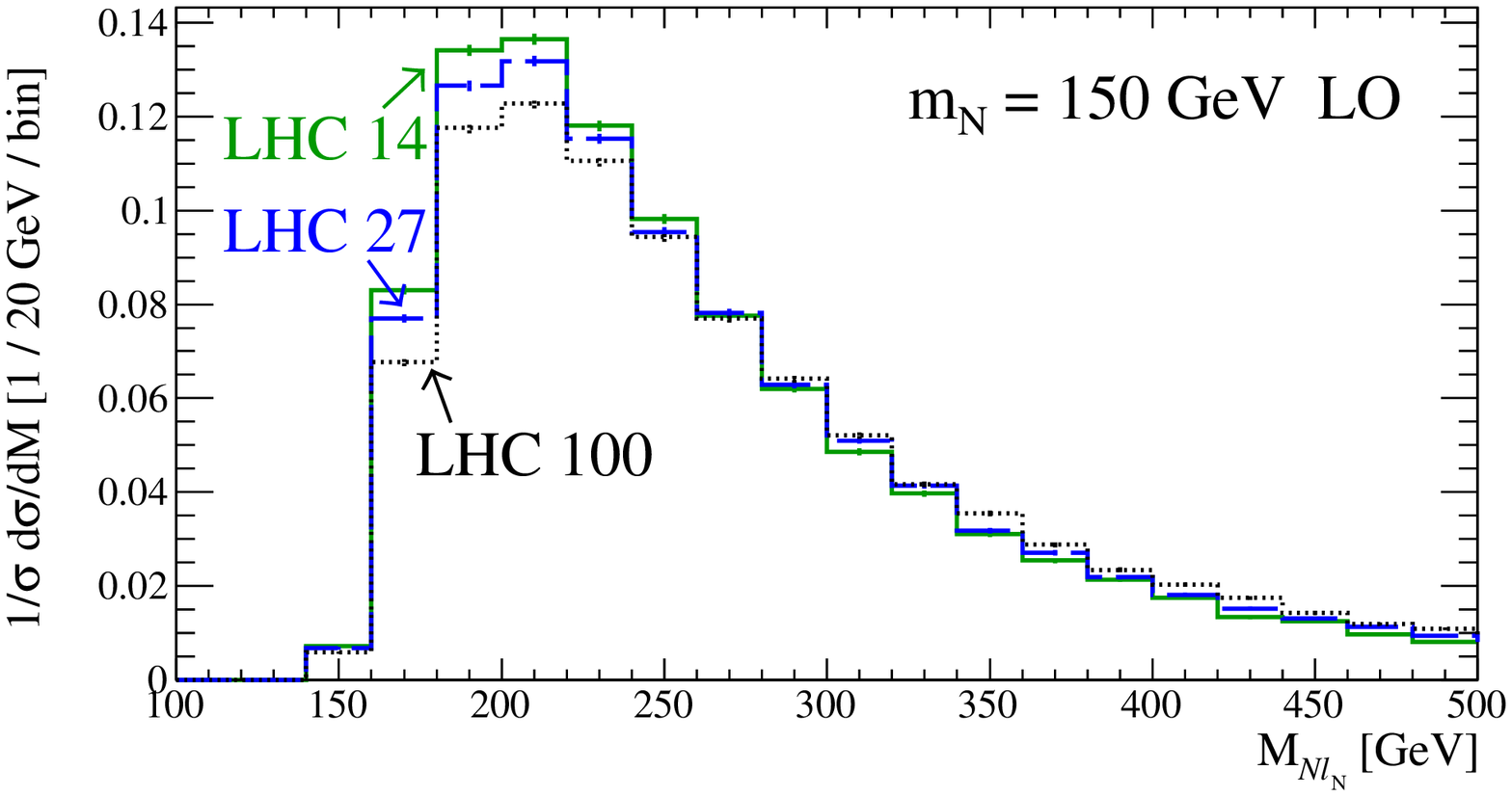}	\label{fig:partonHardQ_mN150_MNlN} }
\subfigure[]{\includegraphics[width=.48\textwidth]{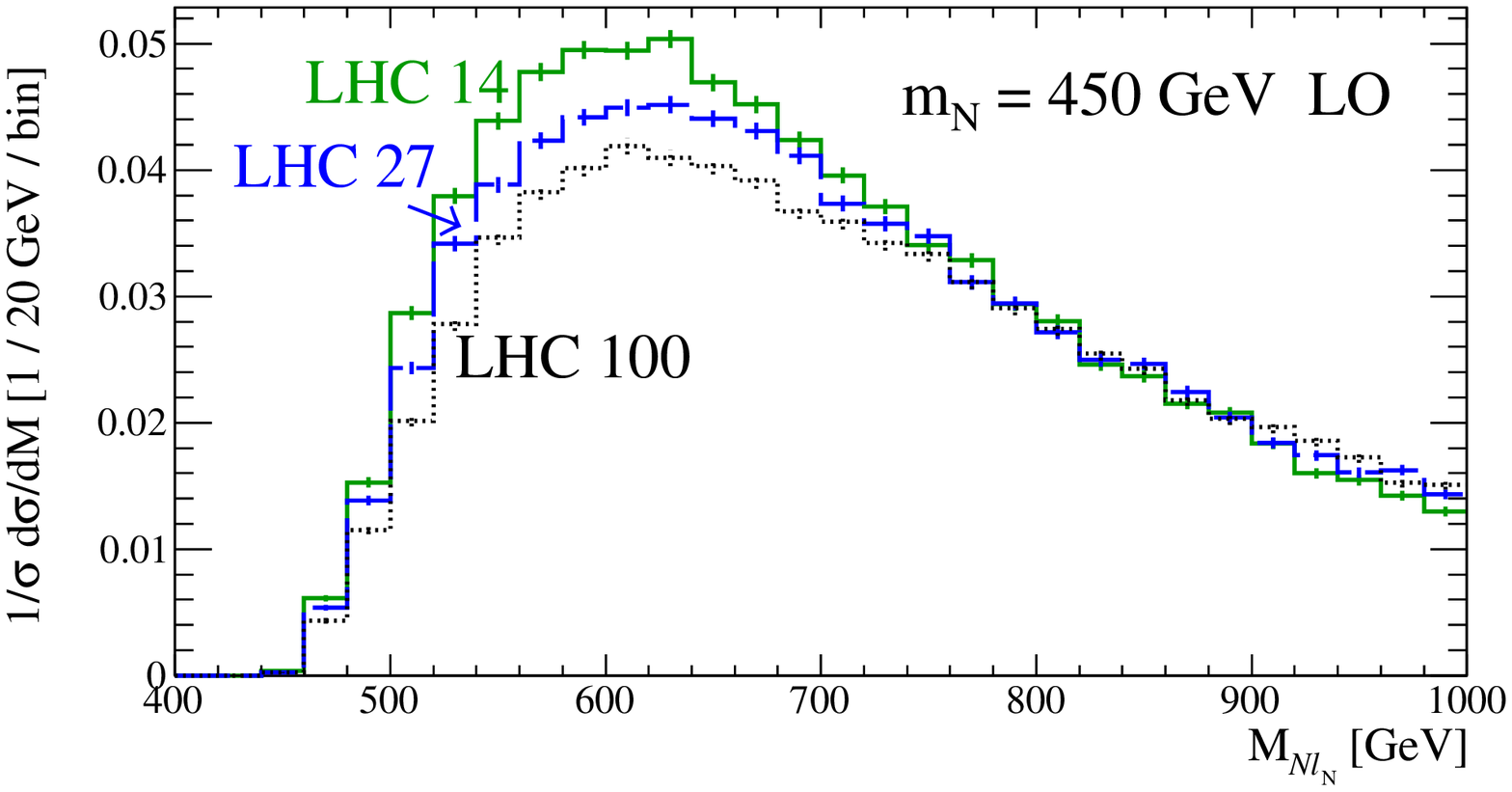}	\label{fig:partonHardQ_mN450_MNlN} }
\end{center}
\caption{Normalized parton-level distribution for the invariant mass of the $(N\ell_N)$ system in the LO DY process $pp\to N l_N \to W l_W l_N \to \nu l_\nu l_W l_N$,
at $\sqrt{s}=14$, 27, and 100 TeV, assuming representative neutrino masses (a) $m_N=$ 150 GeV and (b) 450 GeV.
}
\label{fig:partonHardQ}
\end{figure*}

We begin with a look at the kinematics associated with the hard $(N\ell)$-system itself.
In Fig.~\ref{fig:partonHardQ} we plot the normalized invariant mass distribution of the $(N\ell)$-system for (a) $m_N=150$ and (b) 450 GeV, at $\sqrt{s} = 14$ (solid), 27 (dash), and 100 (dot) TeV.
Immediately, one observes that for a fixed $m_N$ the shape of the invariant  curves are largely the same despite the large differences in collider energies; 
only minor broadening is observed for increasing $\sqrt{s}$.
The reason is that up to kinematic factors related to phase space and angular momentum (e.g., $s$- vs $p$-wave),
the dominant contribution of any matrix element is in the neighborhood of its poles, i.e., when particles are resonant and go onto their mass shell, and is (mostly) independent of collider configuration. 
This is seen explicitly in the parton-level cross section for the $q\overline{q'}\to N\ell$ process, where up to constant factors is
\begin{eqnarray}
\hat{\sigma}(q\overline{q'}\to W^{*}\to N\ell_N) 
&\propto& \cfrac{(Q^2 - m_N^2)^2(2Q^2 + m_N^2)}{Q^4\left[(Q^2-M_W^2)^2 + (M_W\Gamma_W)^2\right]} \\
&\overset{m_N \gg M_W}{\approx}& 
 \cfrac{(Q^2 - m_N^2)^2(2Q^2 + m_N^2)}{Q^8}.
\end{eqnarray}
Here, $Q\gtrsim m_N$ denotes the scale of the hard process, which in this case is equal to the invariant mass of the $(N\ell)$-system.
In the $Q,~ m_N\gg M_W$ limit, one sees clearly the kinematic threshold at $Q = m_N$, below which the process is kinematically forbidden (by momentum conservation).
The kinematical factor of $(2Q^2 + m_N^2)$ acts as a brief ``turn-on'' function before the $\sim1/Q^8$ suppression sinks the rate for larger invariant masses.
Numerically,  the maximum occurs as $Q\sim (4/3)\times m_N$, and indicates that the hard process scale $Q$ is of the same order as the threshold scale $m_N$, independent of collider energy.
Now, as the $2\to2$ process $pp\to N\ell_N$ occurs at a characteristic invariant mass of $Q\sim m_N$, the 4-momenta of $N$ and $\ell_N$ should also carry characteristic values of this order.

\begin{figure*}[!t]
\begin{center}
\subfigure[]{\includegraphics[width=.48\textwidth]{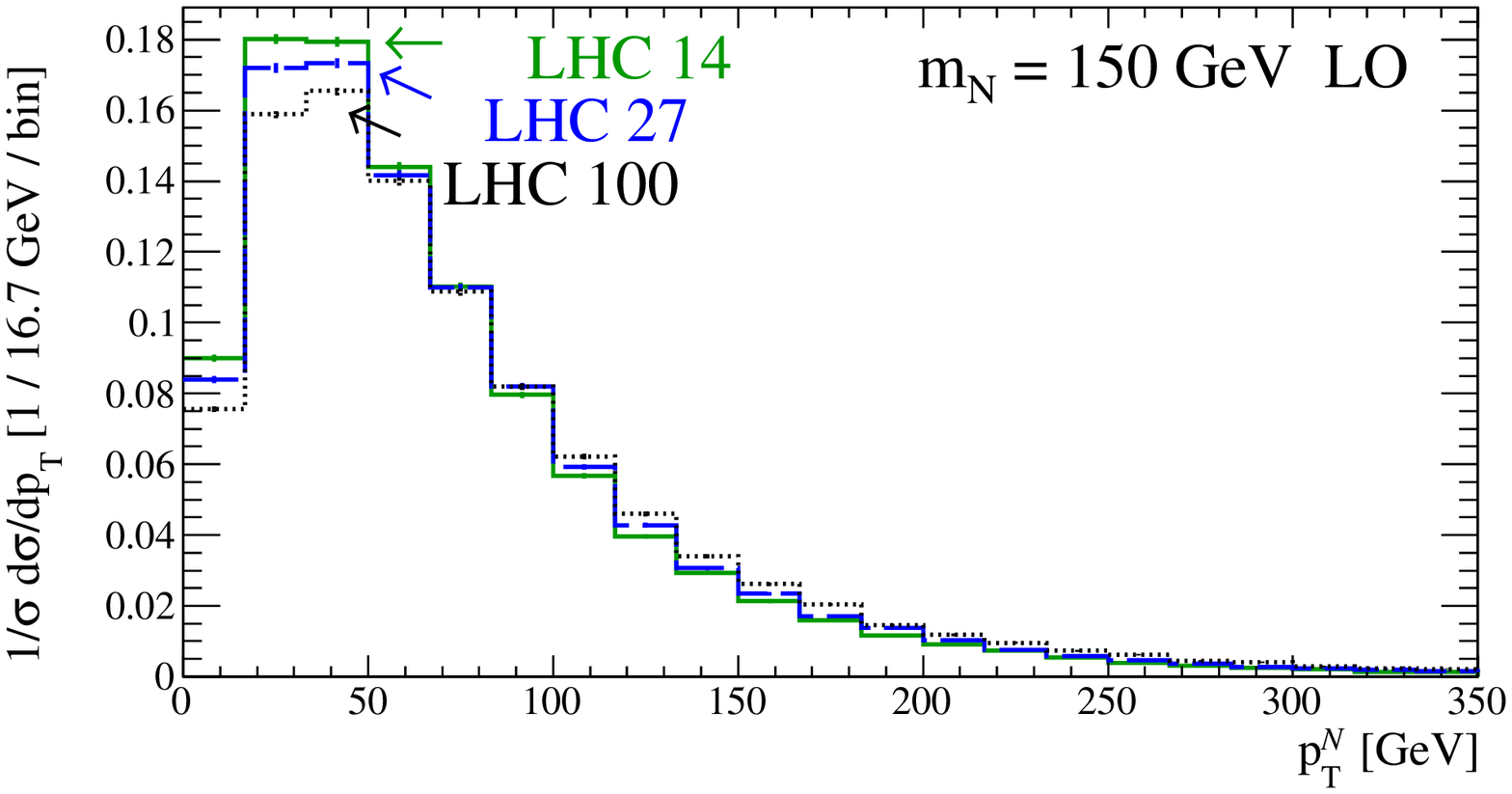}	\label{fig:partonKinNu_mN150_pTN} }
\subfigure[]{\includegraphics[width=.48\textwidth]{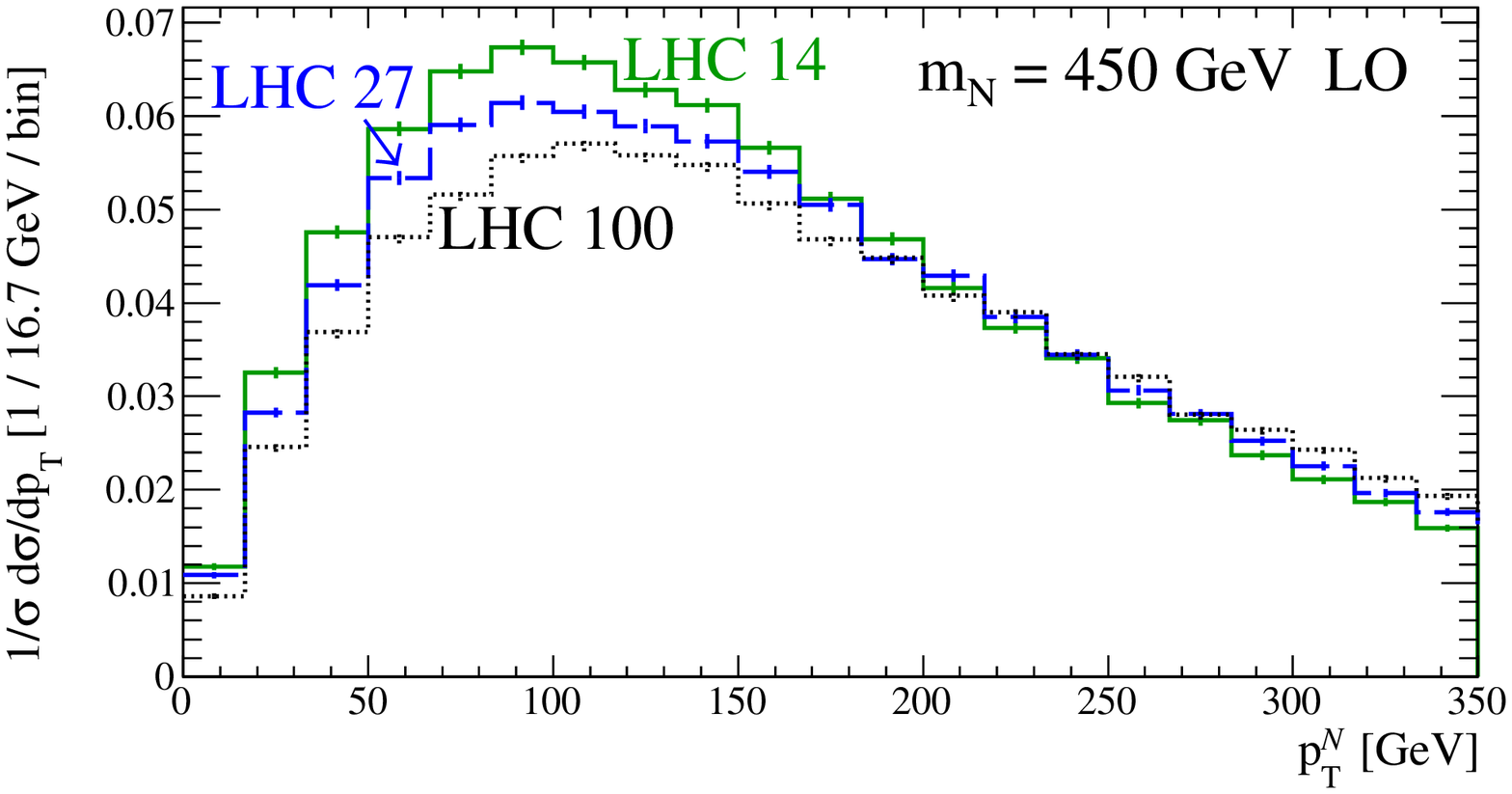}	\label{fig:partonKinNu_mN450_pTN} }
\\
\subfigure[]{\includegraphics[width=.48\textwidth]{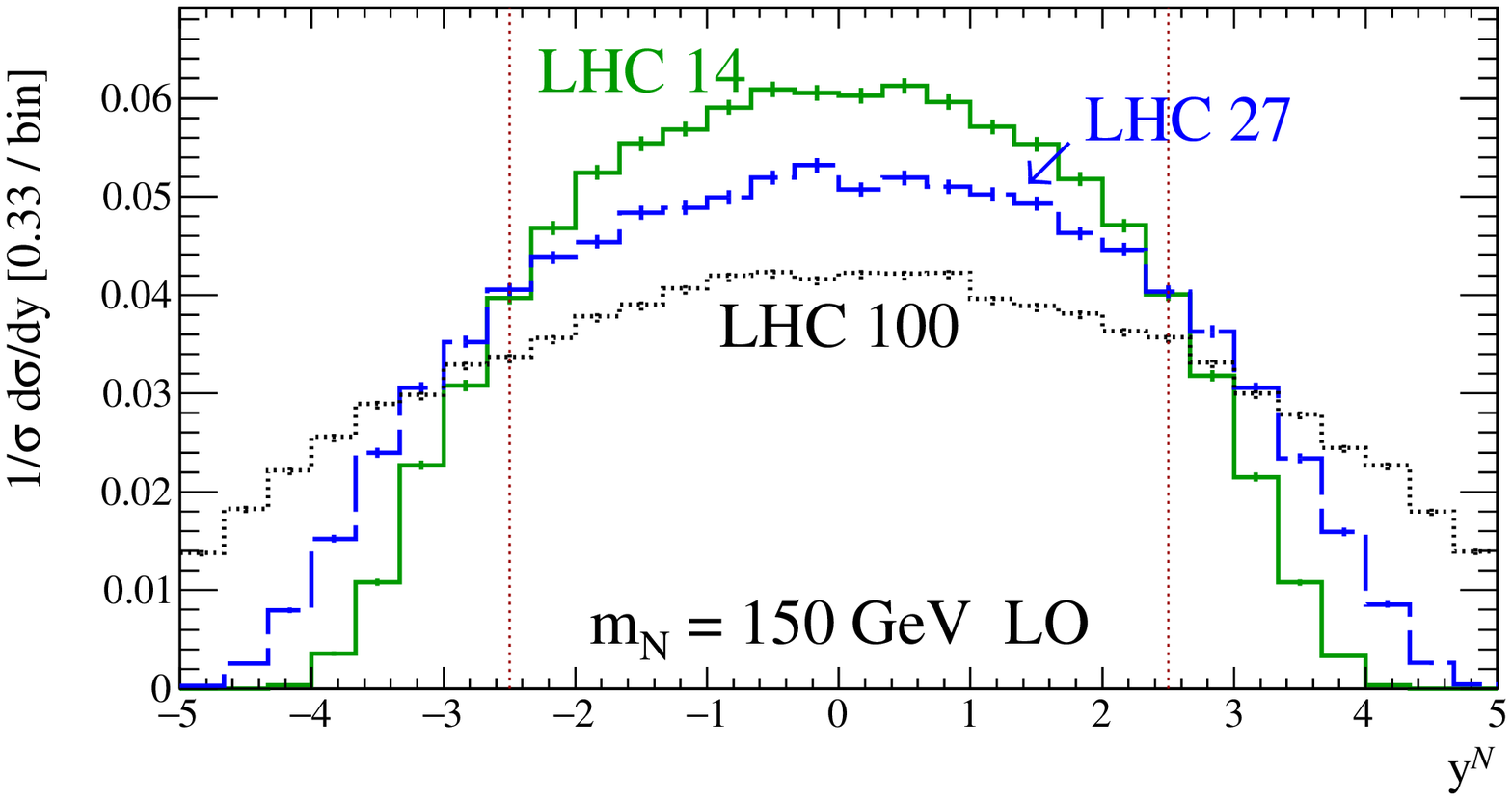}		\label{fig:partonKinNu_mN150_yN} }
\subfigure[]{\includegraphics[width=.48\textwidth]{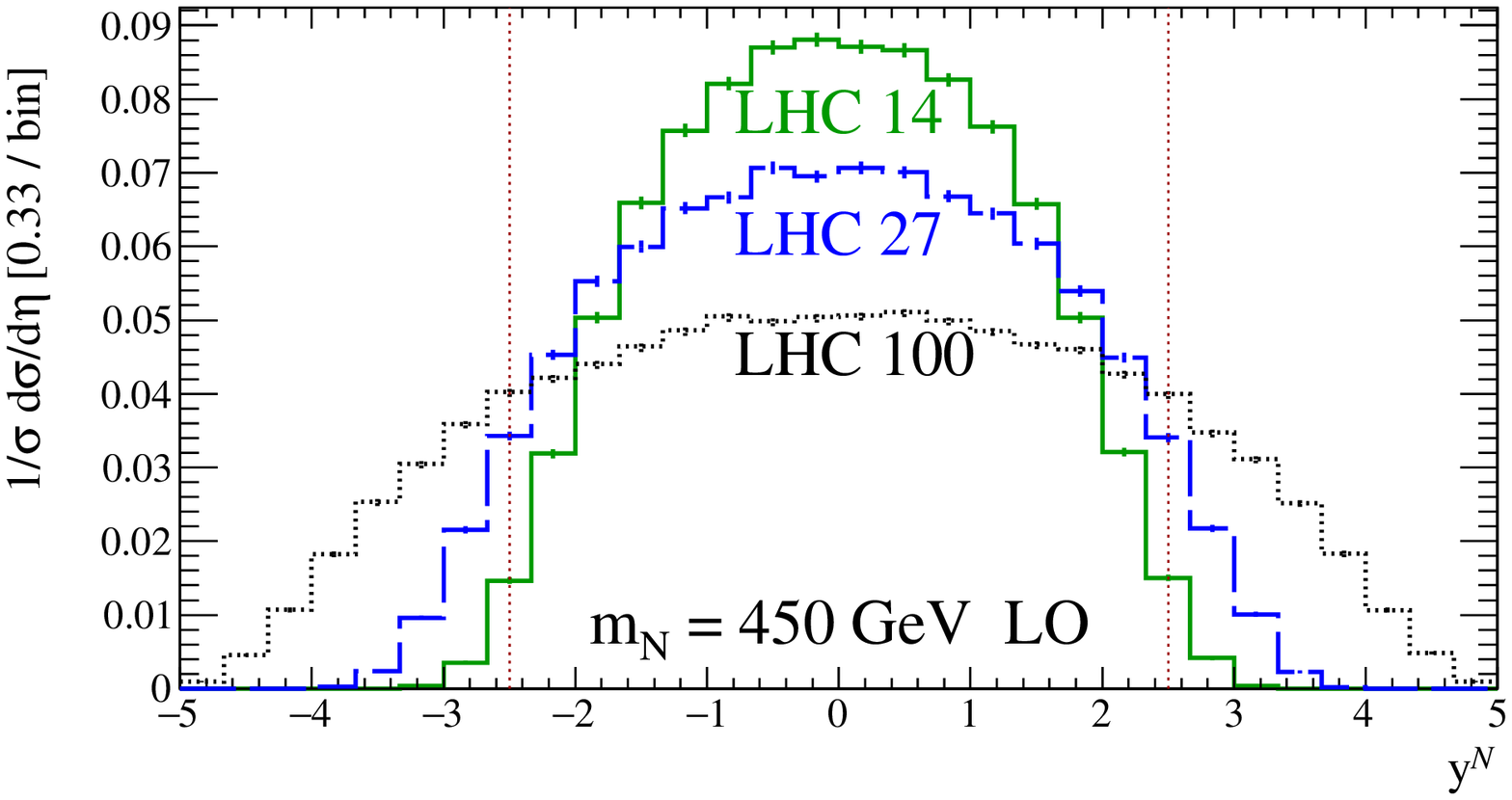}		\label{fig:partonKinNu_mN450_yN} }
\end{center}
\caption{
Normalized parton-level (a,b) transverse momentum $(p_T^N)$ and (c,d) rapidity $(y^N)$ distributions for $N$ produced in the LO DY process $pp\to N l_N \to W l_W l_N \to \nu l_\nu l_W l_N$,
at $\sqrt{s}=14$, 27, and 100 TeV, assuming representative neutrino mass (a,c) $m_N= 150$ and (b,d) 450 GeV.
}
\label{fig:partonKinNu}
\end{figure*}

\begin{figure*}[!t]
\begin{center}
\subfigure[]{\includegraphics[width=.45\textwidth]{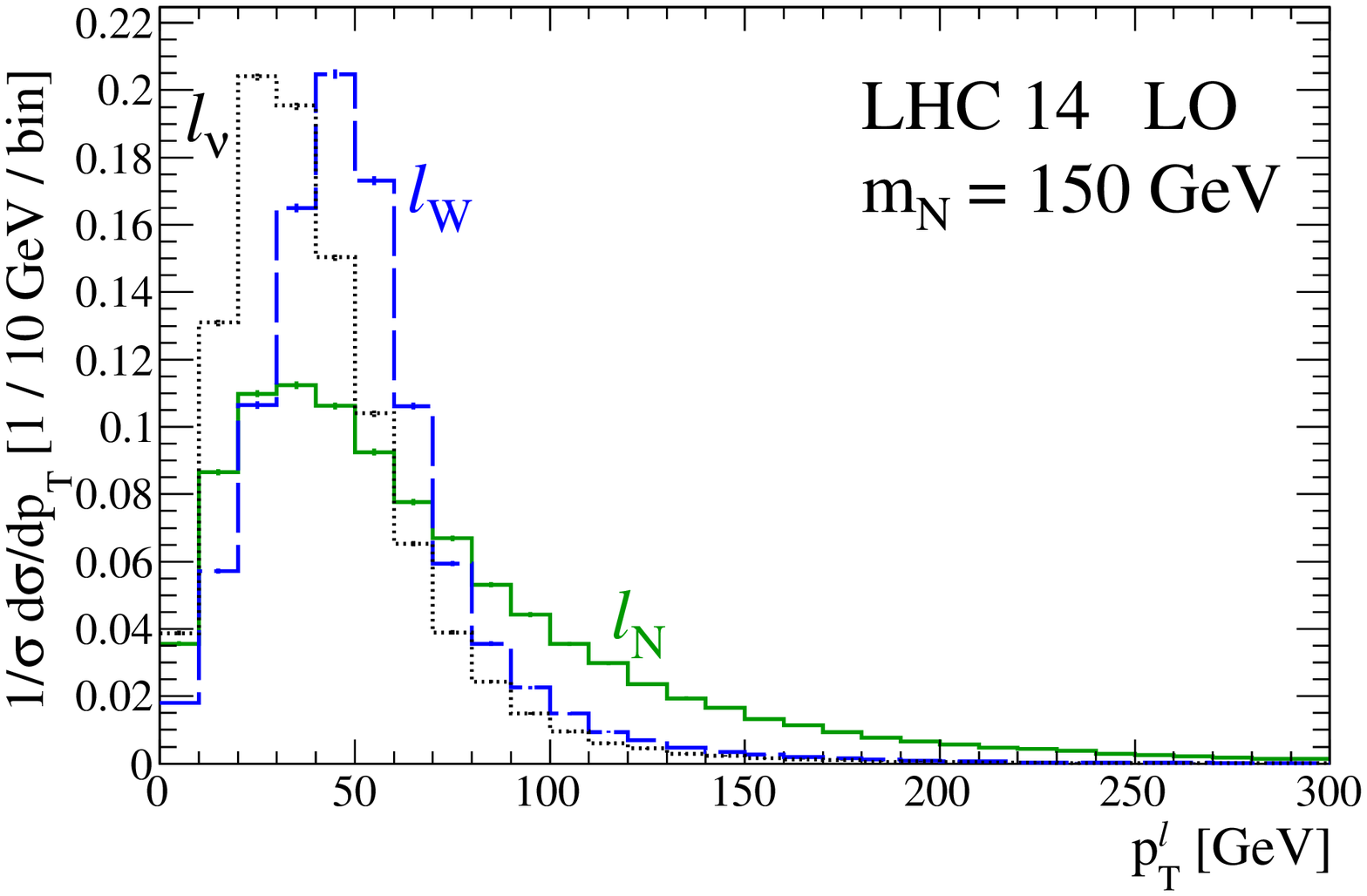}	}
\subfigure[]{\includegraphics[width=.45\textwidth]{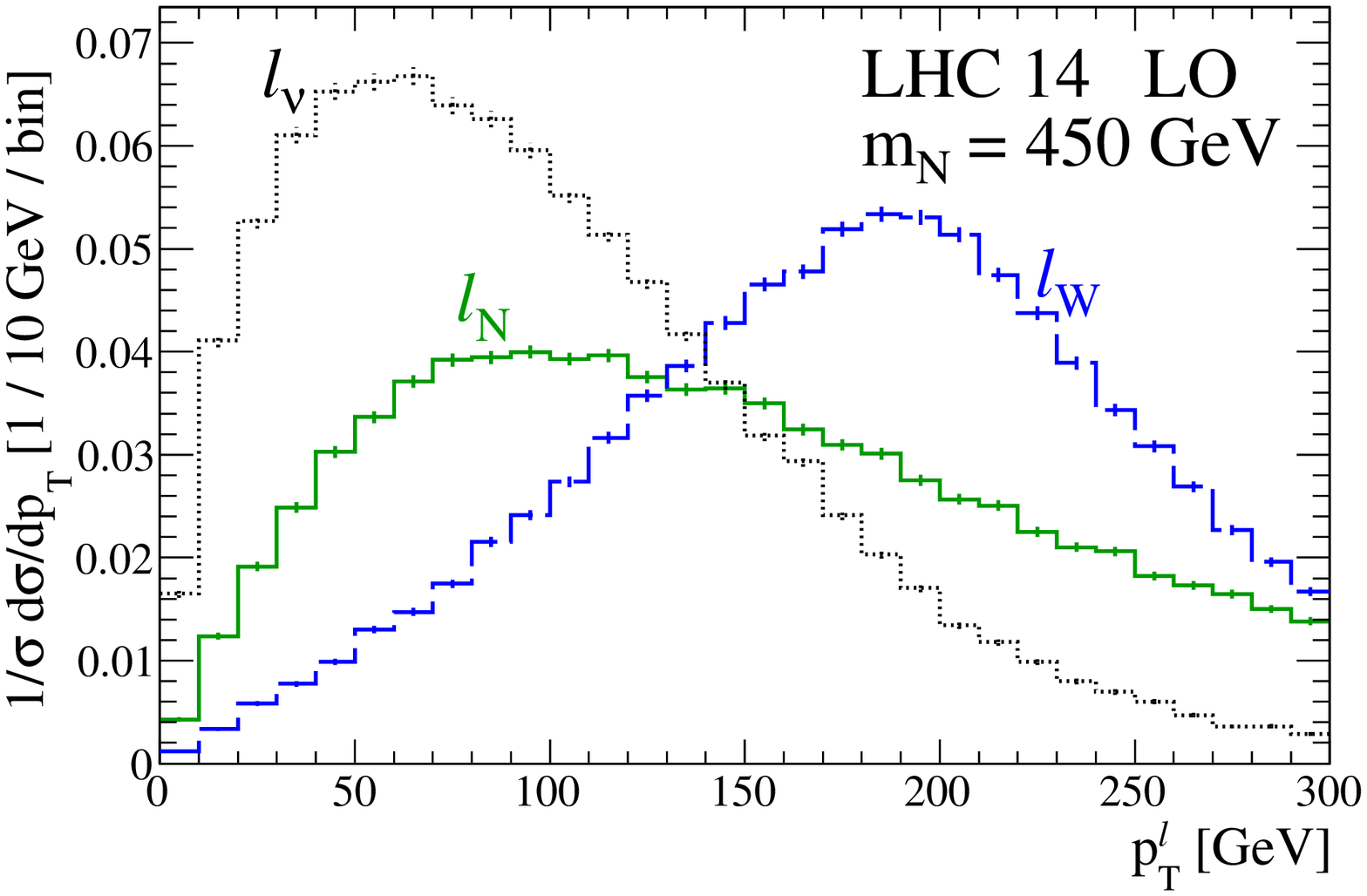}	}
\\
\subfigure[]{\includegraphics[width=.45\textwidth]{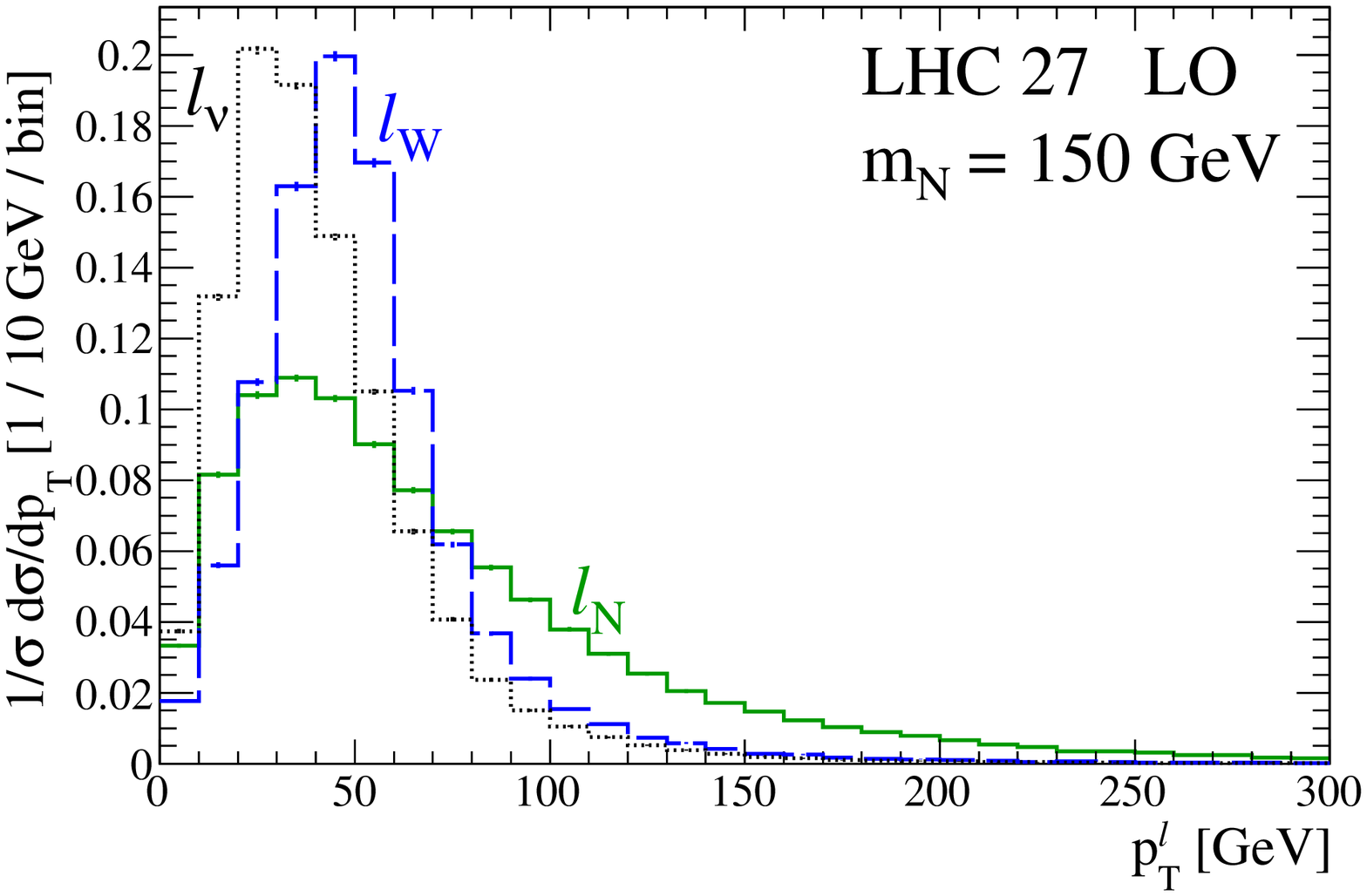}	}
\subfigure[]{\includegraphics[width=.45\textwidth]{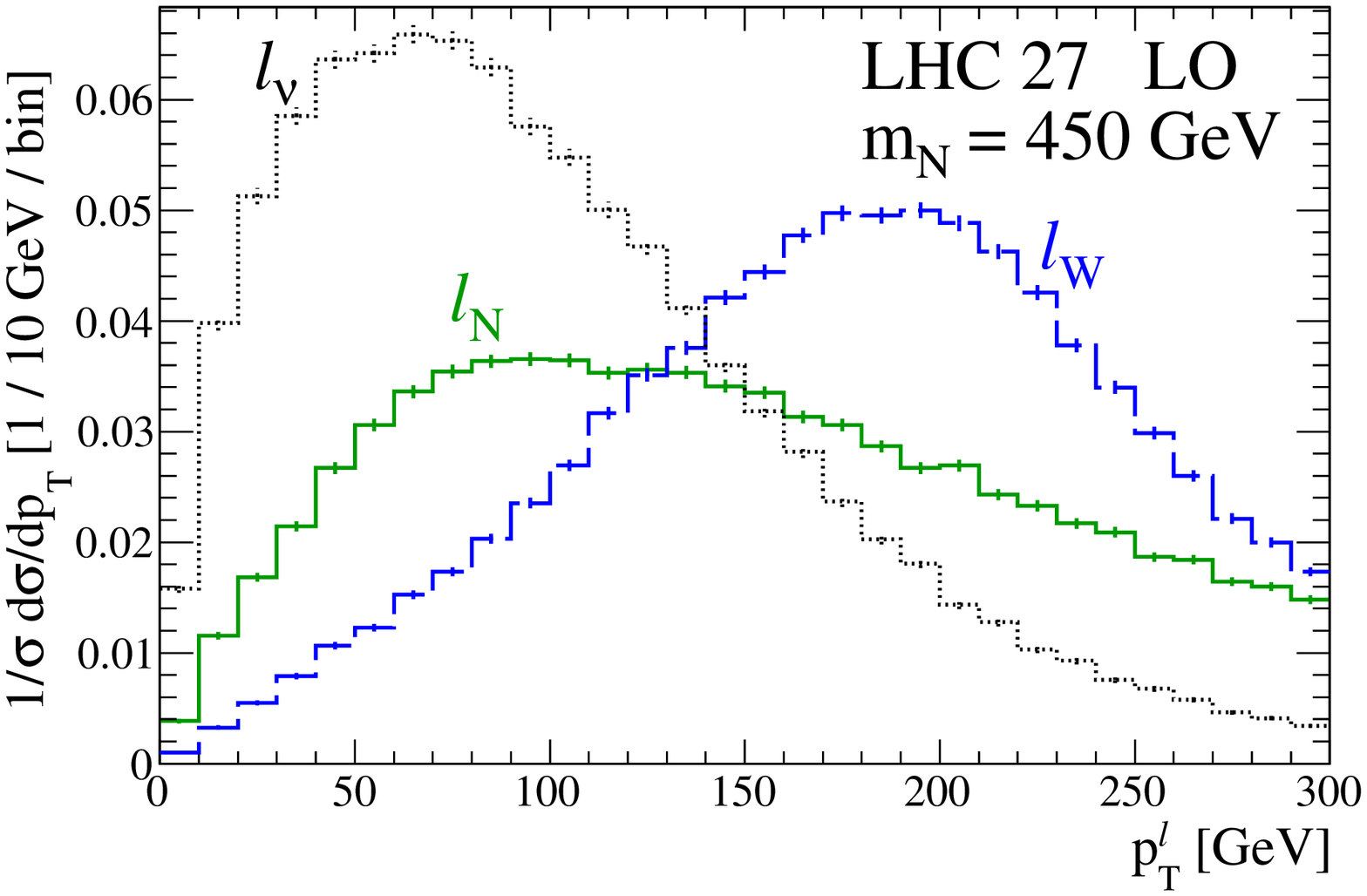}	}
\\
\subfigure[]{\includegraphics[width=.45\textwidth]{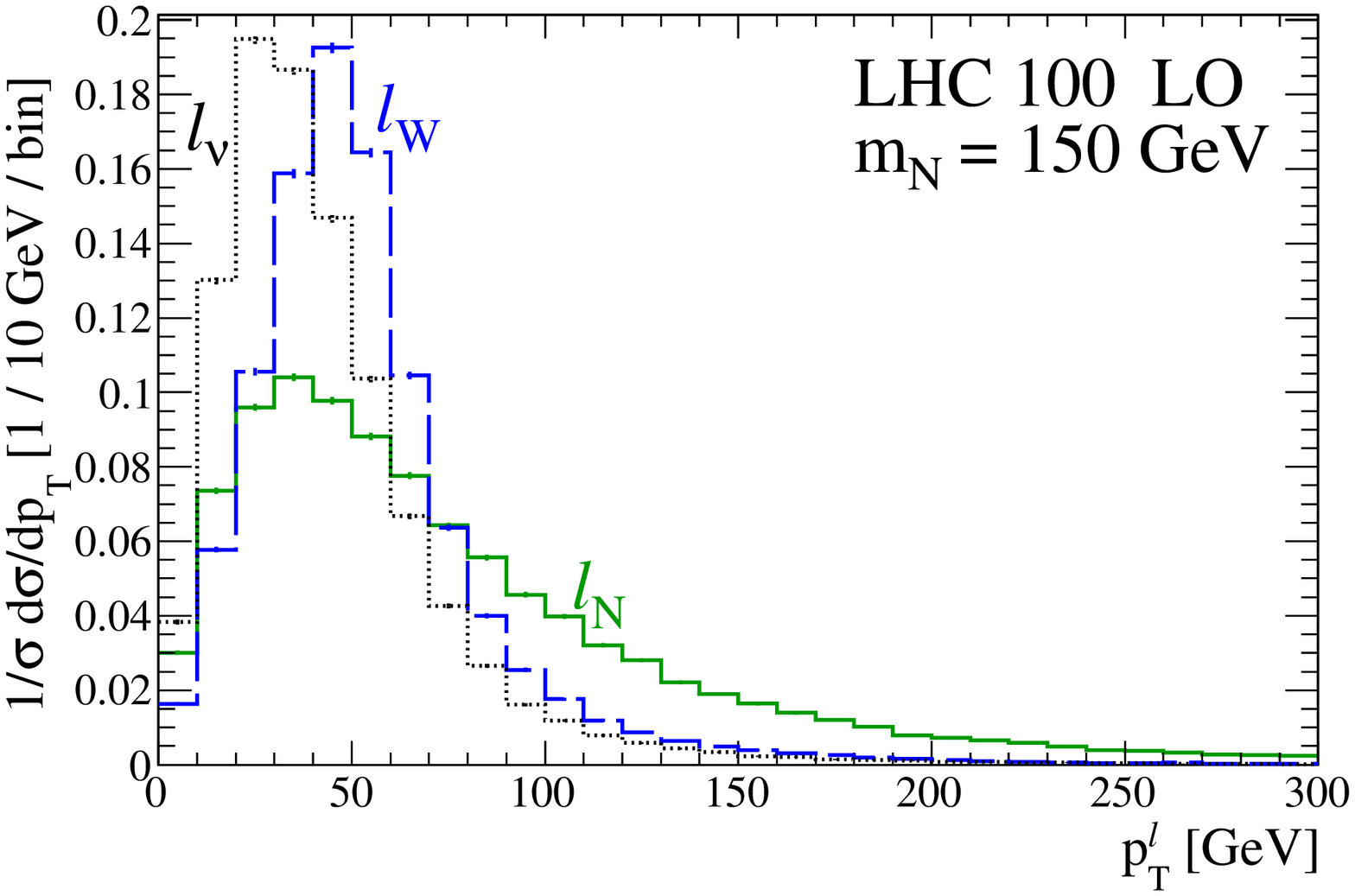}	}
\subfigure[]{\includegraphics[width=.45\textwidth]{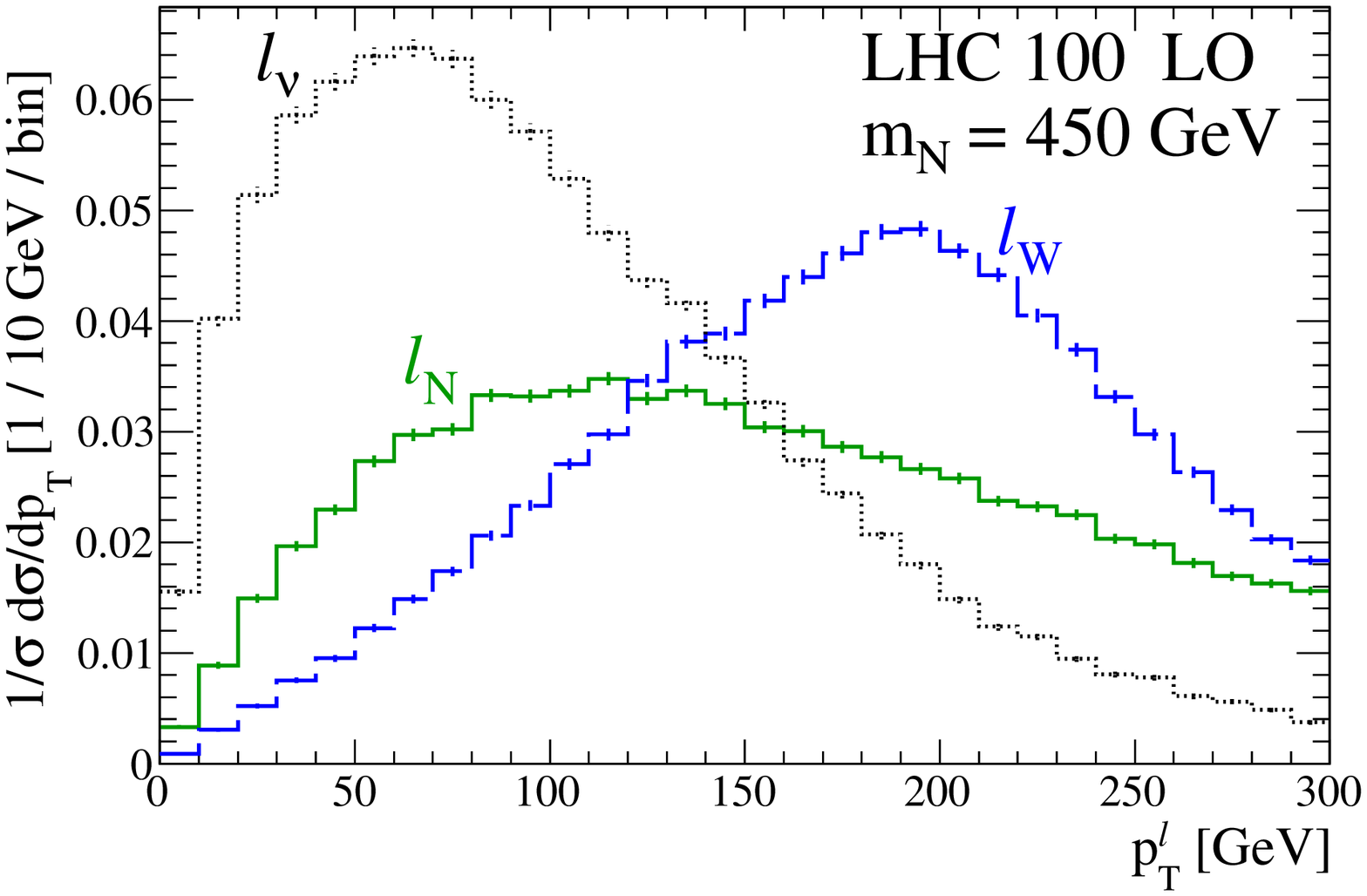}	}
\end{center}
\caption{Normalized parton-level transverse momentum $(p_T)$ distributions for charged leptons produced 
in the process $pp\to N l_N \to W l_W l_N \to \nu l_\nu l_W l_N$ via the DY mechanism
at (a,b)  $\sqrt{s}=14$, (c,d) 27, and (e,f) 100 TeV, for neutrino masses (a,c,e) $m_N=$ 150 GeV and (b,d,f) 450 GeV.
}
\label{fig:partonKinEllpT}
\end{figure*}

To demonstrate this, we show in Fig.~\ref{fig:partonKinNu}, the normalized (a,b) transverse momentum $(p_T^N)$ and (c,d) rapidity
$(y^N)$ distributions of $N$ for (a,c) $m_N=150$ and (b,d) 450 GeV, at $\sqrt{s} = 14$, 27, and 100 TeV.
Like for the invariant mass of the $(N\ell)$-system, for a fixed $m_N$, one observes that the shape of the  heavy neutrino $p_T$ distribution, which scales as 
\begin{equation}
p_T^N \sim \frac{m_N}{5} ,
\end{equation}
 is largely constant despite the large differences in collider energies, confirming the conjecture at LO.
On the other hand, the $y^N$ distribution broadens significantly with increasing collider energy.
This latter feature can be attributed to the fact that, at LO, increasing $\sqrt{s}$ translates directly to increasing the range of longitudinal momentum that initial-state partons may hold, 
and hence the range of longitudinal boosts imparted on the $(N\ell_N)$-system and its descendants.
This suggests that observables that depend on the longitudinal momentum of $N$ or the $(N\ell_N)$-system will tend to broaden with increasing $\sqrt{s}$.
As a function of increasing heavy neutrino mass and for a fixed $\sqrt{s}$, the narrowing $y^N$ distribution reflects the increase in $p_T^N$ observed in (a) and (b).

\begin{figure*}[!t]
\begin{center}
\subfigure[]{\includegraphics[width=.45\textwidth]{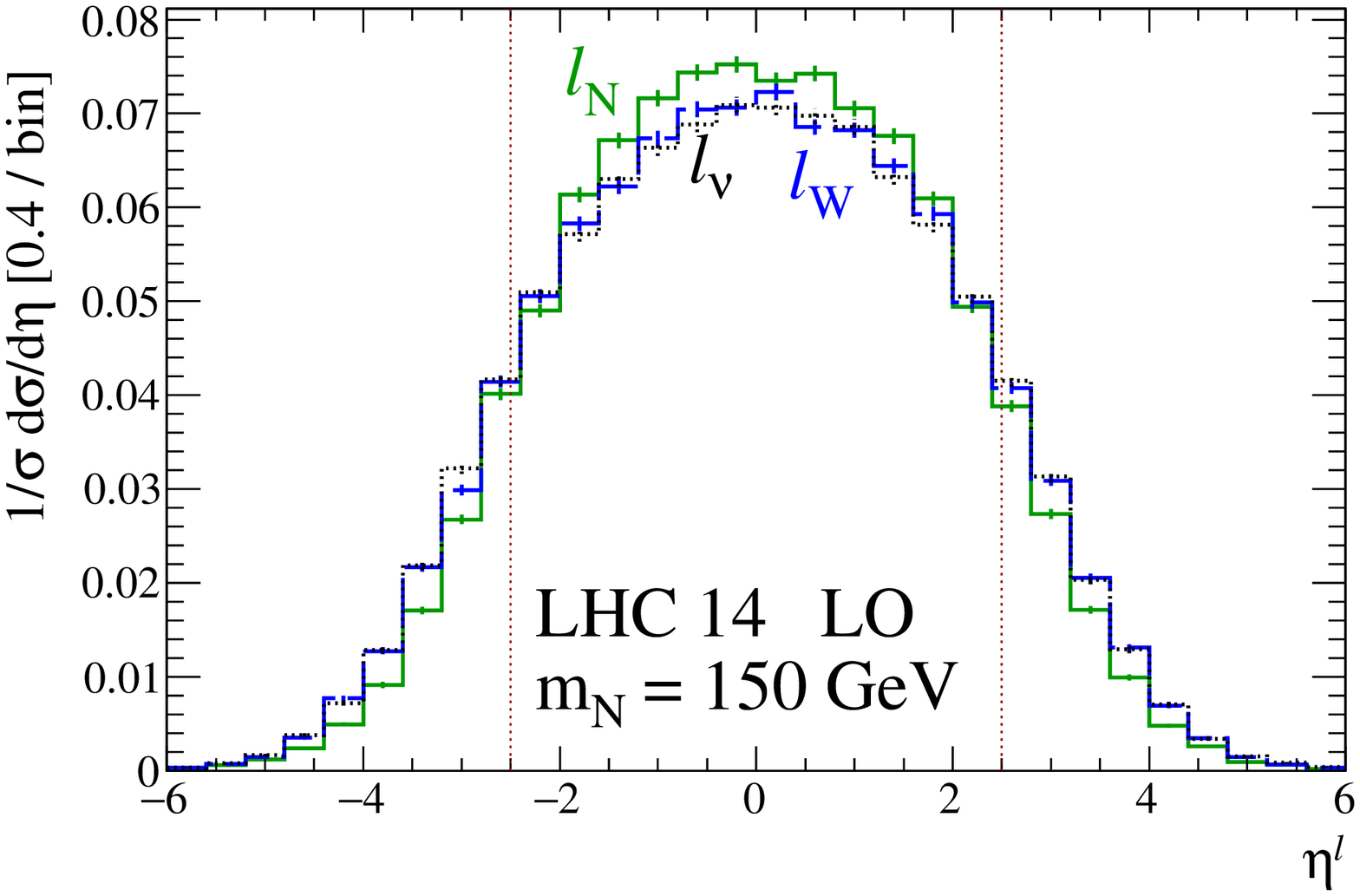}	}
\subfigure[]{\includegraphics[width=.45\textwidth]{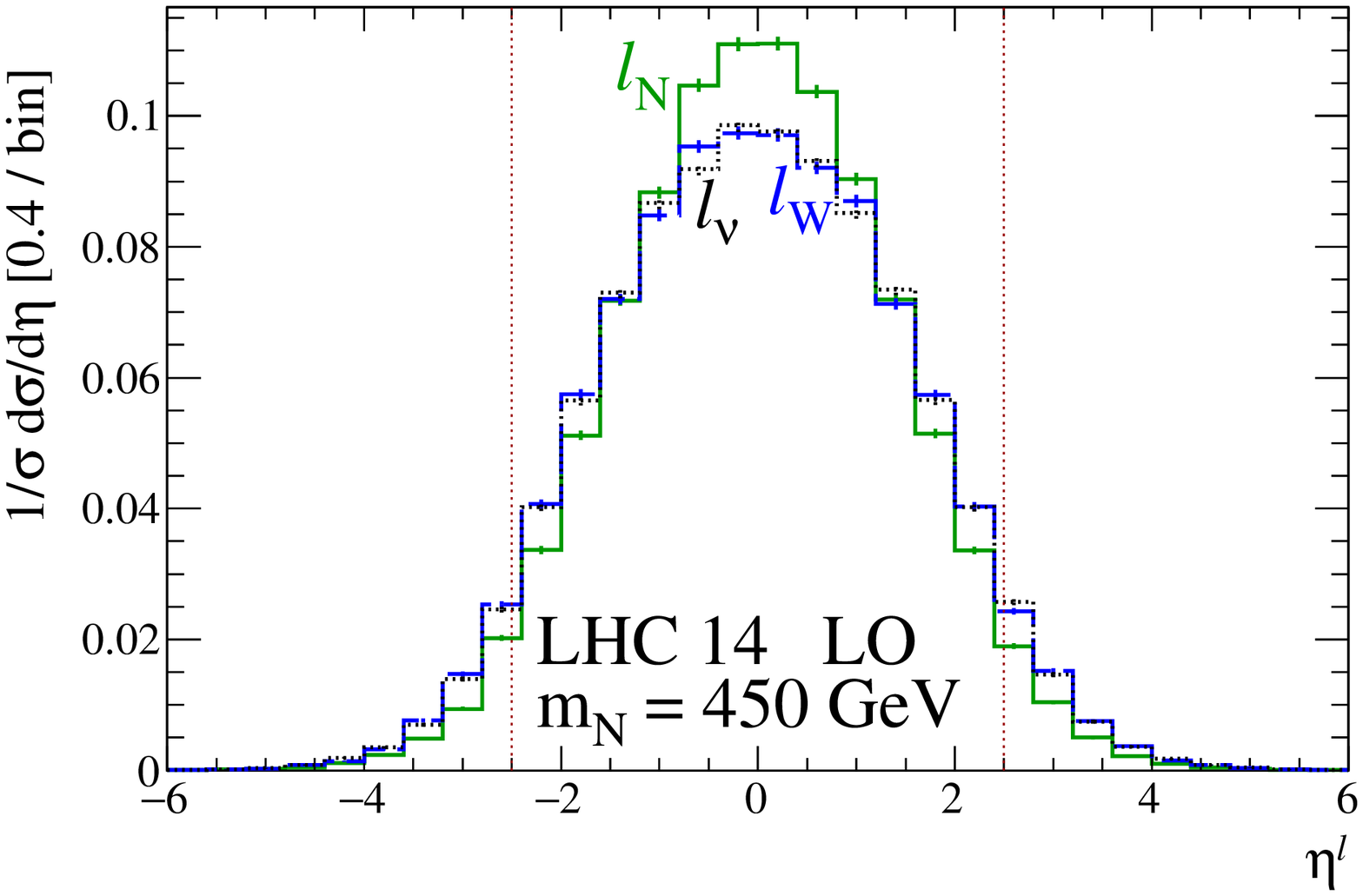}	}
\\
\subfigure[]{\includegraphics[width=.45\textwidth]{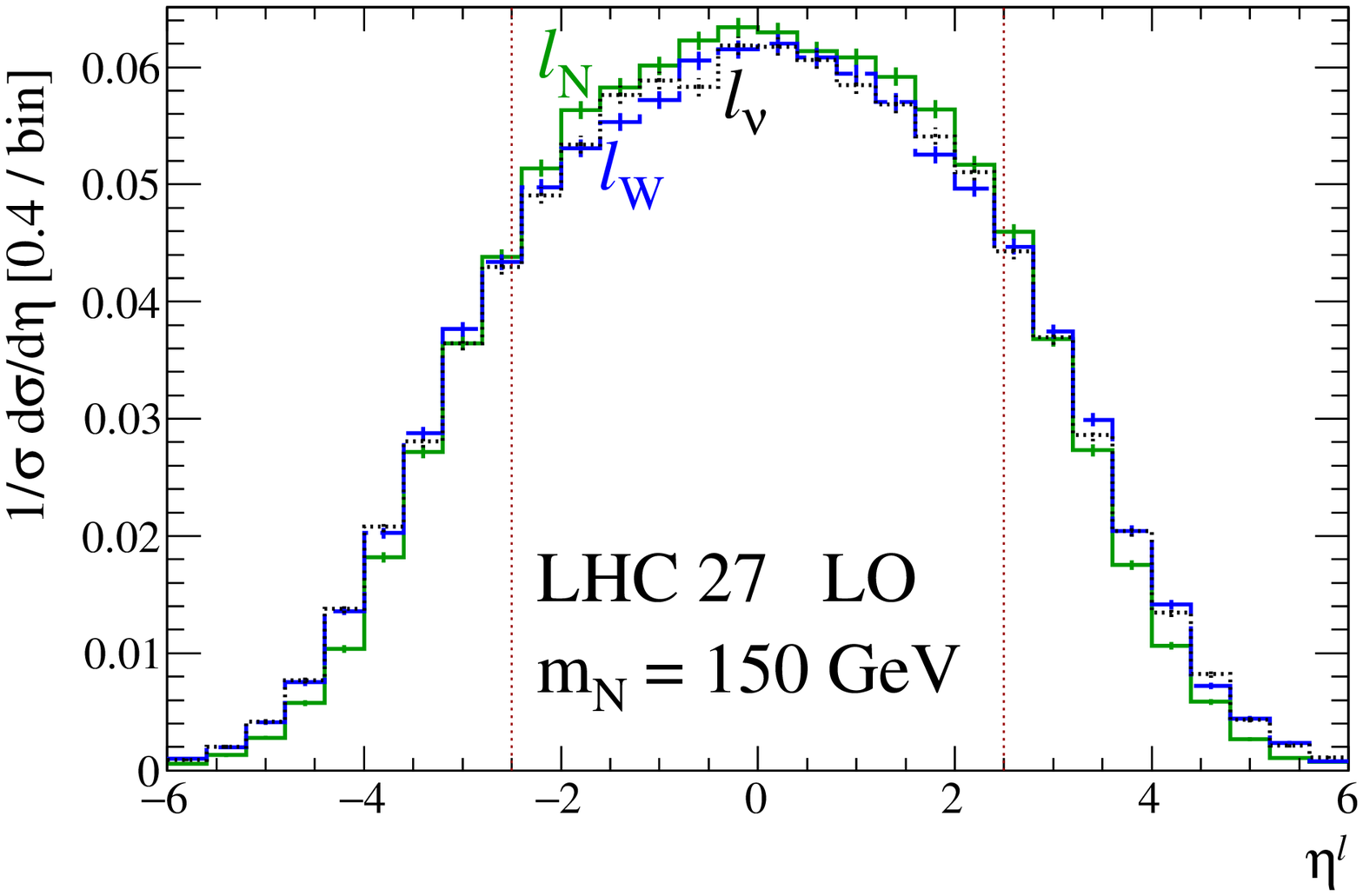}	}
\subfigure[]{\includegraphics[width=.45\textwidth]{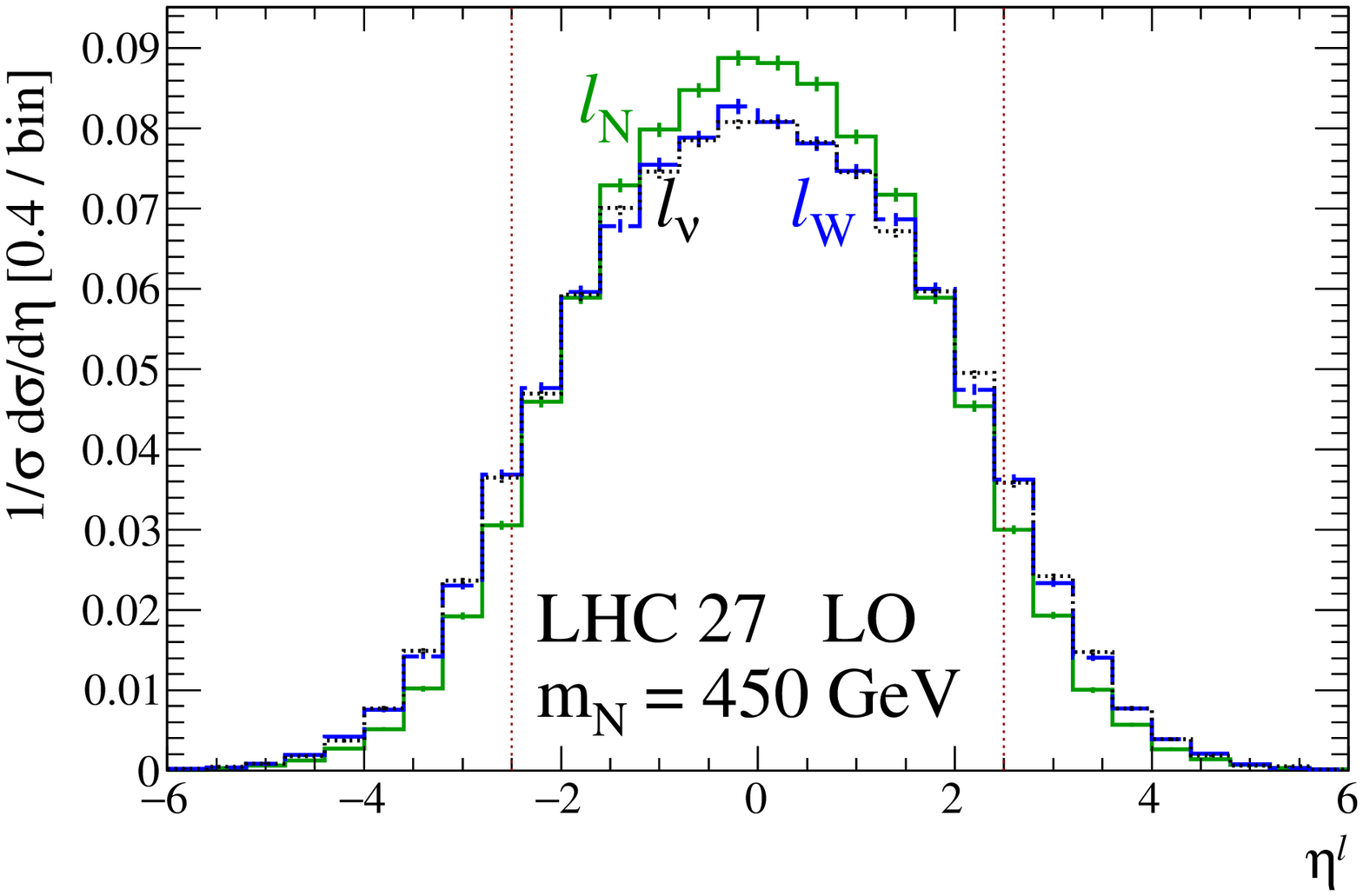}	}
\\
\subfigure[]{\includegraphics[width=.45\textwidth]{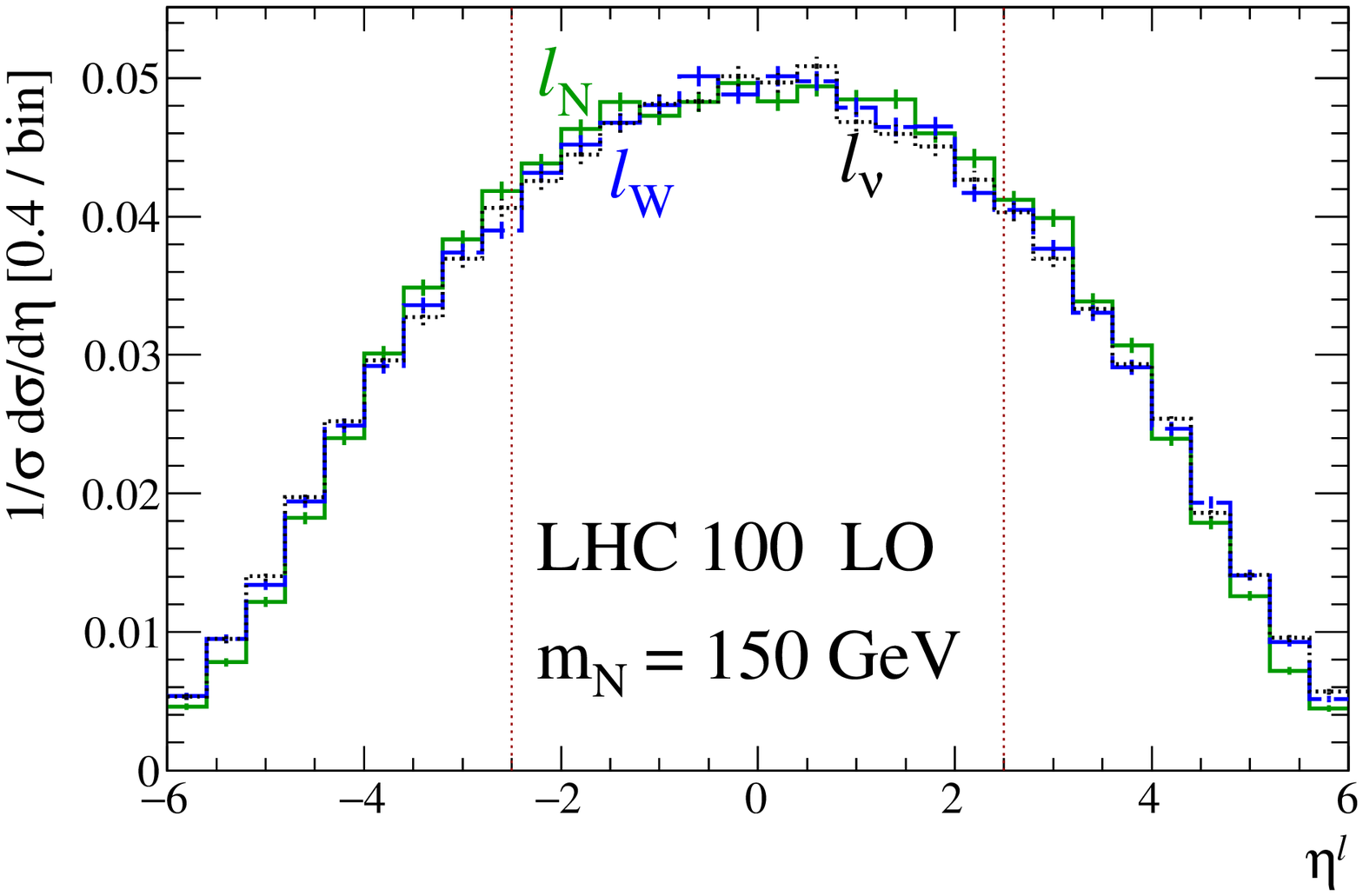}	}
\subfigure[]{\includegraphics[width=.45\textwidth]{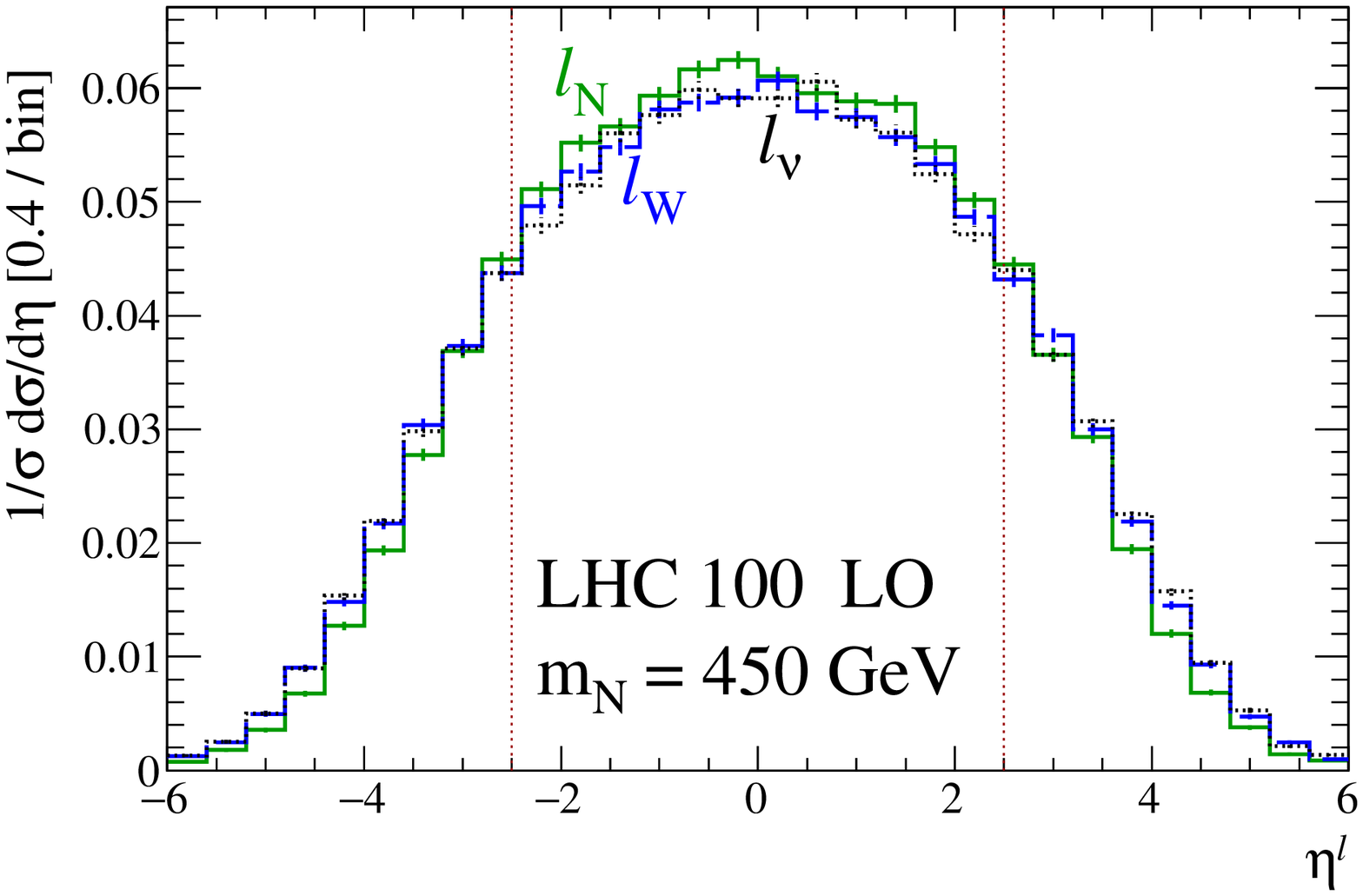}	}
\end{center}
\caption{Same as Fig.~\ref{fig:partonKinEllpT} but for pseudorapidity $(\eta)$ of the charged leptons.}
\label{fig:partonKinElleta}
\end{figure*}

Turning to the final-state charged leptons, Figs.~\ref{fig:partonKinEllpT} and ~\ref{fig:partonKinElleta} depict, respectively,
the normalized $p_T$ and $\eta$ distribution for (a,c,e) $m_N = 150$ and (b,d,e) 450 GeV, at (a,b) $\sqrt{s} = 14$, (c,d) 27, and (e,f) 100 TeV.
For all cases, by momentum conservation, 
the $p_T$ distributions of the charged lepton produced in association with the heavy neutrino $(\ell_N)$ mirror those of $N$ and need not be discussed further. 
The $\eta$ distributions for $\ell_N$ are slightly narrower than the $y^N$ curve due to $\ell_N$'s (approximately) massless nature.

Keeping to Fig.~\ref{fig:partonKinEllpT}, for both low and high mass $N$, one sees that the distribution shapes are largely independent of collider energy.
This follows from the fact that the $p_T$ of $\ell_W$ and $\ell_N$ have characteristic values due to the intermediate resonant structure of the $N\to\ell_W W\to \ell_W\ell_\nu \nu$ decay.
Consequently, up to relative transverse boosts of $N$, the transverse momenta of $\ell_W$ and $\ell_N$ scale as,
\begin{eqnarray}
p_T^{\ell_W} 	&\lesssim&	 E^{\ell_W} \sim \frac{m_N}{2}\left(1-\cfrac{M_W^2}{m_N^2}\right)\sim\frac{m_N}{2}, \quad\text{and} \\
p_T^{\ell_\nu} 	&\lesssim&	 E^{\ell_\nu} \sim \frac{m_N}{4}\left(1+\cfrac{M_W^2}{m_N^2}\right)\sim\frac{m_N}{4},
\end{eqnarray}
where in the last approximation we take the $(M_W/m_N)^2\ll1$ limit. 
As we have argued, however, for the inclusive production of $N$, $p_T^N$ is largely stable against varying beam energy and therefore we expect and observe this independence to be inherited.

For  the $\eta$ distributions in Fig.~\ref{fig:partonKinElleta}, a number of notable features are observed:
The first is that for a fixed $\sqrt{s}$ and $m_N$, the three pseudorapidity curves overlap considerably.
This follows from the combined circumstance that each charged lepton possesses a comparable transverse momentum that scales like $m_N$,
and that all three charged leptons obtain their relative longitudinal momentum from the same source, namely the hard $(N\ell_N)$-system.
As such, the distributions are  indistinguishable and cannot be readily used to help identify a particular lepton's origin.
For the narrowing of the $\eta$ distribution with increasing $m_N$, we reiterate that this simply reflects the relative increase of $p_T$ observed in Fig.~\ref{fig:partonKinEllpT}.

Regarding the $\sqrt{s}$ dependence, again one observes a broadening with increasing collider energy, which is inherited from the parent $(N\ell_N)$-system.
That said, what is important to stress here is the degree of the migration of events to outside the $\vert \eta \vert = 2.5$ boundary.
For $m_N = 150\GeV$, we see that the fraction $(f)$ for each lepton to fall outside this detector boundary grows from 
about \confirm{$f \gtrsim 20\%$, to $\gtrsim30\%$, to $\gtrsim40\%$} as one goes from $\sqrt{s}=14\TeV$ to $27\TeV$, to $100\TeV$.
Neglecting correlation, this means that only about \confirm{$\varepsilon = (1-f)^3 \lesssim 50\%~(35\%)~[25\%]$} of trilepton heavy neutrino events
 fall within the benchmark fiducial coverage at $\sqrt{s}=14~(27)~[100]\TeV$.
 For $m_N = 450\GeV$, the situation is marginally better with  \confirm{$\varepsilon \lesssim 85\%~(60\%)~[35\%]$}, but still discouraging.
 For even larger $m_N$, one can be more hopeful, but at the cost of a smaller cross section (for a fixed active-sterile mixing).
All-in-all, as evident in Fig.~\ref{fig:partonKinElleta}, extending tracker coverage to at least $\vert \eta \vert = 3-4$ 
for future collider detector experiments is vital, if not necessary, to ensure a healthy and competitive search program for anomalous production of charged leptons.

When these distributions are taken together, a fuller picture starts to emerge at LO: 
for symmetric, anti-collinear beams as we consider, 
as one increases $\sqrt{s}$, the incoming initial-state partons entering the $q\overline{q'}\to N\ell$ hard process are allowed to carry larger longitudinal momenta along the beam axis.
As a result, the $(N\ell_N)$-system itself is allowed to carry a larger longitudinal boost, and hence  the observed broadening of the $y^N$ spectra.
On the other hand, to an excellent approximation, the net $p_T$ if the initial proton-proton system is zero.
This does not change as a function of collider energy, and suggest that the $p_T$ of the $(N\ell_N)$-system does not depend on $\sqrt{s}$, 
which is essentially true at LO, as reflected in the $p_T$ distributions of $N$, its descendants, and $\ell_N$.
At the leading log / parton shower level, however, this argument of course begins to break down:
there is soft/collinear ISR that imbues the $(N\ell_N)$-system with nonzero $p_T$ that does indeed depend, to some extent, on the phase space made available by the collider center of mass energy.
This radiation, however, dominantly contributes to the region of phase space where $p_T^{(N\ell)} / Q \ll 1$. 
That is to say, the $p_T$ of the $(N\ell_N)$-system is typically small compared to the invariant mass of the $(N\ell_N)$-system~\cite{Collins:1984kg,Collins:1985ue}.
And as demonstrated in Fig.~\ref{fig:partonHardQ}, the invariant mass of the $(N\ell_N)$-system is driven by the matrix element, not by $\sqrt{s}$.
So while we do expect the $p_T$ of $N$, its descendant leptons, and $\ell_N$ to change with increasing $\sqrt{s}$, this change will be small compared to the hard scale $Q\sim m_N$, 
which sets the initial $p_T$ scale of $N$ and $\ell_N$.

Interestingly, the above argument then suggests that kinematic observables derived strictly from the $p_T$ of $\ell_N,~\ell_W$, and $\ell_\nu$ will inherit this robustness against changing $\sqrt{s}$.
We now turn to particle-level kinematics at NLO+PS(LL) to see how far this holds.

\subsection{Particle-Level Kinematics and Observables at NLO+PS(LL)}\label{sec:ColliderParticleKin}

In this section, we investigate particle-level observables for the same trilepton process in Eq.~\ref{eq:partonDY}, assuming the same representative $m_N$ and $\sqrt{s}$,  but instead at NLO+PS.
We do this by first highlighting the key distinctions between parton- and 
particle-level\footnote[2]{Particle-level events are passed through parton showers, hadronization, and jet clustering algorithms.} objects.
We then point to a feature in MC generation at NLO+PS that enables us to establish a definitive link between parton- and particle-level kinematics.
Building on this correspondence, we are able to see at this significantly more realistic level of modeling how robust the kinematics of $N$ and its descendants are against variable collider energy.
We then demonstrate the existence of an entire class of particle-level observables that exhibits only a weak dependence on $\sqrt{s}$.
This is a main result of this study.

For several reasons operating at the parton shower-level with hadronization modeling is crucial, particularly for this study:
While parton-level kinematics provide important insight and intuition to underlying processes,
not all kinematic observables built at the parton level are physical in hadronic environments~\cite{Collins:1984kg,Collins:1985ue}.
Moreover, detector experiments do not observe partons or bare charged leptons:
at macroscopic distances, high-$p_T$ partons and charged leptons are dressed with collinear and soft QCD and/or QED radiations.
This has a fundamental impact on both exclusive observables, such as jet and lepton multiplicity, 
as well as inclusive ones, such as the transverse momentum of the $(N\ell_N)$-system.
In addition, when the hard scattering process and beam remnant separate beyond a distance of $d_{\rm NP}\sim 1/\Lambda_{\rm NP} \sim 1$ fm, 
color-connection and hadronization must be taken into account.
This gives rise to a cornucopia of (relatively) low-$p_T$ hadrons and such contributions play an important role in this work:
Depending on the largeness of the jet radius parameter $R$, ``out-of-cone'' and ``into-cone'' emission of hadrons can shift jet $p_T$ by up to \confirm{$\mathcal{O}(10)$ GeV}~\cite{Dasgupta:2007wa}.
Weak decays of heavy hadrons are a source of background charged leptons.
And likewise, light neutrinos originating from weak decays of hadrons can cumulatively shift the direction and magnitude of the missing transverse momentum vector  
by \confirm{$\mathcal{O}(\Lambda_{\rm NP}/{\rm decay})$}.
Not to mention, hadronic decays of $\tau$ leptons $(\tau_h)$ require modeling and experimental tuning as their branching rates cannot be derived from first principles. 

Despite the activity of a hadron collider environment, it is possible to make a more or less concrete mapping to parton-level objects, though not necessarily their precise identities.
As done in detector experiments, one must cluster (sum) the momenta of all like-objects into composite objects (jets)
 according to some procedure (a sequential jet clustering algorithm) that is valid at all orders of perturbation theory (is infrared and collinear-safe).
With this additional layer of abstraction, a many-body environment is simplified to a few-body system with correspondence to the partonic event.
This procedure, though, is not entirely free of ambiguities.
A consequence of working at this reconstructed level is the obfuscation of amplitude-/generator-level information about a particle's particular origin.
For example: in Figs.~\ref{fig:partonKinEllpT} and \ref{fig:partonKinElleta}, 
the degree of overlap and similarity of the $p_T$ and $\eta$ curves for $\ell_N,~\ell_W,$ and $\ell_\nu$  is worsened due to recoils against electromagnetic FSR and QCD ISR.
In effect, the exact lineage of the charged leptons are anonymized at the reconstruction level.

\begin{figure*}[!t]
\begin{center}
\subfigure[]{\includegraphics[width=.48\textwidth]{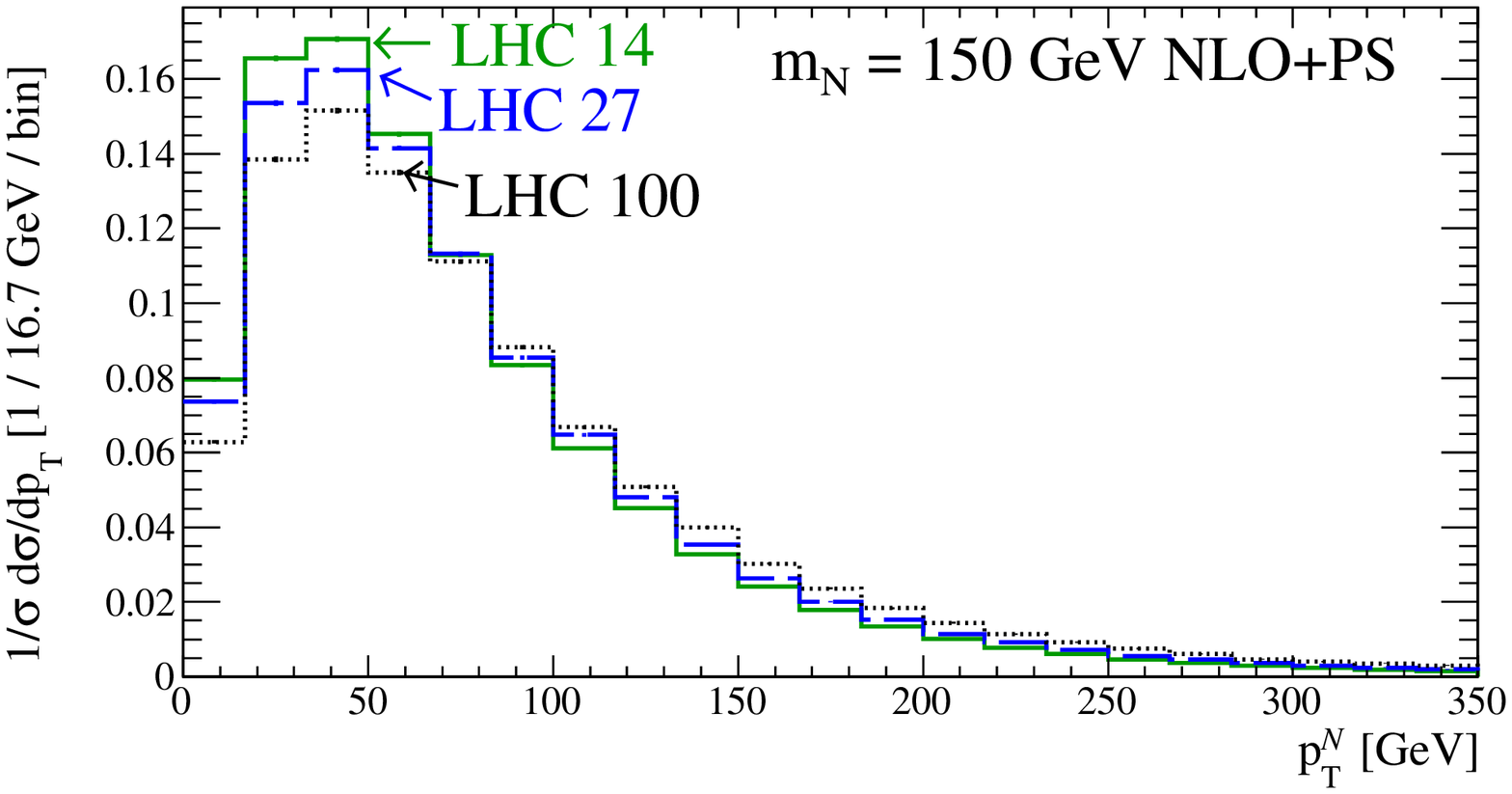}	}
\subfigure[]{\includegraphics[width=.48\textwidth]{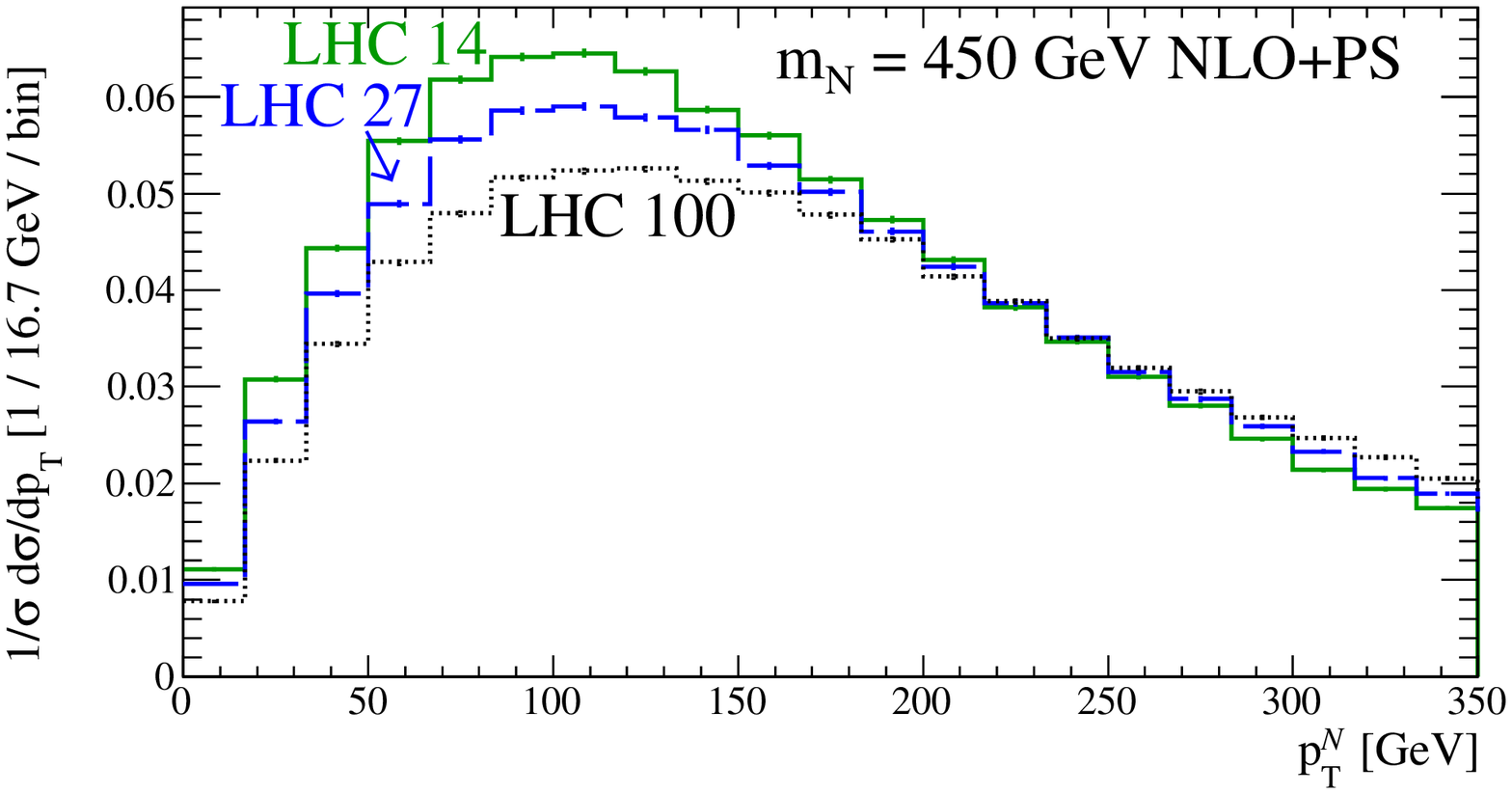}	}
\\
\subfigure[]{\includegraphics[width=.48\textwidth]{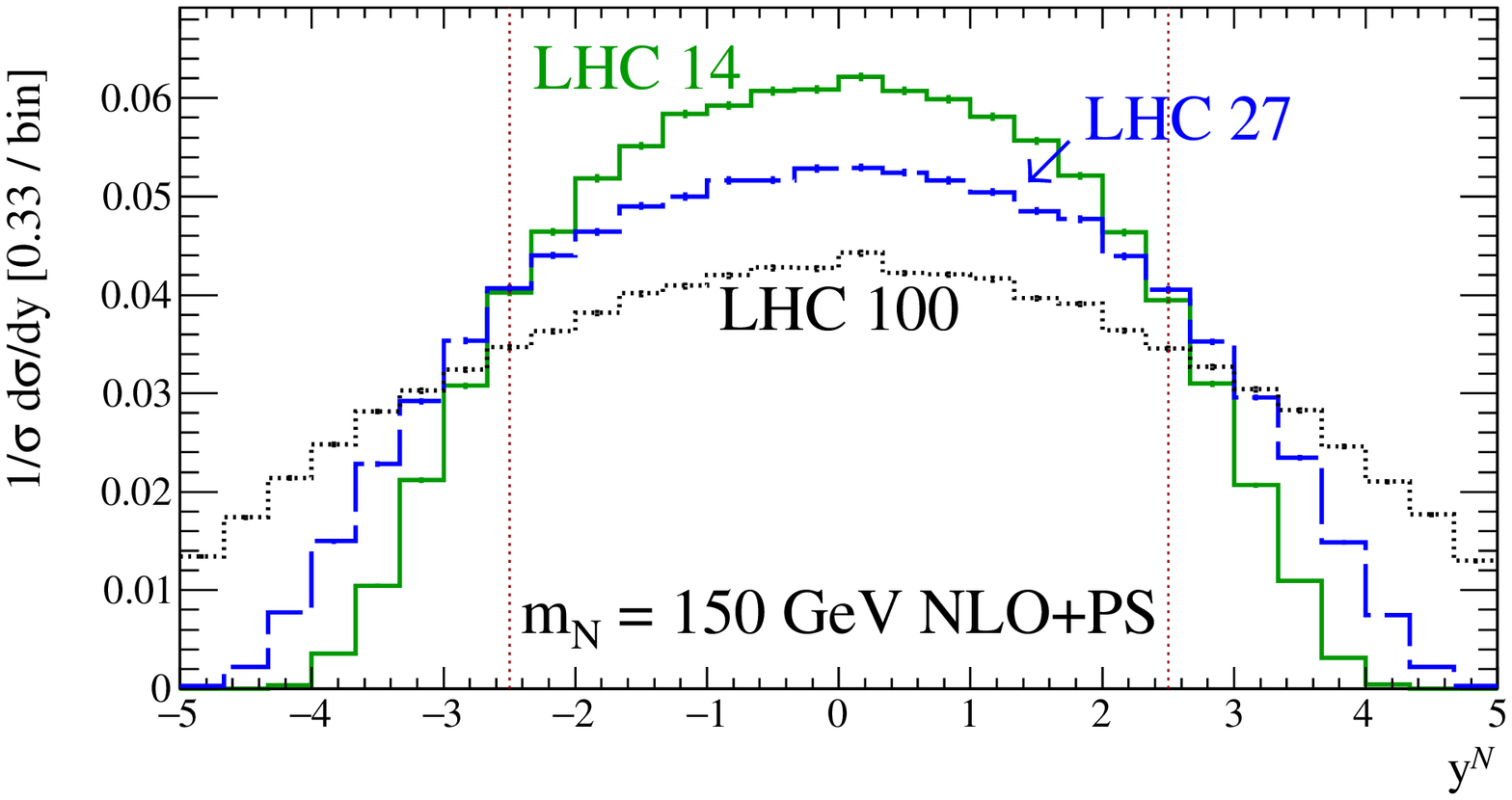}	}
\subfigure[]{\includegraphics[width=.48\textwidth]{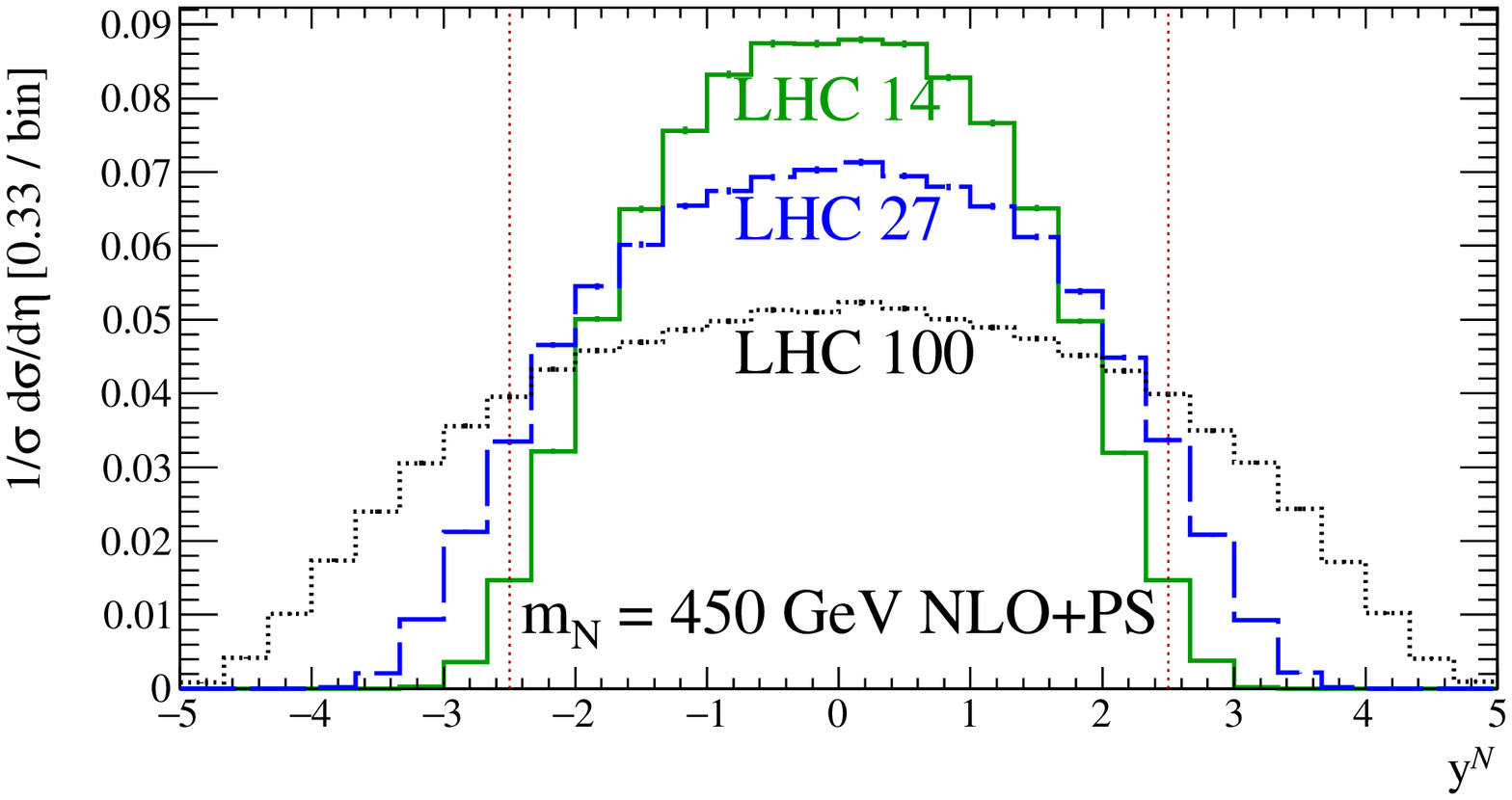}	}
\end{center}
\caption{
Same as Fig.~\ref{fig:partonKinNu} but at NLO+PS accuracy with particle-level reconstruction.
}
\label{fig:particleKinNu}
\end{figure*}

A quirk exist in the present case,  however,  that allows us to establish a stronger link between parton-level and particle-level heavy neutrino events.
As described in Sec.~\ref{sec:ColliderPartonKin}, for event generation with \mgamc, intermediate particles that go on-shell are recorded in the event file independent of the NWA.
This is true even at NLO.
Moreover, the 4-momentum and PID of intermediate, on-shell particles from the original $q\overline{q'}\to N\ell \to \ell\ell W$ hard process are preserved throughout the \mgpyma~ chain, 
thereby allowing us to build kinematic distributions of $N$ at this accuracy.

Subsequently, we show in Fig.~\ref{fig:particleKinNu} the normalized, NLO+PS accurate (a,c) $p_T$ and (b,d) $y$ distributions for the intermediate heavy $N$ in Eq.~\ref{eq:partonDY}, 
for (a,b) $m_N=150$ and (c,d) 450 GeV, at $\sqrt{s}=14$ (solid), 27 (dash), and 100 (dot) TeV.
In comparison to Fig.~\ref{fig:partonKinNu}, one sees very little changes in any of the curves and, crucially, that the insensitivity to $\sqrt{s}$ is retained.
This is due to the fact that QCD corrections for DY-like processes are dominated by  positive virtual corrections~\cite{Altarelli:1979ub,Hamberg:1990np,Anastasiou:2003yy}.
This further indicates that the $p_T^N$ and $y^N$ distributions (and the analogous ones for $\ell_N$) exhibit Born-like features at NLO+PS 
but with a possible enhancement at large $p_T^N$ owing to the opening of partonic channels, e.g., Compton scattering $gq\to N\ell q'$ with $p_T^{q'} > \mu_f$.
However, following Ref.~\cite{Degrande:2016aje}, our choice of factorization and renormalization scales, as given in Eq.~\ref{eq:scale}, 
are intentionally selected to minimize the impact of such channels on distribution shapes.
Thus, such enhancements will be minor, and one may infer that observables directly related to the  kinematics of $N$, 
e.g., $p_T$ of its decay products, remain mostly unchanged relative to LO predictions.

\begin{figure*}[!t]
\begin{center}
\subfigure[]{\includegraphics[width=.45\textwidth]{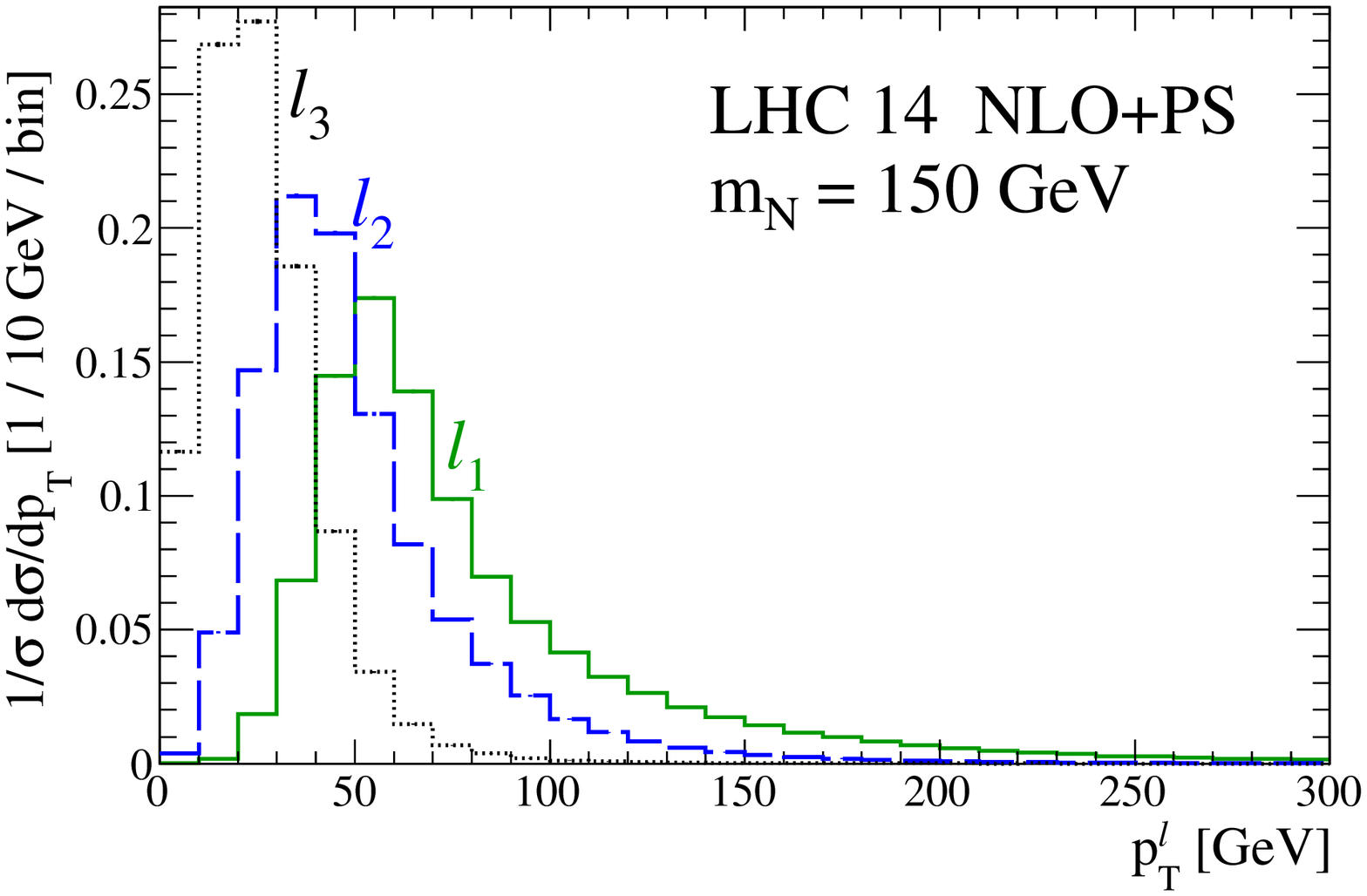}	}
\subfigure[]{\includegraphics[width=.45\textwidth]{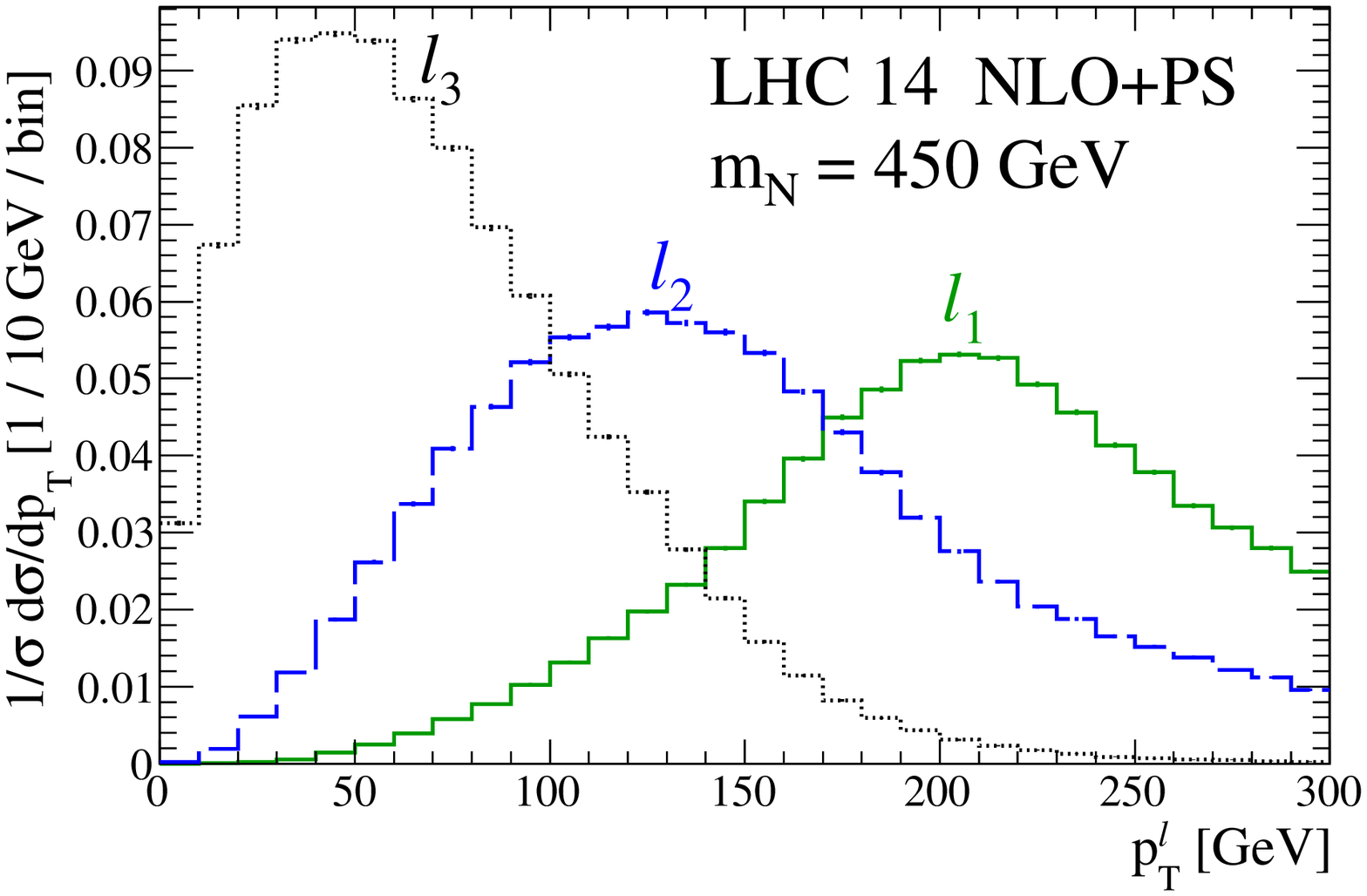}	}
\\
\subfigure[]{\includegraphics[width=.45\textwidth]{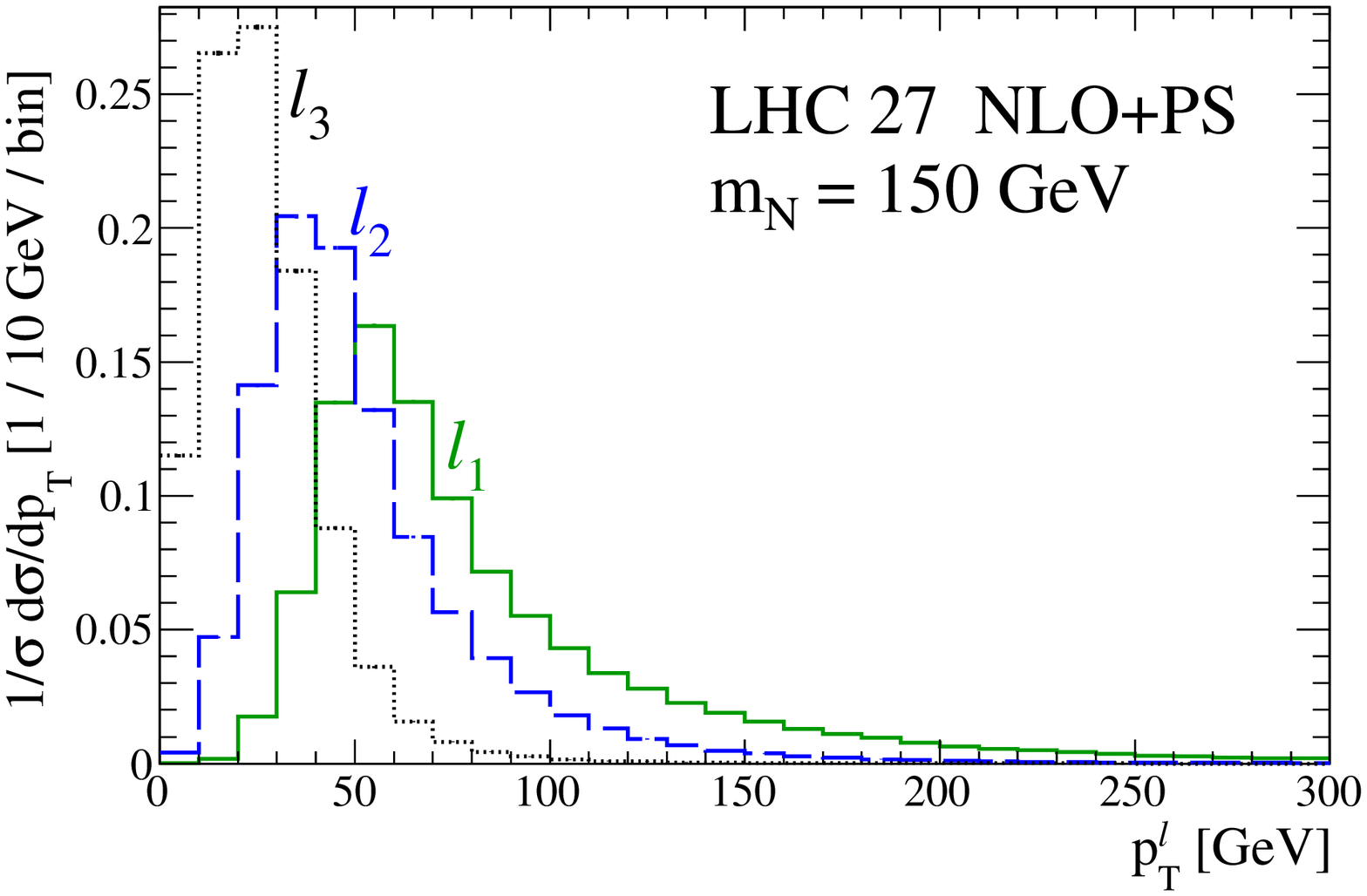}	}
\subfigure[]{\includegraphics[width=.45\textwidth]{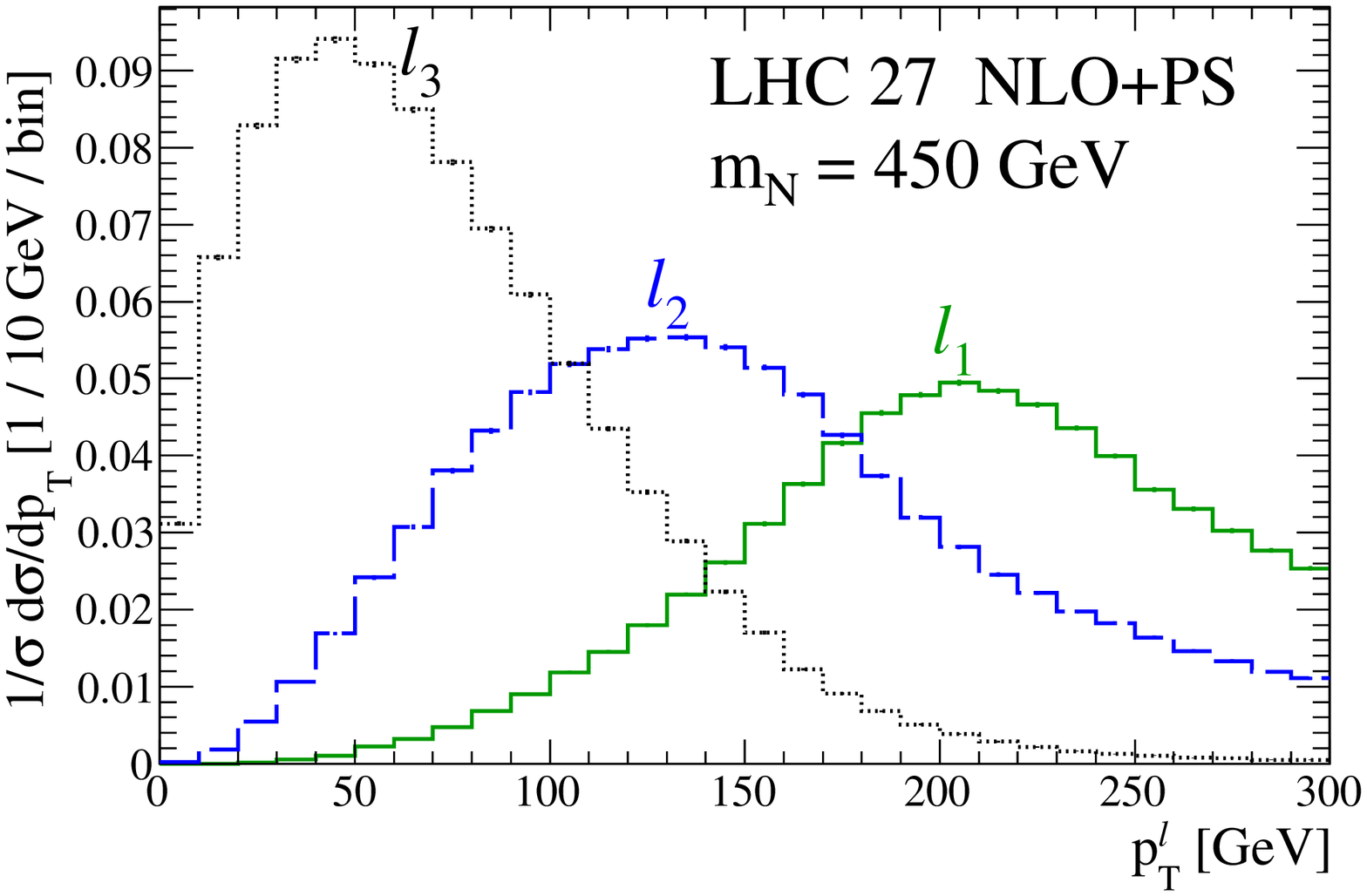}	}
\\
\subfigure[]{\includegraphics[width=.45\textwidth]{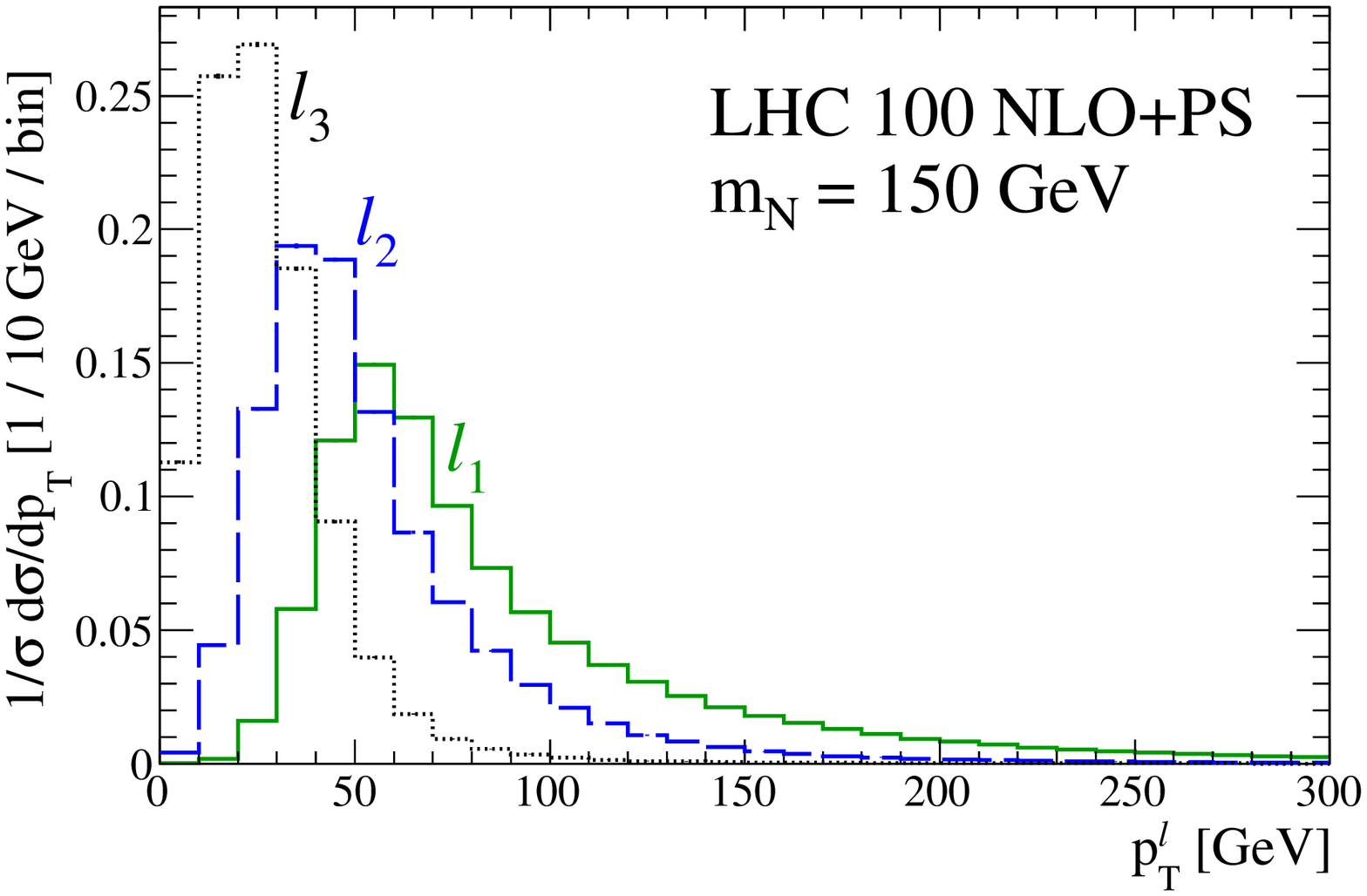}	}
\subfigure[]{\includegraphics[width=.45\textwidth]{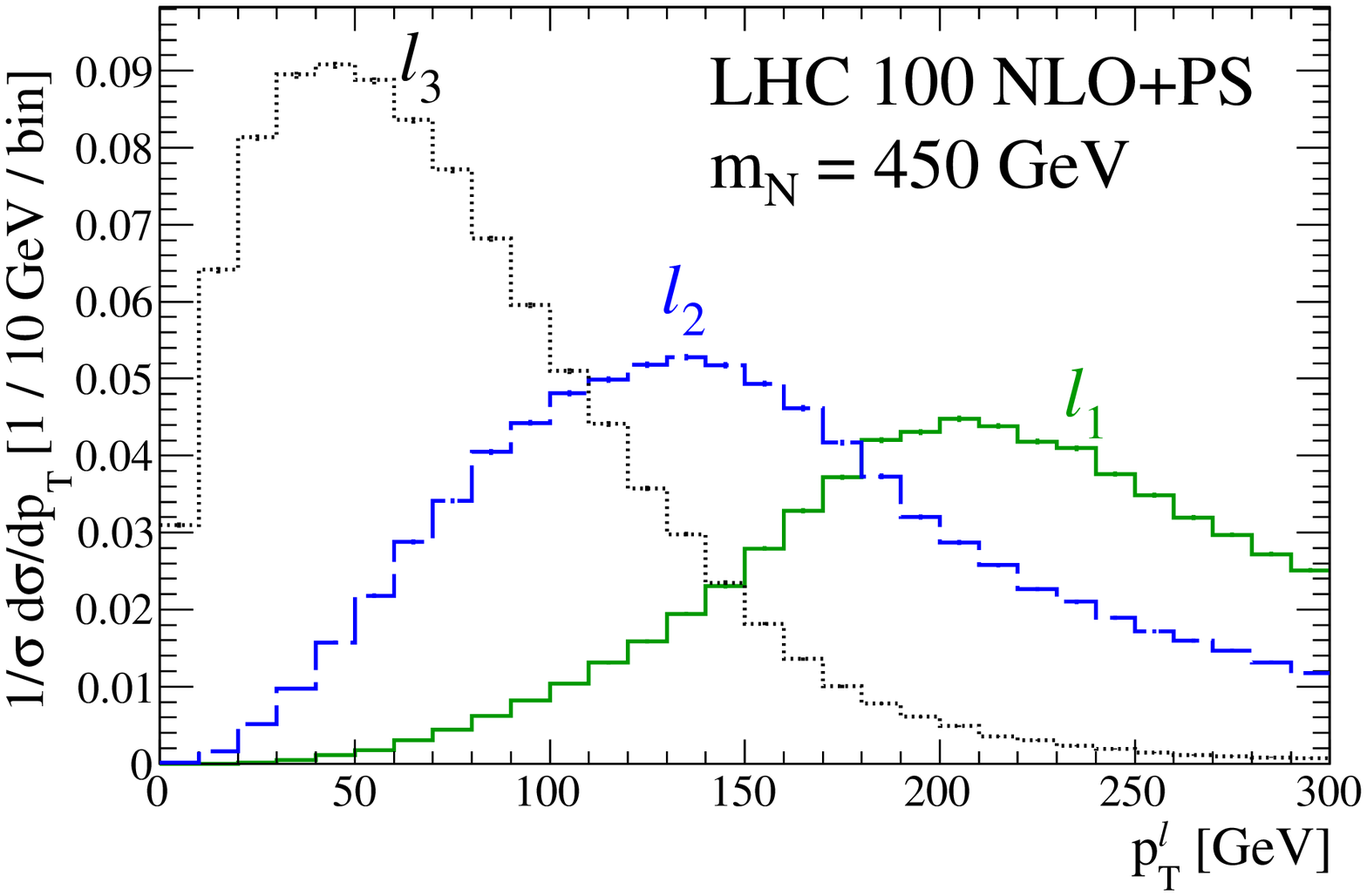}	}
\end{center}
\caption{
Same as Fig.~\ref{fig:partonKinEllpT} but at NLO+PS accuracy with particle-level reconstruction.
}
\label{fig:particleKinEllpT}
\end{figure*}

\begin{figure*}[!t]
\begin{center}
\subfigure[]{\includegraphics[width=.45\textwidth]{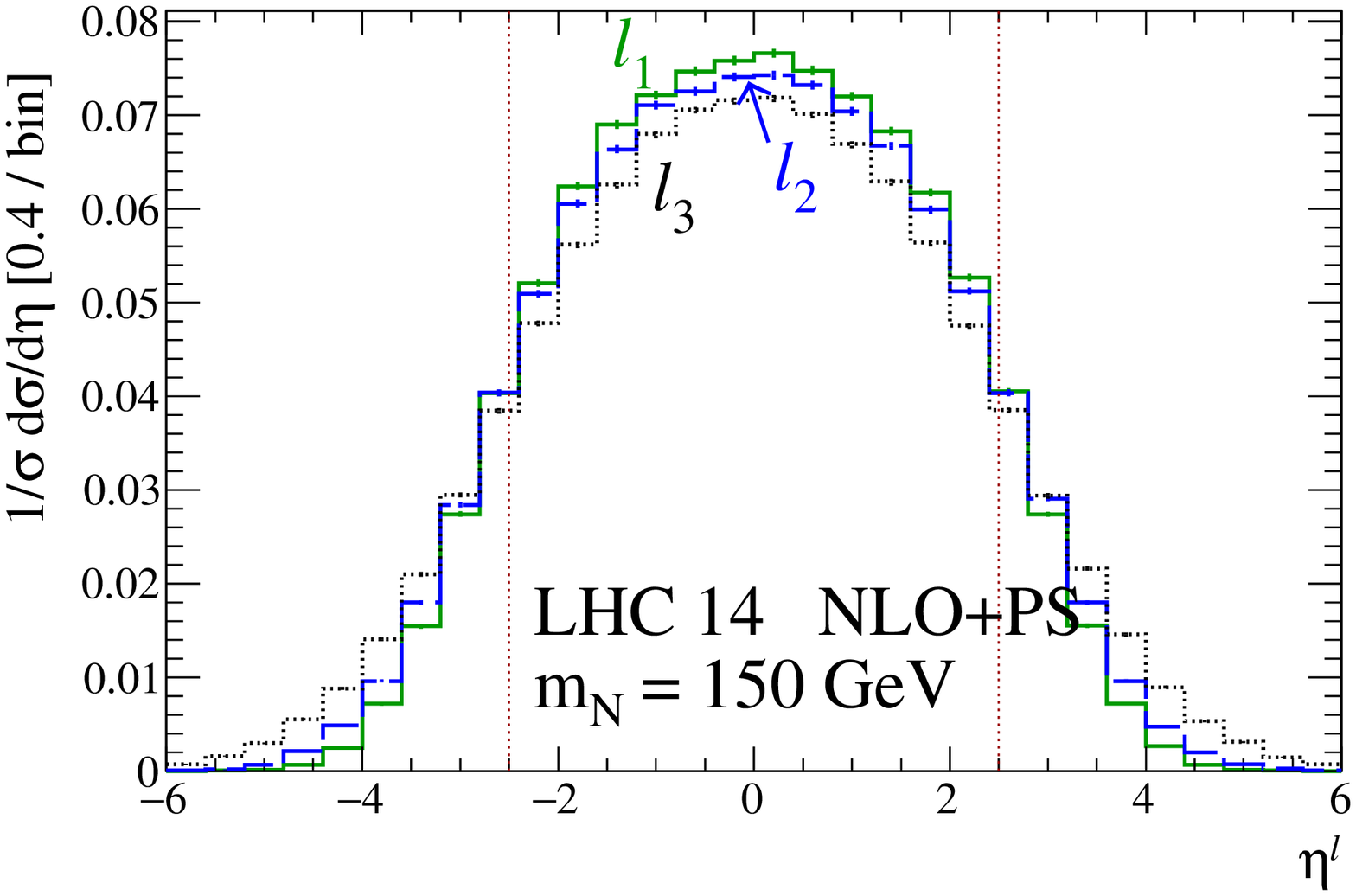}	}
\subfigure[]{\includegraphics[width=.45\textwidth]{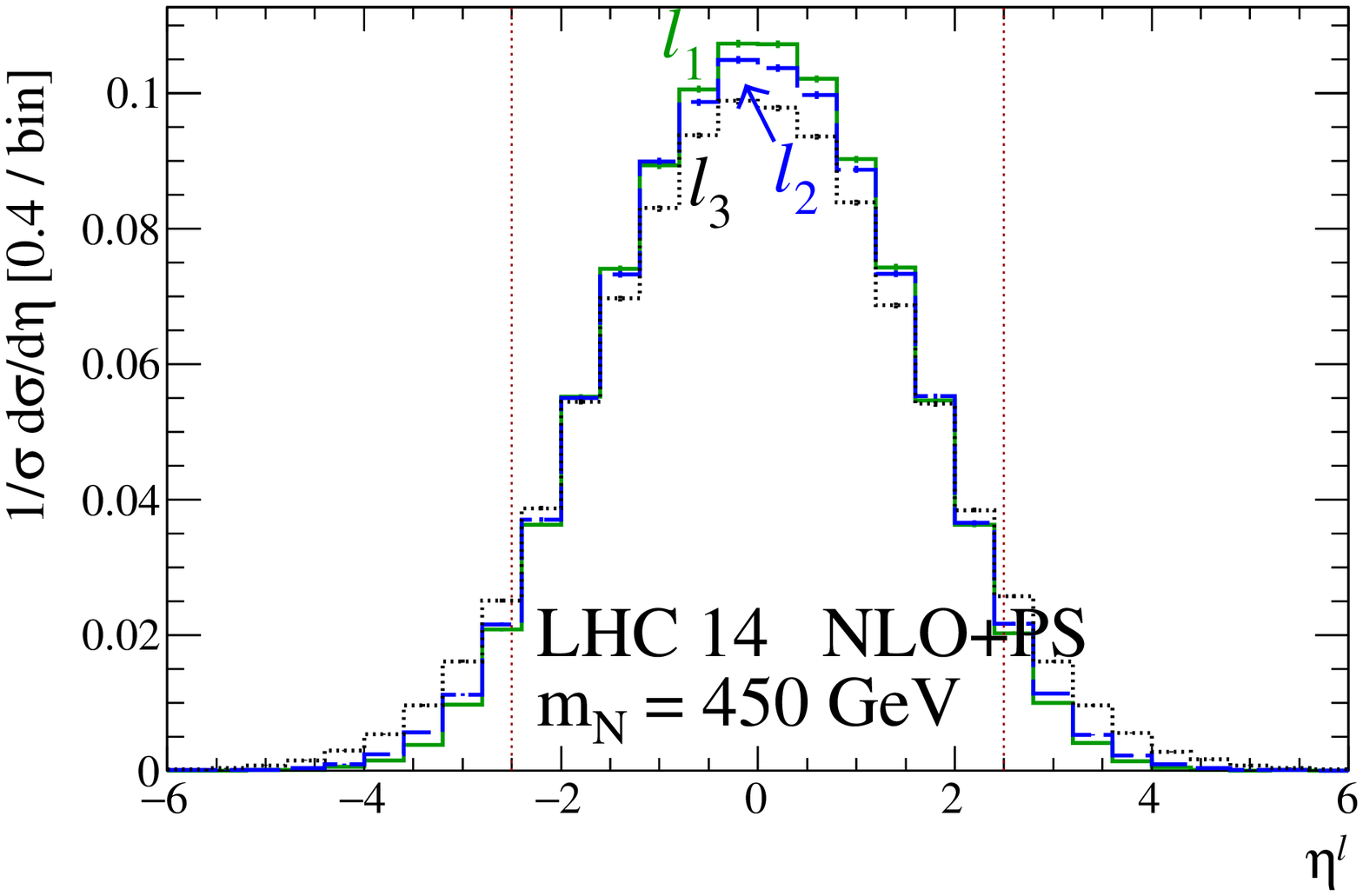}	}
\\
\subfigure[]{\includegraphics[width=.45\textwidth]{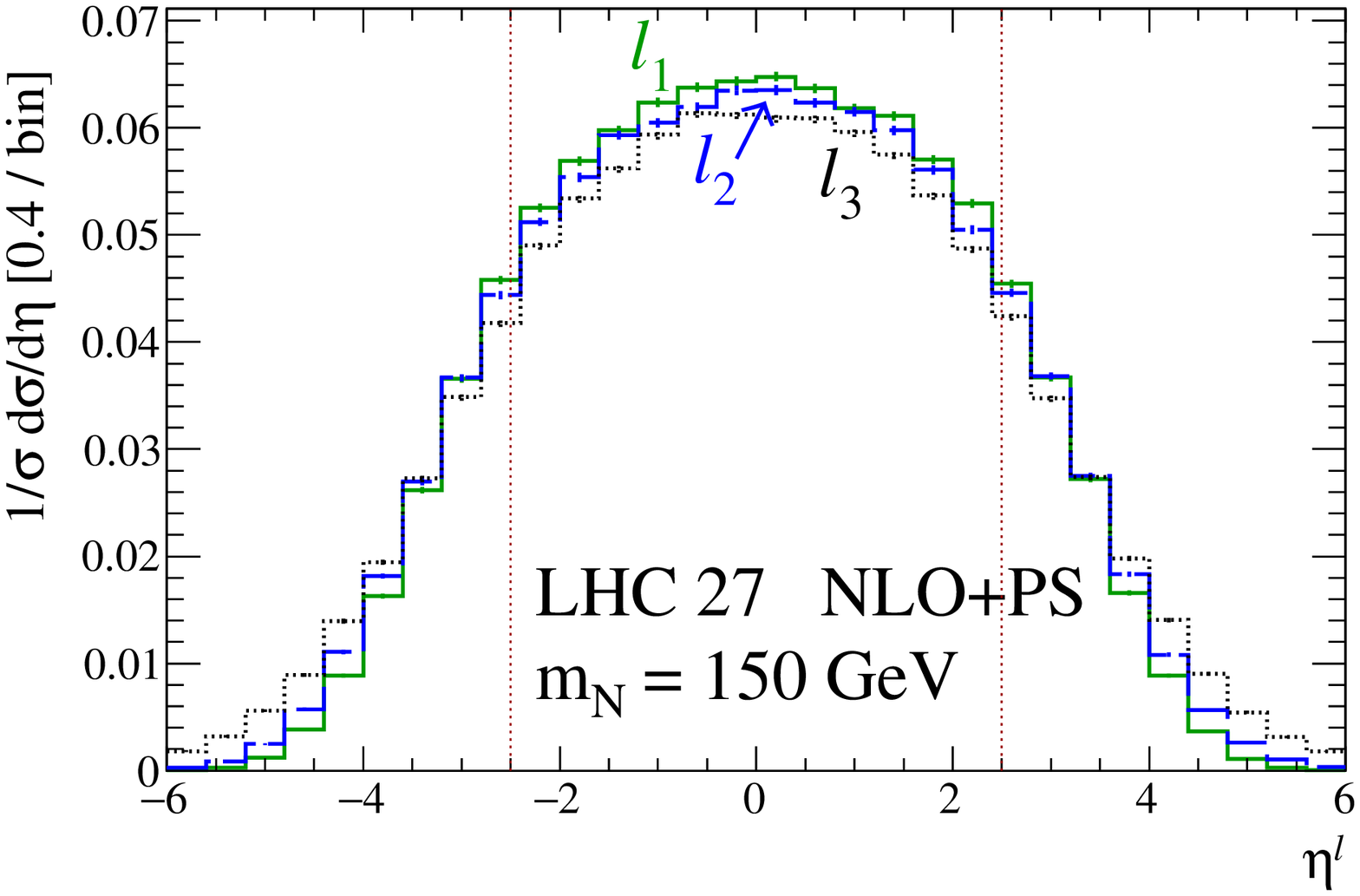}	}
\subfigure[]{\includegraphics[width=.45\textwidth]{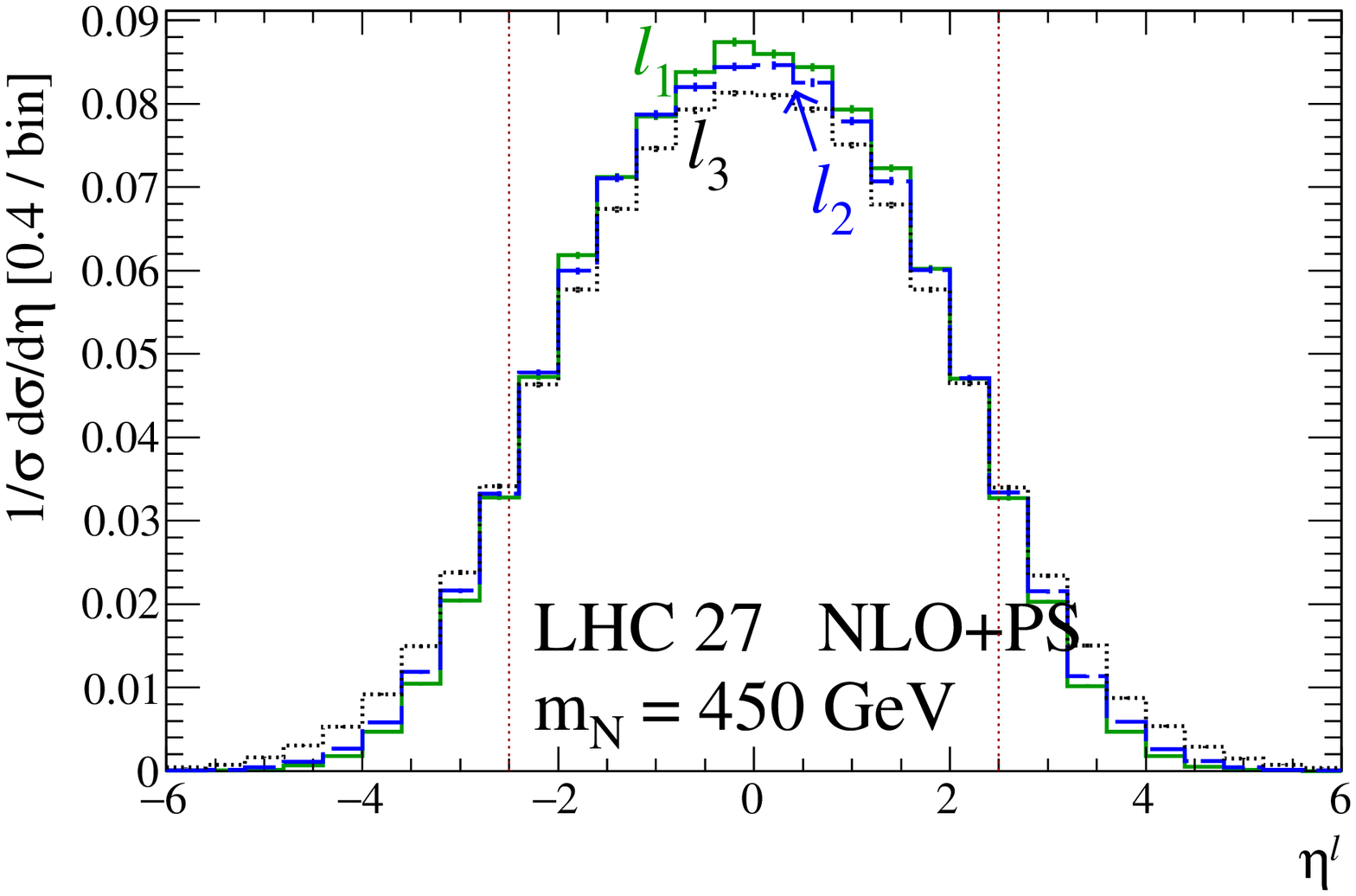}	}
\\
\subfigure[]{\includegraphics[width=.45\textwidth]{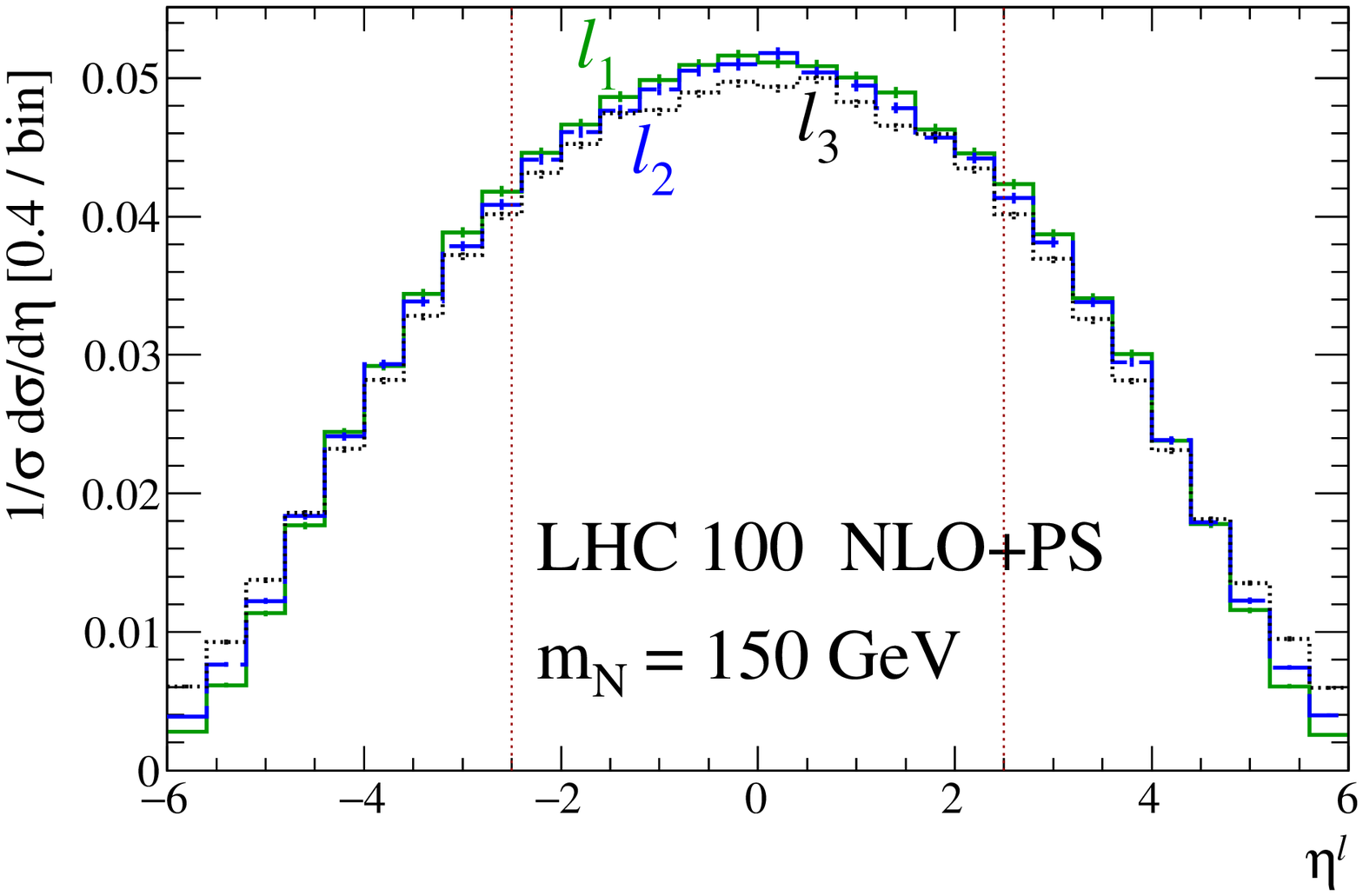}	}
\subfigure[]{\includegraphics[width=.45\textwidth]{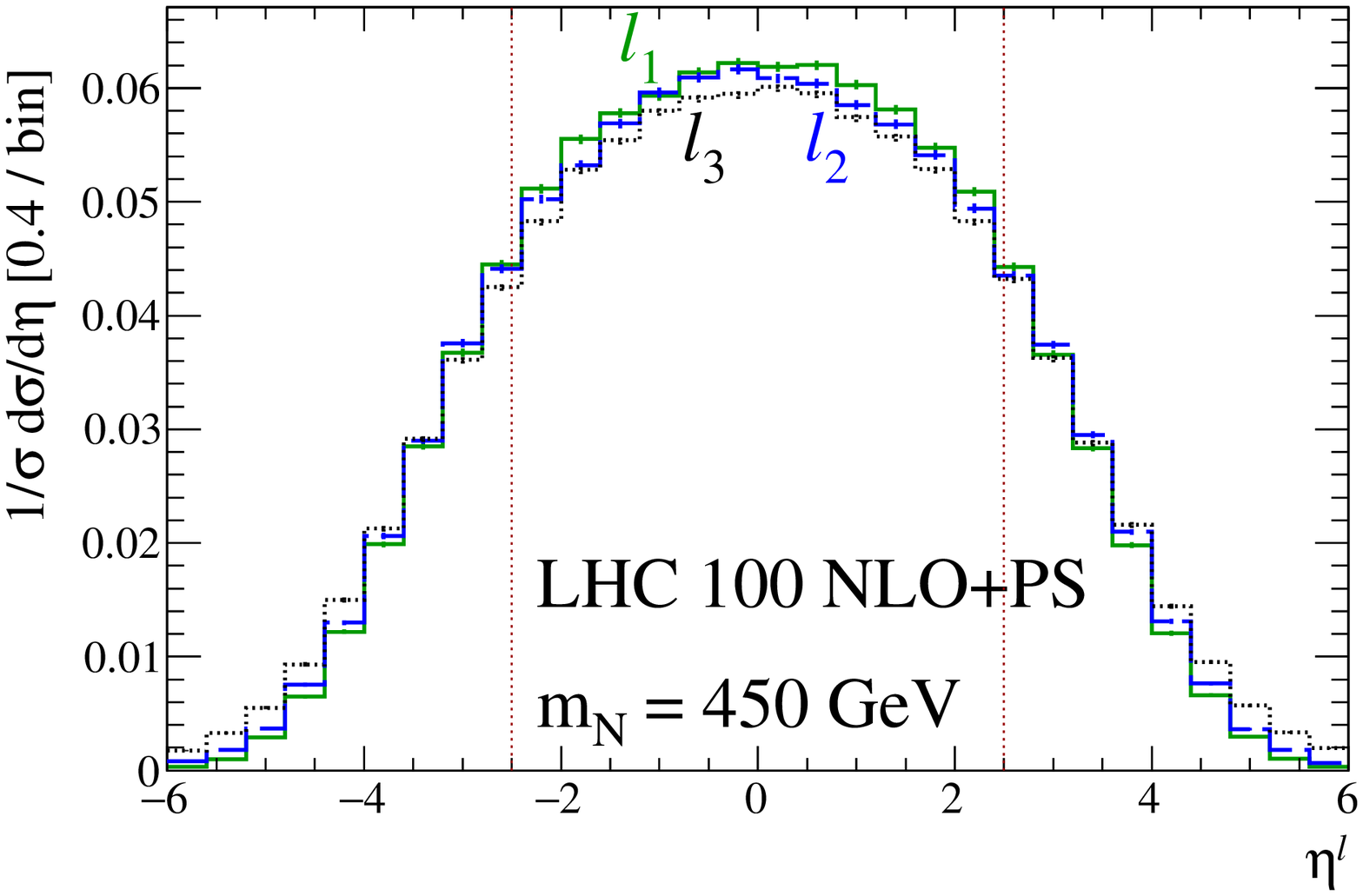}	}
\end{center}
\caption{
Same as Fig.~\ref{fig:partonKinElleta} but at NLO+PS accuracy with particle-level reconstruction.
}
\label{fig:particleKinElleta}
\end{figure*}

Focusing on the decay products of $N$, 
we plot in Figs.~\ref{fig:particleKinEllpT} and ~\ref{fig:particleKinElleta}, respectively, the $p_T$ and $\eta$ distributions of the three hardest, i.e., highest $p_T$, charged leptons,
in each event, at (a,b) $\sqrt{s}=14$, (c,d) 27, and (e,f) 100 TeV, for (a,c,e) $m_N=150$ and (b,d,f) 450 GeV.
Due to QED parton shower effects and leptonic decays of hadrons, more than three charged leptons may be present in each event.
Hence, here and for the remainder of the text, leptons $\ell_k$ with $k=1,\dots$, are ordered such that $p_T^{\ell_k}>p_T^{\ell_{k+1}}$.
(Jets are similarly ranked and labeled according to their $p_T$.)
Despite the potential increase in charged lepton multiplicity, one sees in Fig.~\ref{fig:particleKinEllpT} characteristic structures emerge in the $p_T$ curves.
This indicates that the three leading charged leptons largely originate from heavy, intermediate resonances, and are not low-energy, continuum contributions.
At $m_N=150$ GeV, $\ell_{1,\dots,3}$ cannot be reliably or uniquely associated with $\ell_N,\ell_W,$ or $\ell_\nu$,
which is expected as the LO curves in  Fig.~\ref{fig:partonKinEllpT} indicate that the $p_T$ are all highly comparable for fixed $m_N$.
For heavier $N$, the association is slightly better, with one observing that $p_T^{\ell_1}\sim m_N/2$ and $p_T^{\ell_2}\sim m_N/4$.
This suggests that the leading charged leptons may be more readily identified with $\ell_W$ and $\ell_\nu$, respectively, but still with large uncertainty.
In the pseudorapidity curves of Fig.~\ref{fig:particleKinElleta}, no discerning feature is present to discriminate or associate $\ell_{1,\dots,3}$ with $\ell_N,\ell_W,$ or $\ell_\nu$.
Even the slightly narrower/taller distribution for $\eta^{\ell_N}$ seen at LO is washed out at NLO+PS.
In short, upon discovery of an anomalous trilepton signature consistent with heavy neutrino production and decay,
it will be difficult to readily identify the event's three leading charged leptons with $\ell_N,\ell_W,$ or $\ell_\nu$.
To do so, one must resort to more sophisticated techniques, such as the Matrix Element Method, more complex observables, or additional hypotheses, such as lepton number/flavor conservation or violation.

Regardless of the ability to readily associate $\ell_{1,\dots,3}$ to $\ell_{N,W,\nu}$, 
with respect to the dependence on collider energy, both classes of the $p_T$ and $y/\eta$ distributions reflect the behavior observed in the LO distributions.
Subsequently, for a fixed heavy neutrino mass, one can conclude that since the shapes of the $p_T$ distribution for $N$ remain effectively insensitive to changes of $\sqrt{s}$,
so too do the distributions for $\ell_1,~\ell_2$, and $\ell_3$ at NLO+PS.
We now investigate how far this robustness against varying collider energy holds and 
consider more complex observables that are derived from the $p_T$ of the leading charged leptons.

\begin{figure*}[!t]
\begin{center}
\subfigure[]{\includegraphics[width=.48\textwidth]{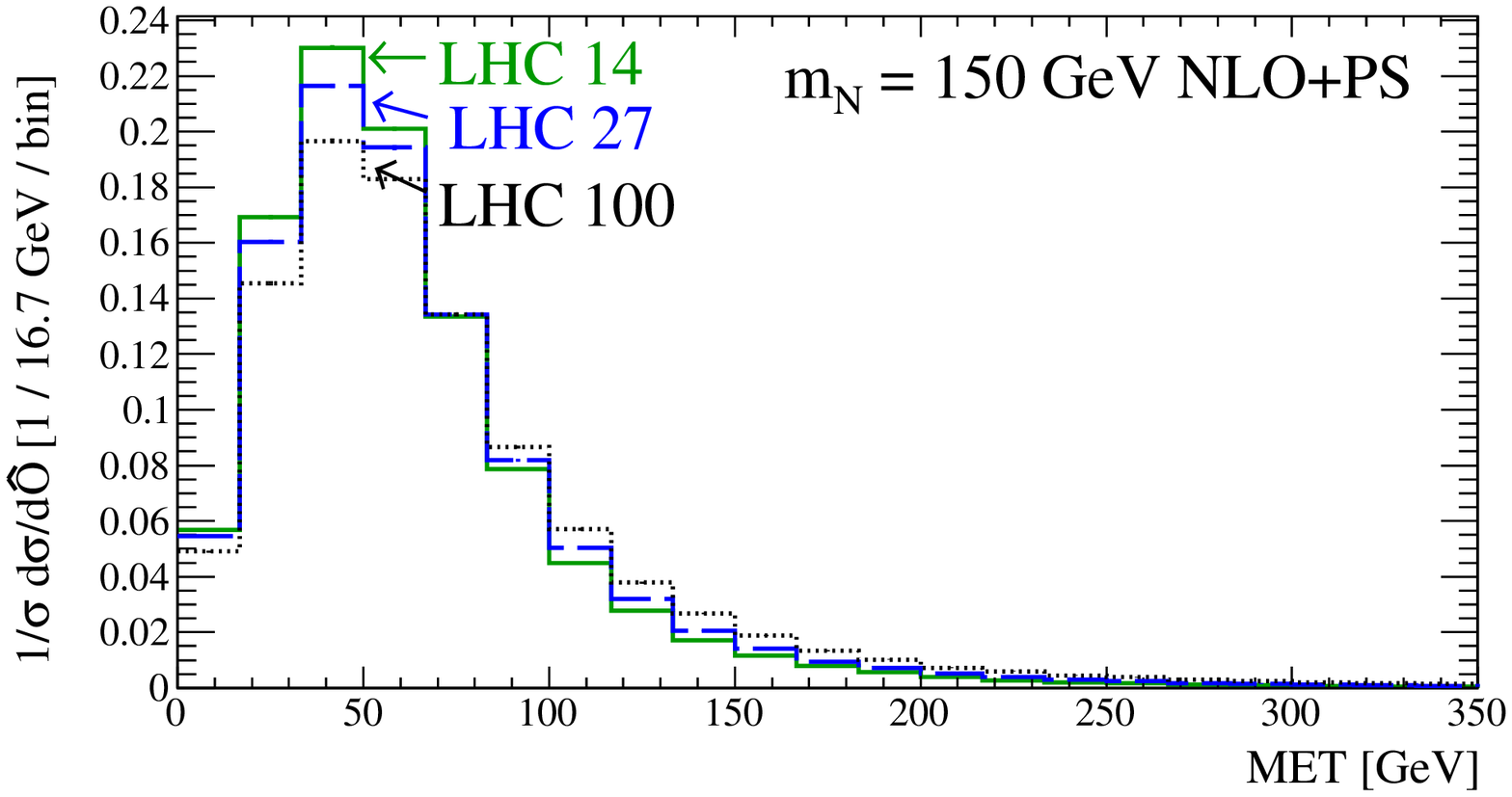}	}
\subfigure[]{\includegraphics[width=.48\textwidth]{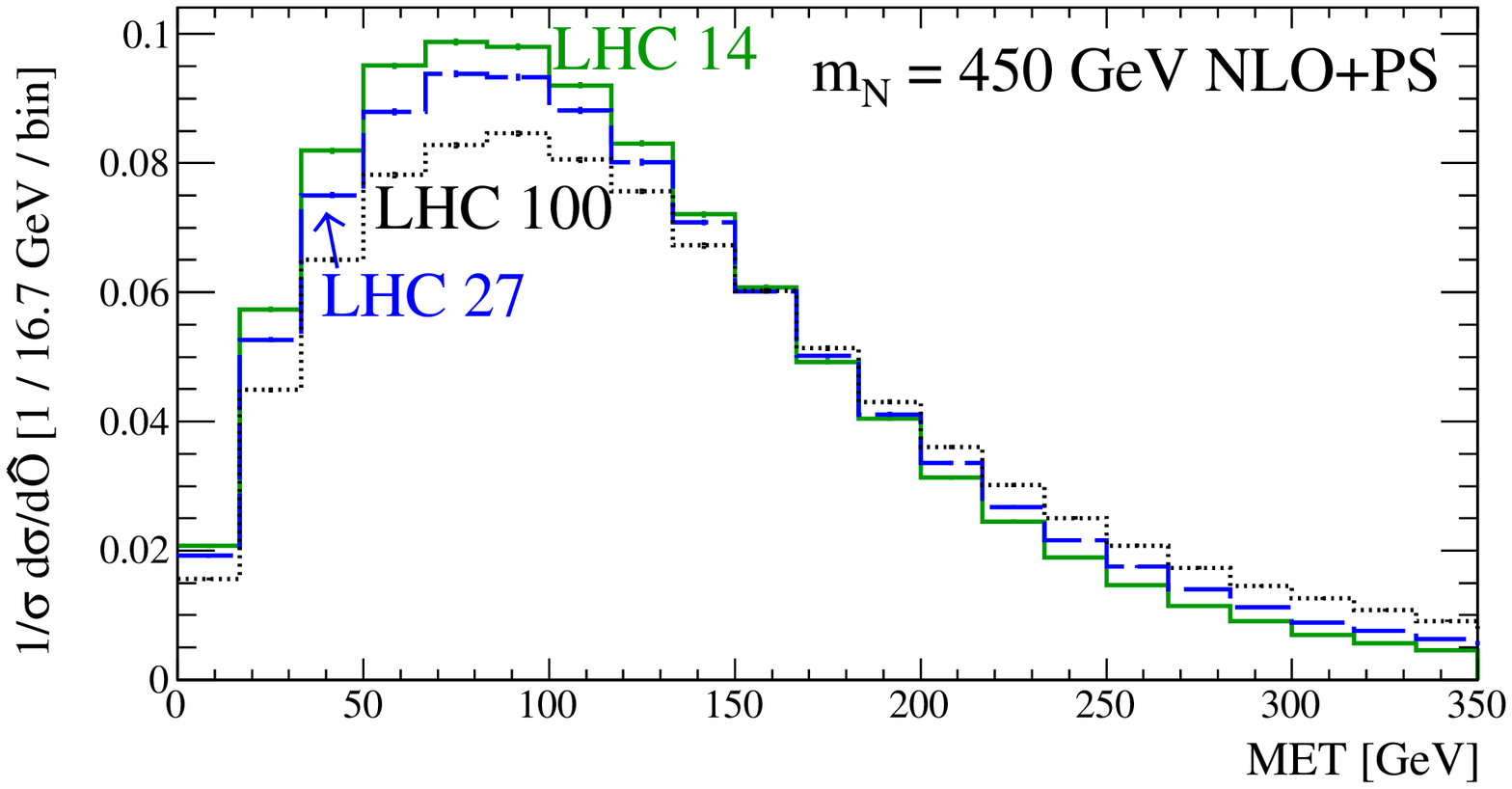}	}
\end{center}
\caption{
Normalized particle-level missing transverse momentum (MET) distributions for the $pp \to N\ell \to 3\ell\nu$ process via the DY mechanism at NLO+PS,
assuming representative neutrino mass (a) $m_N = 150$ and (b) 450 GeV, at $\sqrt{s} = 14$ (solid), 27 (dash), and 100 (dot) TeV.
}
\label{fig:particleKinMET}
\end{figure*}

In addition to the three charged leptons, the collider process of Eq.~\ref{eq:partonDY} also contains final-state light neutrinos originating from the $N\to \ell W \to 2\ell \nu$ decay.
As mentioned above, at our present accuracy, light neutrinos can originate from weak decays of hadrons.
(At the detector-level, there are also sizable contributions from finite detector resolution and, to a lesser extent, finite coverage; see Sec.~\ref{sec:smearing}.)
The collective presence of these particles is inferred from the transverse momentum imbalance of all visible, final-state particles.
Accordingly, the 2-vector $\not\!\vec{p}_T$ and its magnitude $\met$ are defined as
\begin{equation}
 \not\!\vec{p}_T = - \sum_{k\in\{visible\}}\vec{p}_T^{~k}, \quad\text{with}\quad  \met~\equiv \vert \not\!\vec{p}_T \vert,
\end{equation}
where the summation runs over all visible final states regardless of their hardness $(p_T)$.
Disregarding objects below a particular $p_T$ threshold can introduce distortions in the \met~distribution of that same order and worsen the perturbative stability~\cite{Fuks:2017vtl}.

In Fig.~\ref{fig:particleKinMET}, we plot the particle-level \met~for (a) $m_N=150$ and (b) 450 GeV, at $\sqrt{s}=14$, 27, and 100 TeV.
Immediately, one sees that~\met, a quantity built from both leptonic and hadronic transverse momenta, 
more or less aligns with na\"ive $1\to2\to3$ kinematics for $N$ decays, which stipulate that the light neutrino $p_T$ and $E$ scale as
\begin{equation}
\met =  \vert \vec{p}_T^{~\nu} \vert \lesssim E^\nu = \frac{m_N}{4}\left(1 + \frac{M_W^2}{m_N^2}\right) \sim 50~(115)\GeV\quad\text{for}\quad 150~(450)\GeV.
 \label{eq:metExpectation}
\end{equation}
Like the $p_T^{\ell_X}$ distributions in Fig.~\ref{fig:particleKinEllpT}, the structured behavior indicates that the distributions are driven by heavy, intermediate, resonant states.
In contrast to the $p_T$ of the leading charged leptons, we see a stronger collider energy dependence.
The \met~peak drops \confirm{10-to-20\%} from its maximum as one goes from 14 TeV to 100 TeV, for both $m_N =150$ and 450\GeV.
The location of the maximum, however, does not noticeably change.
This broadening can be attributed to two effects:
(i) At higher collider energies, more low-$p_T$ hadrons are generated due to the larger available phase space.
In turn, more hadrons can undergo leptonic weak decays, which in turns generates more light neutrinos.
(ii) Deviations from Eq.~\ref{eq:metExpectation} also originate from neutrinos carrying momentum in the longitudinal direction,
which worsens with increasing $\sqrt{s}$ due to increasingly forward valence quark-sea antiquark annihilations.
This can be seen in the rightward shifts of the \met~tail.
So while the $p_T$ of the light neutrino appearing in the $N \to \ell W \to 2\ell\nu$ decay may remain robust against $\sqrt{s}$, its proxy $(\met)$ is slightly less robust due to hadronic contamination.

One should caution on the strength of the conclusion drawn from Fig.~\ref{fig:particleKinMET}.
Here we only show particle-level \met~and do not (yet) account for detector effects.
Higher collider energies give rise to a higher multiplicity of jets, which can cause additional \met~through momentum mis-measurement and finite fiducial coverage.

\begin{figure*}[!t]
\begin{center}
\subfigure[]{\includegraphics[width=.48\textwidth]{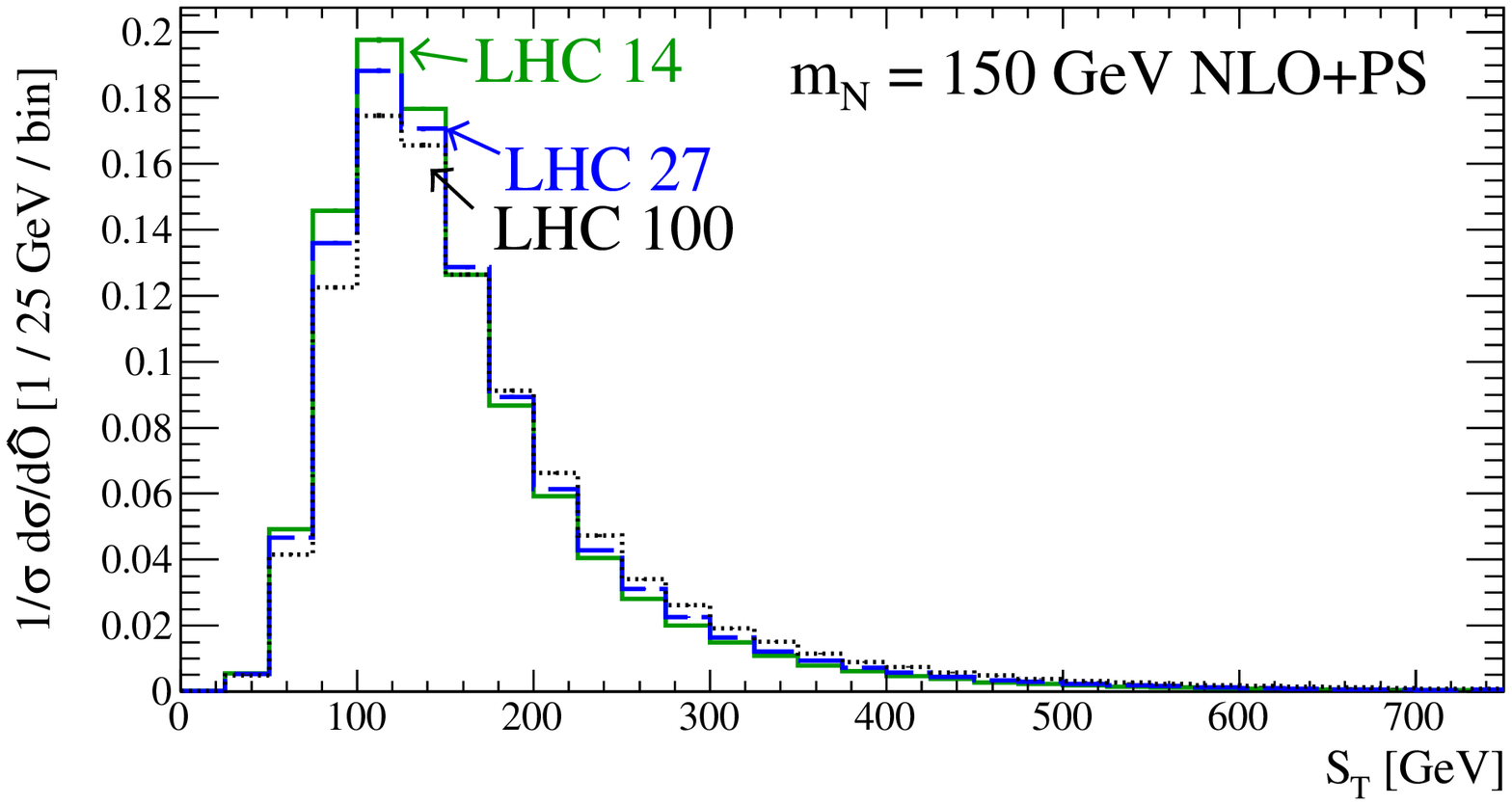} }
\subfigure[]{\includegraphics[width=.48\textwidth]{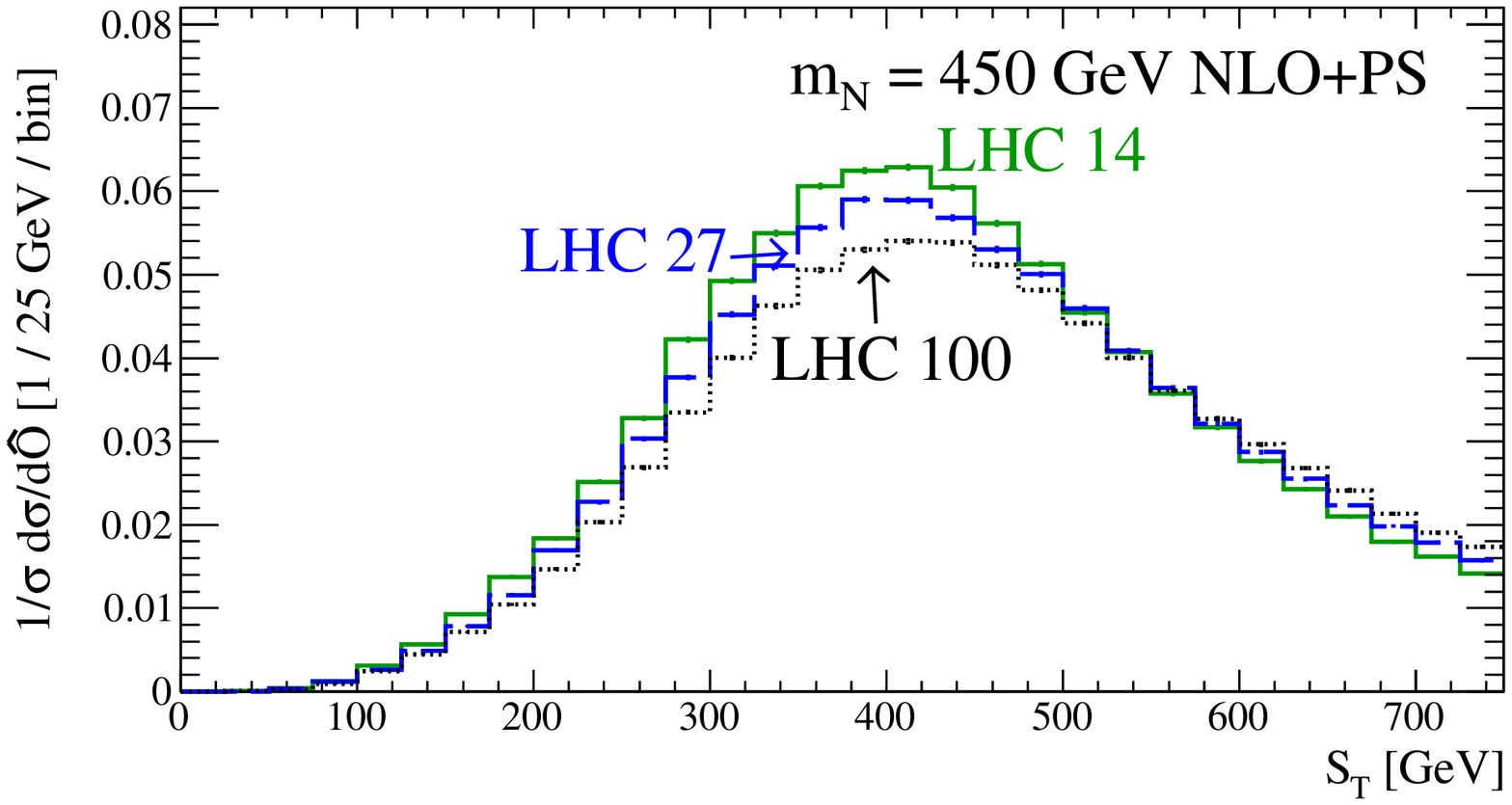} }
\\
\subfigure[]{\includegraphics[width=.48\textwidth]{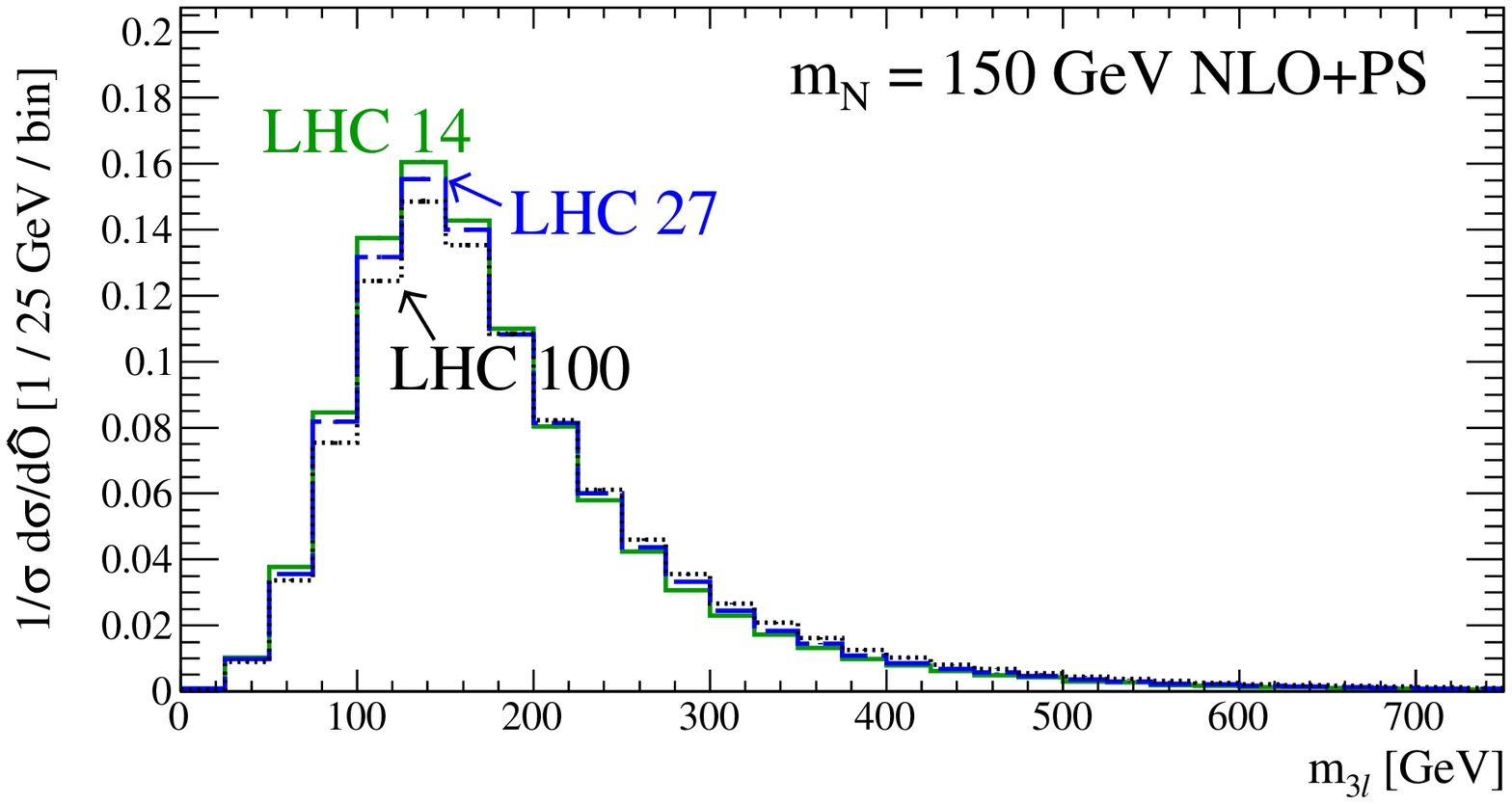} }
\subfigure[]{\includegraphics[width=.48\textwidth]{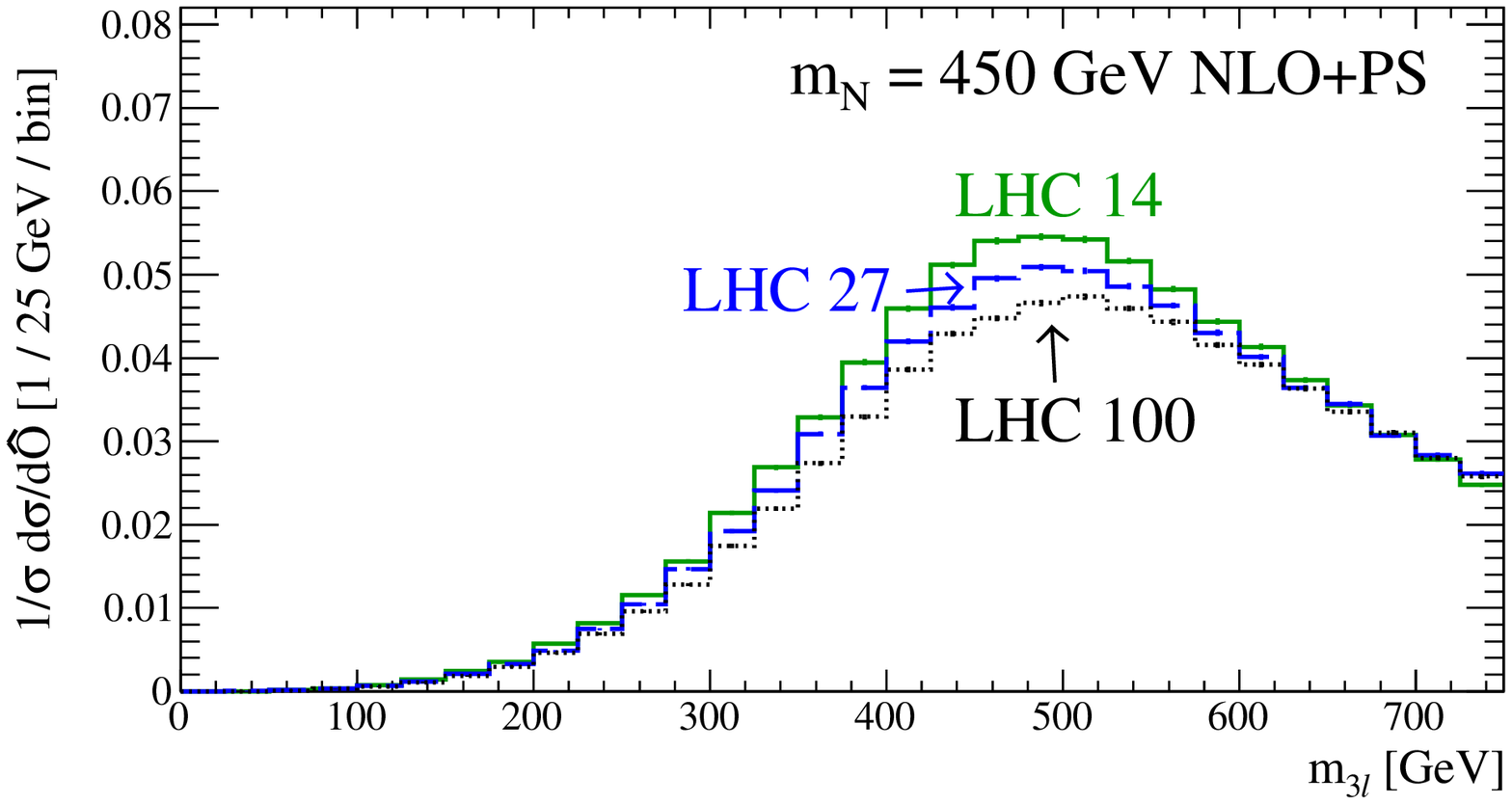} }
\end{center}
\caption{
Same as Fig.~\ref{fig:particleKinMET} but for (a,b) $S_T$ and (c,d) $m_{3\ell}$ of the three leading charged lepton.
}
\label{fig:particleKin_LepMassScale}
\end{figure*}

To investigate the collider energy dependence on global leptonic activity, we consider two representative measures: 
the (exclusive) scalar sum of leptonic $p_T$, $S_T$, defined as
\begin{equation}
 S_T = \sum_{\ell_k\in\{\ell_1,\ell_2,\ell_3\}} \vert \vec{p}_T^{~\ell_k}\vert,
 \label{eq:defST}
\end{equation}
with the sum running only over the three leading charged leptons in the event, as well as the invariant mass of the same three charged leptons:
\begin{equation}
m_{3\ell} = \sqrt{\left(p_T^{\ell_1} + p_T^{\ell_2} + p_T^{\ell_3} \right)^2}.
\label{eq:defM3l}
\end{equation}
In Fig.~\ref{fig:particleKin_LepMassScale}, we show $S_T$ for (a) $m_N=150\GeV$ and (b) 450\GeV, and likewise $m_{3\ell}$, respectively, in (c) and (d).
As anticipated, only a slight broadening of $S_T$ and $m_{3\ell}$ occurs with increasing collider energies.
We observe that all four peaks only reduce  \confirm{5 to 10\%} from their maximum as one goes to 27 or 100 TeV.
Like \met, ~we see that the location of the peak remains unchanged.
The weaker dependence on $\sqrt{s}$ is attributed to the neglect of charged leptons with $p_T < p_T^{\ell_3}$.
As the hard $pp\to N\ell \to 3\ell X$ process only contains three charged leptons in the final state, 
additional charged leptons in the event must necessarily originate from the decays of the low-$p_T$ hadrons in the beam remnant and possibly electromagnetic ISR/FSR.
As we saw in Fig.~\ref{fig:particleKinMET}, these additional contributions have a collider energy dependence, 
and therefore neglecting their contribution to $S_T$ and $m_{3\ell}$ helps fortify these observables' insensitivity to $\sqrt{s}$.
It is noteworthy that both observable peaks near $S_T,~m_{3\ell}\sim m_N$.
This is a somewhat a fortuitous accident and results from the fact that the momenta of the three charged leptons, as observed in Fig.~\ref{fig:partonKinEllpT}, scale as
\begin{equation}
 p_T^{\ell_N}\sim\frac{m_N}{5}, \quad
 p_T^{\ell_W}\sim\frac{m_N}{2}\left(1-\frac{M_W^2}{m_N^2}\right), \quad
 p_T^{\ell_\nu}\sim\frac{m_N}{4}\left(1+\frac{M_W^2}{m_N^2}\right).
 \label{eq:pTlXSummary}
\end{equation}
For a heavy neutrino with masses that we consider, this means for $S_T$ we have
\begin{eqnarray}
 S_T^N &\sim& \frac{m_N}{5} + \frac{m_N}{2}\left(1-\frac{M_W^2}{m_N^2}\right) + \frac{m_N}{4}\left(1+\frac{M_W^2}{m_N^2}\right)  
 \\
& \sim& \left(\frac{19}{20}\right)m_N - \frac{m_N}{4}\left(\frac{M_W}{m_N}\right)^2 \sim m_N.
 \label{eq:STsignal}
\end{eqnarray}
Whereas $S_T$ slightly undershoots $m_N$, we see that $m_{3\ell}$ lurches above it due to the additional longitudinal momentum that goes into $m_{3\ell}$.
This also explains why the $m_{3\ell}$ distributions are broader than the $S_T$ curves.
Otherwise, there is little difference between the observables.
However,  due to this broadening, we prefer to build and limit our signal analysis to quantities constructed only from transverse momenta.
We stress though that while $S_T$ and $m_{3\ell}$ are  numerically comparable to $m_N$ for the CC DY process, this does not hold for all production mechanisms.
The $W\gamma$ fusion channel, for example, kinematically favors lower $p_T^{\ell_N}$ due to an initial $\gamma\to\ell\ell^*$ splitting~\cite{Alva:2014gxa}.
Nonetheless, taking for example the extreme limit of $(p_T^{\ell_N}/m_N)\to0$, one still observes that $S_T^{\rm VBF}\gtrsim 3m_N/4$.
Hence, the more broadly applicable statement is that $S_T$ scales proportionally with $m_N$ in a somewhat universal fashion, a fact that we will exploit in our signal analysis.

From both a search-strategy and a properties-measurement perspective, being able to reconstruct the mass of $N$ is highly desirable.
The capability, however, is complicated by the light neutrino in the $N\to2\ell\nu$ decay chain,
which makes it impossible \textit{a priori} to reconstruct $N$'s 4-momentum, 
and hence its invariant mass, from the charged leptons and MET momentum vectors alone in $pp$ collisions.
And as seen in Fig.~\ref{fig:particleKin_LepMassScale}, leptonic observables such as $S_T$ and $m_{3\ell}$ 
do not appear to be sufficiently robust estimates of heavy neutrino masses due to their broadness.
Though interestingly, like the top quark and Higgs boson, it is possible to build a more reliable measure of  the invariant mass of $N$.
We do this by exploiting simultaneously the $1\to2\to3$ decay structure of $N$ and that the intermediate $W$ boson is largely on its mass shell for the range of $m_N$ we consideration.
For $1\to3$ and $1\to4$-body decays, a class of transverse mass observes exist~\cite{Barger:1983wf,Barger:1988mr,Han:2005mu} 
that are essentially multi-body extensions of the canonical, two-body transverse mass variable used in $W$ boson mass measurements~\cite{Barger:1983wf}.
Differences in the observables are based on the multiplicity of final-state neutrinos and whether decays are sequential or in parallel.

\begin{figure*}[!t]
\begin{center}
\subfigure[]{\includegraphics[width=.45\textwidth]{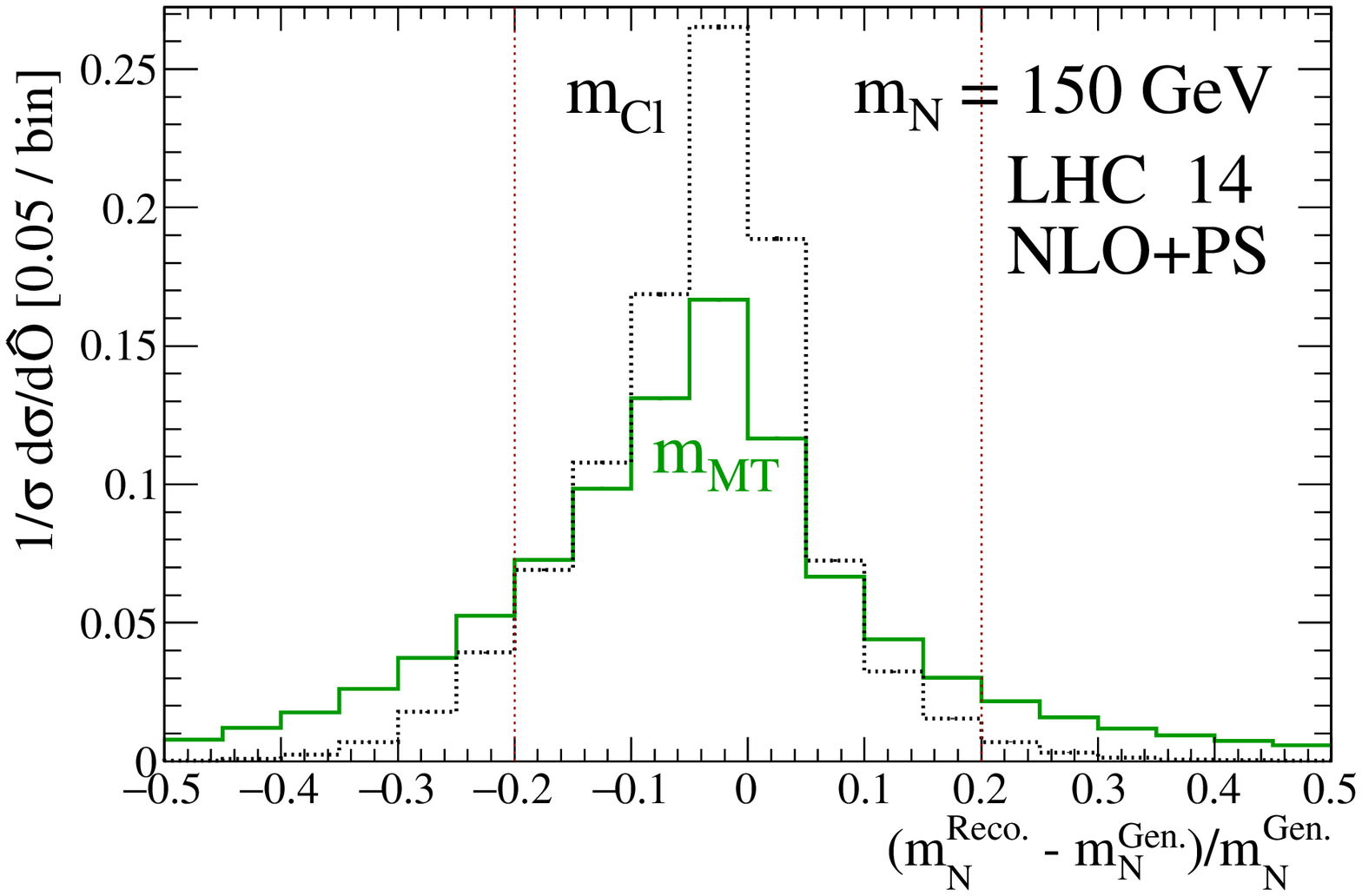}	}
\subfigure[]{\includegraphics[width=.45\textwidth]{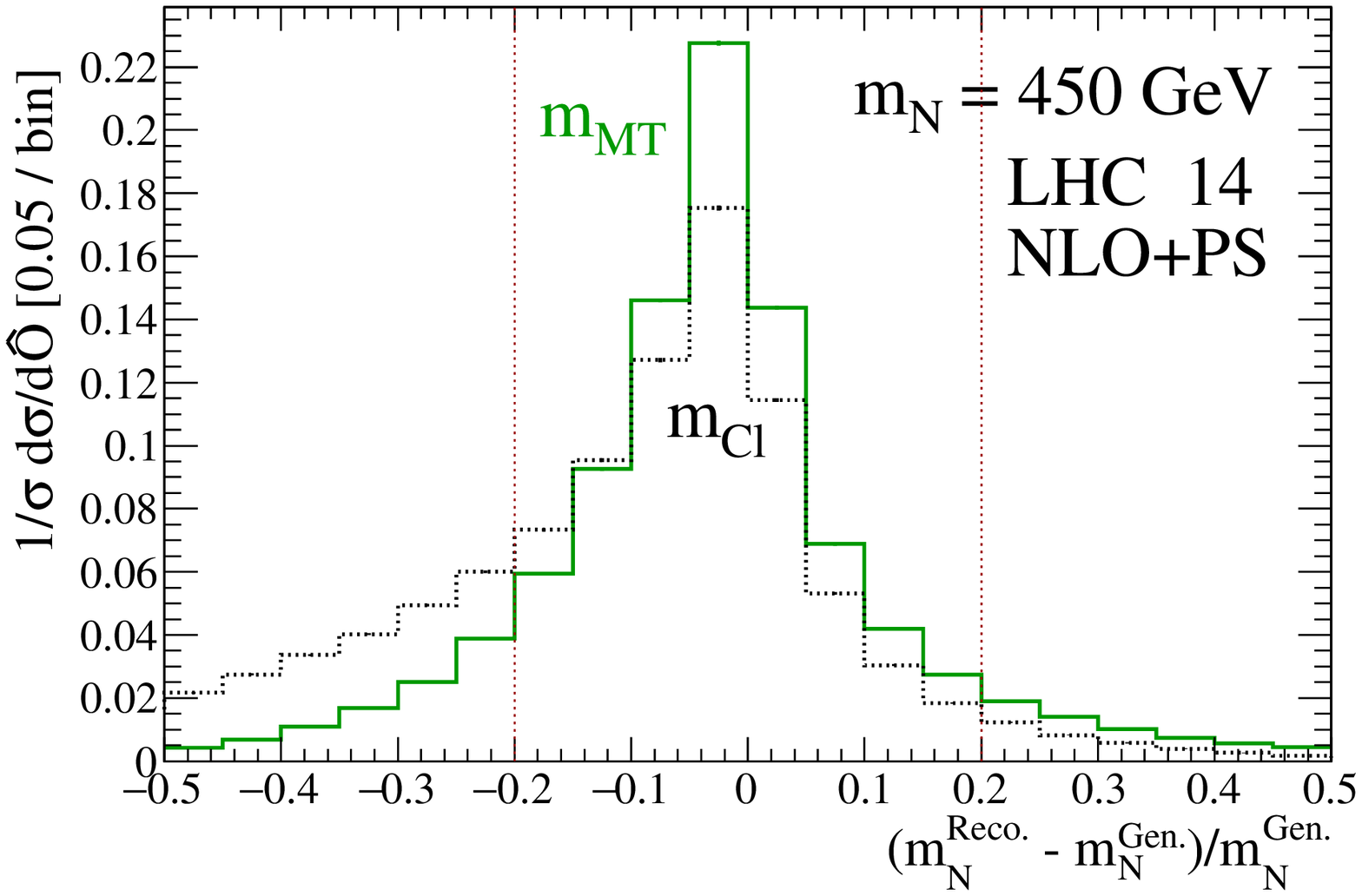}	}
\\
\subfigure[]{\includegraphics[width=.45\textwidth]{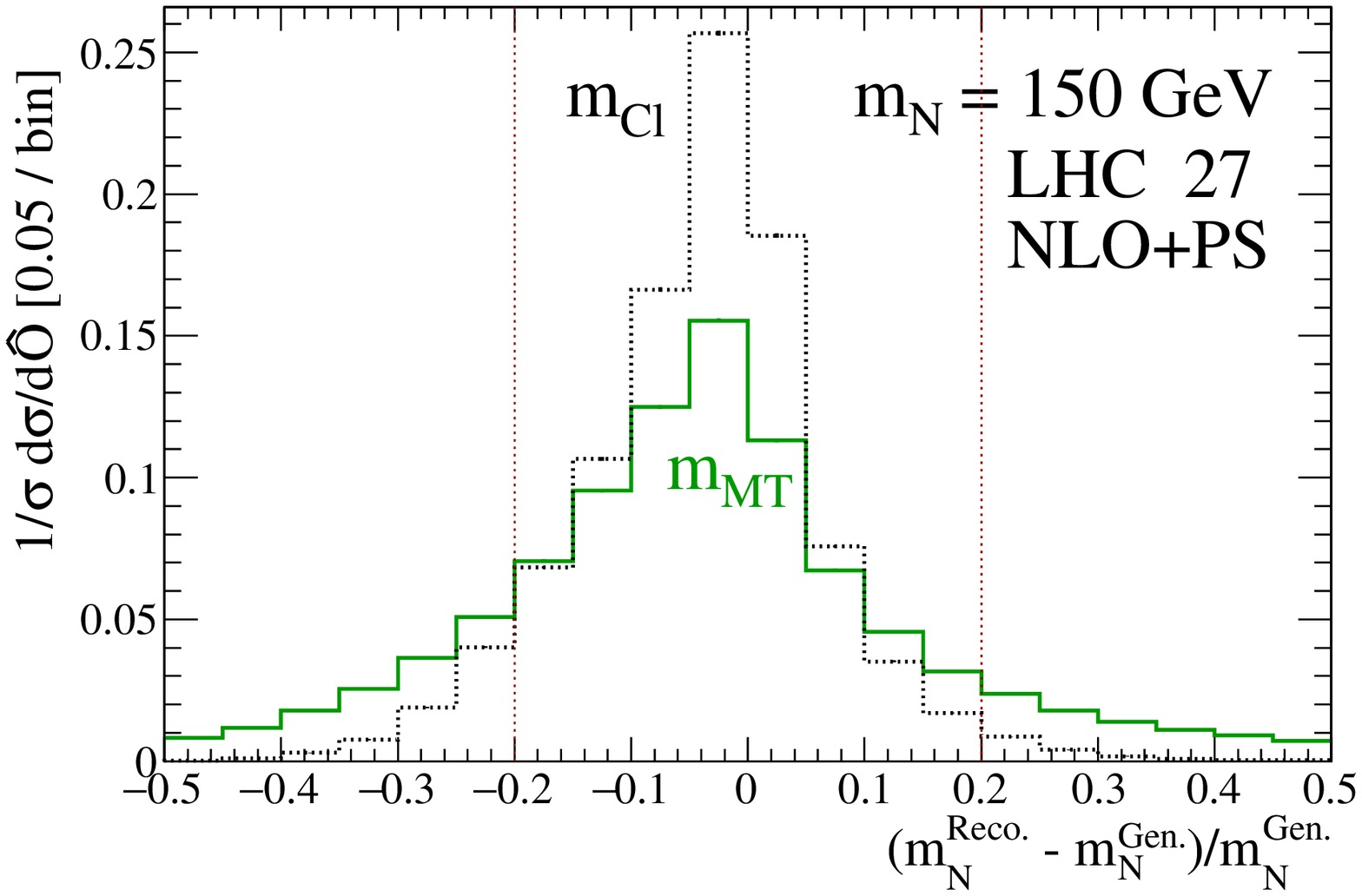}	}
\subfigure[]{\includegraphics[width=.45\textwidth]{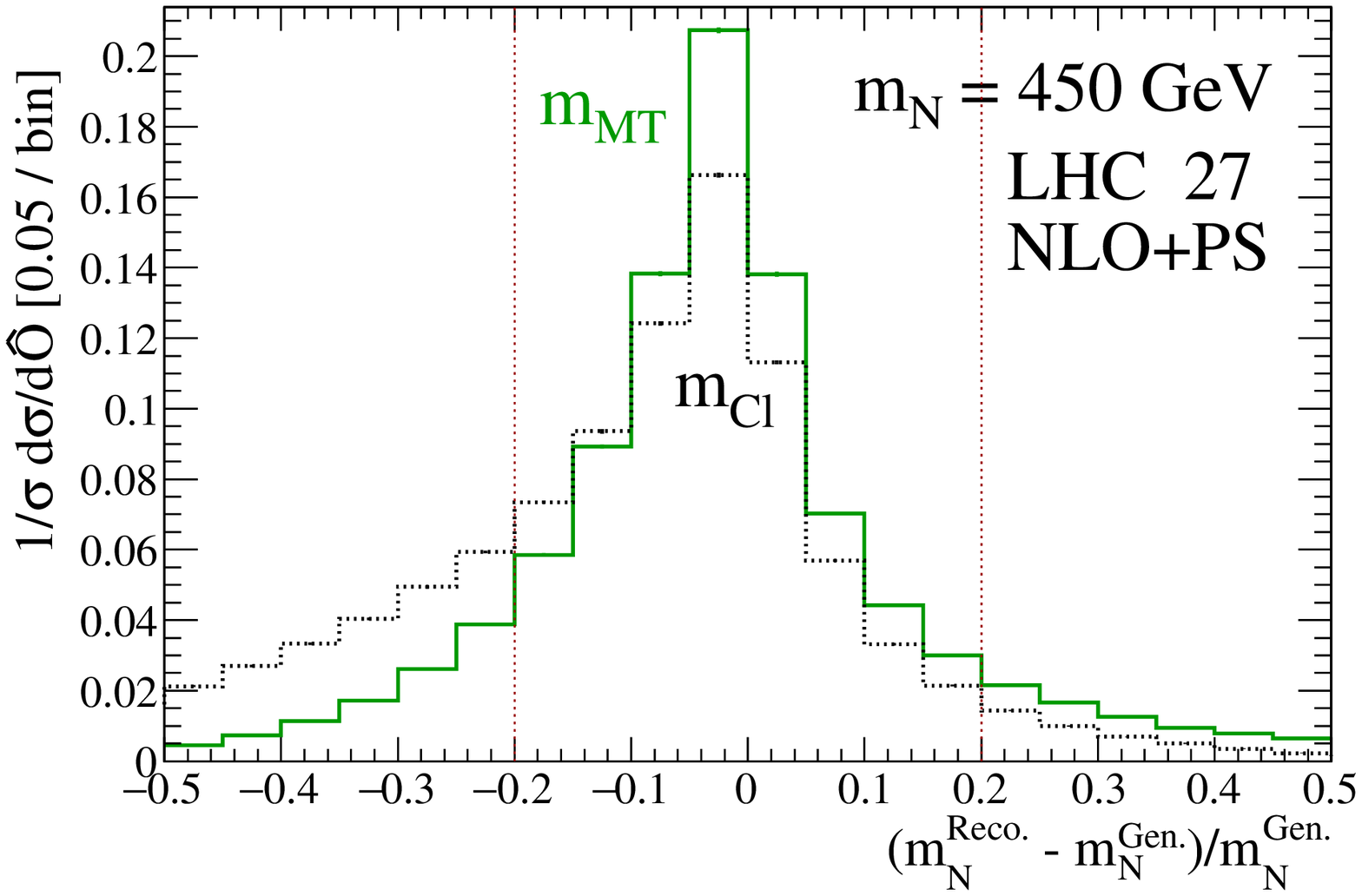}	}
\\
\subfigure[]{\includegraphics[width=.45\textwidth]{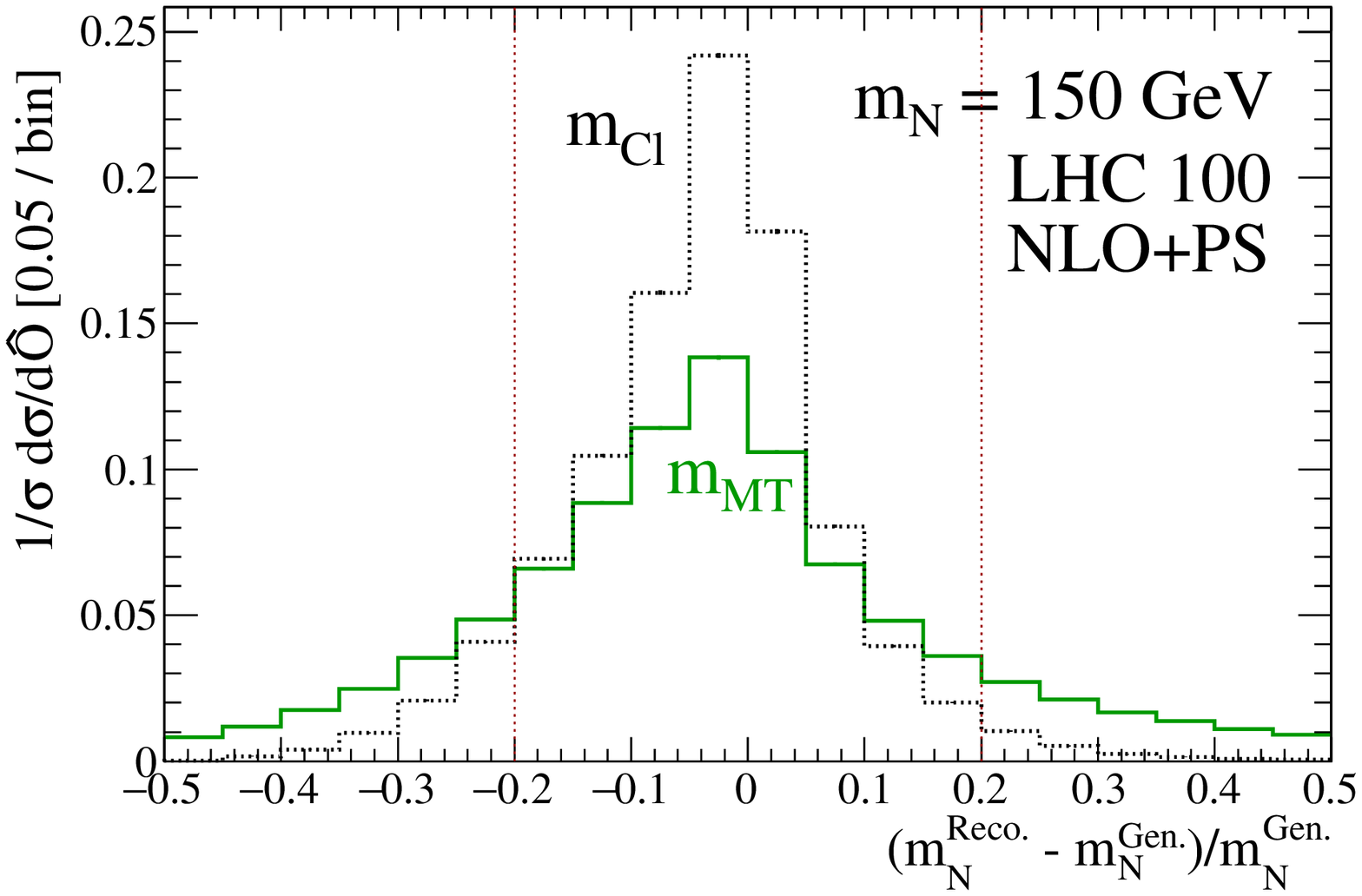}	}
\subfigure[]{\includegraphics[width=.45\textwidth]{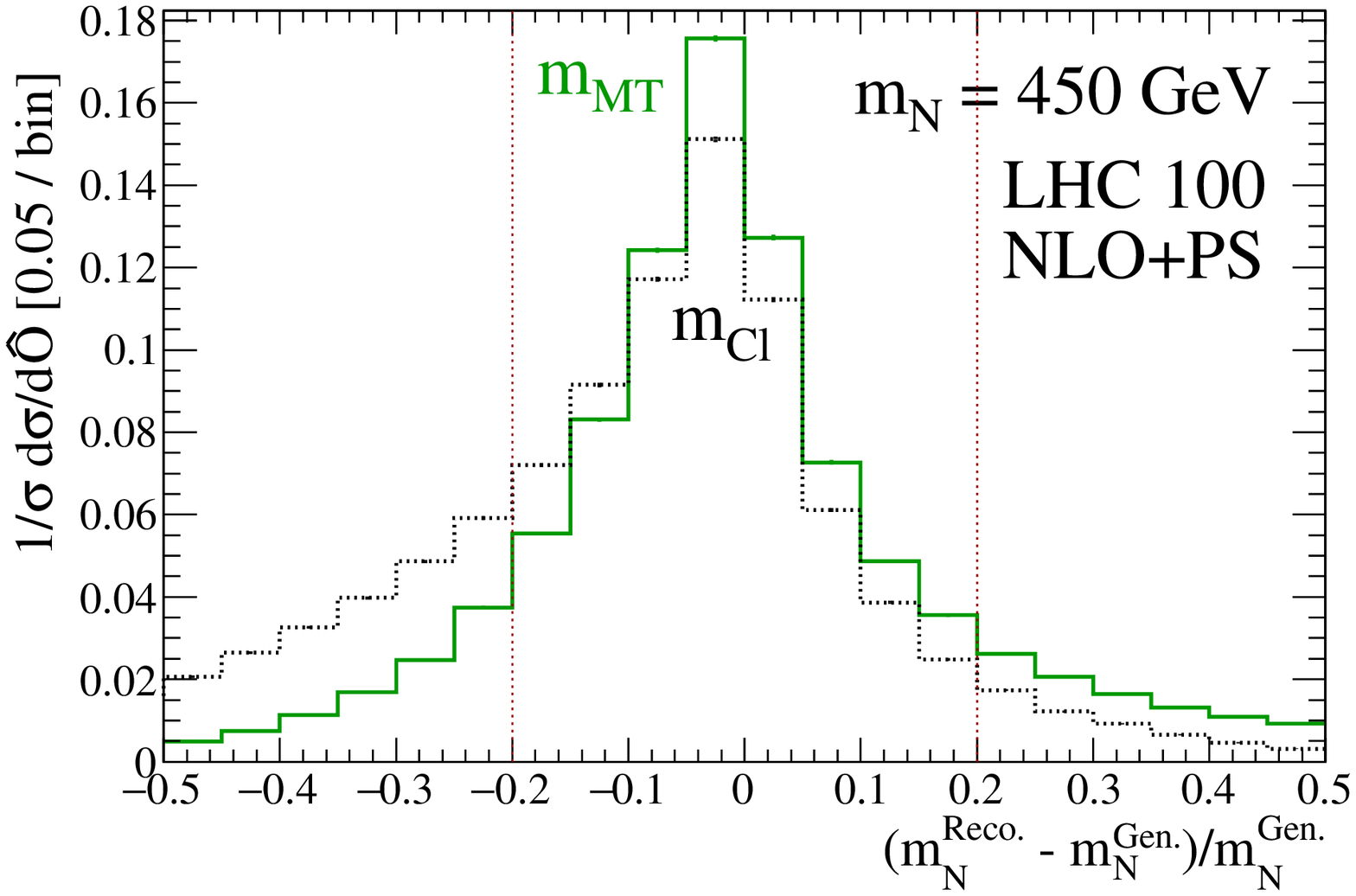}	}
\end{center}
\caption{
The relative difference $(\delta m/m)$ between the value of the reconstructed mass $m_N^{\rm Reco.} \in\{\hat{m}_{MT}~\text{(solid)}, \hat{m}_{Cl}~\text{(dot)}\}$ 
and $m_N^{\rm Gen.}$m for (a,c,e) $m_N=150\GeV$ and (b,d,f) 450\GeV, at (a,b) $\sqrt{s}=14$, (c,d) 27, and (e,f) 100\TeV, and at NLO+PS.
Guide lines placed at $\pm20\%$.
}
\label{fig:particleKin_dMass}
\end{figure*}

For our $N \to \ell_W W \to \ell_W \ell_\nu \nu$ configuration, two\footnote[2]{
Using comparable assumptions, techniques pioneered for leptonic and dileptonic decays of top quark 
pairs~\cite{Dalitz:1991wa,Sonnenschein:2005ed,Sonnenschein:2006ud,Betchart:2013nba} 
may also be applicable to heavy neutrinos. Such investigations are encouraged.
} 
possible mass-estimates can serve our purpose:
the multi-body transverse mass $m_{MT}$ (labeled as $M_{T,WW}'$ in Ref.~\cite{Han:2005mu}),
\begin{equation}
 m_{MT}^2 = \left(\sqrt{p_T^2(\ell_W) + m(\ell_W)^2} + \sqrt{p_T^2(\ell_\nu,\not\!\!\vec{p}_T)^2 + M_W^2} \right)^2 
 - \left(\vec{p}_T(\ell_W,\ell_\nu) + \not\!\vec{p}_T\right)^2,
 \label{eq:DefMultibodyMT}
\end{equation}
and the multi-body cluster mass $m_{Cl}$ (labeled as $M_{C,WW}$ in Ref.~\cite{Han:2005mu}),
\begin{equation}
 m_{Cl}^2 = \left(\sqrt{p_T^2(\ell_W,\ell_\nu) + m(\ell_W,\ell_\nu)^2} + \not\!\! p_T\right)^2 
 - \left(\vec{p}_T(\ell_W,\ell_\nu) + \not\!\!\vec{p}_T\right)^2.
\end{equation}
Considering the assumptions that go into constructing these quantities, it not at all obvious whether one observable is (or should be) a better estimate of $m_N$.
Therefore, we compute both: 
At the particle-level, charged leptons ranked according to $p_T$ cannot readily be associated with $\ell_N,\ell_W,$ or $\ell_\nu$ (see Fig.~\ref{fig:particleKinEllpT}).
By electric charge conservation, however, 
we know that the sum of the three's electric charges must add to $Q_{3\ell} = \sum_\ell^3 Q_\ell = \pm 1$ and that the charges of $\ell_W$ and $\ell_\nu$ must cancel. 
Hence, without guidance, there are in principle two (four) permutations of $\ell_{1,\dots,3}$ that one could use to build either mass variable for Dirac (Majorana) $N$.
One choice should concentrate near the true invariant mass of $N$ whereas the others should follow a continuum distribution.
Due to the low multiplicity of choices, we (in an unsophisticated manner) brute force a guess-and-test determination.
More specifically, for a particular mass hypothesis $(m_N^{\rm Hypo.})$, we build all permutations of $m_{MT}$ and $m_{Cl}$ allowed by 
electric charge conservation and choose the one closest to $m_N^{\rm Hypo.}$ for each variable, labeled $\hat{m}_{MT}$ and $\hat{m}_{Cl}$.

To help gauge whether one transverse mass better approximates $m_N$, 
we take advantage of our access to generator-level information on $N$ and set $m_N^{\rm Hypo.}$ 
to the generator-level value of $m_N$ $(m_N^{\rm Gen.})$, on an event-by-event basis.
This accounts for shifts and variations in the virtuality of $N$ due to its finite width and is done only for exploratory means.
(We do not access such information in the detector-level analysis of Sec.~\ref{sec:Observability}.)

We quantify the ability of particle-level $\hat{m}_{MT}$ (solid) and $\hat{m}_{Cl}$ (dot) to estimate $m_N^{\rm Gen.}$ by calculating  
the relative difference $(\delta m/m)$ between the value of the reconstructed mass $m_N^{\rm Reco.} \in\{\hat{m}_{MT},\hat{m}_{Cl}\}$ 
and the value of $m_N^{\rm Gen.}$ found in the event file:
\begin{equation}
 \frac{\delta m_N^{\rm Reco.}}{m_N^{\rm Reco.}} = \cfrac{\left(m_N^{\rm Reco.} -  m_N^{\rm Gen.}\right)}{m_N^{\rm Gen.}}.
\end{equation}
In Fig.~\ref{fig:particleKin_dMass}, we plot this mass resolution for both observables, 
for (a,c,e) $m_N=150\GeV$ and (b,d,f) 450\GeV, and at (a,b) $\sqrt{s}=14$, (c,d) 27, and (e,f) 100\TeV.
Guide lines are inserted at $\delta m / m = \pm 20\%$.
Remarkably, though perhaps no longer surprisingly, both $\hat{m}_{MT}$ and $\hat{m}_{Cl}$ retain their shape and sharpness with increasing $\sqrt{s}$.
As repeatedly argued, this stems from the two objects only being a function of transverse momenta and elementary (constant) mass scales.
While both transverse masses give a reasonable estimation for $m_N^{\rm Gen.}$, a qualitative difference is noticeable.
Whereas $\hat{m}_{MT}$ appears to better describe higher mass $N$, i.e., give  a sharper, narrower distribution, $\hat{m}_{Cl}$ is found to do better for lower $N$ masses.
We do not investigate this difference further, but note that such behavior may be useful input in constructing multivariate analyses.
In light of the better performance for higher mass $N$ as well as our goal of quantifying the discovery potential of EW- and TeV-scale $N$ at higher energy $pp$ colliders,
we focus on employing the multibody transverse mass $m_{MT}$ for the remainder of this study.

\subsection{Summary}
In this section we have reported in exhaustive detail production- and decay-level properties of high-mass $(m_N > m_h)$ heavy neutrinos produced in hadron colliders, 
at various perturbative accuracies and collider energy of mass energies.
Due to the amount of detail, it is appropriate to briefly summarize the immediate findings:
We report that the leading production mechanisms of heavy neutrinos across a range of masses varies enormously as a function of   $\sqrt{s}$.
While the $W\gamma$ fusion process inevitably dominates for the heaviest mass scales, 
the largely studied CC DY process becomes subleading to neutral current processes at collider energies immediately beyond those reached at the LHC.
For a fixed heavy neutrino mass $m_N$, we also report the existence of an entire class of observables,
namely inclusive (with respect to hadronic activity) quantities built (dominantly) from the transverse momenta of the leading charged leptons,
 whose distribution shapes display only a weak sensitivity to changes in collider energy.

We now turn to the jet activity in heavy neutrino production at $pp$ colliders.

\section{Heavy Neutrinos and Dynamic Jet Vetoes}\label{sec:jetVeto}

We now investigate the hadronic activity in heavy neutrino events when they are produced through the CC DY and $W\gamma$ mechanisms.
We consider this for a range of neutrino masses and hadron collider energies, as well as for representative SM backgrounds to the inclusive
\begin{equation}
pp \to N\ell + X \to 3\ell  + X',
\label{eq:inclTrilepton}
\end{equation} 
collider signature.
It is important to reiterate that here and throughout the text we mean ``inclusive'' with respect to hadronic activity and, potentially, 
additional charged leptons that may appear in an event.
The goal of this section is to quantify the amount of hadronic activity in these various processes 
and to establish to what extent the relative amount of hadronic and leptonic activities in these processes changes with $\sqrt{s}$.
As noted in Sec.~\ref{sec:ColliderParticleKin}, for a fixed heavy neutrino masses, the amount of leptonic activity, 
as measured by leading charged lepton $p_T$ or exclusive $S_T$, does not change appreciably with $\sqrt{s}$.
The main result of the present section is to show that discriminating events according to the \text{relative} amounts of leptonic and hadronic activities, 
on an event-by-event basis, is a powerful means to improve the signal-to-background ratio $(\mathcal{S}/\mathcal{B})$ in multilepton searches for heavy $N$.

We reach this conclusion in the following manner:
In Sec.~\ref{sec:jetVetoSignal}, we discuss the characteristic behavior of QCD activity in the CC DY and $W\gamma$ signal processes as a function of $m_N$ and $\sqrt{s}$.
We present how jet vetoes, which exploit this information, can be extended to account for the relative amount of leptonic activity on an event-by-event basis, i.e., a dynamic jet veto.
In Sec.~\ref{sec:jetVetoBkg}, we look into how a dynamic jet veto impacts leading backgrounds.
We briefly discuss alternative, complementary measures of hadronic and leptonic activities that may be fruitful for future studies in Sec.~\ref{sec:jetVetoNewVetoes}.
Lastly, in Sec.~\ref{sec:jetVetoLL}, we examine the uncertainties associated with static and dynamic jet vetoes at NLO+PS and LO+PS.

\subsection{Signal Processes and Jet Vetoes at NLO+NNLL}\label{sec:jetVetoSignal}

As shown in Fig.~\ref{fig:xsecVsMass}, the inclusive trilepton process of Eq.~\ref{eq:inclTrilepton} is driven by the CC DY mechanism, i.e., quark-antiquark annihilation,
\begin{equation}
 q \overline{q'} \to W^{*} \to N \ell_1 \to \ell_1 \ell_2 \ell_3 \nu_\ell,
\end{equation}
at lower neutrino masses and by $W\gamma$ fusion for larger masses,
\begin{equation}
 W \gamma \to N \ell_1 \to  \ell_1 \ell_2 \ell_3 \nu_\ell.
\end{equation}
These are both color-singlet processes where initial-state partons scatter into a colorless final state. 
This often leads to the na\"ive (and incorrect) suggestion that there is no QCD radiation in hadronic production of heavy neutrinos, especially at leading/lowest order.
Protons, of course, are color-singlet states.
Hence, for DY-like process, whatever remains after removing the initial-state quark or antiquark (a ${\bf 3}$ or ${\bf \overline{3}}$ under QCD) 
must also be collectively charged under QCD due to color conservation (as a  ${\bf \overline{3}}$ or ${\bf 3}$, respectively).
As a result, initial partons and the beam remnant are connected via color fields that spontaneously give rise to jets that are characteristically collinear with the beam axes.
For the $W\gamma$ process, the argument is less subtle: 
initial-state $W$ bosons are generated perturbatively in $pp$ collisions from $q\to Wq'$ splitting.
As a result, the $W\gamma\to N\ell$ process is typically associated with a forward jet possessing $p_T^{j}\sim M_W/2$ that 
remains spatially close and color-connected to the rest of its parent proton's constituents.
(A subleading, high-$\vert\eta\vert$ jet with $p_T\ll M_W/2$ in backward direction is also usually present.)
Beyond LO, this behavior remains much the same for both 
SM processes~\cite{Altarelli:1979ub,Hamberg:1990np,Anastasiou:2003yy,Han:1992hr,Bolzoni:2010xr,Bolzoni:2011cu,Cacciari:2015jma,Dreyer:2016oyx} 
and heavy neutrino production~\cite{Ruiz:2015zca,Ruiz:2015gsa,Degrande:2016aje}.

Regardless of production mode, 
color conservation and mass scales 
lead to the fact that heavy neutrinos are accompanied by jets that are predominantly forward (large $\vert\eta\vert$) and/or soft (low $p_T$). 
This is in contrast to QCD processes, such as top quark production, where jets are typically central (small $\vert\eta\vert$) and hard (high $p_T$). 
This qualitative difference provides the rationale for central jet vetoes~\cite{Barger:1990py,Barger:1991ar,Bjorken:1992er,Fletcher:1993ij,Barger:1994zq},
wherein events that contain central $(\vert\eta\vert < 2.5-3)$ jets with a transverse momentum above $\pTVeto$ are classified as background and rejected in measurements and searches.

\begin{figure}[!t]
\begin{center}
\subfigure[]{\includegraphics[width=.45\textwidth]{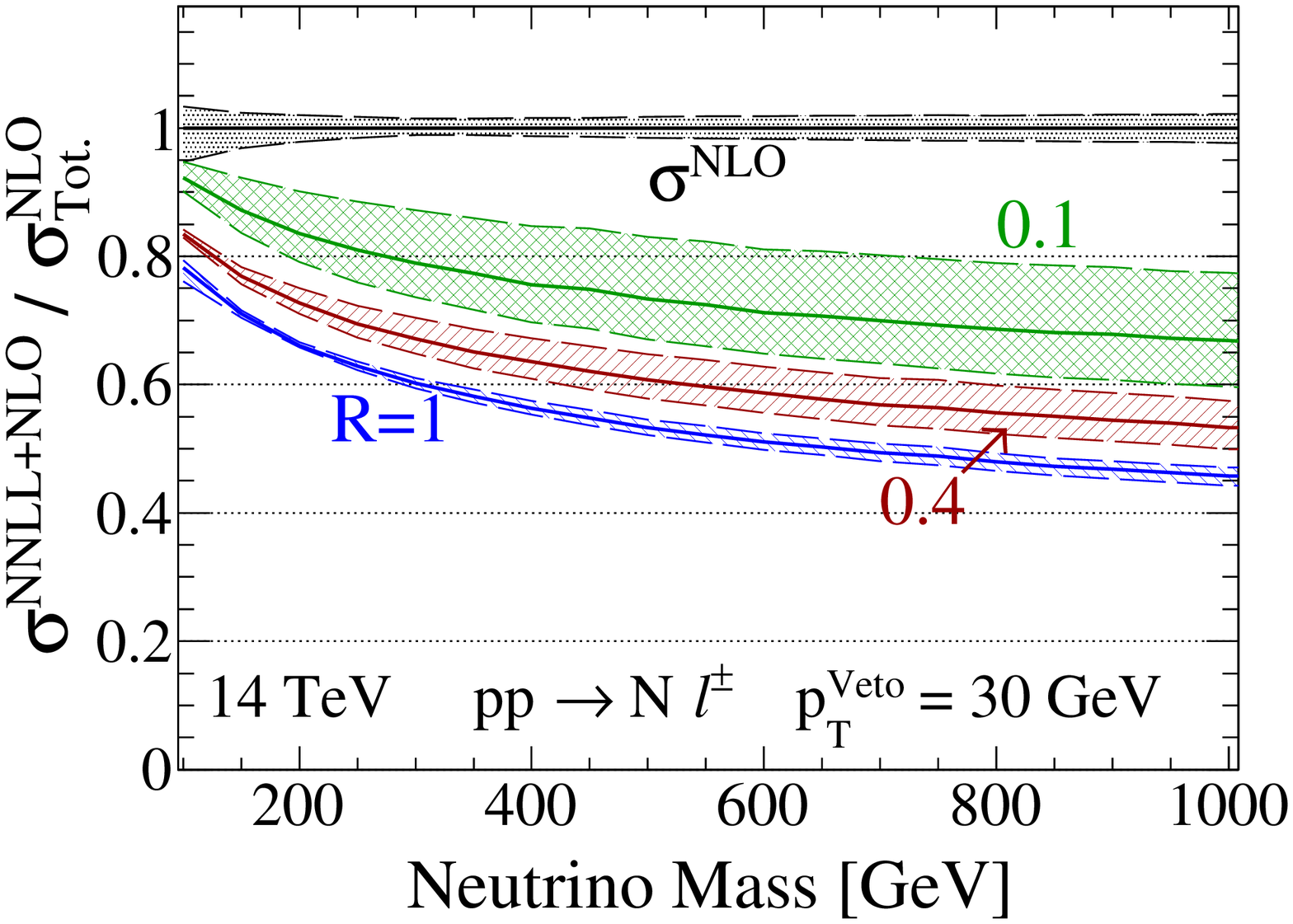}		}
\subfigure[]{\includegraphics[width=.45\textwidth]{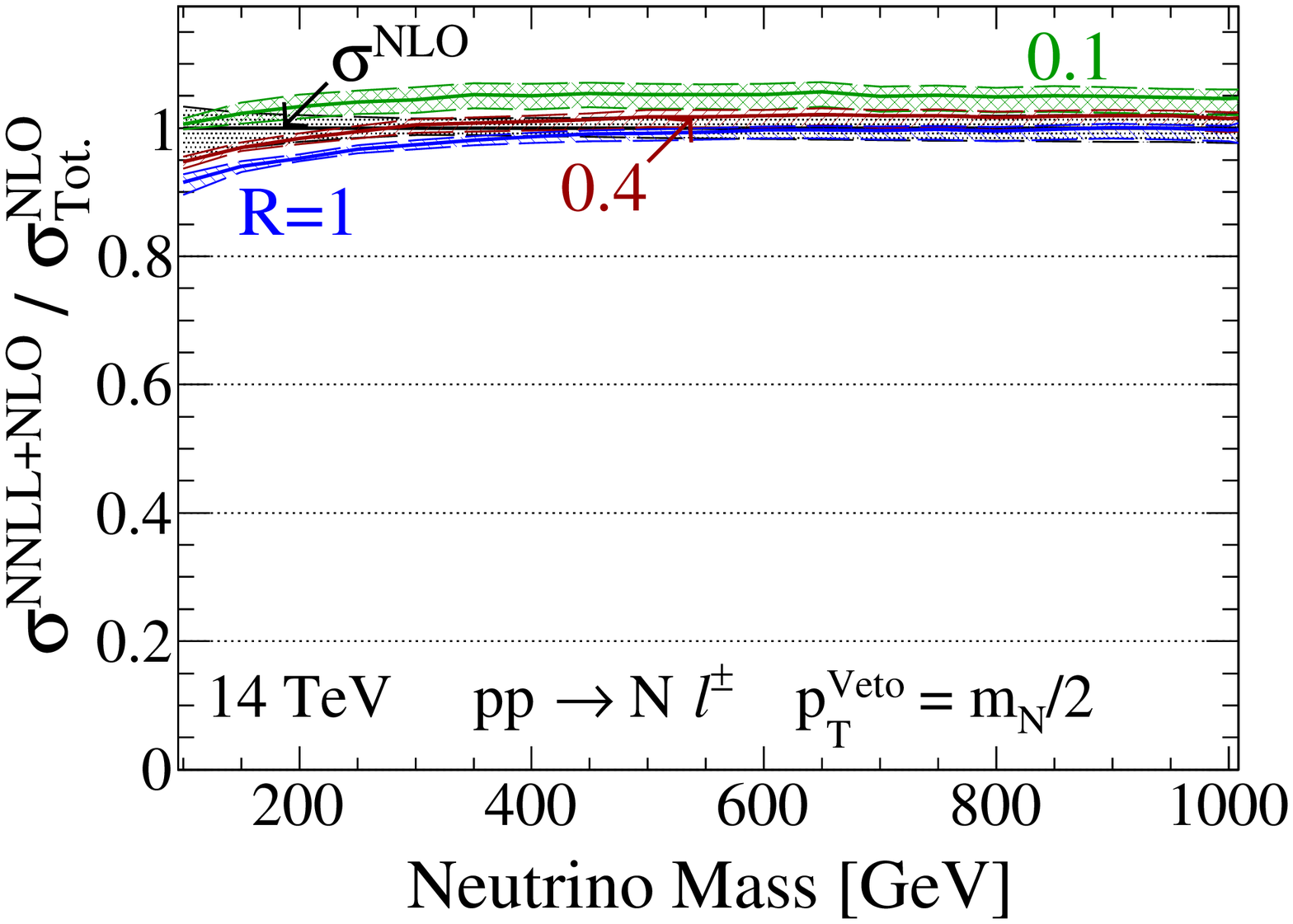}	}
\\
\subfigure[]{\includegraphics[width=.45\textwidth]{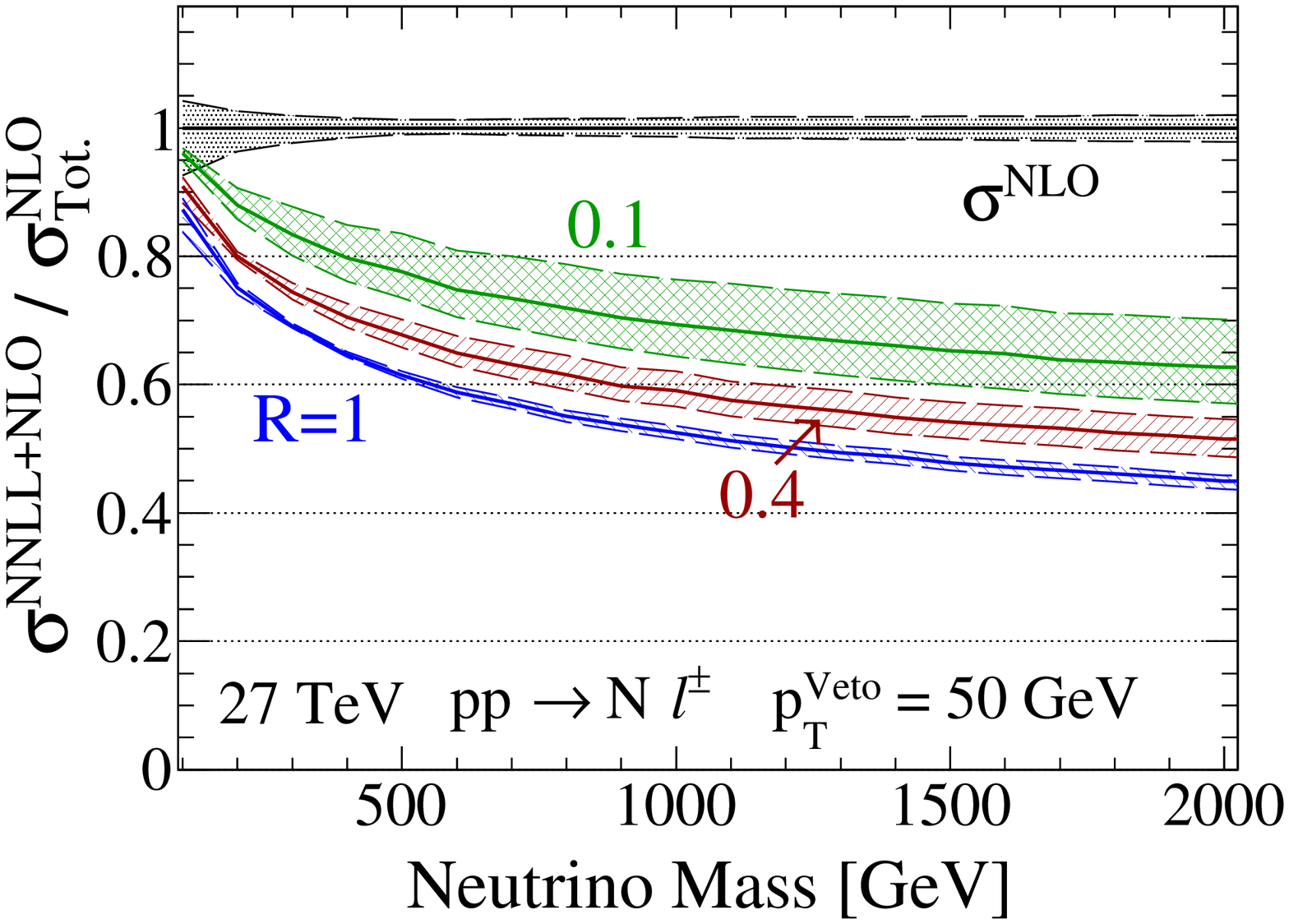}		}
\subfigure[]{\includegraphics[width=.45\textwidth]{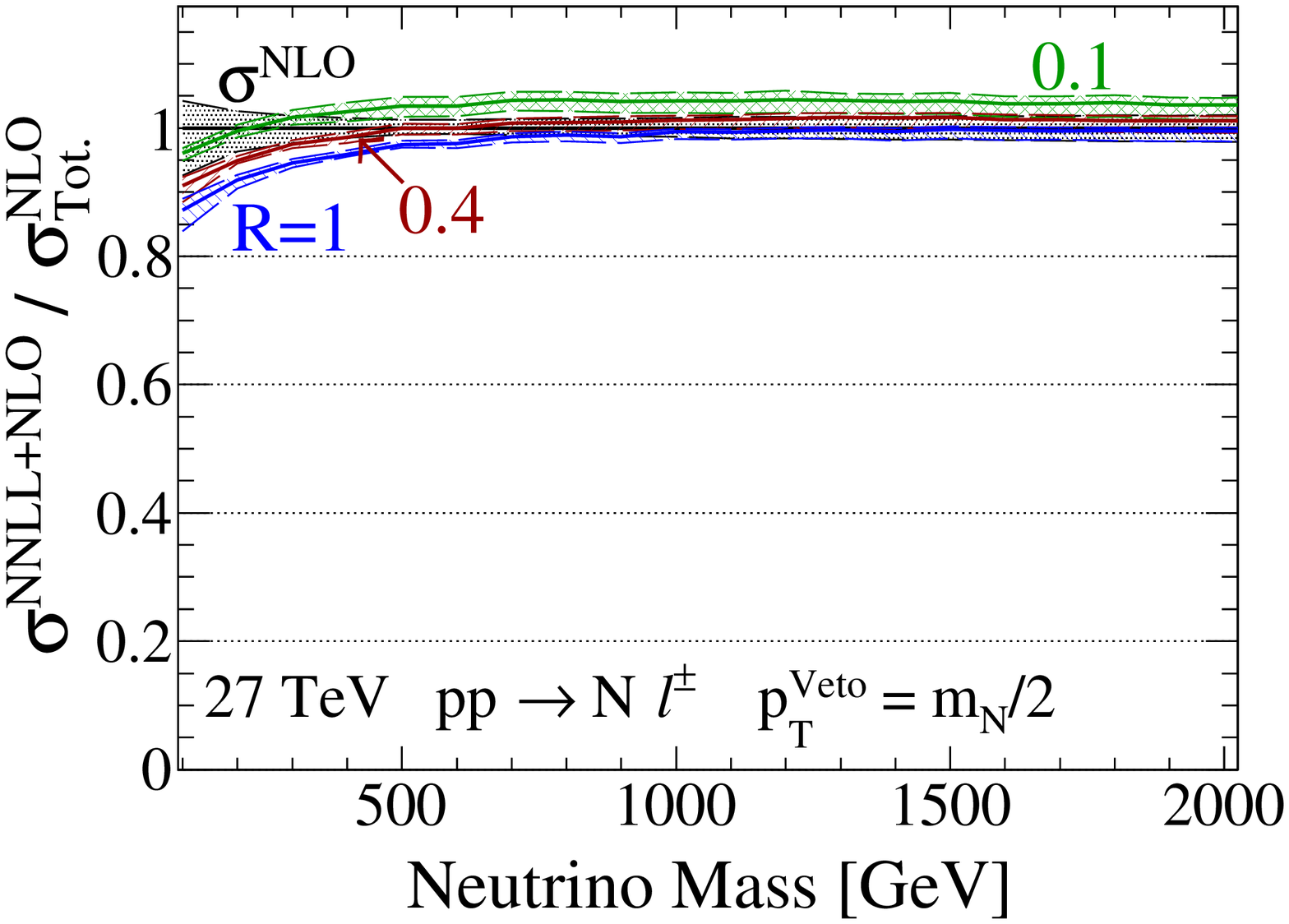}	}
\\
\subfigure[]{\includegraphics[width=.45\textwidth]{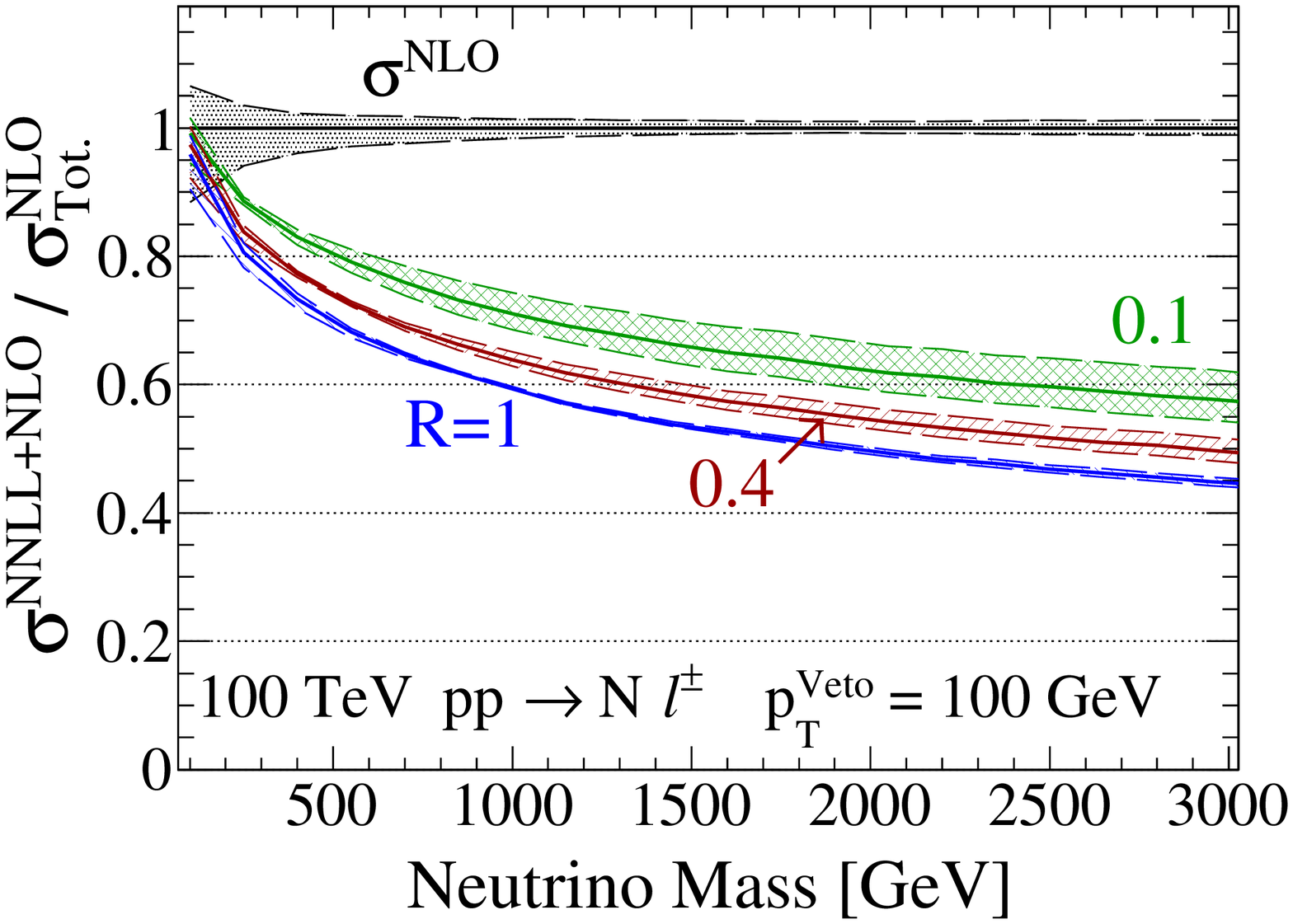}	}
\subfigure[]{\includegraphics[width=.45\textwidth]{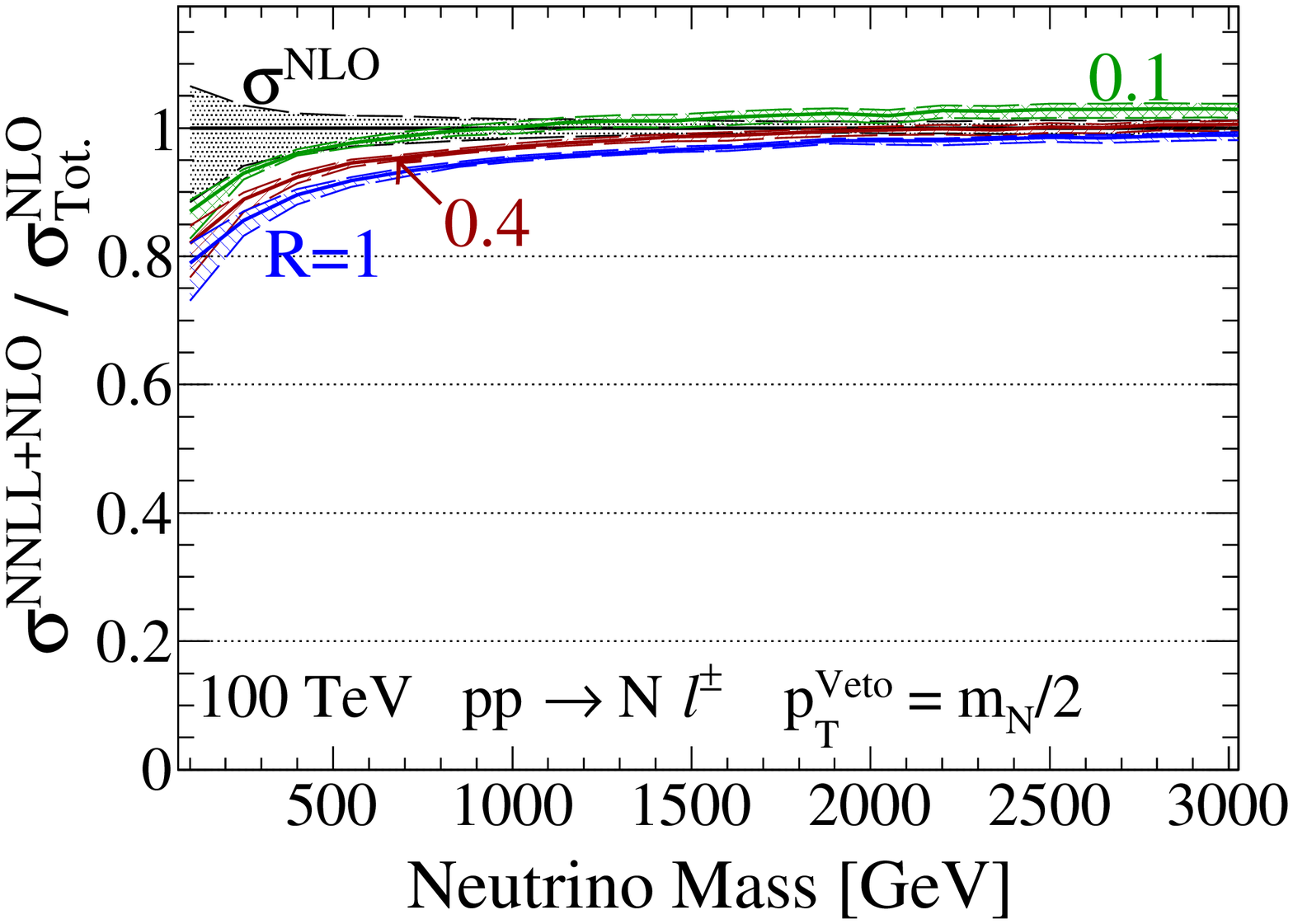}	}
\end{center}
\caption{
Jet veto efficiency at NLO+NNLL(veto) and residual scale dependence for the CC DY process $pp \to N\ell$,
as a function of heavy neutrino mass for representative jet radii $R$, at (a,b) $\sqrt{s} = 14$, (c,d) 27, and (e,f) 100 TeV,
assuming a static jet veto of (a) $\pTVeto = 30$, (c) 50, and (e) 100\GeV, and a representative dynamic veto of (b,d,f) $m_N/2$.
}
\label{fig:jetVetoEffdyxNLONNLL}
\end{figure}

While qualitatively sound, in reality, application of traditional jet vetoes with static values of $\pTVeto=20-50\GeV$ do much to hurt signal efficiency
in searches for heavy, new colorless particles ~\cite{Tackmann:2016jyb,Ebert:2016idf,Fuks:2017vtl,Pascoli:2018rsg,Fuks:2019iaj}.
To quantify this statement, we plot in Fig.~\ref{fig:jetVetoEffdyxNLONNLL}(a) the jet veto efficiency $\varepsilon(\pTVeto)$ and uncertainty for the CC DY process, defined as the ratio:
\begin{equation}
\varepsilon^{\rm NLO+NNLL(veto)}(\pTVeto) = \frac{\sigma^{\rm NLO+NNLL(veto)}(pp \to N\ell + X ~; ~p_T^j < \pTVeto)}{\sigma^{\rm NLO}_{\rm Tot.}(pp \to N\ell +X)},
\label{eq:vetoEffDef}
\end{equation}
as a function of heavy neutrino mass, for jet radii $R=0.1,~0.4$, and $1.0$, and a (static) jet veto threshold of $\pTVeto=30\GeV$.
Here, $\sigma^{\rm NLO+NNLL(veto)}(pp \to N\ell + X ~; ~p_T^j < \pTVeto)$ is the NLO+NNLL(veto)-accurate signal process cross section 
with a phase space cut to remove jets with $p_T^j > \pTVeto$.
 $\sigma^{\rm NLO}_{\rm Tot.}(pp \to N\ell +X)$ is the totally inclusive NLO cross section without the jet veto.
We use the computation formalism of Refs.~\cite{Alwall:2014hca,Becher:2014aya},
which implicitly applies a jet veto within the rapidity range $\vert \eta^{j_1} \vert < \eta^{\rm Veto}$, where $\eta^{\rm Veto} \to \infty$ is assumed.
This approximation nevertheless provides an excellent estimate of the total normalization and uncertainty for when $\eta^{\rm Veto} = 4.5-5$~\cite{Michel:2018hui}, 
which is the rapidity gap we consider at the analysis level in Sec.~\ref{sec:Observability}.
Conversely, for more restrictive jet veto rapidity windows, such as the commonly used $\eta^{\rm Veto} = 2.5$, 
there is a considerable increase in sensitivity to higher order radiative corrections~\cite{Michel:2018hui},
and hence a much larger $(\gtrsim1.5-2\times)$ scale uncertainty than what we report below.
 For additional details, see Ref.~\cite{Michel:2018hui} and references therein. 

We immediately see several features:
(i) For increasing neutrino mass and $R=1.0~(0.4)~[0.1]$, 
the veto efficiency drops precipitously from about ~$90\%~(85\%)~[80\%]$ to about ~$65\%~(55\%)~[45\%]$ over the mass range considered.
The poor acceptance can be attributed to the structure of initial-state radiation of gluons:
after phase space integration, one finds that such contributions are of the form $\delta\sigma/\sigma \sim \alpha_s(\mu_r=\pTVeto)\log^2(Q/\pTVeto)^2$,
for a hard scale $Q\sim m_N$. This indicates a propensity for heavier $N$ to be accompanied by higher $p_T$ gluons.
For example: 
the likelihood for a (relatively) soft gluon with momentum $\hat{p}_T$ to accompany a neutrino of mass $\hat{m}_N$ 
is essentially the same for a gluon of momentum $2\hat{p}_T$ to accompany a neutrino with mass $2\hat{m}_N$.
Hence, heavier $N$ are accompanied by jets with $p_T$ that are more and more likely to exceed (and thus fail) the $\pTVeto$ threshold.
(ii) In addition to this, one also sees that larger $R$ jets uniformly fail the veto more readily than smaller $R$ jets.
This is due to larger $R$ jets containing more constituents that are roughly traveling in parallel motion, and therefore contribute constructively to the jet's net $p_T$.
(iii) For this same reason, i.e., that larger $R$ jets are more inclusive than smaller $R$ jets, we observe a smaller scale dependence for large-$R$ jets than small-$R$ jets.
However, this is also impacted by formally $\mathcal{O}(\alpha_s^2)$ terms in the NNLL resummation that scale as $\log(R)$, 
and therefore numerically vanish in the $R\to1$ limit, minimizing the jet radius dependence.
Some unmatched scale dependence is also inserted by the matching of the NLO calculation to an NNLL resummation, and not a NLL resummation.
This scale dependence behavior is quasi-universal in that it is largely dependent on color structure and relevant mass scales~\cite{Fuks:2017vtl},
and hence can be found when jet vetoes are applied to slepton/electroweakino pair production~\cite{Tackmann:2016jyb,Fuks:2019iaj} 
or Sequential Standard Model $W'/Z'$ boson production~\cite{Fuks:2017vtl}.

For higher collider energies, which is shown in Fig.~\ref{fig:jetVetoEffdyxNLONNLL}(c) for $\sqrt{s}=27$ and (d) 100 TeV, we see a similar, if not slightly worsening, picture.
To avoid vanishing efficiencies for heavier $m_N$, we relax the veto threshold to $\pTVeto = 50$ and $100\GeV$ respectively.
For a fixed $(m_N/\pTVeto)$ ratio, we see, with for example $(m_N,\pTVeto)=(600\GeV,30\GeV)$ at 14 TeV and $R=0.1$, 
 that the efficiency roughly is constant at $\confirm{\varepsilon(\pTVeto)\sim70\%}$ between $\sqrt{s}=14$ and 27 TeV, but drops below \confirm{$65\%$} for 100 TeV.
A similar behavior is observed for other $(m_N/\pTVeto)$ ratios and $R$ choices.
In the end, this shows that independent of collider, if one wishes to apply a jet veto with a fixed $\pTVeto$ threshold and retain sensible efficiencies, 
i.e., $\varepsilon(\pTVeto) \gtrsim 80-95\%$,
then one must restrict their searches to heavy neutrino mass scales to at most a few hundred GeV, or relax $\pTVeto$ to effectively act as if no veto is applied.

While $m_N$  is an unknown quantity (assuming $N$ exists),
the premise of relaxing $\pTVeto$ with increasing $m_N$ to act as if no veto is applied is not exceptional nor impossible.
As discussed extensively in Secs.~\ref{sec:ColliderPartonKin} and ~\ref{sec:ColliderParticleKin}, 
the three leading charged leptons in the CC DY $pp\to N\ell \to 3\ell X$ process all possess momenta that scale with $m_N$ (see Eq.~\ref{eq:pTlXSummary}). 
At the reconstructed level, they do not appreciably change with variable collider energy and are distinguishable from continuum contributions (see  Fig.~\ref{fig:particleKinEllpT}).
A similar argument for the leading charged leptons in the $W\gamma \to N\ell$ fusion process also holds.
Hence, in a genuine heavy neutrino event, the $p_T$ of the leading charged leptons scale proportionally with the neutrino mass and can act as a proxy for $m_N$.
Therefore, by setting $\pTVeto$ on an event-by-event basis, for example, to the leading charged lepton in an event, one would expect an increase in the jet veto efficiency for 
genuine heavy neutrino trilepton events.

Variable jet vetoes are not new, per se. 
In precision SM calculations~\cite{Denner:2009gj,Nhung:2013jta,Frye:2015rba}, 
comparable choices for $\pTVeto$ have been made as a computational convenience to alleviate large veto logarithms of the form $\log^2(Q/\pTVeto)^2$, 
rendering resummation unnecessary.
Independently, the lepton-over-jet $p_T$ ratio $(p_T^\ell/p_T^j)$ has also been used experimentally~\cite{Khachatryan:2017qgo}  
at the LHC for particle identification  to help distinguish leptons associated with hadronic / jet activity.
However, as first reported in the companion letter of this report~\cite{Pascoli:2018rsg}, 
we find that dynamic jet vetoes can be successfully used in a much broader class of experimental measurements and searches, 
notably searches for new, high-mass colorless particles.

We investigate the impact of a dynamic jet veto on the CC DY production of heavy neutrinos by first noting 
that the leading charged lepton in such events possesses a transverse momentum that scales as $p_T^{\ell_1}\sim m_N/2$.
Hence, we set\footnote[2]{
It is not  presently possible to set $\pTVeto$ on an event-by-event basis in the public NLO+NNLL(veto) code of Refs.~\cite{Alwall:2014hca,Becher:2014aya}.
Therefore, we set the veto threshold here explicitly to $\pTVeto=m_N/2$. 
In all following parton shower-level discussions, $\pTVeto$ is set on an event-by-event basis to the appropriate quantity.
}
the veto threshold here to
\begin{equation}
\pTVeto =   p_T^{\ell_1} \sim \frac{m_N}{2},
\label{eq:dynamicVetoBenchmark}
\end{equation}
 and plot in Fig.~\ref{fig:jetVetoEffdyxNLONNLL}(b,d,f) the veto efficiency respectively for collider energies (b) $\sqrt{s}=14,$ (d) $27$, and (f) $100\TeV$.
Dramatically, one sees that veto efficiencies jump to $\varepsilon(\pTVeto) \gtrsim 80-99\%$ for \confirm{$m_N \gtrsim 100\GeV$,} for all collider configurations.
Depending on the precise collider beam energy, $\varepsilon(\pTVeto)$ remains above the $90\%$ threshold for all {$m_N \gtrsim 100-430\GeV$}.
The lower efficiency at $m_N \lesssim 100-200\GeV$, can be tied precisely to the scaling of $p_T^{\ell_1}$, 
which in absolute terms corresponds to $\pTVeto = m_N/2 \lesssim 50-100\GeV$, and hence reproduces the behavior of the static jet veto case.
We observe that the veto efficiency $\varepsilon(\pTVeto)$ remains essentially independent of $m_N$  {for larger $m_N$}, 
indicating that the structure of jet veto Sudakov logarithms $\alpha_s(\mu_r=\pTVeto)\log^2(Q/\pTVeto)^2$ is captured and matched 
by associating $\pTVeto$ with the characteristic $p_T$ of the signal event's leading charged lepton.
Qualitatively, this differs from static/fixed $\pTVeto$ choices, where $\varepsilon(\pTVeto) \to 0$ as $m_N$ increases.

Two additional features are observed:
The first is a significant decrease in the residual scale dependence, which spans from {$\delta\varepsilon \sim \pm1\% $ to $\pm5$ for $\pTVeto = m_N/2$.}
Remarkably, the uncertainty exhibits little-to-no sensitivity to the heavy neutrino mass scale, jet radius, or collider energy under this veto scheme.
 This is attributed simply to the fact that tying $\pTVeto$ to $p_T^\ell \propto m_N\sim Q$ outright avoids the generation of large jet veto Sudakov logarithms.
 For $m_N$ much larger than the static veto thresholds of $\pTVeto = 30-100\GeV$, 
 one obtains a more inclusive cross section and therefore recovers an uncertainty comparable to the totally inclusive one.
As a consequence of this reduction, we are able to observe the appearance of an excess in all three $R=0.1$ curves, 
where for some starting $m_N$, $\varepsilon(\pTVeto)$ surpasses  unity.
Upon further investigation~\cite{Pascoli:2018rsg}, we find that this originates from $\mathcal{O}(\alpha_s^2)$ terms in the NNLL(veto) resummation that scale as $\alpha_s^2\log R$.
We confirm this by setting $R=0.01$ and finding that  \confirm{$\varepsilon(\pTVeto)\approx 100-115\%$ for $m_N\gtrsim 150\GeV$},
clearly illustrating a breakdown of the perturbative calculation and the need for a dedication resummation of $\log R$ terms as done in Refs.~\cite{Becher:2013xia,Dasgupta:2014yra}.
We stress that this does not imply a breakdown of the dynamic jet veto scheme, only a breakdown of the NLO+NNLL(veto) computation in the $R\to0$ limit.
The picture that emerges from Fig.~\ref{fig:jetVetoEffdyxNLONNLL}(b,d,f) may indeed hold for $R\ll 0.4$, but further investigation is required and encouraged.
For additional details, see Ref.~\cite{Pascoli:2018rsg}.

\subsubsection*{VBF and Jet Vetoes}

It is worth discussing how the $W\gamma$-fusion process, which possesses at least one energetic forward jet with $p_T^{j_{\rm VBF}} \gtrsim M_W/2\sim40\GeV$, 
is affected by a dynamic jet veto.
An important conceptual point to stress is that jet vetoes do not require events to have zero hadronic activity.
They select out only events with (potentially central) jets possessing a transverse momentum above $p_T^\mathrm{Veto}$ but are inclusive  with respect to jet activity outside this threshold. 
 As a consequence, the VBF process will survive the veto as long as $p_T^{\rm Veto} > p_T^{j_{\rm VBF}}$.
 While this might not always be true for static vetoes of $p_T^{\rm Veto}=20-40\GeV$, 
  for the neutrino mass scales where VBF is relevant, a dynamic jet veto with $p_T^{\rm Veto}= p_T^{\ell_1} \sim m_N/2$ will readily satisfy this condition.
More specifically, as seen in Fig.~\ref{fig:xsecVsMass}, $W\gamma$ fusion becomes a numerically important channel for $m_N \gtrsim 500\GeV$.
 Setting $\pTVeto=p_T^{\ell_1},~p_T^{\ell_2},$ or even $p_T^{\ell_3},$ translates to (see Eq.~{\ref{eq:pTlXSummary}) $\pTVeto\sim 100-250\GeV$,
 and is well above the $p_T^{j_1}\gtrsim M_W/2$ threshold.
 Hence, with very high efficiency, 
 one expects $W\gamma$ fusion events to characteristically survive a jet veto based on the leading charged leptons in an event,  down to $m_N \sim 250\GeV$.
 For explicit demonstration of this, see Sec.~\ref{sec:jetVetoLL}.

\subsubsection*{$\tau_h$ and Jets Vetoes}

Another point worth examining is the treatment of hadronically decaying $\tau$ leptons $(\tau_h)$ in the presence of a jet veto.
Experimentally, $\tau_h$ are reconstructed first as jets before $\tau$-tagging is applied~\cite{Khachatryan:2015dfa,Aad:2015unr}.
Hence, one must justify whether a $\tau_h$ can be excluded from the veto procedure,
which requires demonstrating that $\tau_h$ are objects totally independent QCD jets.
As argued in Ref.~\cite{Pascoli:2018rsg}, under the NWA, 
a final-state, on-shell $\tau$ is color-disconnected at a perturbative level from the rest of the hard $pp$ collision, to all orders in $\alpha_s$.
At a non-perturbative level, the $\tau \to \tau_h\nu$ decay occurs on a macroscopic distance of
\begin{equation}
d_\tau = c \tau_\tau \gamma_\tau \sim \left(\frac{1}{\Gamma_\tau}\right)\left(\frac{E_\tau}{m_\tau}\right) \sim \frac{m_N} {\Gamma_\tau m_\tau}
\approx 5.6~{\rm mm}\times\left(\frac{m_N}{100\GeV}\right),
\end{equation}
away from the $pp$ interaction point.
Here $\gamma_\tau$ is the $\tau$ lepton's Lorentz factor, and $\Gamma_\tau \sim 2\times10^{-12}\GeV$ is its total width.
This is a much later transition than hadronization, which for a non-perturbative scale of $\Lambda_{\rm NP} \sim 1-2$ GeV, 
occurs at a distance of $d_{\rm NP} = 1-2$ fm away from the $pp$ collision.
Consequently, $\tau$ leptons outlive primary hadronization of high-$p_T$ $pp$ collisions, and the remaining coupling to the $(pp)$-system is 
through long-range, color-singlet exchanges~\cite{Sjostrand:1993rb,Khoze:1994fu,Khoze:1998gi,Collins:2011zzd}, i.e., beyond twist-$2$ in the operator product expansion.
Contributions of this kind are beyond the formal accuracy of the Collinear Factorization Theorem in Eq.~\ref{eq:factThm},
and hence can be consistently neglected, thereby demonstrating a decoupling of $\tau_h$ from QCD jets in an event.

\subsubsection*{Dynamic Jet Vetoes and PDF Uncertainties}

\begin{figure}[!t]
\begin{center}
\subfigure[]{\includegraphics[width=.47\textwidth]{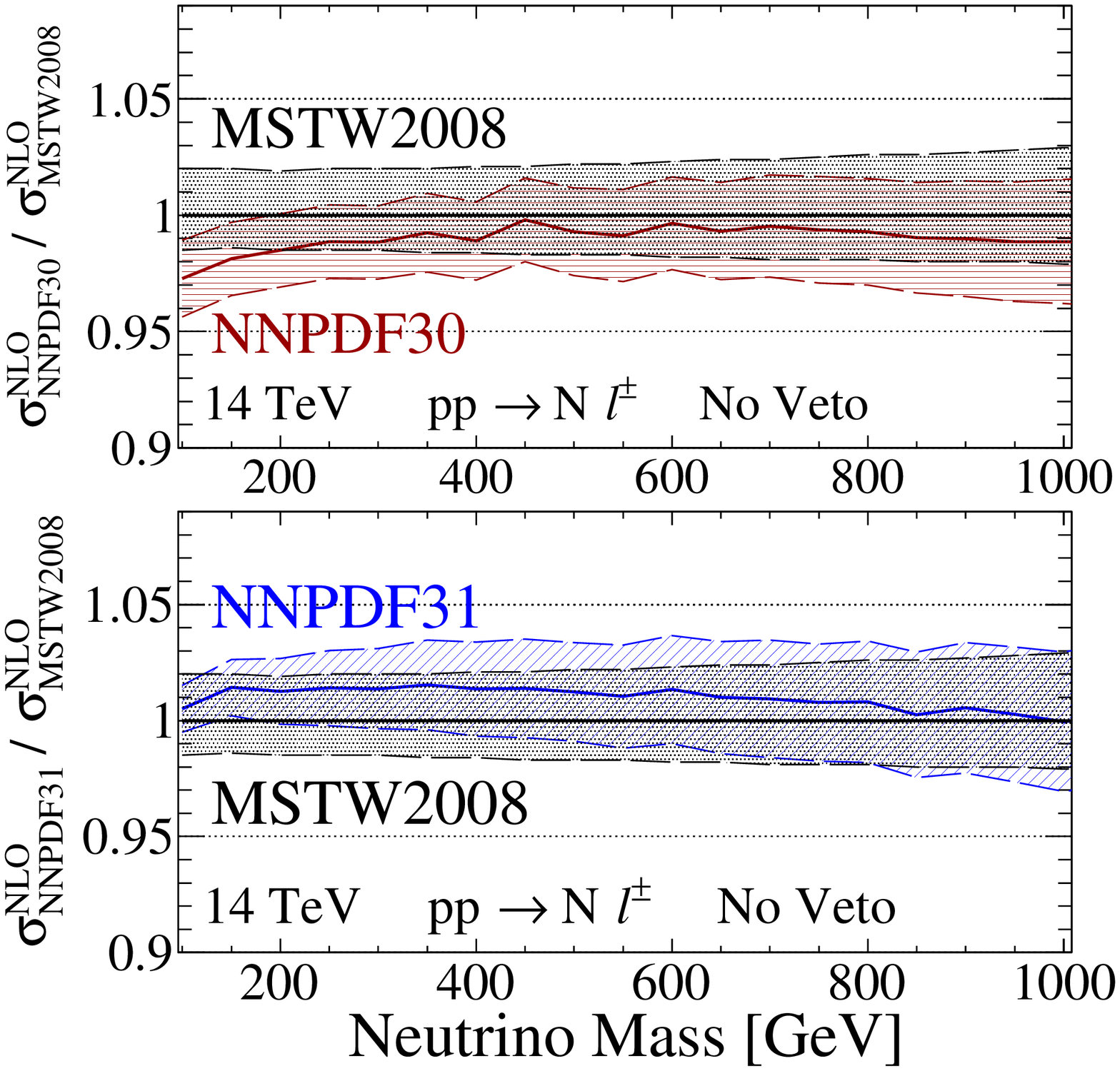}	}
\subfigure[]{\includegraphics[width=.47\textwidth]{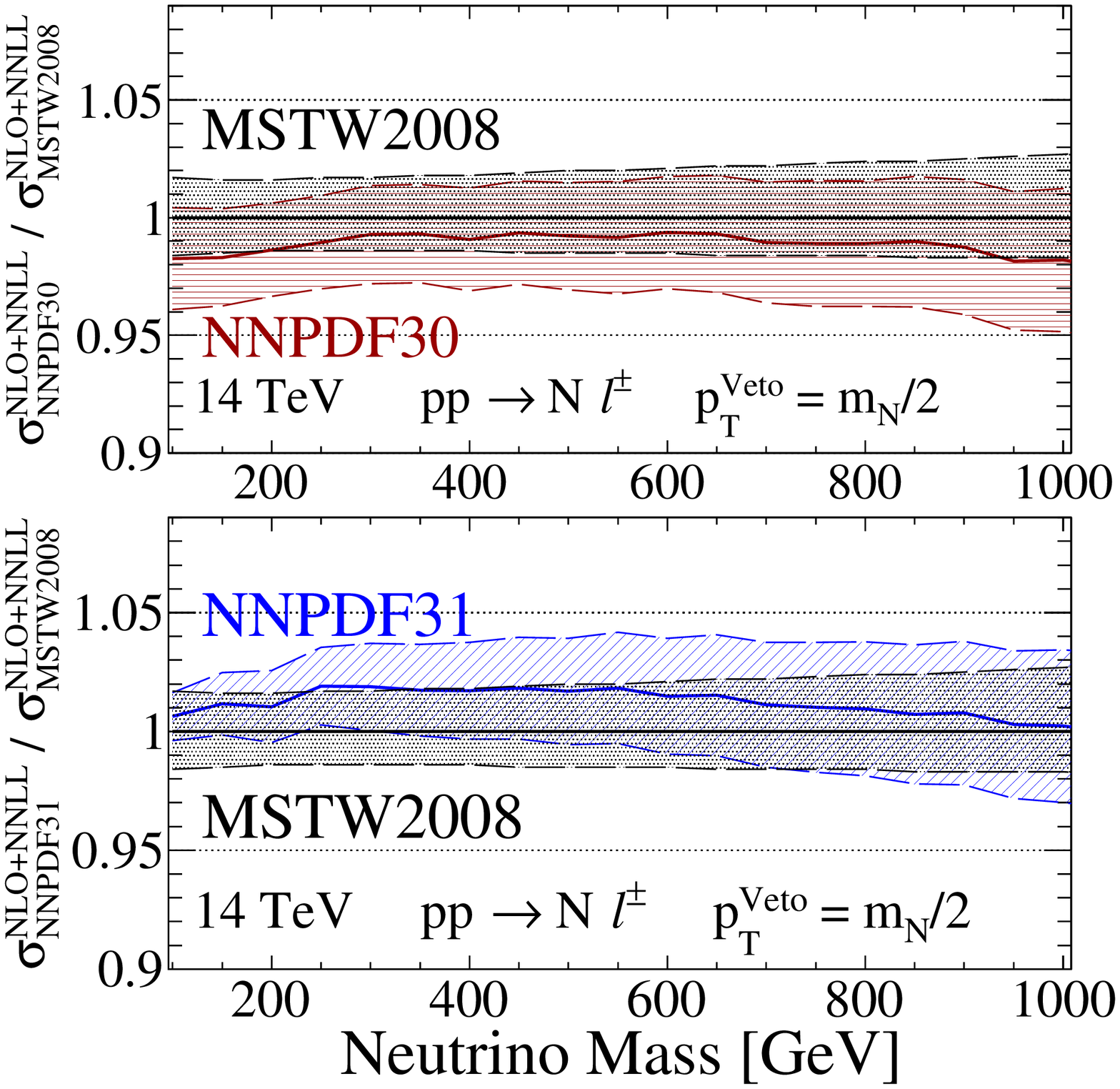}	}
\end{center}
\caption{
PDF uncertainty of the charged current Drell-Yan heavy neutrino production cross section at $\sqrt{s}=14$ TeV
 (left) without a jet veto and (right) with a jet veto threshold $p_T^\mathrm{Veto}=m_N/2$.
}
\label{fig:jetVetoPDFunc}
\end{figure}

Owing to the complexity of jet vetoes, there are numerous sources of theoretical uncertainty.
One such uncertainty stems from fitting PDF normalizations to data
and the subsequent impact on $\sigma^{\rm NLO+NNLL(veto)}$ cross section predictions.
For the CC DY $pp\to N\ell+X$ signal process, we attempt to quantify this by considering the ratio of cross sections,
\begin{eqnarray}
\mathcal{R}^{\rm NLO}  &=& \cfrac{\sigma^{\rm NLO}(pp\to N\ell+X ~;   ~ {\rm PDF_1})}{\sigma^{\rm NLO}(pp\to N\ell+X ~;  ~ {\rm PDF_{Ref.}})},
\\
\mathcal{R}^{\rm NLO+NNLL(veto)}  &=& 
\cfrac{\sigma^{\rm NLO+NNLL(veto)}(pp\to N\ell+X ~; \pTVeto, ~ {\rm PDF_1})}{\sigma^{\rm NLO+NNLL(veto)}(pp\to N\ell+X ~; \pTVeto , ~ {\rm PDF_{Ref.}})},
\end{eqnarray}
in which the scattering rates in the numerators and denominators are evaluated with a different PDF for fixed values of $m_N$, $\pTVeto$ (if applied), etc.
As reference (Ref.) PDF, we use the NNLO  \texttt{MSTW~2008} PDF set~\cite{Martin:2009iq};
we also consider the NNLO \texttt{NNPDF 3.0} set~\cite{Ball:2014uwa} as a representative ``PDF4LHC Run II'' recommendation~\cite{Butterworth:2015oua}.
Uncertainties are derived from PDF replicas in the appropriate statistical manner~\cite{Buckley:2014ana}.

In Fig.~\ref{fig:jetVetoPDFunc} and as a function of heavy neutrino mass, we show the ratios 
(a) $\mathcal{R}^{\rm NLO}$ and (b) $\mathcal{R}^{\rm NLO+NNLL(veto)}(\pTVeto=m_N/2)$,
with PDF uncertainties, at $\sqrt{s}=14\TeV$.
In the upper (lower) plots, the \texttt{MSTW~2008} reference curve is overlaid with the \texttt{NNPDF 3.0} ~ (\texttt{NNPDF 3.1+LUXqed}) curve.
Notably, for all three PDF sets, we observe few differences, qualitatively and quantitatively,  between the (a) no-veto and (b) jet-veto uncertainties. 
For the mass range $m_N = 100-1000\GeV$, we see that the uncertainty for all three PDFs span $\pm1\%-\pm3\%$.
Qualitatively, we see that central value of the \texttt{NNPDF 3.0} ~ (\texttt{NNPDF 3.1+LUXqed}) PDF uniformly undershoots (overshoots) 
the reference PDF's central value but essentially always stays within its $1\sigma$ band.
The slight exception is for the \texttt{NNPDF 3.0} case at $m_N<200\GeV$, where the central value exceeds the $1\sigma$ lower band by $\sim1\%$.

For the uncertainty at higher collider energies, we first note that the range of Bjorken-$x$ considered here spans roughly $x \sim m_N / \sqrt{s} \approx 0.01-0.07$.
This corresponds roughly to $m_N \approx 300-1900\GeV$ at $\sqrt{s}=27\TeV$ and $m_N=1-7\TeV$ at $\sqrt{s}=100\TeV$,
and hence covers the bulk of the investigated parameter space.
Over these mass ranges, one expects a similar PDF uncertainties as those reported in Fig.~\ref{fig:jetVetoPDFunc} due to PDF scale invariance.
For smaller $x$ down to $x\sim (150\GeV/100\TeV) = 1.5\times10^{-3}$, the \texttt{NNPDF 3.1+LUXqed} uncertainly is approximately unchanged~\cite{Bertone:2017bme}.
For larger $x$ up to $x\sim (20\TeV/100\TeV) = 0.2$, the uncertainty grows larger but remains under \confirm{$\pm5\%$}~\cite{Bertone:2017bme}.

\subsection{Backgrounds and Jet Vetoes at NLO+PS}\label{sec:jetVetoBkg}
In light of the improved veto efficiency of the dynamic jet veto for the CC DY and VBF signal processes relative to the static veto,
it is necessary to explore how background processes fare.
For the inclusive $pp \to N\ell +X \to 3\ell +X$ signal process, the leading SM processes fall into three categories:
(i) top quarks, 
(ii) EW diboson and multi-boson production, and
(iii) fake charged leptons,
which we now discuss.
We note that to emulate minimal, analysis-level selection cuts,
throughout Sec.~\ref{sec:jetVetoBkg}, all signal and background processes are evaluated at NLO+PS according to Sec.~\ref{sec:Setup}
and the following cuts are applied to jets and the three leading charged leptons after reconstruction:
\begin{equation}
\vert \eta^j \vert < 4.5,  \quad
p_T^{\ell_k} > 15\GeV, 
\quad\text{and}\quad \vert \eta^{\ell_k} 
\vert < 2.5, \quad\text{for}\quad k=1,\dots3.
\label{eq:toyFidKinCuts}
\end{equation}

\subsubsection*{Top Quark Background}
 
Associated top quark production processes at multi-TeV hadron colliders, e.g.,
\begin{eqnarray}
  pp \to t\overline{t}\ell\ell,                             	&\quad\text{with}\quad&  t \to Wb \to \ell\nu_\ell b,\\
  pp \to t\overline{t}\ell \nu,                             	&\quad\text{with}\quad&  t \to Wb \to \ell\nu_\ell b, \quad\text{and} \\
  \overset{(-)}{b}q \to \overset{(-)}{t}q'\ell\ell,        &\quad\text{with}\quad&  t \to Wb \to \ell\nu_\ell b,
\end{eqnarray}
are major background for any multilepton measurement and search due to their large cross sections, intrinsic mass scales, and diversity of final states.
Since ${\rm BR}(t\to Wb)\approx100\%$, multilepton search strategies for heavy neutrinos make use of $b$-tagging,
and hence $b$-jet vetoing, to help suppress these backgrounds~\cite{delAguila:2008cj,Sirunyan:2018mtv}. 
However, even for high-efficiency taggers, such as the CMS \texttt{CSVv2} algorithm~\cite{Chatrchyan:2012jua,Sirunyan:2017ezt} used by Ref.~\cite{Sirunyan:2018mtv},
which possesses a tagging efficiency of $\varepsilon^{\rm b-tag}\approx 70-80\%$, at least $4-9\%$ of top quark pairs survive single- and double-b-tagging.
Of the total number of top quarks, the fraction is actually larger when one takes into account that only jets within $\vert \eta \vert \lesssim 2.5$, i.e., within tracker coverage, can be tagged.
It has been reported~\cite{Fuks:2017vtl} that relaxing the $b$-tagging requirement and simply vetoing central jets with $\pTVeto = 20-50\GeV$, independent of flavor composition,
can improve top quark rejection in searches for DY-like processes, even with an ideal efficiency of $\varepsilon^{\rm b-tag}=100\%$.
We improve upon this by allowing $\pTVeto$ to be set on an event-by-event basis to the $p_T$ of the leading charged lepton in our multilepton final state.

\begin{figure*}[!t]
\begin{center}
\subfigure[]{\includegraphics[width=.48\textwidth]{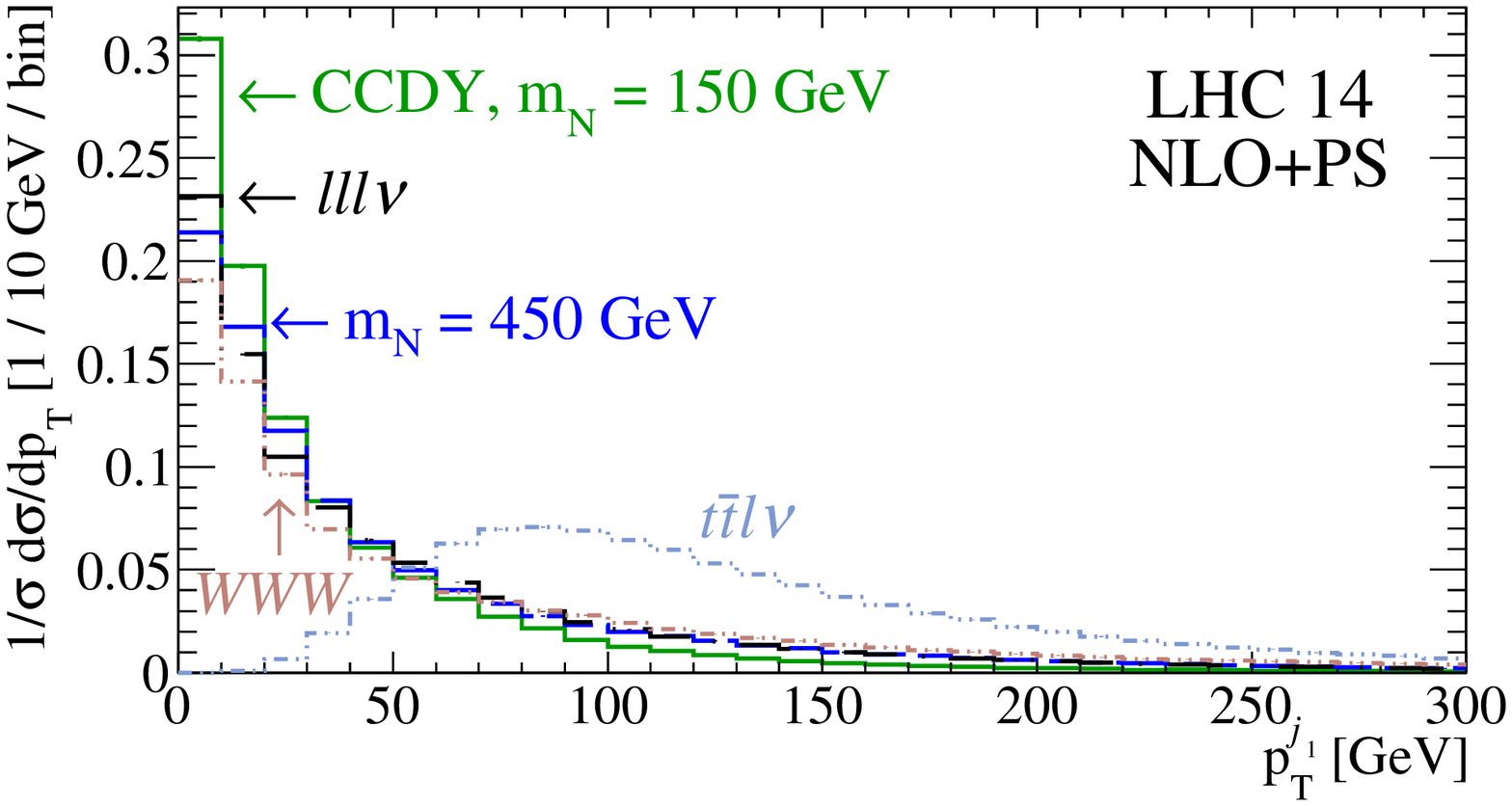}       	}
\subfigure[]{\includegraphics[width=.48\textwidth]{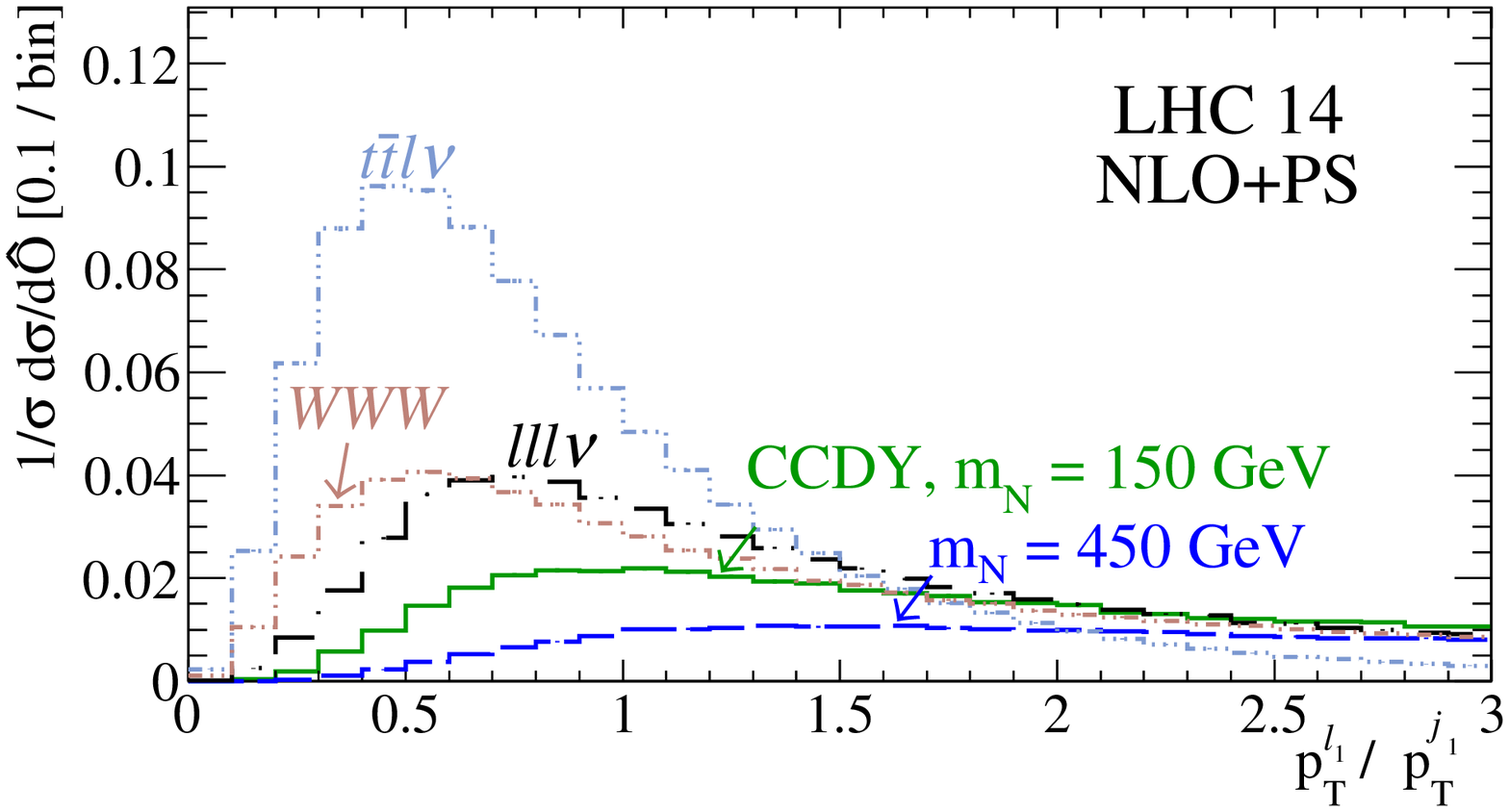}       	}
\\
\subfigure[]{\includegraphics[width=.48\textwidth]{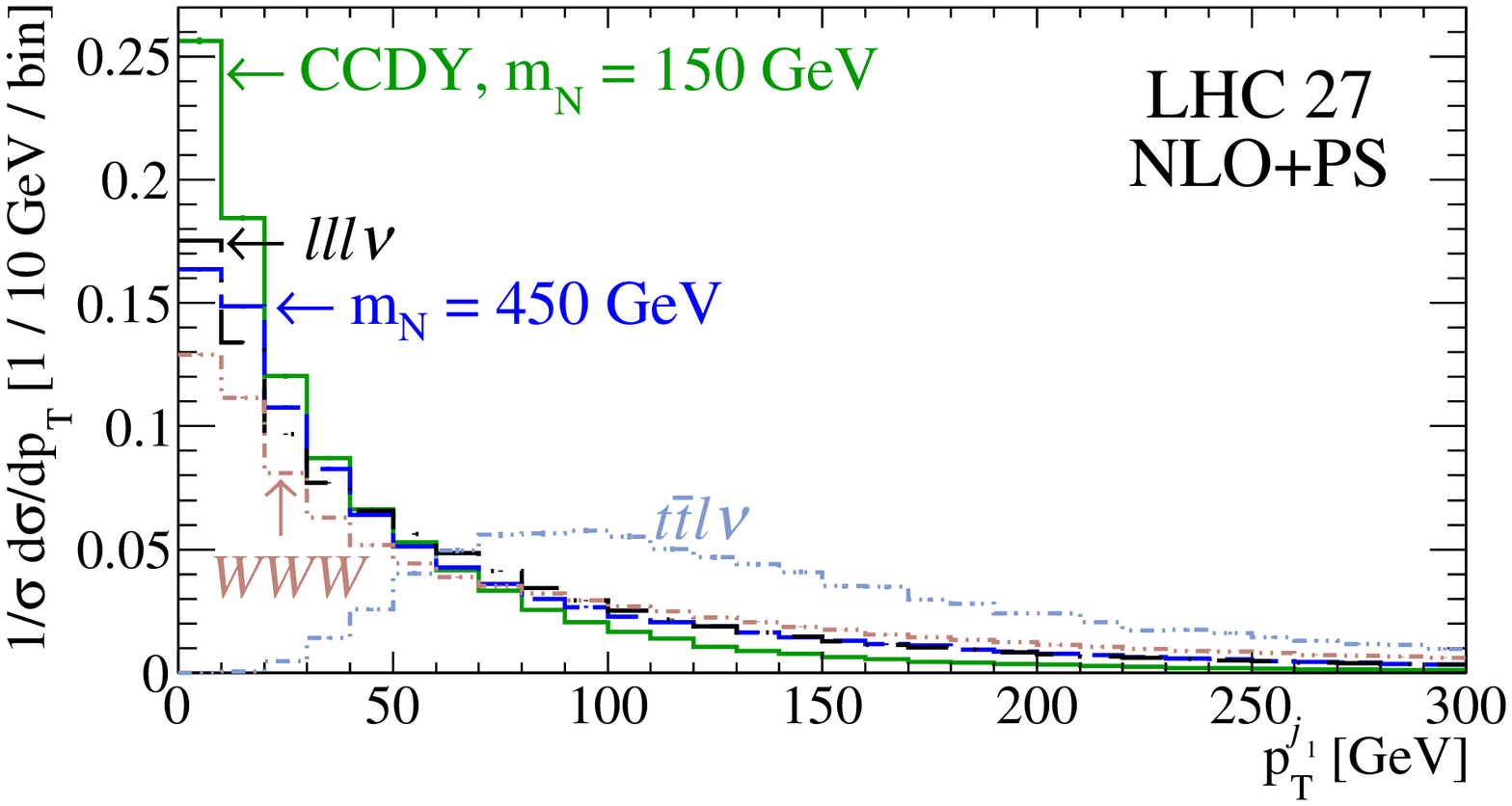}     		}
\subfigure[]{\includegraphics[width=.48\textwidth]{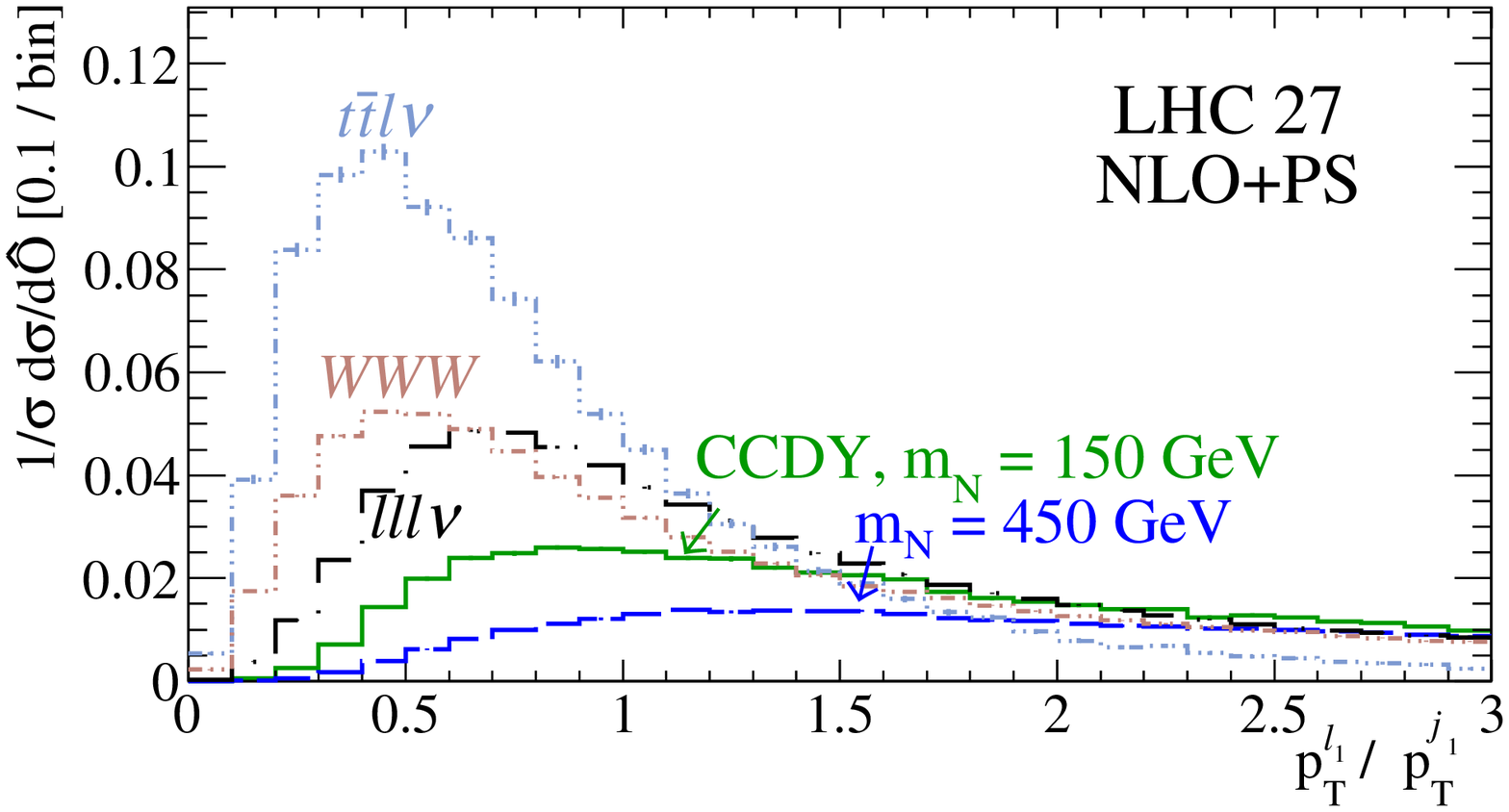}	}
\\
\subfigure[]{\includegraphics[width=.48\textwidth]{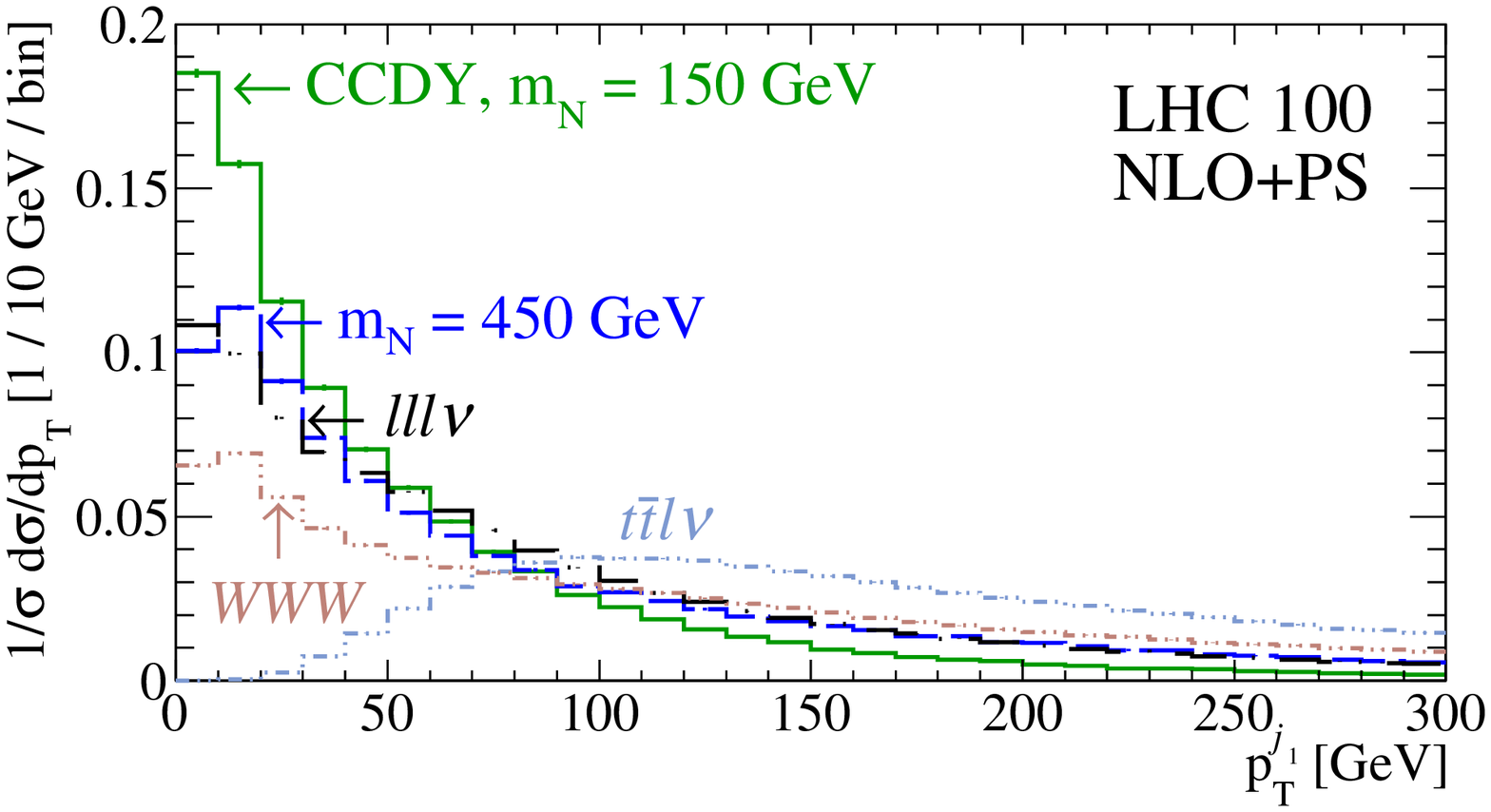}     	}
\subfigure[]{\includegraphics[width=.48\textwidth]{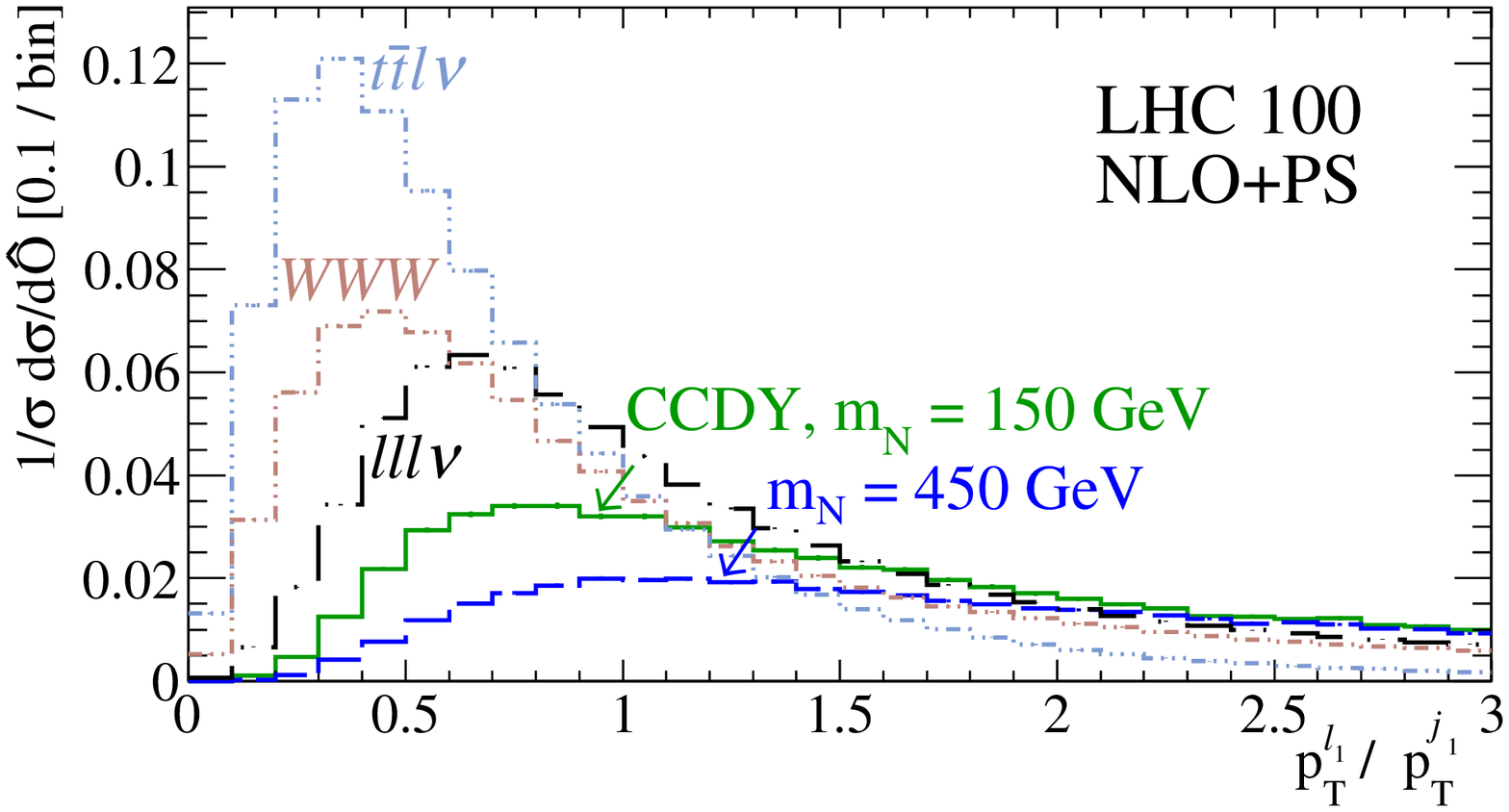}      }
\end{center}
\caption{
 Normalized NLO+PS distributions with respect to (a,c,e) the $p_T$ of leading jet $(p_T^{j_1})$ and (b,d,f) the ratio of leading lepton $p_T$ ~$(p_T^{l_1})$ to $p_T^{j_1}$,
 i.e., $\hat{\mathcal{O}}=p_T^{\ell_1}/p_T^{j_1}$,   at (a,b) $\sqrt{s}=14$ TeV, (c,d) 27 TeV, and (e,f) 100 TeV,
  for the charged current Drell-Yan (CCDY) $pp\to N \ell \to 3\ell\nu$ signal process with $m_N = 150\GeV$ (solid) and $450\GeV$ (dash),
  as well as the $pp\to 3\ell\nu$ (dash-1 dot), $pp\to WWW\to 3\ell\nu$ (dash-2 dots), and $pp\to t\overline{t}\ell\nu\to3\ell X$ (dash-3 dots)
  background processes.
  }
\label{fig:vetoKinLeadingJet}
\end{figure*}

The manner in which a dynamic jet veto such as $\pTVeto = p_T^{\ell_1}$ functions
largely follows from the resonant structure of the $t\overline{t}V\to 2WV$ production and decay chain for $V\in\{W,Z\}$.
For such processes, the leading charged leptons possess the characteristic momenta, 
\begin{equation}
p_T^{\ell_1} \sim \frac{m_t}{4}\left(1 + \frac{M_W^2}{m_t^2}\right) \approx 50-55\GeV, 
\quad\text{and}\quad
p_T^{\ell_2},~p_T^{\ell_3} \sim \frac{M_V}{2} \sim 40-45\GeV.
\end{equation}
On the other hand, the $b$-jet momenta, independent of being tagged, scale as
\begin{equation}
p_T^{b_1},p_T^{b_2} \sim \frac{m_t}{2} \left(1 - \frac{M_W^2}{m_t^2}\right) \approx 60-70\GeV.
\label{eq:bkgTopPTb}
\end{equation}
Therefore, we see that the top quark backgrounds in multilepton searches for heavy $N$ inherently fail a dynamic veto that requires $p_T^{\ell_1} > p_T^{j_1}$.
In principle, top quark processes fail regardless of choosing the leading charged lepton $\ell_1$, the subleading charged lepton $\ell_2$, or even the trailing charged lepton $\ell_3$.

To show this, we plot in Fig.~\ref{fig:vetoKinLeadingJet} 
the (a,c,e) $p_T$ of the leading jet $(p_T^{j_1})$ and (b,d,f) leading lepton-to-leading jet $p_T$ ratio 
\begin{equation}
r^{\ell_1}_{j_1} = p_T^{\ell_1} ~/~ p_T^{j_1}
\label{eq:defDynamicVeto}
\end{equation}
at (a,b) $\sqrt{s} = 14$, (c,d) 27, and (e,f) 100 TeV
for representative CC DY signal processes with $m_N = 150$ (solid) and $450\GeV$ (dash),
the $t\overline{t}\ell\nu$ (3-dot-dash) background, as well as the $WWW$ and $3\ell\nu$ backgrounds to the inclusive $pp\to3\ell+X$ collider process.

At $\sqrt{s}=14\TeV$, we see in that the CC DY signal and $t\overline {t}\ell\nu$ background processes both exhibit their characteristic $p_T^{j_1}$ behavior at low- and high-$p_T$ respectively.
Due to binning effects, the Sudakov shoulder in the signal process is unobservable.
For $t\overline {t}\ell\nu$, the wide bump near $p_T^{j_1}\sim 70-90\GeV$ corresponds to the anticipated $p_T$ for $b$-jets as given in Eq.~\ref{eq:bkgTopPTb}.
The higher value for the maximum is attributed to hard ISR and FSR that appears at $\mathcal{O}(\alpha_s)$.
Focusing now on the ratio plot, we see crucially a qualitative difference in the top quark and signal process:
 The signal process has an incredibly broad, continuum-like spectrum that appreciably starts at $r^{\ell_1}_{j_1} \sim 0.25$,
 possesses a very shallow peak at $r^{\ell_1}_{j_1} \sim1~(1.5)$ for $m_N=150~(450)\GeV$, and readily spans rightwards by several units.
 The $t\overline{t}\ell\nu$ background, on the other hand, peaks at a ratio of $r^{\ell_1}_{j_1} \sim 0.5$, with over \confirm{75\%} of events sitting below $r^{\ell_1}_{j_1}<1$.
While not shown, we report that ratio for the subleading lepton $r^{\ell_2}_{j_1} = (p_T^{\ell_2} / p_T^{j_1})$ features a slightly stronger separation 
between signal and background.

For higher collider energies, we observe encouraging behavior:
While the $p_T^{j_1}$ distribution CC DY signal process broadens, with the lowest bin occupancy for $m_N=150~(450\GeV)$ dropping from 
about \confirm{$30\%~(22\%)$ at $\sqrt{s}=14$ TeV, to $25\%~(17\%)$ at 27 TeV, down to $20\%~(10\%)$ at 100 TeV,}
the $t\overline{t}\ell\nu$ rightward shift toward higher values of $p_T^{j_1}$ is more significant.
This seen clearest in the $r^{\ell_1}_{j_1}$ ratios at 27 and 100 TeV, where about \confirm{35\% and 40\%} of events, respectively, have
a leading lepton-to-leading jet $p_T$ ratio less than $r^{\ell_1}_{j_1}< 0.5$.
This is in comparison to the roughly \confirm{25\%} of $t\overline{t}\ell\nu$ events at 14 TeV.
For the CC DY process, the migration of $r^{\ell_1}_{j_1}$ to values below unity is found to be only slight for $m_N =150\GeV$ and mostly negligible for heavier neutrino masses.

At first,  the dynamic jet veto choice of $\pTVeto=p_T^{\ell_1}$ does not appear  to improve top quark discrimination over $b$-tagging central jets.
We will see next, though, that in conjunction with additional information on charged lepton activity, it is sufficient.

\subsubsection*{EW Diboson Continuum and Resonant Multiboson Production}
In multilepton searches for heavy neutrinos with masses at or above the EW scale, the production of two or more EW bosons, either non-resonantly or resonantly,
\begin{eqnarray}
 pp \to 3\ell\nu, 				&\quad&	 pp \to 4\ell, \\
  pp \to WWW \to 3\ell3\nu,	&\quad&    pp \to WW\ell\ell \to 3\ell\nu X+4\ell2\nu,
\end{eqnarray}
represent the leading EW backgrounds at $pp$ colliders~\cite{delAguila:2008cj,Sirunyan:2018mtv}.
However, despite their color-singlet nature, they are significantly less immune to jet vetoes than the CC DY and $W\gamma$ fusion signal processes.
Due to the presence of radiation zeros in Born-level amplitudes, a large fraction of the inclusive diboson and triboson processes 
contain at least one high-$p_T$ jet~\cite{Mikaelian:1977ux,Brown:1979ux,Mikaelian:1979nr,Zhu:1980sz,Brodsky:1982sh,Brown:1982xx}.
For example: 
NLO studies of the inclusive $pp\to3W+X$ process reveal that $\mathcal{O}(30\%)$ of the cross section is comprised of the
subprocess $pp \to 3W+1j$ with $p_T^j > 50\GeV$~\cite{Campanario:2008yg,Binoth:2008kt}.
As a result, NLO and NNLO corrections to EW 
diboson~\cite{Smith:1989xz,Ohnemus:1991gb,Ohnemus:1992jn,Baur:1993ir,Ohnemus:1994qp,
Baur:1995uv,Baur:1997kz,Catani:2011qz,Cascioli:2014yka,Gehrmann:2014fva,Grazzini:2015nwa,Grazzini:2016swo} 
and triboson~\cite{Lazopoulos:2007ix,Hankele:2007sb,Binoth:2008kt} contribute to large increases 
to the total normalization of inclusive cross sections $(>+100\%)$  and multiplicity of high-$p_T$ jets.

\begin{figure*}[!t]
\begin{center}
\subfigure[]{\includegraphics[width=.48\textwidth]{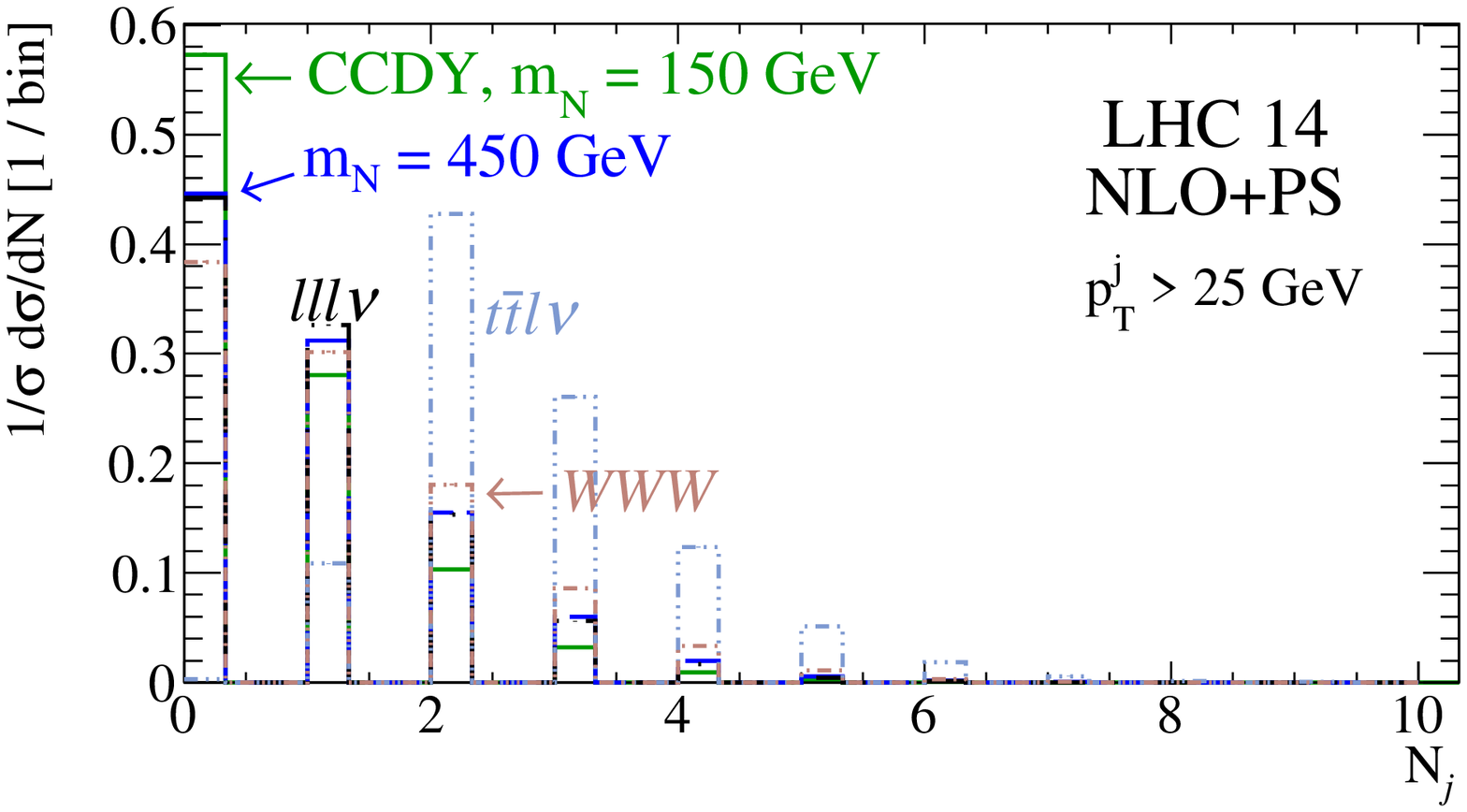}   }
\subfigure[]{\includegraphics[width=.48\textwidth]{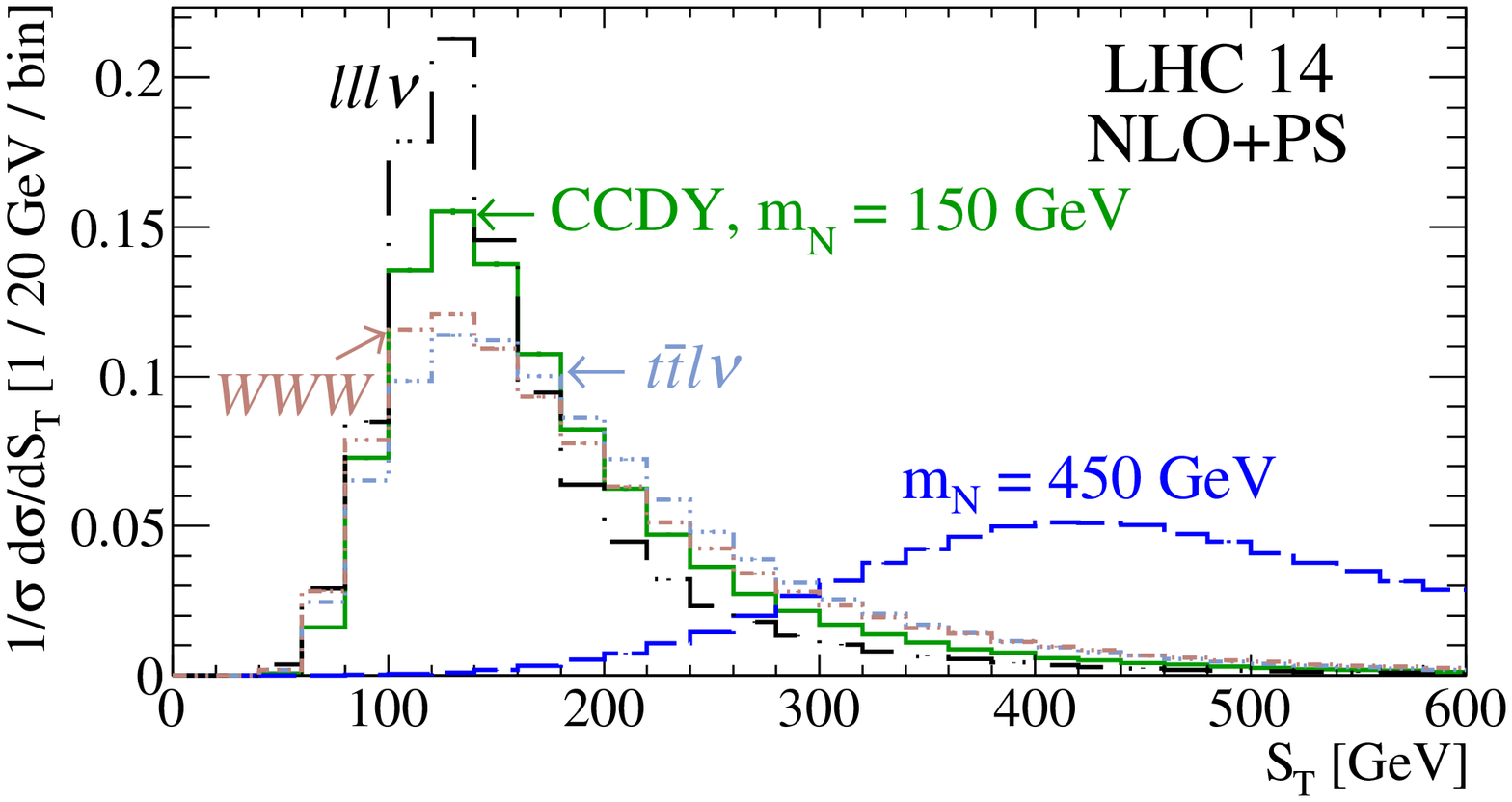}       }
\\
\subfigure[]{\includegraphics[width=.48\textwidth]{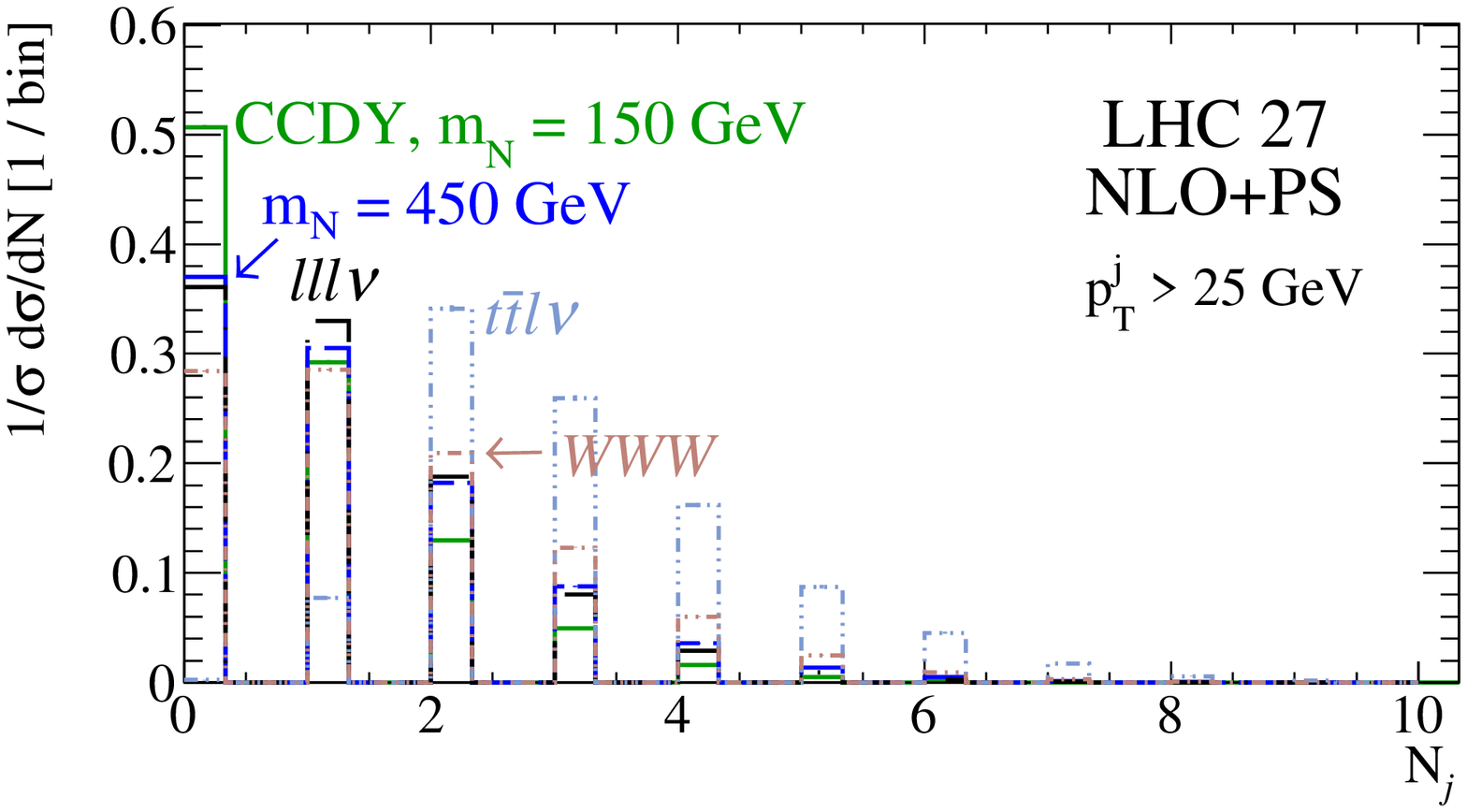}	}
\subfigure[]{\includegraphics[width=.48\textwidth]{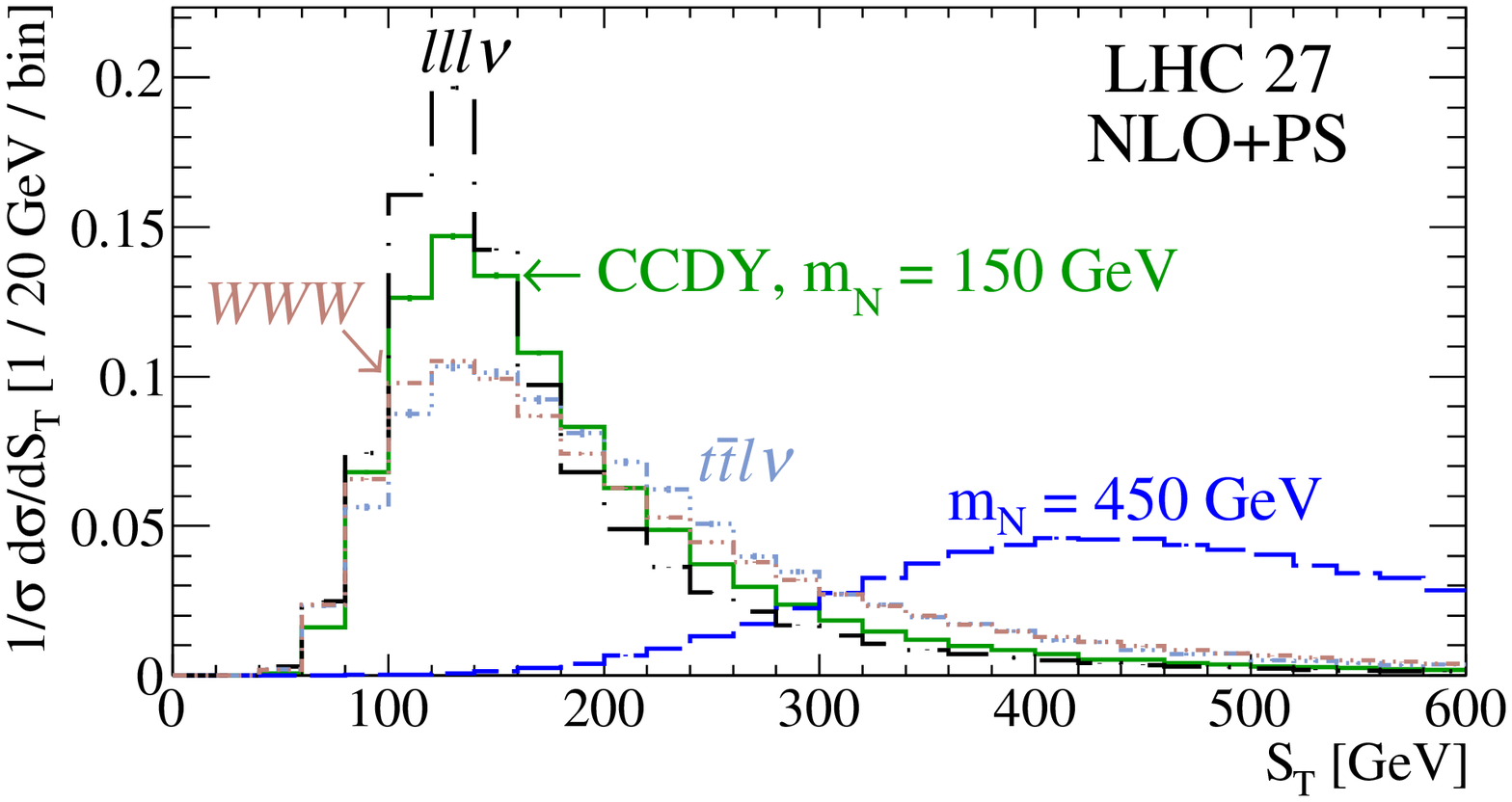}       }
\\
\subfigure[]{\includegraphics[width=.48\textwidth]{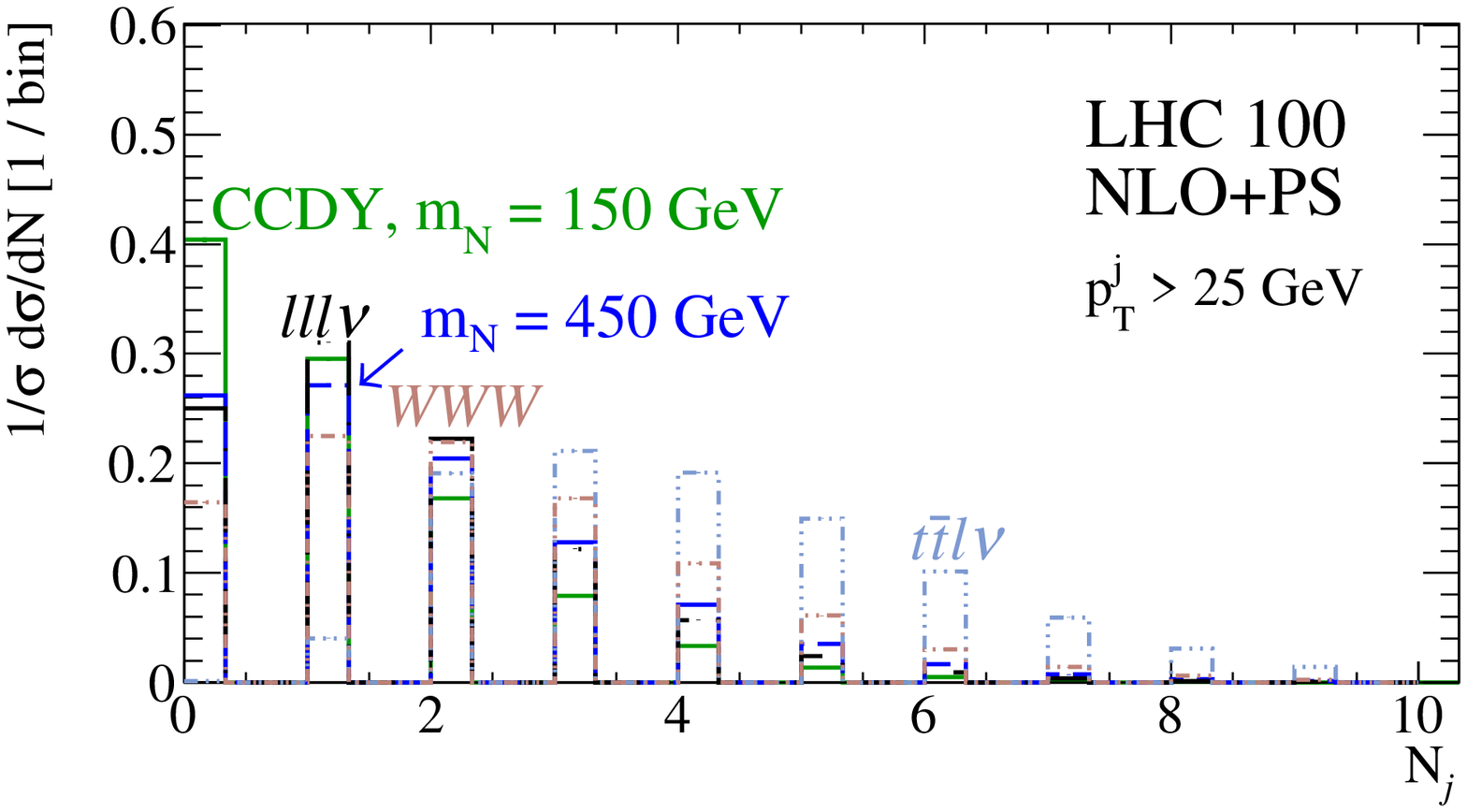}   }
\subfigure[]{\includegraphics[width=.48\textwidth]{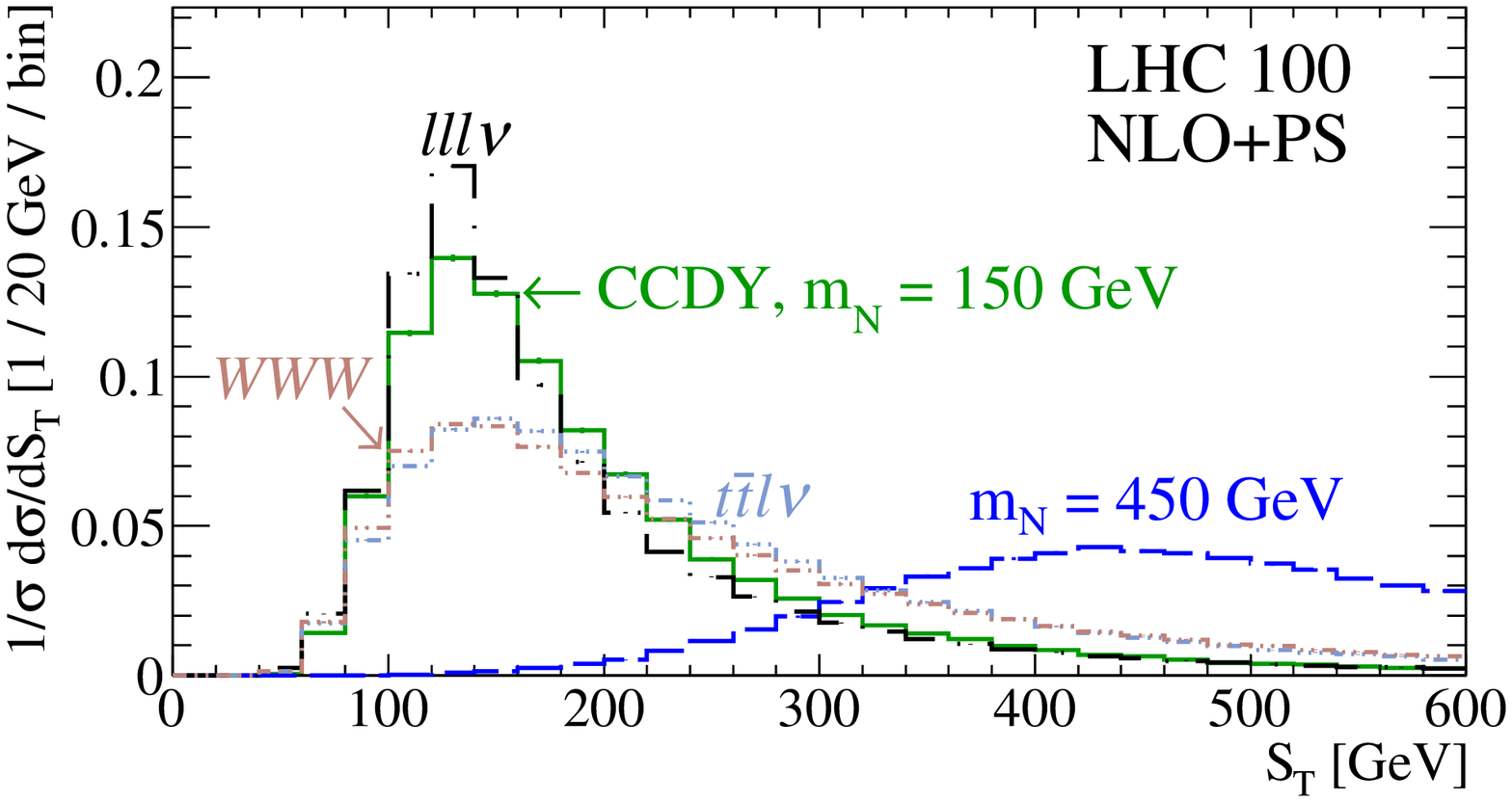}       }
\end{center}
\caption{
Same as Fig.~\ref{fig:vetoKinLeadingJet} but for (a,c,e) the number of jets with $p_T^j > 25\GeV$ and (b,d,f) $S_T$.
}
\label{fig:vetoKinHadLepActivity}
\end{figure*}

This can be effectively observed in Fig.~\ref{fig:vetoKinHadLepActivity},
 where we plot the (normalized) number of jets with $p_T>25\GeV$ at (a) $\sqrt{s} = 14$, (c) 27, and (e) 100 TeV, 
for the CCDY signal process assuming $m_N = 150$ (solid) and $450\GeV$ (dash) as well as the 
$3\ell\nu$ (dash-1 dot), $WWW$ (dash-2 dots), and $t\overline{t}\ell\nu$ (dash-3 dots)  background processes.
At $\sqrt{s}=14\TeV$, one sees that fewer than $40-45\%$ of $pp\to3\ell\nu$ and $3W$ events fall into the zero-jet bin, i.e., events with only jets possessing $p_T^j < 25\GeV$.
This is comparable to the CC DY signal process at $m_N=450\GeV$ but unlike the $m_N=150\GeV$ process, which is closer to the 55-60\% threshold.
Conversely, the $pp\to3\ell\nu$ and $3W$ processes more readily populate higher jet multiplicities, with \confirm{$\gtrsim10\%$} of events possessing at least three jets, 
whereas only 5\% of CC DY events at $m_N=150\GeV$ do.
At higher $\sqrt{s}$, one clearly sees that the lowest (highest) multiplicity bins deplete (grow) faster for the EW backgrounds than the CC DY samples.
Concretely, for the signal process at $m_N = 150~(450)\GeV$, one observes that the fraction of events with zero jets above 25 GeV 
drop from about \confirm{60\%~(45\%) at $\sqrt{s}=14\TeV$} to roughly \confirm{50\%~(35\%)} at 27 TeV, to approximately \confirm{40\%~(25\%)} at 100 TeV.
Similarly, for the $t\overline{t}\ell\nu$ process, at least \confirm{10\% of events have four or more jets at 14 TeV};
at 27 TeV, at least \confirm{10\% of events have five or more jets};
and at 100 TeV, \confirm{over 10\% of events have at least six jets}.
Roughly speaking, at our level of modeling, \confirm{about $5\%$ of $t\overline{t}\ell\nu$ events have eight or nine jets} with $p_T>25\GeV$.
One should caution in interpreting these results:
For smaller jet radii, one expects the level of migration (and multiplicity) to increase dramatically.
In addition, the accuracy here is only NLO+PS, indicating that the highest jet multiplicity bins are populated by the parton shower.

To see how the  EW background for the $pp\to N\ell+X \to 3\ell+X$ process copes with a dynamic jet veto in light of this jet activity, 
we revisit Fig.~\ref{fig:vetoKinLeadingJet}, which additionally shows the representative $pp\to 3\ell\nu$ (dash-1-dot) and $pp\to3W\to3\ell 3\nu$ (dash-2-dots).
As the EW processes above are driven by $(q\overline{q})$-annihilation, we see much the same qualitative behavior in the $p_T^{j_1}$ distribution as we do for the signal processes.
Quantitatively, however, the larger jet activity of the multiboson channels causes both the $3\ell\nu$ and $3W$ processes,
which have mass scales of the order $M_{VV}\sim 2M_V \sim 160-180\GeV$ and $M_{VVV}\sim 240-270\GeV$,
 to behave significantly more like the $m_N=450\GeV$ channel than the $m_N\sim 150\GeV$.
 As a result, in the leading charged lepton-to-leading jet $p_T$ ratio $r^{\ell_1}_{j_1} = (p_T^{\ell_1} / p_T^{j_1})$,
 we see that most of the EW events possess $r^{\ell_1}_{j_1}<1$.
 This is very unlike the CC DY process and in fact much more the like $t\overline{t}\ell\nu$ process. 
 As a function of collider energy, we observe a similar (though perhaps slightly milder) $\sqrt{s}$ dependence for  $r^{\ell_1}_{j_1}$ as we do the $t\overline{t}\ell\nu$ case, 
 i.e., an increase separation of signal and background, and need not be discussed further.

 An interesting consequence of the high jet activity in the EW $pp\to 3\ell\nu$ and $3W$ processes is the impact on the $p_T$ of the leading leptons.
 Unlike the CC DY and VBF signal processes, which feature low-$p_T$ ISR, the EW backgrounds feature high-$p_T$ ISR that recoil against the leading charged leptons.
 The impact of this can be seen in Fig.~\ref{fig:vetoKinHadLepActivity}, 
 where we show for (b) $\sqrt{s}=14$, (d) 27, and (f) 100 TeV,  exclusive $S_T$ as defined in Eq.~\ref{eq:defST}, 
  for the several processes presently under discussion. 
 For the EW processes, the $p_T$ of the charged leptons scale as $p_T^\ell \sim M_V/2$, leading to a characteristic $S_T$ of
\begin{equation}
 S_T^{2V},~S_T^{3V} = \sum_{\ell} \vert \vec{p_T}^\ell \vert \sim 3\frac{M_V}{2} \sim 120-135\GeV.
 \label{eq:STewBkg}
\end{equation}
This distribution is observed at $\sqrt{s}=14\TeV$, but with a particular broadness.
We attribute the dispersion in $S_T^{2V}$ and $S_T^{3V}$, which would otherwise be much narrower distributions since they result from discrete transitions / decays,
to the hard ISR.
For reference, the broadness of the CC DY signal processes originate from the prompt charged lepton in the  initial $pp \to N\ell_N$ scattering process,
which leads to a continuum distribution for $p_T^{\ell_N}$ (see Fig.~\ref{fig:partonKinEllpT}).
At higher collider energies, this broadening is worsened, which can be attributed to the associated increase in jet multiplicity.

Now, with Fig.~\ref{fig:vetoKinHadLepActivity} in mind, imposing a dynamic jet veto of $r^{\ell_1}_{j_1} = (p_T^{\ell_1} / p_T^{j_1})>1$ 
on the EW backgrounds would result in two outcomes:
First is the removal of a sizable fraction of  background events, as evident in Fig.~\ref{fig:vetoKinLeadingJet}.
Second, is the reduction of hard ISR that presently kicks the multilepton systems and is responsible for broadening  the $S_T^{3\ell\nu}$ and $S_T^{3W}$ distributions.
This implies that the EW boson systems would largely be at rest and, crucially, 
that their decay products would possess Born-like kinematics more in line with Eq.~\ref{eq:STewBkg}.
This should be contrasted with $S_T^N \gtrsim m_N$, for the heavy neutrinos in the mass range we are considering ($m_N > 150\GeV$); see Eq.~\ref{eq:STsignal}.
(For the top quark background, similar arguments also hold for $S_T^{t\overline{t}\ell\nu} \sim 140-155\GeV$.)
Hence, imposing a selection of
\begin{equation}
S_T > 125-175\GeV,
\end{equation}
\textit{in conjunction} with a dynamic jet veto would eradicate the EW (and top quark) backgrounds, while remaining resilient as a function of collider energy.

\subsubsection*{Fake Leptons}

Non-prompt leptons from hadron decays and light-flavored QCD jets mis-identified as electrons or hadronically decaying $\tau$ leptons $(\tau_h)$,
objects collectively known as ``fake leptons,'' represent a non-negligible source of backgrounds in multilepton searches for heavy neutrinos 
at hadron colliders~\cite{Abachi:1995yi,delAguila:2007qnc,Alva:2014gxa,Aad:2015xaa,Sirunyan:2018mtv,Sirunyan:2018xiv,Aaboud:2018spl,Pascoli:2018rsg}. 
Fake leptons satisfying selection criteria, however, must necessarily originate from high-$p_T$ hadronic activity.
By color conservation, this implies~\cite{Pascoli:2018rsg} a high likelihood of additional hadronic activity,  i.e., a second jet, with comparable $p_T$ elsewhere in the detector,
and suggests that a jet veto can improve the rejection rate of fake lepton backgrounds.

The reason for this is that high-$p_T$ jets and hadrons that give rise to fake leptons are seeded by high-$p_T$ partons that are charged under QCD.
As the initial protons colliding form precisely a color-singlet state with zero, net transverse momentum, 
the color charge and transverse momentum of this parton must be balanced by one or more recoiling partons, which separately form hadrons / jets.
For the specific case of hadron decays, e.g., $\mathcal{B}\to\mathcal{D}\ell\nu$, 
this may simply be the spectator hadron, which is known to have similar momenta as the outgoing charged lepton~\cite{Isgur:1988gb,Scora:1995ty}. 
Independent of this, rates for mis-identifying light jets as electrons and $\tau_h$ are highest for the lowest $p_T$ jets
in an event due to poorer resolution~\cite{Alvarez:2016nrz,CMS-DP-2017-036,CMS-DP-2018-009}.
Hence, when taken together, the presence of a fake lepton implies not only the existence of one or more additional clusters of hadronic activity, 
but that these additional clusters have a slightly higher likelihood to possess a $p_T$ greater than the fake lepton itself.
As a consequence, the use of a dynamic jet veto of $r^{\ell_1}_{j_1}  = (p_T^{\ell_k} / p_T^{j_1})>1$ for $k=1,\dots,3$, can improve the rejection rate of this class of background.

\subsection{Dynamic Vetoes Beyond $p_T$ Ratios}\label{sec:jetVetoNewVetoes}

At its heart, the dynamic jet veto we employ in Eq.~\ref{eq:defDynamicVeto}, in conjunction with $R=1$ jets,  
functions as a discriminant between leptonic and hadronic activities.
The $p_T$ of an event's leading lepton and jet, however, are not the sole measures of such behavior.
Other observables can also quantify this in complementary aspects due to their varying levels of inclusiveness.
For leptons, this includes exclusive $S_T$, as defined in Eq.~\ref{eq:defST}.
For hadronic activity, there is inclusive $H_T$, which is defined as the scalar sum of all hadronic $p_T$,
\begin{equation}
 H_T = \sum_{k~\text{s.t.}~\vert \eta^{j_k} \vert < \eta^{\rm max}} \vert \vec{p}_T^{~j_k} \vert, \quad\text{with}\quad \eta^{\rm max}=4.5.
\label{eq:defIncHT}
\end{equation}
In practice, such generalizations of jet vetoes can be employed by requiring, for example, 
that the $H_T$ of an event is below some threshold $H_T^{\rm Veto}$, or by setting $\pTVeto$ proportionally to $S_T$, on an event-by-event basis.
In light this perspective,  we briefly explore if alternative dynamic jet veto criteria can fulfill our main intent of improving sensitivity of multi-lepton searches for heavy neutrinos.
As in Sec.~\ref{sec:jetVetoBkg}, signal and background processes are evaluated at NLO+PS according to Sec.~\ref{sec:Setup}
and the nominal fiducial and kinematic cuts of Eq.~\ref{eq:toyFidKinCuts} are applied to jets and the three leading charged leptons after reconstruction.

\begin{figure*}[!t]
\begin{center}
\subfigure[]{\includegraphics[width=.48\textwidth]{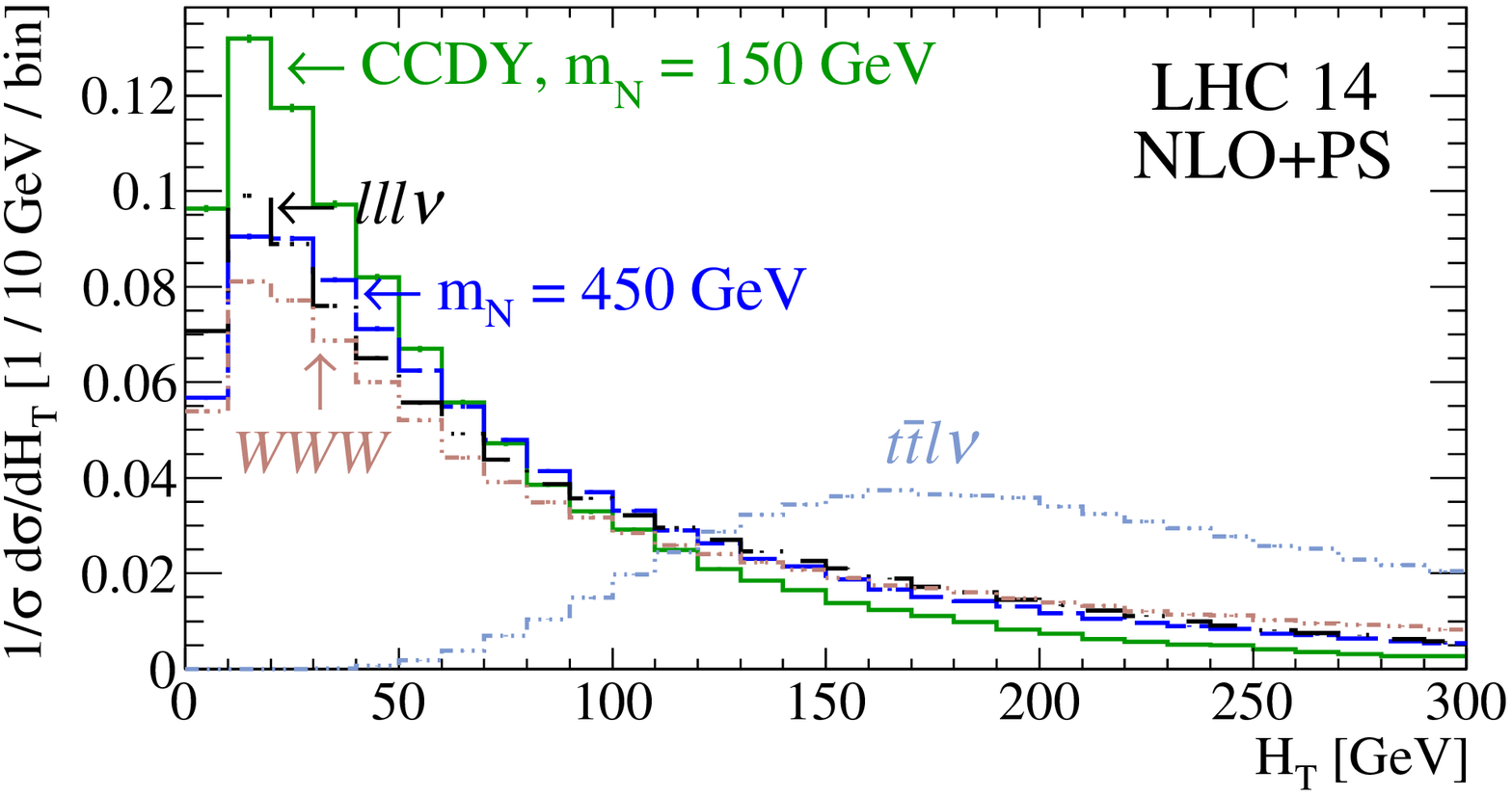}	}
\subfigure[]{\includegraphics[width=.48\textwidth]{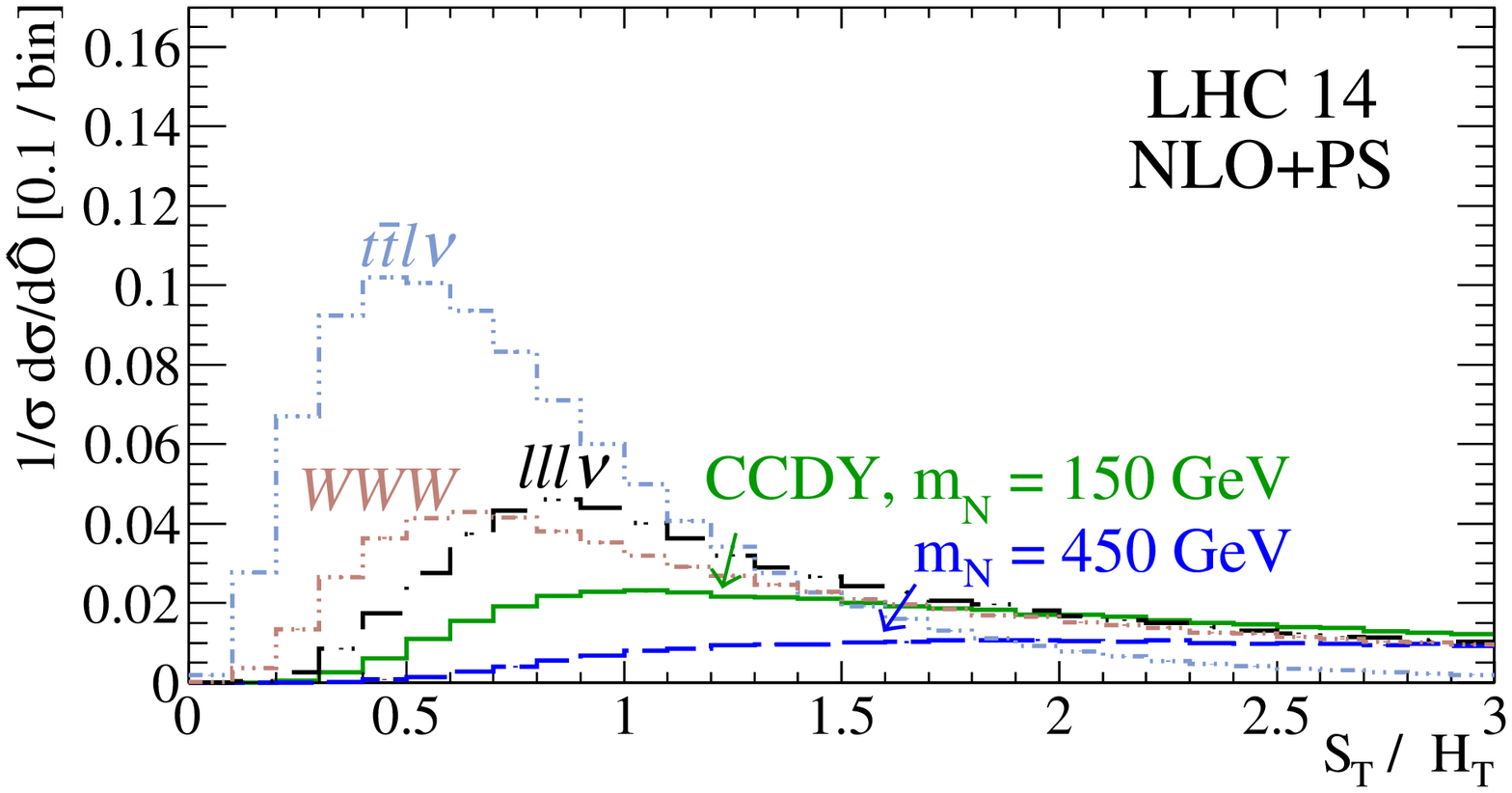}       }
\\
\subfigure[]{\includegraphics[width=.48\textwidth]{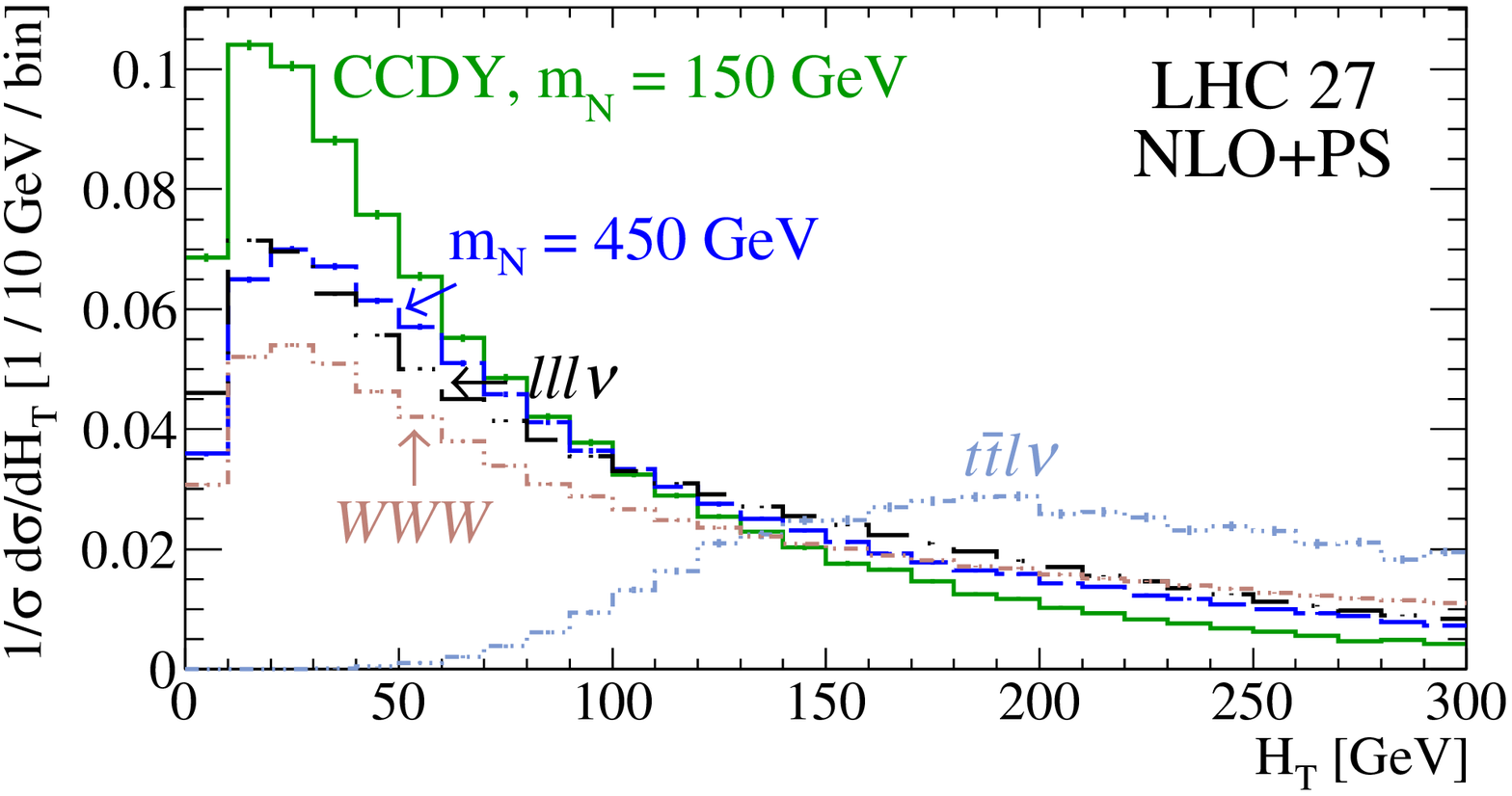}	}
\subfigure[]{\includegraphics[width=.48\textwidth]{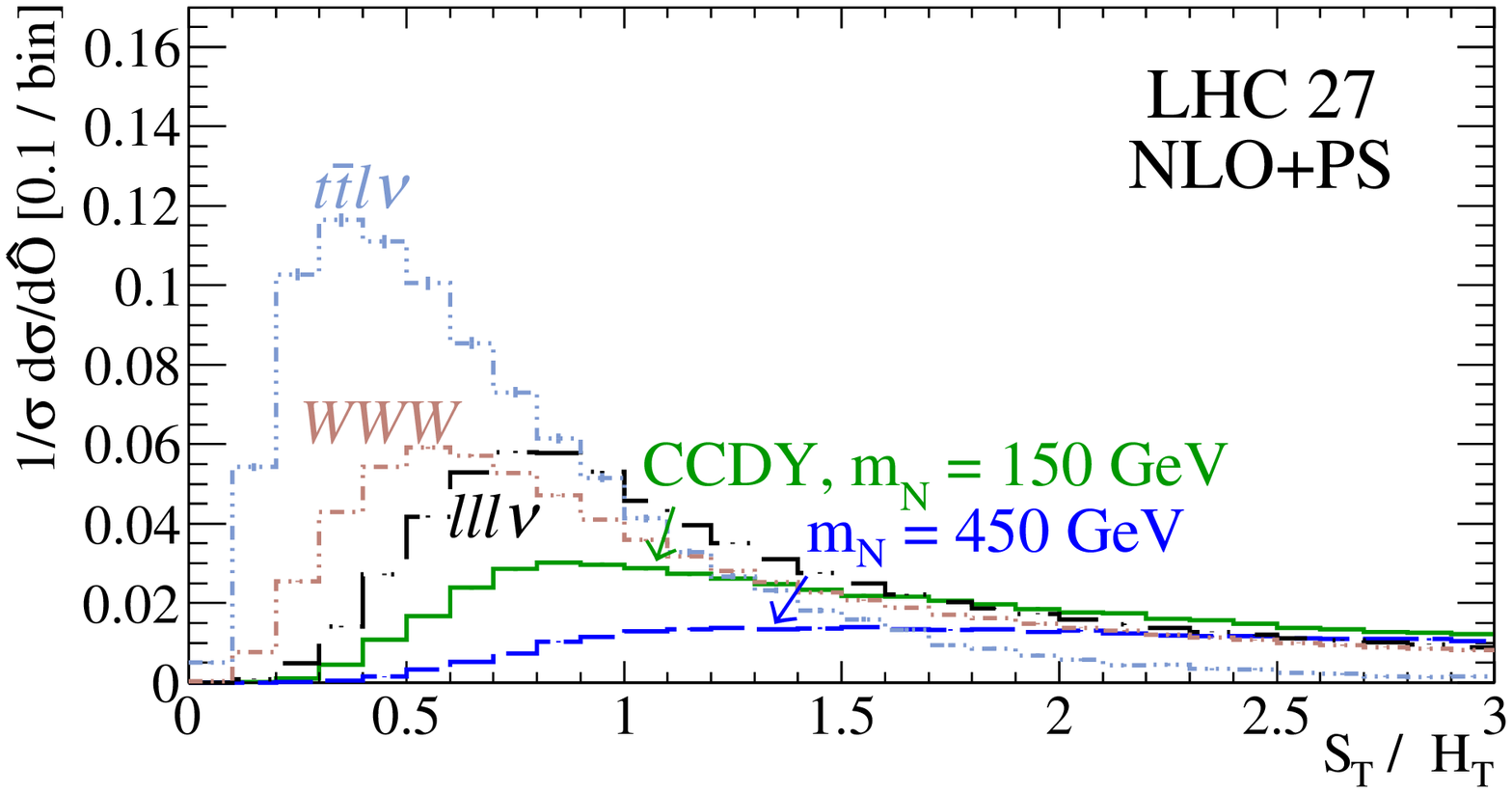}       }
\\
\subfigure[]{\includegraphics[width=.48\textwidth]{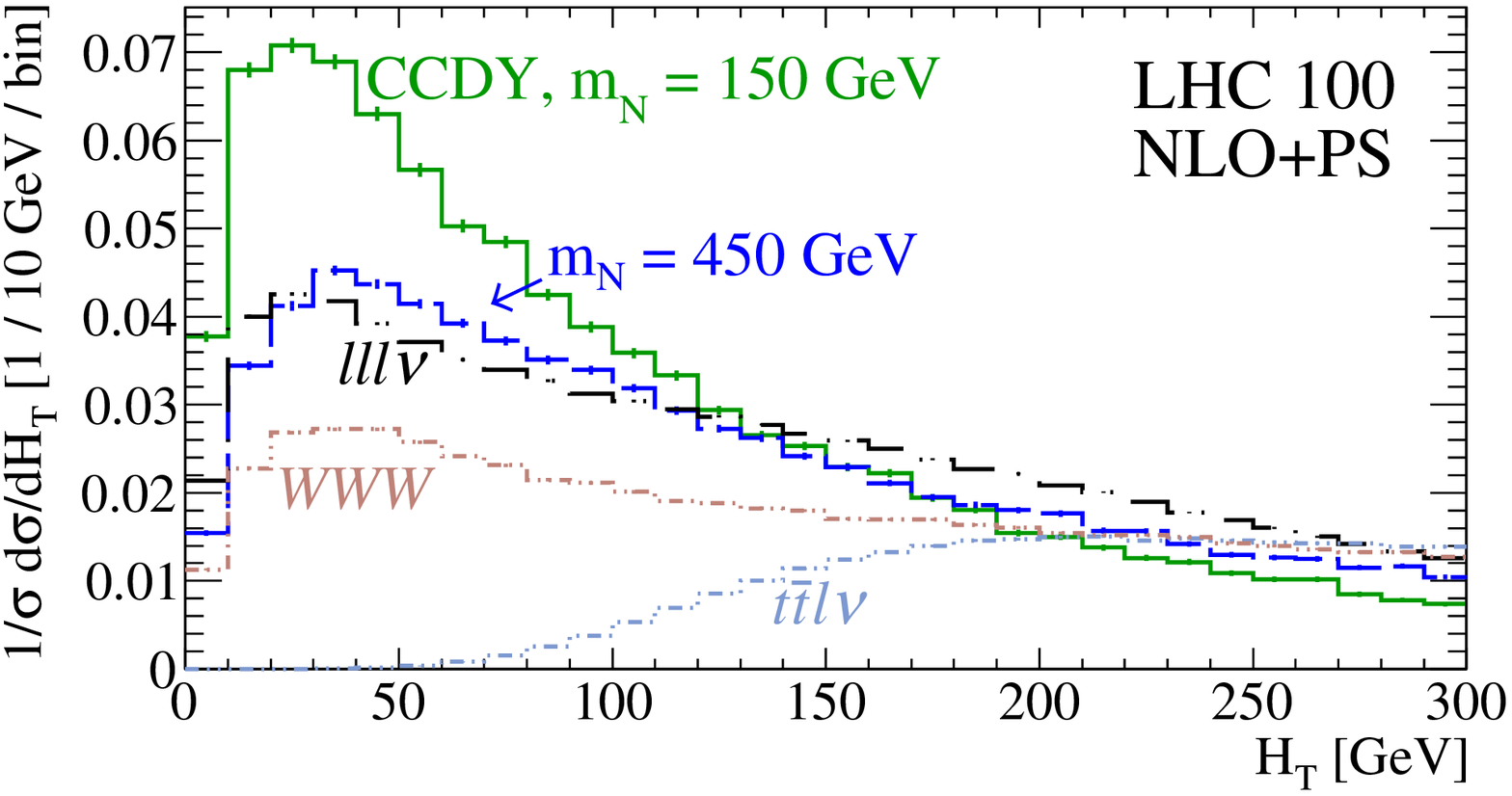}	}
\subfigure[]{\includegraphics[width=.48\textwidth]{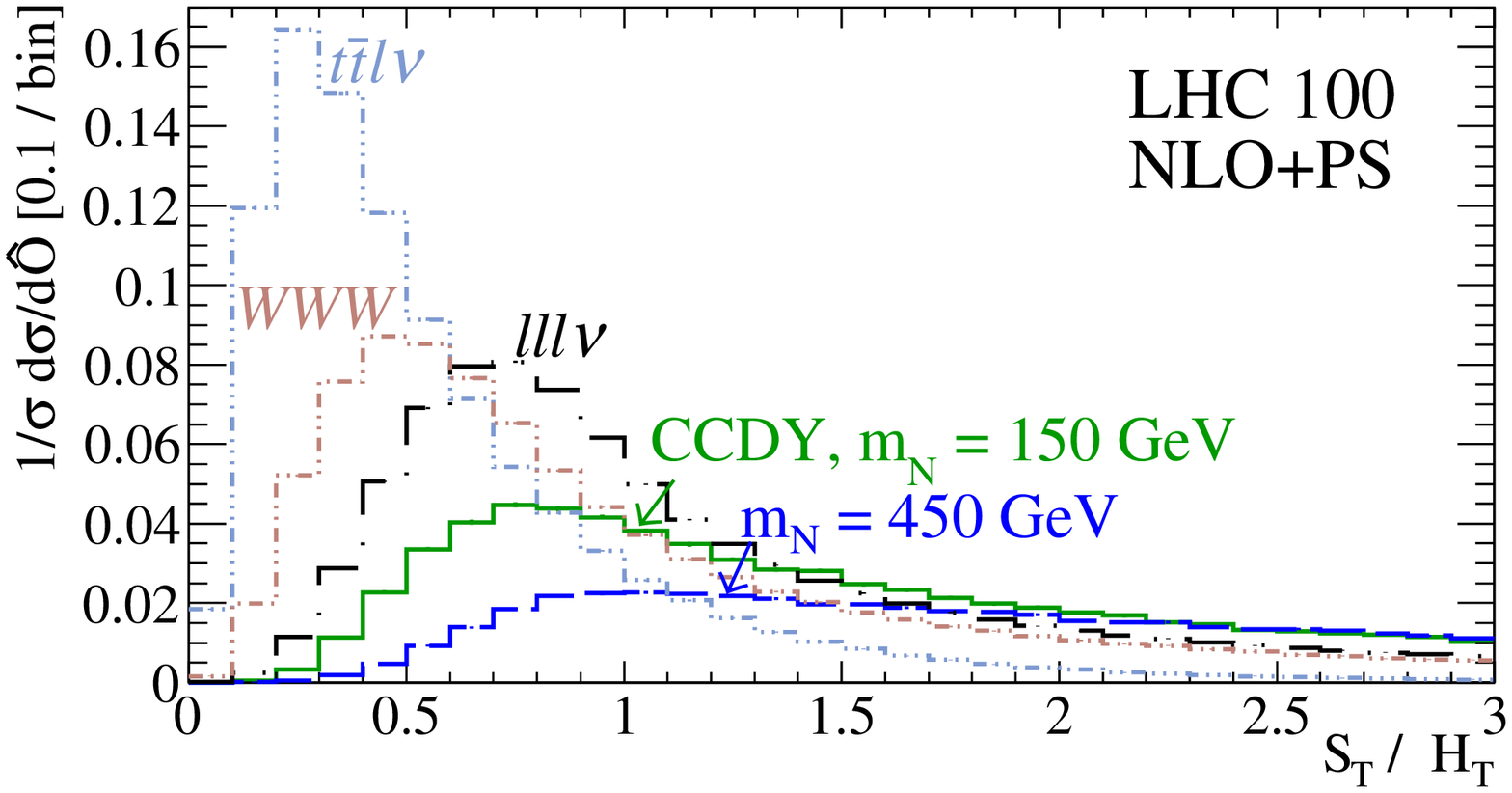}       }
\end{center}
\caption{
Same as Fig.~\ref{fig:vetoKinLeadingJet} but for (a,c,e) $H_T$ as well as (b,d,f) the ratio $S_T/H_T$.
}
\label{fig:vetoHTSTratio}
\end{figure*}

The motivation for considering alternative measures of leptonic or hadronic activity stems from the observation in Sec.~\ref{sec:ColliderParticleKin} 
that there exists an entire class of leptonic observables, not just $p_T^{\ell_k}$ for $k=1,\dots,3$, whose distributions are largely insensitive to varying collider energies.
Such a property does not hold in general, as seen throughout Secs.~\ref{sec:ColliderParticleKin} and ~\ref{sec:jetVetoBkg}.
In Fig.~\ref{fig:vetoKinHadLepActivity}, for example, we observed an uptick in the jet multiplicity with increasing $\sqrt{s}$ for all processes due, in part, to the opening of phase space.

Now, a consequence of increasing jet multiplicity is the increase in hadronic energy carried away from the hard scattering process to the detector experiments.
To quantify this, we show in Fig.~\ref{fig:vetoHTSTratio} the normalized inclusive $H_T$ distribution, as defined in Eq.~\ref{eq:defIncHT}, at (a) $\sqrt{s}=14$, (c) 27, and (e) 100 TeV,
for the C CDY signal process assuming $m_N = 150$ (solid) and $450\GeV$ (dash) as well as the 
$3\ell\nu$ (dash-1 dot), $WWW$ (dash-2 dots), and $t\overline{t}\ell\nu$ (dash-3 dots)  background processes.
Interestingly, while one sees a considerable broadening of the $H_T$ for both signal and background processes, 
the background processes broaden at a slightly faster rate and  is most pronounced for the $3\ell\nu$ and $t\overline{t}\ell\nu$ channels.
Moreover, for increasing $\sqrt{s}$, a rightward lurch to larger $H_T$ can be observed in the $t\overline{t}\ell\nu$.
The acute sensitivity of $H_T$ to collider energy suggests that, like the ratio $r^{\ell_1}_{j_1}=p_T^{\ell_1} / p_T^{j_1}$,
the ratio $p_T^{\ell_1} / H_T$ may serve as  discriminant to base a veto on hadronic activity.
On the other hand, the robustness of $S_T$ over varying collider energy for the signal process
suggests that the ratio $S_T / H_T$ may too serve as a comparable, if not better, discriminant of leptonic and hadronic activity.
To investigate this, in Fig.~\ref{fig:vetoHTSTratio}(b,d,f), we show the ratio $r^{S_T}_{H_T}=S_T / H_T$ for $\sqrt{s} = 14,~27,$ and 100 TeV, respectively.
Remarkably, as $\sqrt{s}$ increases, the background processes grow significantly more narrow than the CC DY signal processes
and shifts leftward to smaller $S_T/H_T$,
suggesting a potentially powerful means to reject backgrounds at future colliders that inherits the single-scale properties of the $p_T^{\ell_1}/p_T^{j_1}$ discriminant.

\begin{figure*}[!t]
\begin{center}
\subfigure[]{\includegraphics[width=.48\textwidth]{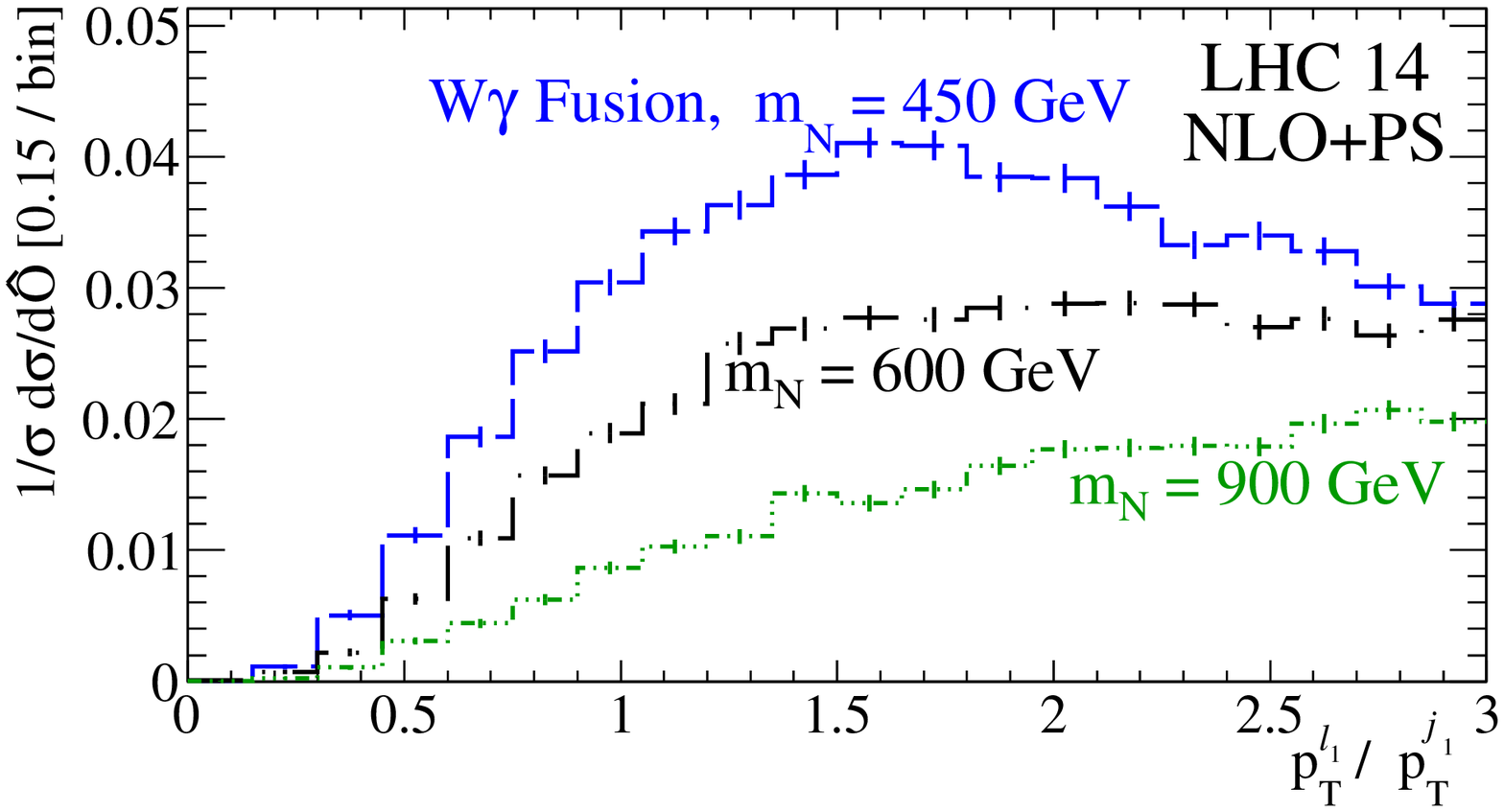}	}
\subfigure[]{\includegraphics[width=.48\textwidth]{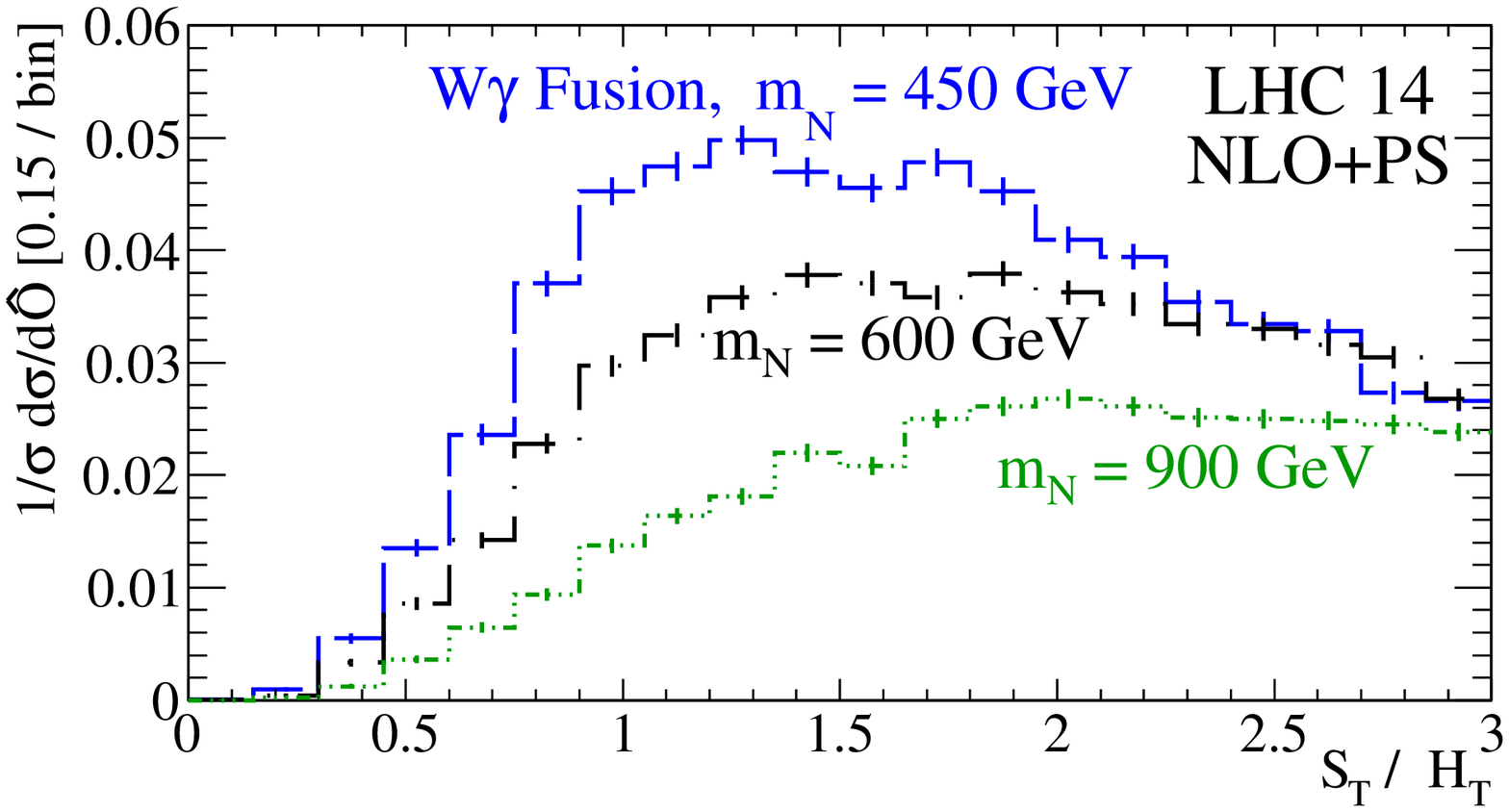}       }
\\
\subfigure[]{\includegraphics[width=.48\textwidth]{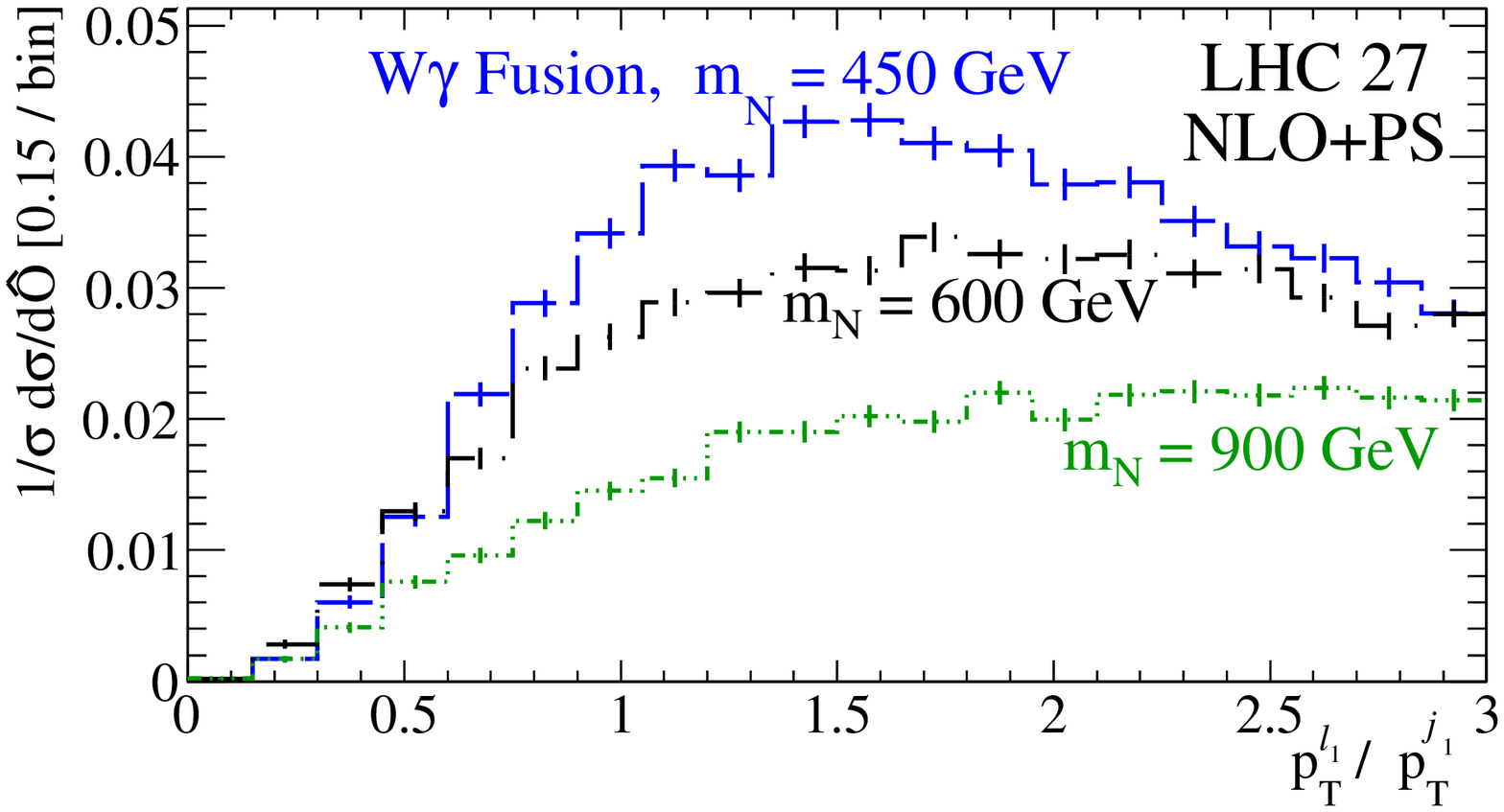}	}
\subfigure[]{\includegraphics[width=.48\textwidth]{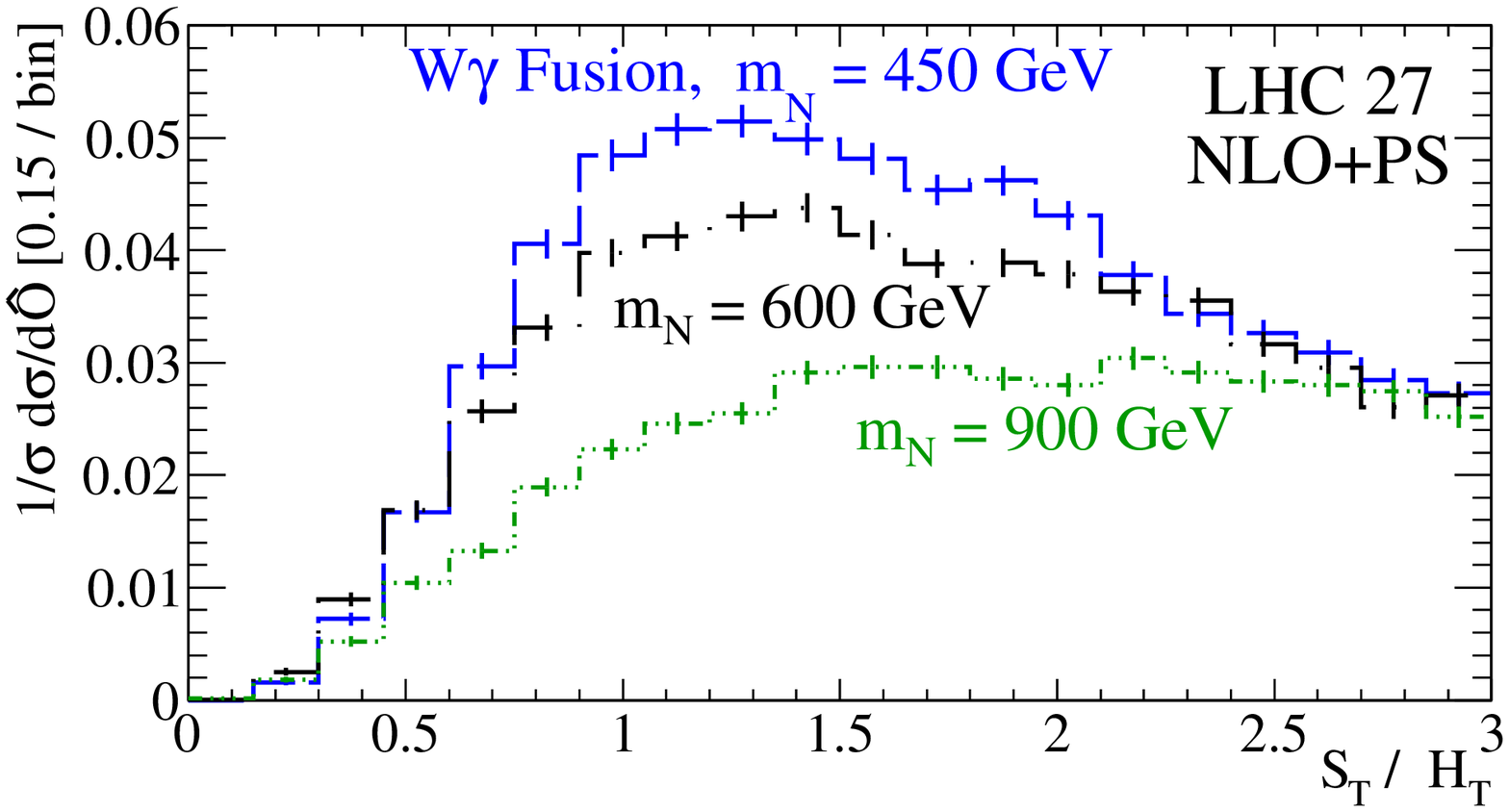}       }
\\
\subfigure[]{\includegraphics[width=.48\textwidth]{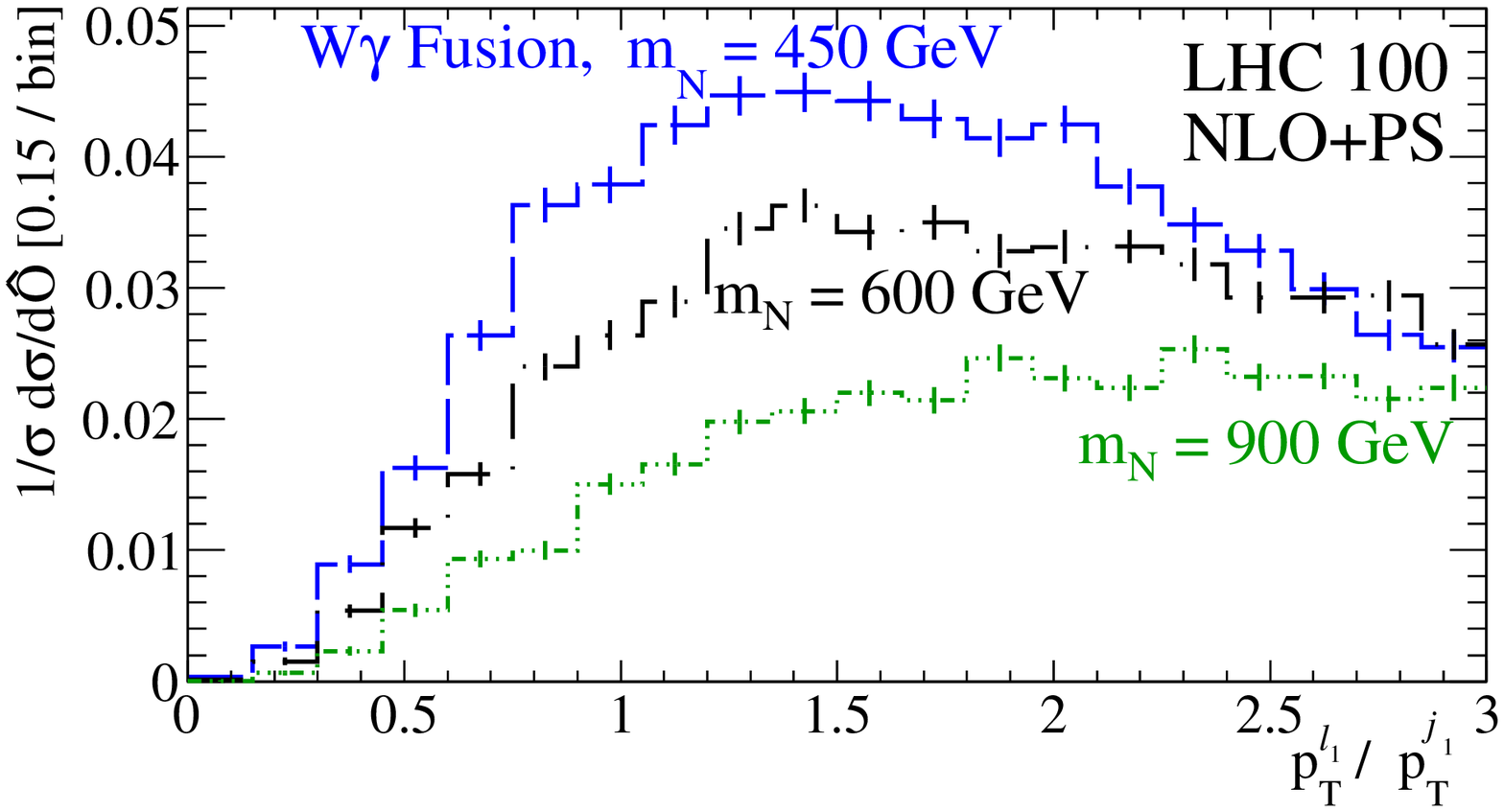}	}
\subfigure[]{\includegraphics[width=.48\textwidth]{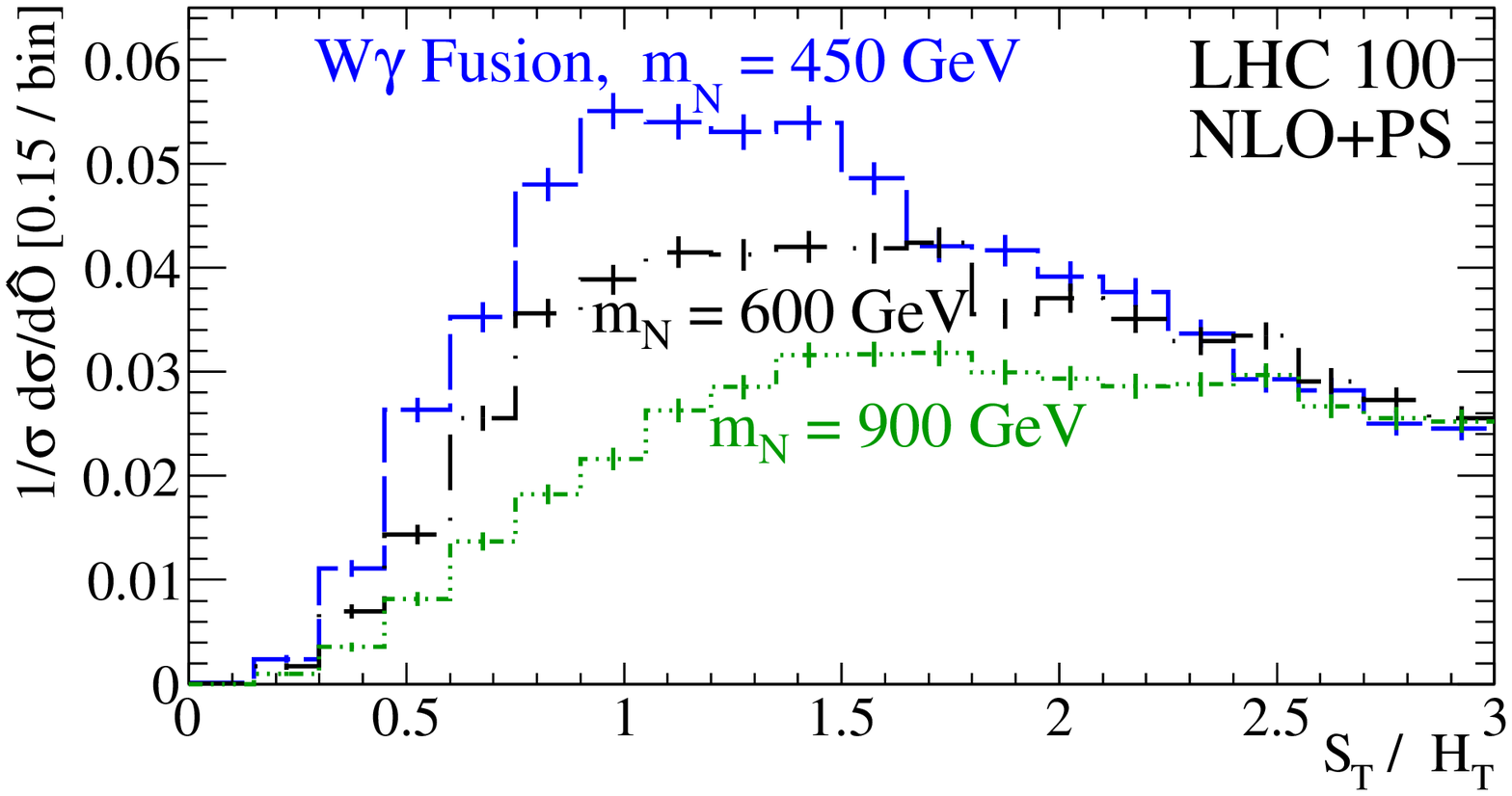}       }
\end{center}
\caption{
Normalized NLO+PS distributions with respect to the ratios (a,c,e) $ p_T^{\ell_1} / p_T^{j_1}$ and (b,d,f) $S_T/H_T$,
at (a,b) $\sqrt{s}=14$ TeV, (c,d) 27 TeV, and (e,f) 100 TeV,
  for the $W\gamma$ fusion (VBF) signal process, with $m_N = 450\GeV$ (dash), $600\GeV$ (dash-dot), and $900\GeV$ (dash-3 dots).}
\label{fig:vetoHTSTratioVBF}
\end{figure*}

As the $W\gamma \to N\ell$ fusion process, as shown in Fig.~\ref{fig:feynmanMulti}(c), 
becomes an increasingly important production vehicle for heavy neutrinos with masses $m_N \gtrsim 500\GeV$ (see Fig.~\ref{fig:xsecVsMass}),
it is worth exploring briefly to what extent generalizations of dynamic jet vetoes would impact sensitivity in the VBF channel.
To this extent, we plot in Fig.~\ref{fig:vetoHTSTratioVBF} the
normalized NLO+PS distributions with respect to the ratios (a,c,e) $r^{\ell_1}_{j_1}=p_T^{\ell_1} / p_T^{j_1}$ and (b,d,f) $r^{S_T}_{H_T}=S_T/H_T$,
at (a,b) $\sqrt{s}=14$ TeV, (c,d) 27 TeV, and (e,f) 100 TeV, for the VBF signal process, with $m_N = 450\GeV$ (dash), $600\GeV$ (dash-dot), and $900\GeV$ (dash-3 dots).
Like in the CC DY signal case, we see broad distributions with  most of the phase space favoring lepton-to-hadron ratios greater than unity.
However, for $m_N = 450\GeV$, we observe a larger concentration of VBF events near $r^{\ell_1}_{j_1},~r^{S_T}_{H_T} \sim 1-1.5$, than for the CC DY channel at the same mass.
This is due to the VBF process inherently containing at least one jet with $p_T^j \gtrsim M_W/2$.
Though as noted in Sec.~\ref{sec:jetVetoSignal}, the characteristic  $p_T^j$ scale is a property of the VBF mechanism and is largely independent of heavy neutrino mass.
This is unlike $p_T^{\ell_1}$, which scales with $m_N$.
Hence, as $m_N$ increases, both $r^{\ell_1}_{j_1}$ and $r^{S_T}_{H_T}$ quickly dampen and shift rightward.
The ratio $r^{S_T}_{H_T}$ tends to be lower than $r^{\ell_1}_{j_1}$ due to the sizable likelihood of resolving the forward VBF jet associated with the 
initial $q\to q\gamma^*$ splitting, which thereby increases $H_T$.
As a function of collider energy, we see much the same leftward shifts to smaller ratios as observed in the CC DY and background processes and need not be discussed further.

Further optimization of jet veto selection criteria, 
such the relative gain (or lack thereof) of choosing $S_T / H_T > 1$ as a veto criterion rather than $p_T^{\ell_1}/p_T^{j_1} > 1$,
are beyond the scope of this work and is deferred to future studies, e.g., Ref.~\cite{Fuks:2019iaj}.
Similarly, investigations into whether the precise requirement that $r^{\ell_1}_{j_1} = p_T^{\ell_1}/p_T^{j_1} > 1$ is better or worse than choosing,
for example, $r^{\ell_1}_{j_1} > 0.5$ or $r^{\ell_1}_{j_2} > 2$, are strongly encouraged, particularly in the context of a multivariate analysis.
Regardless, for the remainder of this study, we set as our dynamic jet veto threshold $r^{\ell_1}_{j_1}  > 1$.

\subsection{Dynamical Jet Vetoes at Leading Logarithmic Accuracy}\label{sec:jetVetoLL}

\begin{figure*}[!t]
\begin{center}
\subfigure[]{\includegraphics[width=.48\textwidth]{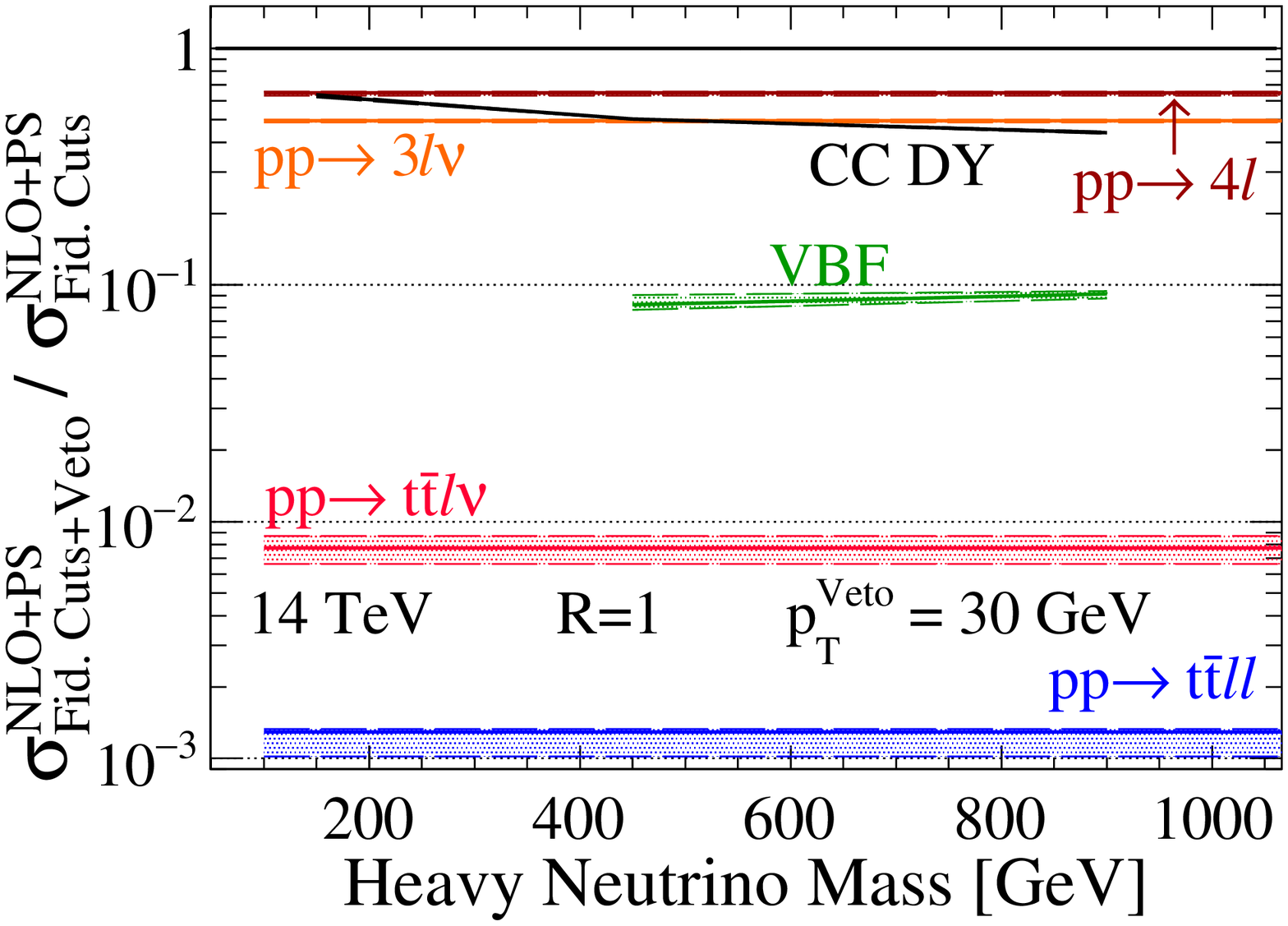}	}
\subfigure[]{\includegraphics[width=.48\textwidth]{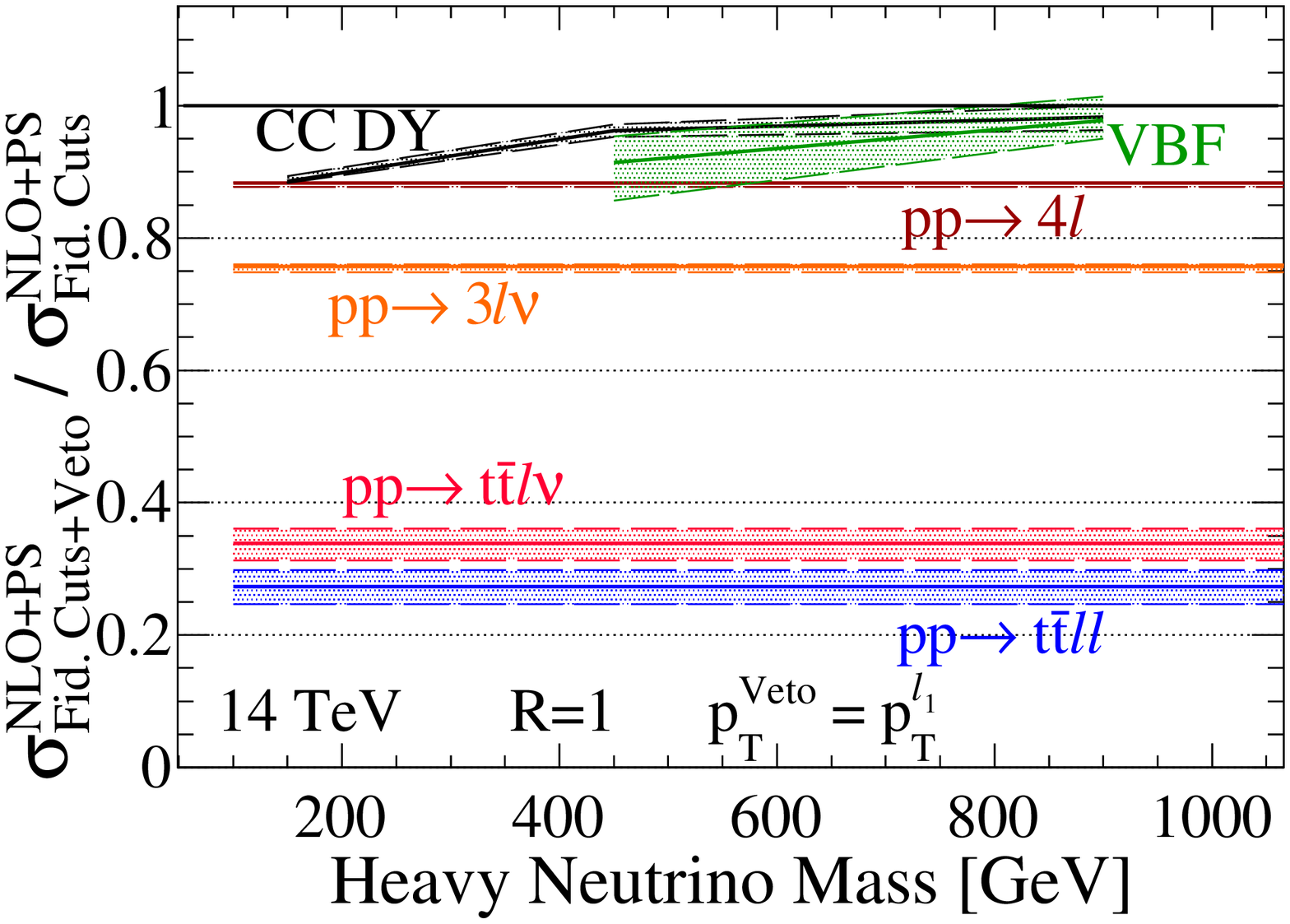}	}
\end{center}
\caption{
As a function of $m_N$, the NLO+PS jet veto efficiency (relative to the fiducial cross section) for the DY and VBF signal processes
as well as representative background processes at 14 TeV,
assuming (a) a static jet veto of $\pTVeto=30\GeV$ and (b) a dynamic jet veto of $\pTVeto=p_T^{\ell_1}$.
}
\label{fig:VetoEffScaleUncNLOPS}
\end{figure*}

As discussed in Sec.~\ref{sec:jetVetoSignal}, 
a principle consequence of choosing a dynamic $\pTVeto$ in lieu of a static veto is a reduced dependence 
of jet veto cross sections on IR and UV cutoff scales, e.g., $\mu_f,~\mu_r,~\mu_s$.
This is clear at NLO+NNLL(veto) in Fig.~\ref{fig:jetVetoEffdyxNLONNLL}, 
where scale dependencies reduce from the $1$-to-$15\%$ level to the sub-percent level for DY production of $N$.
This subsequently raises two questions: 
(i) Does such a reduction also hold for other color structures and scattering topologies (and hence different radiation patterns)?
(ii) How large is the scale dependence at lower logarithmic accuracies, particularly at LL, which is universally calculable with parton showers?
Finding a universal reduction in scale uncertainties for jet veto cross sections of processes with alternative structures  
and at formally lower precision would facilitate a broad application of dynamic jet vetoes in experimental searches.
Such a broad, systematic investigation, however, is beyond the present study but encouraged.
Instead, we take a more limited but highly illustrative first step.

\begin{table}[!t]
\begin{center}
\resizebox{\columnwidth}{!}{
 \begin{tabular}{ c | c | c | c | c | c }
 \hline
                \multicolumn{6}{ c }{Charged Current Drell-Yan: $pp \to N\ell+X \to 3\ell\nu+X$}                 \tabularnewline
\hline \hline
\multirow{2}{*}{Mass $(m_N)$}   & \multirow{2}{*}{Observable}   & Inclusive     & Fiducial  & Fid.~Cuts+     & Fid.~Cuts+ \tabularnewline
                                &                               & (Generator)   & Cuts      & Dynamic~Veto & Static~Veto                                
                \tabularnewline\hline\hline
          &  $\sigma^{\rm NLO+PS}$ [ab] & $125^{+0.6\%}_{-<0.5\%}$  & $58.1^{+0.8\%}_{-<0.5\%}$ & $51.5^{+0.9\%}_{-<0.5\%}$ & $36.6^{+2\%}_{-2\%}$ \tabularnewline
          &  $\sigma^{\rm LO+PS}$  [ab] & $104^{+4\%}_{-6\%}$       & $48.2^{+5\%}_{-6\%}$      & $46.2^{+4\%}_{~-6\%}$       & $34.9^{+5\%}_{-6\%}$ \tabularnewline
 150 GeV  &  $K=\cfrac{\sigma^{\rm NLO+PS}}{\sigma^{\rm LO+PS}}$  & $1.19$ & $1.21$ & $1.11$ & $1.05$ \tabularnewline
          & & & & & \tabularnewline
          &  $\vareps=\cfrac{\sigma^{\rm NLO+PS}_{\rm All~Cuts}}{\sigma^{\rm NLO+PS}_{\rm Fid.~Cuts}}$  & ---  & 100\%  & 89\%  & 63\%  \tabularnewline
          \hline
          &  $\sigma^{\rm NLO+PS}$ [ab] & $1.74^{+2\%}_{-1\%}$ & $1.25^{+2\%}_{-1\%}$ & $1.20^{+1\%}_{-1\%}$ & $0.628^{-<0.5\%}_{-0.7\%}$ \tabularnewline
          &  $\sigma^{\rm LO+PS}$  [ab] & $1.50^{+2\%}_{-2\%}$ & $1.08^{+2\%}_{-2\%}$ & $1.06^{+2\%}_{-2\%}$ & $0.578^{~+2\%}_{-2\%}$ \tabularnewline
 450 GeV  &  $K=\cfrac{\sigma^{\rm NLO+PS}}{\sigma^{\rm LO+PS}}$  & $1.16$ & $1.16$ & $1.13$ & $1.09$ \tabularnewline
                    & & & & & \tabularnewline
          &  $\vareps=\cfrac{\sigma^{\rm NLO+PS}_{\rm All~Cuts}}{\sigma^{\rm NLO+PS}_{\rm Fid.~Cuts}}$  & ---  & 100\%  & 96\%  &  50\% \tabularnewline
          \hline
          &  $\sigma^{\rm NLO+PS}$ [zb] & $98.4^{+2\%}_{-2\%}$ & $83.1^{+2\%}_{-2\%}$ & $81.6^{+2\%}_{-2\%}$ & $36.4^{+1\%}_{-0.9\%}$ \tabularnewline
          &  $\sigma^{\rm LO+PS}$  [zb] & $85.0^{+7\%}_{-6\%}$ & $71.9^{+6\%}_{-6\%}$ & $71.3^{+6\%}_{-6\%}$ & $32.5^{+7\%}_{-5\%}$ \tabularnewline
 900 GeV  &  $K=\cfrac{\sigma^{\rm NLO+PS}}{\sigma^{\rm LO+PS}}$  & $1.16$ & $1.16$ & $1.15$ &  $1.12$ \tabularnewline
                    & & & & & \tabularnewline
          &  $\vareps=\cfrac{\sigma^{\rm NLO+PS}_{\rm All~Cuts}}{\sigma^{\rm NLO+PS}_{\rm Fid.~Cuts}}$  & ---  & 100\%  & 98\%  &  44\% \tabularnewline
\hline\hline
\end{tabular}
} 
\caption{
For the CC DY $pp\to N\ell+X \to 3\ell\nu+X$ signal process, with $e-\mu/$no-$\tau$ mixing as given in Eq.~(\ref{eq:mcMixingLFV}) and representative $m_N$,
the total inclusive cross section and QCD scale uncertainty [\%] at $\sqrt{s}=14\TeV$, 
the cross section after the nominal kinematical and fiducial cuts of Eq.~(\ref{eq:toyFidKinCuts}),
the same with a dynamic jet veto of $\pTVeto=p_T^{\ell_1}$,
and the same but with an alternative, flavor-agnostic static jet veto of $\pTVeto=30\GeV$,
at NLO+PS(LL), LO+PS(LL).
Also shown are the QCD NLO+PS $K$-factor and the veto efficiency $(\vareps)$ relative to the fiducial cross section.
}
\label{tb:VetoScaleUncDY}
\end{center}
\end{table}

For the DY and VBF signal processes at representative heavy $N$ masses 
and for representative background processes ($pp\to~4\ell,~3\ell\nu,~t\overline{t}\ell\ell,$ and $t\overline{t}\ell\nu$),
we report the scale dependence of jet veto cross sections  and efficiencies at $\sqrt{s}=14\TeV$, up to NLO+PS(LL),
respectively, in Tables~\ref{tb:VetoScaleUncDY}, ~\ref{tb:VetoScaleUncVBF}, and ~\ref{tb:VetoScaleUncBkg}, and graphically in Fig.~\ref{fig:VetoEffScaleUncNLOPS}.
Specifically, we compute the factorization $(\mu_f)$ and renormalization $(\mu_r)$ scale dependence in cross sections  
when jointly scanned over a three-point scale 
variation\footnote[2]{That is, we evaluate $(\mu_f,\mu_r)$ at $(2\times,2\times)$, $(1\times,1\times)$, and $(0.5\times,0.5\times)$ their default/nominal values.}, at LO+PS and NLO+PS.
As a normalization, we define a generic particle-level, fiducial cross section (4th column) obtained from
applying the same requirements as given in Eq.~\ref{eq:toyFidKinCuts} on jets and the three leading charged leptons in an event.
The total inclusive rate, i.e., without kinematic selection criteria beyond necessary generator-level cuts, is also reported (third column) for completeness.
In addition to Eq.~\ref{eq:toyFidKinCuts}, 
we consider the case of a dynamic veto with $\pTVeto=p_T^{\ell_1}$ (fifth column) 
as well as the case of a static veto with $\pTVeto=30\GeV$ (sixth column).
For alternative $\sqrt{s}$, we conjecture that one should observe comparable changes in scale dependence as observed in Fig.~\ref{fig:jetVetoEffdyxNLONNLL} and Ref.~\cite{Fuks:2017vtl}.
To quantify the impact of QCD corrections, we report the NLO+PS $K$-factor at each cut iteration. 
The statistical confidence correspond to 500k generated events, except for the $t\overline{t}\ell\ell$ samples, which correspond to 1M events.

\begin{table}[!t]
\begin{center}
\resizebox{\columnwidth}{!}{
 \begin{tabular}{ c | c | c | c | c | c }
\hline\multicolumn{6}{ c }{$W\gamma$~Fusion: $pp \to N\ell j+X \to 3\ell\nu j+X$}                 \tabularnewline\hline\hline
\multirow{2}{*}{Mass $(m_N)$}   & \multirow{2}{*}{Observable}   & Inclusive     & Fiducial  & Fid.~Cuts+     & Fid.~Cuts+ \tabularnewline
                                &                               & (Generator)   & Cuts      & Dynamic~Veto & Static~Veto
                \tabularnewline\hline\hline 
          &  $\sigma^{\rm NLO+PS}$ [zb] & $408^{+5\%}_{-6\%}$ & $271^{+5\%}_{-6\%}$ & $248^{+4\%}_{-6\%}$  & $22.4^{+10\%}_{-5\%}$         \tabularnewline
          &  $\sigma^{\rm LO+PS}$  [zb] & $391^{+3\%}_{-3\%}$ & $259^{+2\%}_{-2\%}$ & $235^{+3\%}_{-2\%}$  & $23.2^{+5\%}_{-7\%}$          \tabularnewline
450 GeV   &  $K=\cfrac{\sigma^{\rm NLO+PS}}{\sigma^{\rm LO+PS}}$  & $1.04$ & $1.05$ & $1.05$ & $0.97$          \tabularnewline
                    & & & & & \tabularnewline
          &  $\vareps=\cfrac{\sigma^{\rm NLO+PS}_{\rm All~Cuts}}{\sigma^{\rm NLO+PS}_{\rm Fid.~Cuts}}$  &  --- & 100\%  & 91\%  &  8\% \tabularnewline
          \hline
          &  $\sigma^{\rm NLO+PS}$ [zb] & $104^{+3\%}_{-3\%}$        & $75.9^{+4\%}_{-3\%}$ & $74.2^{+4\%}_{-3\%}$ &  $6.94^{+3\%}_{-4\%}$  \tabularnewline
          &  $\sigma^{\rm LO+PS}$  [zb] & $94.4^{+<0.5\%}_{-<0.5\%}$ & $68.9^{+<0.5\%}_{-<0.5\%}$ & $67.2^{+<0.5\%}_{-<0.5\%}$ & $6.49^{+0.7\%}_{-2\%}$   
 \tabularnewline
 900 GeV  &  $K=\cfrac{\sigma^{\rm NLO+PS}}{\sigma^{\rm LO+PS}}$  & $1.10$ & $1.10$ & $1.10$ & $1.07$          \tabularnewline
                    & & & & & \tabularnewline
          &  $\vareps=\cfrac{\sigma^{\rm NLO+PS}_{\rm All~Cuts}}{\sigma^{\rm NLO+PS}_{\rm Fid.~Cuts}}$  &  --- & 100\%  & 98\%  & 9\%  \tabularnewline
\hline\hline
\end{tabular}
} 
\caption{
Same as Table~\ref{tb:VetoScaleUncDY} but for the VBF $pp\to N\ell j+X \to 3\ell\nu j+X$ signal process.
}
\label{tb:VetoScaleUncVBF}
\end{center}
\end{table}

For the DY signal process (Table~\ref{tb:VetoScaleUncDY}), a number of qualitative features can be discerned:
Foremost is that the total inclusive, fiducial, and both jet veto scale uncertainties at NLO+PS are all comparable in size, and span at most $\pm2\%$;
at LO+PS, the uncertainties are $2$-to-$10\times$ larger, but do not exceed $\pm7\%$.
At NLO+PS, efficiencies span approximately \confirm{90-98\%} for the dynamic veto and \confirm{45-65\%} for the static veto.
For increasing $m_N$, veto efficiencies increase (decrease) under the dynamic (static) veto.
Importantly, this is qualitatively and quantitatively comparable to both the central values and uncertainties observed at NLO+NNLL(veto), in Fig.~\ref{fig:jetVetoEffdyxNLONNLL}.
This suggests that NLO+PS provides a good description of dynamic and static jet vetoes, and sufficient for discovery purposes.
For static vetoes, this is consistent with the findings of Refs.~\cite{Monni:2014zra,Fuks:2017vtl}. 
For dynamic vetoes, this is the first such quantitative comparison between parton showers and higher logarithmic jet veto resummations.
In addition, the spectrum of $K$-factors indicate the relative contribution of real and virtual radiation in heavy lepton production via the DY mechanism:
At larger $(m_N/\sqrt{s})$, the inclusive, fiducial, and both veto $K$-factors are comparable, consistent with QCD corrections being dominantly virtual.
At smaller $(m_N/\sqrt{s})$, more exclusive / less inclusive final states possess smaller $K$-factors, indicating the increasing presence of real radiation.
Owing to the uniformity of the NLO+PS $K$-factors for the dynamic veto,
it appears that modeling the DY signal processes at LO+PS using the scale scheme of Eq.~\ref{eq:scale}
and normalizing by a multiplicative $K$-factor of $K\approx1.1$ is a reasonable procedure.
Such a prescription does not hold for the static veto.

\begin{table}[!t]
\begin{center}
\resizebox{\columnwidth}{!}{
 \begin{tabular}{ c | c | c | c | c | c }
\hline \hline
\multirow{2}{*}{Process}   & \multirow{2}{*}{Observable}   & Inclusive      & Fiducial  & Fid.~Cuts+     & Fid.~Cuts+ \tabularnewline
                           &                               &  (Generator)   & Cuts      & Dynamic~Veto & Static~Veto
                \tabularnewline\hline\hline
                    &  $\sigma^{\rm NLO+PS}$ [fb] & $1600^{+2\%}_{-2\%}$ & $631^{+3\%}_{-3\%}$ & $478^{+<0.5\%}_{-1\%}$ & $311^{+1\%}_{-2\%}$ \tabularnewline
 $pp \to 3\ell\nu$  &  $\sigma^{\rm LO+PS}$  [fb] &  $941^{+8\%}_{-9\%}$ & $351^{+5\%}_{-6\%}$ & $335^{+5\%}_{-7\%}$    & $263^{+6\%}_{-7\%}$ \tabularnewline
                    &  $K=\cfrac{\sigma^{\rm NLO+PS}}{\sigma^{\rm LO+PS}}$  & $1.70$ & $1.80$ & $1.42$ & $1.18$ \tabularnewline
                              & & & & & \tabularnewline
                    &  $\vareps=\cfrac{\sigma^{\rm NLO+PS}_{\rm All~Cuts}}{\sigma^{\rm NLO+PS}_{\rm Fid.~Cuts}}$  &   & 100\%  & 76\%  & 49\%  \tabularnewline
                    \hline
                    &  $\sigma^{\rm NLO+PS}$ [fb] & $219^{+<0.5\%}_{-0.5\%}$ & $111^{+0.5\%}_{-0.7\%}$ & $97.8^{+<0.5\%}_{-0.7\%}$ & $71.5^{+1\%}_{-2\%}$ \tabularnewline
 $pp \to 4\ell$     &  $\sigma^{\rm  LO+PS}$ [fb] & $159^{+10\%}_{-11\%}$ & $81.1^{+7\%}_{-8\%}$ & $78.8^{+7\%}_{-8\%}$ & $63.5^{+7\%}_{-8\%}$
                                    \tabularnewline
                    & $K=\cfrac{\sigma^{\rm NLO+PS}}{\sigma^{\rm LO+PS}}$ & $1.37$ & $1.37$ & $1.24$ & $1.13$ \tabularnewline
                              & & & & & \tabularnewline
                    &  $\vareps=\cfrac{\sigma^{\rm NLO+PS}_{\rm All~Cuts}}{\sigma^{\rm NLO+PS}_{\rm Fid.~Cuts}}$  &   & 100\%  & 88\%  &  65\% \tabularnewline
                                    \hline
                    &  $\sigma^{\rm NLO+PS}$ [fb] & $24.0^{+11\%}_{-10\%}$ & $11.9^{+10\%}_{-10\%}$ & $4.04^{+7\%}_{-8\%}$   & $92.1^{+12\%}_{-14\%}\times10^{-3}$ \tabularnewline
 $pp \to t\overline{t}\ell\nu$  
                    &  $\sigma^{\rm LO+PS}$  [fb] & $15.2^{+23\%}_{-18\%}$ & $7.69^{+23\%}_{-17\%}$ & $2.81^{+23\%}_{-17\%}$ & $62.7^{+19\%}_{-19\%}\times10^{-3}$ \tabularnewline
                    &  $K=\cfrac{\sigma^{\rm NLO+PS}}{\sigma^{\rm LO+PS}}$  & $1.58$ & $1.55$ & $1.44$ & $1.47$ \tabularnewline
                              & & & & & \tabularnewline
                    &  $\vareps=\cfrac{\sigma^{\rm NLO+PS}_{\rm All~Cuts}}{\sigma^{\rm NLO+PS}_{\rm Fid.~Cuts}}$  &   & 100\%  & 34\%  & 0.8\%  \tabularnewline
                    \hline
                    &  $\sigma^{\rm NLO+PS}$ [fb] & $108^{+7\%}_{-11\%}$   & $35.4^{+9\%}_{-10\%}$  & $9.68^{+9\%}_{-10\%}$   & $45.8^{+3\%}_{-21\%}\times10^{-3}$ \tabularnewline
 $pp \to t\overline{t}\ell\ell$  
                    &  $\sigma^{\rm LO+PS}$  [fb] & $72.7^{+30\%}_{-21\%}$ & $24.2^{+30\%}_{-21\%}$ & $4.85^{+28\%}_{-11\%}$ & $24.1^{+35\%}_{-16\%}\times10^{-3}$ \tabularnewline
                    &  $K=\cfrac{\sigma^{\rm NLO+PS}}{\sigma^{\rm LO+PS}}$  & $1.48$ & $1.47$ & $1.56$ & $1.75$ \tabularnewline
                              & & & & & \tabularnewline
                    &  $\vareps=\cfrac{\sigma^{\rm NLO+PS}_{\rm All~Cuts}}{\sigma^{\rm NLO+PS}_{\rm Fid.~Cuts}}$  &   & 100\%  & 27\%  & <0.5\%  \tabularnewline
                    \hline
\hline
\end{tabular}
} 
\caption{Same as Table~\ref{tb:VetoScaleUncDY} but for representative SM background processes.}
\label{tb:VetoScaleUncBkg}
\end{center}
\end{table}

For the $W\gamma$ signal process (Table~\ref{tb:VetoScaleUncVBF}), the overall behavior is comparable to the DY channel.
First we observe that the inclusive, fiducial, and dynamic veto rates exhibit a common $\mathcal{O}(5)\%$ scale uncertainty at NLO+PS, 
for both intermediate (450 GeV) and large (900 GeV) $m_N$.
As discussed in Sec.~\ref{sec:ColliderXSec},  this is larger than the scale dependence at LO+PS and is driven by subprocesses sensitivity to the gluon PDF.
For the static veto at large $m_N$, a similar scale dependence is observed at NLO+PS and LO+PS;
for intermediate $m_N$, the scale dependence increases by about $2\times$, reaching values as large as $\mathcal{O}(10)\%$ with a $K$-factor of $K\lesssim 1$.
Like the DY channel, the dynamic jet veto efficiency is about $90-98\%$.
For the static veto, the situation is more dismal, with efficiencies of $\mathcal{O}(10)\%$ due to the $p_T$ of the leading jet scaling as $p_T^{j_1}\sim M_W/2$, which is above the threshold.
While the precise choice of $\pTVeto=30\GeV$ and the $\vert \eta \vert$-window are somewhat arbitrary, the principle at hand remains: 
for the mass scales under consideration, VBF topologies are not robust against arbitrary static jet vetoes due to the presence of forward, high-$p_T$ jets.
We conclude that modeling the signal process at LO+PS with a QCD $K$-factor of $1-1.05$ for a static veto and $1.05-1.1$ for dynamic veto 
is a sufficient description of the VBF processes at NLO+PS.

Turning to our representative backgrounds (Table~\ref{tb:VetoScaleUncBkg}), the situation is more complex.
Beginning with the $pp\to3\ell\nu$ process (first row), 
one notices first the large difference between the NLO+PS and LO+PS cross sections for the inclusive, fiducial, and dynamic veto rates.
The  $K$-factors span $K\approx 1.4-1.8$ and are much larger than what the LO+PS scale dependencies, which span $\mathcal{O}(5-10)\%$, would suggest.
The uncertainties at NLO+PS span $\mathcal{O}(1-3)\%$ for all four rates, and are in-line with fiducial predictions up to 
NNLO~\cite{Smith:1989xz,Ohnemus:1991gb,Ohnemus:1992jn,Baur:1993ir,Grazzini:2015nwa,Grazzini:2016swo}.
The huge increase is due to a radiation zero~\cite{Mikaelian:1977ux,Brown:1979ux,Mikaelian:1979nr,Zhu:1980sz,Brodsky:1982sh,Brown:1982xx} 
in the $q\overline{q'}\to W\gamma^{(*)}$ process, which suppresses the Born rate,
and is broken in the $q\overline{q'}\to W\gamma^{(*)} g$ and $q\overline{q'}\to W\gamma^{(*)} gg$ sub-processes.
For the static veto, the $K$-factor is $K\approx 1.2$, indicating the modest size of virtual and unresolved real radiation.
For the dynamic and static vetoes, the efficiencies are about $75\%$ and $50\%$, respectively, showing the utility of jet vetoes even on EW background processes.
Due to a much weaker radiation zero, 
the $pp\to4\ell$ process (second row) exhibits a smaller $K$-factor but much the same scale uncertainties at NLO+PS and LO+PS.
Therefore, aside from a slightly larger veto efficiency of \confirm{$\vareps\approx90\%$ and 65\%} for the dynamic and static veto, respectively, 
the channel needs not to be discussed further.

Qualitatively, the top quark channels $pp\to t\overline{t}\ell\nu$ (third row) and $t\overline{t}\ell\ell$ (fourth row)
exhibit a very different scale dependencies and veto efficiencies than the $3\ell\nu$ and $4\ell$ processes.
The $pp\to t\overline{t}\ell\nu~(t\overline{t}\ell\ell)$ process reveals $K$-factors of $K\approx1.45-1.6~(1.45-1.75)$ across all rates,
with  $\mathcal{O}(20-30)\%$ uncertainties at LO+PS that reduce to $\mathcal{O}(10)\%$ at NLO+PS.
At NLO+PS, the dynamic (static) veto cross section possesses a scale dependence a bit lower (larger) than this average; the difference between the two is about $1.7-2\times$.
Notably, in the absence of other selection criteria, the dynamic and static veto efficiencies differ radically, with about $30\%$ and $<1\%$, respectively.
The very low static veto efficiency is due almost entirely to the existence of two high-$p_T$ $b$ quarks in the decays of the $t\overline{t}$ pair.
The much higher efficiency (but still low in absolute terms) observed for the dynamic veto is due to the prompt $(\ell\nu)$ system is recoiling against a larger multi-body system.
Roughly, boosting this prompt charged lepton by an additional $10\GeV$ translates to only a $5\GeV$ recoil to each $b$ quark, 
and therefore permits the low-$p_T^b$ tail of events to pass the veto.
This suggests, however, that alternative dynamic choices for $\pTVeto$, such at $\pTVeto=p_T^{\ell_2}$ or $p_T^{\ell_3}$, could further lower the dynamic veto efficiency.
(We have verified this but find the improvement unnecessary due to other cuts applied.)
Numerically, the $t\overline{t}\ell\ell$ process contains larger $K$-factors after cuts than the $t\overline{t}\ell\nu$.
This is attributed to the presences of contributions like $gg\to t\overline{t}\ell\ell$, which are present at the Born-level, 
and receive larger virtual corrections than the $t\overline{t}\ell\nu$ process, which is strictly initiated by $(q\overline{q'})$-annihilation at LO.

Collectively, the results of this section are summarized and illustrated in Fig.~\ref{fig:VetoEffScaleUncNLOPS} for both the (a) static  and (b) dynamic jet vetoes.


\section{Observability of Heavy Neutrinos at Hadron Colliders}\label{sec:Observability}
In  Secs.~\ref{sec:ColliderParticleKin} and \ref{sec:jetVeto}, we established the existence of measures for lepton and hadronic activities
that remained robust to changes in collider energies.
We also argued that when such measures are used together, namely a dynamic jet veto in conjunction with exclusive $S_T$,
the combination can appreciably improve signal acceptance and background rejection.
We now turn to quantifying the impact of a dynamic jet veto on the discovery potential of trilepton searches for heavy neutrinos.
The remainder of this section continues as follows:
In Sec.~\ref{sec:DetModeling}, we describe our collider detector modeling, including object definition requirements and tagging/misidentification rates.
We continue in Sec.~\ref{sec:obsSigDef} with defining our proposed dynamic jet veto analysis and benchmark trilepton analysis, 
which is based closely on the $\sqrt{s}=13\TeV$ CMS analysis reported in Ref.~\cite{Sirunyan:2018mtv}.
Finally, in Sec.~\ref{sec:Results}, we report our results for search prospects for Dirac and Majorana neutrinos at the $\sqrt{s}=14\TeV$,
as well as projections for a 27 TeV HE-LHC and a hypothetical 100 TeV VLHC.

\subsection{Hadron Collider Detector Modeling}~\label{sec:DetModeling}
In this section, we describe our modeling of a generic LHC detector experiment.
The fiducial volume and segmentation are based on 
the ATLAS~\cite{ATLAS:1999uwa,ATLAS:1999vwa,ATLAS:1502664} and CMS~\cite{Bayatian:2006nff,Ball:2007zza,CMSCollaboration:2015zni} detector experiments.
Detector response modeling, particle identification (PID) tagging and mistagging rates, as well as kinematic and isolation acceptances/thresholds,
are based on published physics analyses, 
dedicated calibration and detector performance studies, where publicly available,
and published trigger menus, all of which we now summarize.

\subsubsection*{Global Fiducial Volume}\label{sec:globalVolume}
Our analysis starts from the particle-level event samples described in Sec.~\ref{sec:Setup}.
Particle-level objects are stable on a detector's length scale and assumed, momentarily, 
to be reconstructed with 100\% efficiency if they fall within a subdetector's fiducial volume.
All objects with ultra forward rapidities, i.e., $\vert y \vert > 5$, are ignored outright.
Electrons, muons, hadronically decayed tau leptons ($\tau_h$), and photons that fall outside the electromagnetic calorimeter (ECAL) coverage, 
i.e., $\vert y \vert > 3$, are relabeled categorically as light jets.

\subsubsection*{Detector Response}\label{sec:smearing}
We next smear the momenta of these objects according to their (potentially new) PID.
For all objects we employ a Gaussian smearing profile.
Explicitly, this means that for some kinematic observable $\hat{\mathcal{O}}$, e.g.,  $\hat{\mathcal{O}} = p_T$ or $E$,  
its value is perturbed randomly according to a normal distribution centered on the original value of $\hat{\mathcal{O}}$ 
with a spread of $\sigma_{\hat{\mathcal{O}}}$, i.e.,
\begin{equation}
 \hat{\mathcal{O}} \rightarrow \hat{\mathcal{O}}' = \hat{\mathcal{O}} + \text{Gaus}[\mu=0,\sigma_{\hat{\mathcal{O}}}].
\end{equation}
In this language, the resolution $(\delta)$ of $\hat{\mathcal{O}}$ is the (dimensionless) quantity, 
$\delta = \sigma_{\hat{\mathcal{O}}}/\hat{\mathcal{O}}$.
The 4-momentum is then recalculated assuming that the direction of a relativistic particle always remains unchanged.
This is done because the direction of an infinitely energetic stable object can be measured in hermetic detectors with high certainty, unlike its energy.
Momentum reconstruction for electrons, photons, jets, and hadronic taus is determined largely through calorimetry.
Hence, we smear their momenta via shifts in their energies.

For electrons and photons, the energy smearing is parameterized by~\cite{CMS:2015kjy,Khachatryan:2016zqb,Khachatryan:2016jww}
\begin{equation}
 \sigma_{E^e} = b_{E}^e \times E^e  \quad\text{and}\quad    \sigma_{E^\gamma}  = b_{E}^\gamma \times E^\gamma.
\end{equation}
Here $b_{E}^e = 2\%$ for $E_T < 500\GeV$, at all $\eta$.
For larger $E_T$, $b_{E}^e = 4~(6)\%$ for objects inside (outside) the central barrel, which extends to $\vert \eta \vert < 1.444$.
For simplicity, electrons and photons are treated identically and so we set $b_E^\gamma = b_{E}^e$.

For jets, we combine the energy smearing adopted in the 13 TeV $pp\to t\overline{t}+nj$ analysis of Ref.~\cite{Khachatryan:2015mva},
which exploits energy resolution measurements for $R=0.5-0.7$ jets~\cite{Chatrchyan:2011ds,Khachatryan:2015mva},
with $p_T$ smearing based on a dedicated calibration of $R=1.0$ jets~\cite{Aaboud:2018kfi}.
Jet energies and $p_T$ are varied independently so that changes in a jet's momentum are translated into a shift in the jet's mass, 
again leaving the 3-momentum direction unmodified.
Like electrons and photons, the jet energy smearing is given by 
\begin{equation}
 \sigma_{E^j} = b_{E}^j \times E^j,
\end{equation}
where the coefficient is $b_{E}^j = 3\%$ (5\%) for central (forward) jets 
with rapidities satisfying $\vert y\vert < 3$ ($\vert y\vert > 3$)~\cite{Chatrchyan:2011ds}.
From Ref.~\cite{Aaboud:2018kfi}, one can extract the jet $p_T$-smearing function
\begin{equation}
 \sigma_{p_T^j} = \alpha  \times \left(\frac{p_T}{\GeV}\right)^\beta \times 1\GeV,
 \quad\text{where}\quad \alpha\approx110\% ~\text{and}~ \beta\approx0.52.
 \label{eq:jetPtResolution}
\end{equation}
For jets with $p_T^j \in [250\GeV,1.5\TeV]$, Ref.~\cite{Aaboud:2018kfi} reports
that the power-law expression and the values of the coefficients are nearly uniform over the most central region of the ATLAS detector's barrel.
We therefore extrapolate this to all jets with $\vert y \vert < 3$.
Outside this $p_T$ window, we assume a linear function with coefficients set by the boundaries of Eq.~(\ref{eq:jetPtResolution}), 
\begin{equation}
 \sigma_{p_T^j} = b_{p_T}^j \times p_T^j,
\end{equation}
where $b_{p_T}^j \approx 7.6\%~(3.2\%)$ for $p_T^j < 250\GeV ~(p_T^j > 1.5\TeV)$.
For forward jets with $\vert y \vert > 3$, we use a blanket, $y$-independent coefficient of $b_{p_T}^j \approx 7.6\%$.
%
As $\tau_h$'s are experimentally a special class of jets, we apply the same smearing protocol to them as we do to QCD jets.

The momenta of muons are determined via track curvature.
Hence, their smearing is modeled correspondingly through shifts in $p_T$, 
with deviations $(\sigma_{p_T^\mu})$  parameterized by.
\begin{equation}
 \sigma_{p_T^\mu} = a_{p_T^\mu} ~p_T^2\ .
\end{equation}
Here, $a_{p_T^\mu} = 10\%~\TeV^{-1}$ and 20\%~TeV$^{-1}$ for central ($\vert\eta^\mu\vert < 0.9$) 
and forward ($\vert\eta^\mu\vert>0.9$) muons, respectively~\cite{CMS:2015kjy,Khachatryan:2016jww}.

\subsubsection*{PID Tagging and Mistagging}

Summarily, PID tagging $(\epsilon^{\rm Tag.})$ and mistagging $(\epsilon^{\rm Mis-Tag.})$ probabilities 
are estimated from published ATLAS and CMS physics analyses and detector performance studies.
For a given object, PID tagging/mistagging is implemented simply 
by comparing a randomly generated number (with \texttt{TRandom3::Uniform()}) to the probability, which may depend on $p_T$.

We first check if $b$-jets survive the tagging cull using the \texttt{DeepCSV (loose)} efficiencies  as a function of jet $p_T$, 
as reported in Ref.~\cite{Sirunyan:2017ezt}; those that do not survive are henceforth identified as light jets.
Next, $\tau_h$ are checked using the $p_T$-dependent \confirm{$e\tau_h$-\texttt{Era-2017-F}} tagging efficiencies of Ref.~\cite{CMS-DP-2018-009}; 
those $\tau_h$ that are not tagged are identified as light jets.
$\tau_h$-tagging rates span (roughly) $\epsilon^{\tau_h\to\tau_h} \sim 35-95\%$ for $p_T \gtrsim 30\GeV$~\cite{CMS-DP-2018-009}.
After, genuine light-flavor QCD jets (and charged leptons and photons that are outside the ECAL coverage) are checked if they are misidentified first as
(a) $b$-jets according to Ref.~\cite{Sirunyan:2017ezt},
the rates of which span $\epsilon^{j\to b}\sim1-3\%$ for $p_T^j \gtrsim 20\GeV$;
then as 
(b) $\tau_h$  assuming the $p_T$-dependent \texttt{Tight MVA Discriminant} rates of Ref.~\cite{CMS-DP-2017-036}, 
which span $\epsilon^{j\to\tau_h}\sim0.01-1\%$ for $p_T^j \gtrsim 20\GeV$.
As a consistency check, 
any $b$-tagged jets with $\vert y^j\vert > 2.4$ and $p_T^j < 25\GeV$ is reclassified as a light-flavored jet.
The electric charge of jets misidentified as $\tau_h$ is assigned with equal probability.
Genuine $b$-jets and $\tau_h$ that are misidentified as light jets are grouped with light-flavored jets that survived mistagging.

We also take into account the possibility of a light QCD jet being misidentified as an electron/positron,
which is a main component of the ``fake lepton'' background in heavy neutrino trilepton searches~\cite{delAguila:2008cj,Sirunyan:2018mtv}.
For those background processes where such an occurrence is potentially important (see Sec.~\ref{sec:mcBkgGen}), on an event-by-event basis
we randomly choose a jet from the event (using \texttt{TRandom->Integer()}) and relabel it either an electron or positron (with equal probability).
The likelihood (weight) for the event itself is then re-weighted uniformly by a factor of $\epsilon^{j\to e} = 7.2\times10^{-5}$~\cite{Alvarez:2016nrz}, independent of jet kinematics.
Throughout the analysis, electron/positron charge mis-measurement is neglected;
for dedicated searches for LNV, this is a poor

\subsubsection*{Isolation and Analysis-Object Definitions}\label{sec:AnaObjectDef}

Stable electrons and muons $(\ell^\pm)$ are considered hadronically isolated when the sum of the total 
hadronic $E_T$ within a distance of $R_{\rm max}$ centered on the lepton candidate is less than a fraction $\varepsilon_{\rm Had.~Iso.}^{\ell}$ of its $E_T$.
Symbolically, this is given by
\begin{equation}
I_{\rm Had.~Iso.}^{\ell} \equiv 
\sum_{k\in\{\text{had.}\}} E_{T}^{k} / E_{T}^{\ell} 
\approx 
\sum_{k'\in\{\text{jets}\}} E_{T}^{k'} / E_{T}^{\ell} 
< \varepsilon_{\rm Had.~Iso.}^{\ell}~\quad\text{for}\quad~\Delta R_{\ell k'} < R_{\rm max} .
\label{cut:HadIso}
\end{equation}
Following Ref.~\cite{Sirunyan:2018mtv}, we use a loose isolation criterion and set $\varepsilon_{\rm Had.~Iso.}^{\ell}=30\%,~R_{\rm max}=0.3$. 
Reconstructed jets, and hence $\tau_h$~\cite{Sirunyan:2018pgf}, are inherently isolated from other pockets of hadronic activity 
since they are built up from sequential jet clustering algorithms.
Furthermore, two leptons $(\ell_1,\ell_2)$ are considered leptonically isolated if they satisfy the separation,
\begin{equation}
 \Delta R_{\ell_1,\ell_2} > 0.3.
 \label{cut:LepIso}
\end{equation}

At $\sqrt{s}=14\TeV$, analysis-quality charged leptons and jets are subsequently defined as isolated objects satisfying the following fiducial and kinematic criteria:
\begin{eqnarray}
 p_T^{e,~(\mu),~[\tau_h],~\{j\}} > 15~(15)~[30]~\{25\}\GeV, \quad\text{with}\\
 \vert \eta^{\mu,\tau_h} \vert < 2.4, \quad  \vert \eta^{j} \vert < 4.5,  \quad\text{and}\quad \vert \eta^e \vert < 1.4 ~\text{or}~ 1.6 < \vert \eta^e \vert < 2.4.
 \label{cut:AnaObjDef}
\end{eqnarray}
We stress that in hadron collisions, not all isolated objects satisfy the above requirements.
In particular, stray QED emission off electrically charged leptons and partons can give rise to central, soft electron and muon pairs;
relatively soft muon pairs and $\tau_h$ can also originate from decays of hadronic resonances;
and hard, non-diffractive hadron collisions inherently give rise to forward, low-$p_T$ jets due to color conservation and confinement.

Independent of the multiplicity of analysis-quality objects
and as a proxy for the net impact of energetic, final-state, light neutrinos,
we define the transverse momentum-imbalance vector $(\not\!\vec{p}_T)$ and its magnitude $(\met)$ by the following:
\begin{equation}
 \met~\equiv \vert \not\!\vec{p}_T \vert, \quad\text{where}\quad \not\!\vec{p}_T = - \sum_{k\in\{Vis.\}}\vec{p}_T^{~k}.
\end{equation}
The summation here is after smearing and over all visible objects within the fiducial volume of the detector. 
The exception to this are reconstructed jets with $p_T < 1\GeV$.

\subsection{Signal Process and Collider Signature Definitions}\label{sec:obsSigDef}
We now describe our proposed dynamic jet veto and benchmark analyses at $\sqrt{s}=14\TeV$.

\subsubsection*{Dynamic Jet Veto Trilepton Analysis at the $\sqrt{s}=14$ TeV LHC}

\begin{table}[!t]
\begin{center}
\resizebox{\columnwidth}{!}{
 \begin{tabular}{ c || c || c | c | c | c }
\hline \hline
Signal Category  &  cLFC/V & $\vert V_{eN}\vert$ &   $\vert V_{\mu N}\vert$  &   $\vert V_{\tau N}\vert$ & Signal Process  \tabularnewline\hline
\hline
\texttt{EE}      	&  cLFC & $\neq0$                     &   $=0$                                    &   $=0$                    &  $pp\to e^+ e^- \ell_X$ \tabularnewline
  \hline    
\texttt{EMU-I}   & cLFV & \multirow{2}{*}{ $\neq0$ }  &   \multirow{2}{*}{ $=\vert V_{eN}\vert$ }                 &   \multirow{2}{*}{ $=0$ }  
        & $pp\to  e e \ell_X + e^\pm \mu^\mp \tau_h$ \tabularnewline    
\texttt{EMU-II}  & cLFV   &                             &                                           &                           
        & $pp\to \mu \mu \ell_X$  \tabularnewline
\hline
\texttt{ETAU-I}   & cLFV  & \multirow{2}{*}{ $\neq0$ }  &   \multirow{2}{*}{ $=0$ }                 &   \multirow{2}{*}{ $=\vert V_{eN}\vert$ }  
        & $pp\to e^\pm \tau_h^\mp  \ell_X + e^+ e^- l_i  $ \tabularnewline    
\texttt{ETAU-II}  & cLFC   &                             &                                           &                           
        & $pp\to \tau_h^+ \tau_h^- \tau_h + \tau_h^+ \tau_h^- \mu $  \tabularnewline
\hline
\texttt{MUTAU-I}   & cLFV & \multirow{2}{*}{ $=0$ }  &   \multirow{2}{*}{ $\neq0$ }                 &   \multirow{2}{*}{ $=\vert V_{\mu N}\vert$ }  
        & $pp\to \mu^\pm \tau_h^\mp \ell_X + \mu^+ \mu^- l_i $ \tabularnewline    
\texttt{MUTAU-II}  &  cLFC  &                             &                                           &                           
        & $pp\to \tau_h^+ \tau_h^- \tau_h + \tau_h^+ \tau_h^- e $  \tabularnewline                
\hline
\texttt{MUMU}    &  cLFC & $=0$                        &   $\neq0$                                 &   $=0$                    &  $pp\to \mu^+ \mu^- \ell_X$ \tabularnewline
\hline
\texttt{TAUTAU-I}  & cLFC  & \multirow{2}{*}{ $=0$ }     &   \multirow{2}{*}{ $=0$ }              &   \multirow{2}{*}{ $\neq0$ }  
        & $pp\to\tau_h l_i^+ l_j^- $  \tabularnewline
\texttt{TAUTAU-II} &  cLFC &                             &                                           &                           
        & $pp\to \tau_h^+ \tau_h^- \ell_X + \tau_h^\pm \tau_h^\pm l_i$  \tabularnewline
  \hline
  \hline
\end{tabular}
} 
\caption{Signal categories for trilepton signal processes mediated by a heavy Dirac neutrino $(N)$,
 underlying mixing hypotheses,  
whether the signal process is charged lepton flavor-conserving (cLFC) or -violating (cLFV).
Here $\ell_X \in \{e,\mu,\tau_h\}$, $l_i,l_j \in \{e,\mu\}$, and no $\pm$ indicates that both lepton charges are permitted.
}
\label{tb:SignalRegions}
\end{center}
\end{table}

As our underlying signal process, we consider the inclusive production of a single heavy neutrino $N$ and a charged lepton $\ell_N$ through the 
CCDY and $W\gamma$ VBF processes, with $N$ decaying through a SM charged current to a fully leptonic final state, given by
\begin{equation}
pp \to N ~\ell_N ~+~ X, \quad\text{with}\quad ~N ~\to ~\ell_W ~W ~\to~ \ell_W ~\ell_\nu ~\nu.
\end{equation}
In accordance to the benchmark active-sterile flavor mixing scenarios, given in Eqs.~\ref{eq:mcMixingLFC} and ~\ref{eq:mcMixingLFV},
we define several flavor permutations for our signal process,
which we categorize by flavor-hypothesis and summarize in Tab.~\ref{tb:SignalRegions}.
For a particular signal category, 
we define our collider signature as precisely three analysis-quality charged leptons and $\met$,
\begin{equation}
 p p \to ~\ell_1 ~\ell_2 ~\ell_3 ~+~ \met ~+~ X, \quad\text{for}\quad \ell_k\in\{e,\mu,\tau_h\}.
 \label{cut:SigDef}
\end{equation}
Here the charged leptons $\ell_k$ (as well as all other objects) are ordered according to their $p_T$, with $p_T^k > p_T^{k+1}$,
and, as above, $X$ denotes an arbitrary number (including zero) of high-$p_T$ jets.
To minimize contamination from multi-EW boson processes, we reject events with four or more analysis-quality charged leptons,
but remain inclusive with respect to charged leptons that do not meet the analysis object definitions in Sec.~\ref{sec:AnaObjectDef}.

We remove  $Z$ pole events and the low-mass SM Drell-Yan spectrum by requiring the following invariant mass cuts on sets of analysis-quality charged leptons:
\begin{equation}
 m_{\ell_i \ell_j} > 10\GeV,~\quad \vert m_{\ell_i\ell_j}-M_Z\vert > 15\GeV,~\quad \vert m_{3\ell}-M_Z\vert > 15\GeV.
 \label{cut:Zpole}
\end{equation}
Dilepton invariant masses are built from all possible $(\ell_i\ell_j)$ permutations, independent of flavor or electric charge,
to suppress electric charge mis-measurement and fake leptons.
The trilepton invariant mass suppresses contributions from rare, but nonzero $Z\to4\ell$ decays.

To curb EW, top quark, and fake backgrounds, we impose a dynamic central jet veto, namely the so-called ``safe jet veto''~\cite{Pascoli:2018rsg}, 
and set on an event-by-event basis
\begin{equation}
\pTVeto = p_T^{\ell_1}.
\label{cut:dynamicVeto}
\end{equation}
More precisely, we require that all analysis-qualitatively jets possess $p_T$ less than the $p_T$ of the event's leading charged lepton.
Events with any number of analysis-quality jets possessing a $p_T$ greater than the event's leading charged lepton are cut.
We reiterate that the jet veto does not eliminate all jet activity in the event.
We remain inclusive with respect to soft and forward hadronic activity:
events may contain an arbitrary number of jets with $p_T<\max(25\GeV,p_T^{\ell_1})$  and/or $\vert y \vert > 4.5$.
(We briefly report that a slight improvement in the $S/B$ ratio was observed when setting $\pTVeto=p_T^{\ell_2}$;
the effect was less pronounced for the trailing charged lepton.
We encourage further investigation into the matter.)

To further repress EW and top quark production, we demand for $\ell_1,\dots,\ell_3$, that
\begin{equation}
 S_T > 125\GeV.
 \label{cut:ST}
\end{equation}
This requirement severely impacts the survival of continuum $3\ell\nu$ and $4\ell$ processes
since such cross sections scale inversely with $S_T$, with $\sigma(pp\to n\ell+X)\sim 1/M_{n\ell}^2 \sim 1/S_T^2$.

As a proxy to the mass of $N$, we build the multi-body transverse mass variable $(\tilde{M}_{MT})$,
\begin{eqnarray}
\tilde{M}_{MT,i}^2 &=& \left[\sqrt{p_T^2(\ell^{\rm OS}) + m_{\ell^{\rm OS}}^2} + \sqrt{p_T^2(\ell_i^{\rm SS},\not\!\!\vec{p}_T) + M_W^2}\right]^2
\nonumber\\
&-& \left[\vec{p}_{T}(\ell^{\rm OS},\ell_i^{\rm SS}) + \not\!\vec{p}_T \right]^2, \quad i=1,2
\label{eq:DefSigMultibodyMT}
\end{eqnarray}
for both opposite-sign (OS) and same-sign (SS) charge lepton permutations allowed for a Dirac neutrino.
Of the two $\tilde{M}_{MT,i}$, 
we choose the one $(\hat{M}_{T})$ closest to our mass hypothesis $(m_N^{\rm hypothesis})$ and select for events satisfying
\begin{equation}
 -0.15 \times m_N^{\rm hypothesis} < (\hat{M}_{MT} - m_N^{\rm hypothesis}) < 0.1 \times m_N^{\rm hypothesis}.
 \label{cut:NuMass}
\end{equation}
The dependence of the mass-window on $m_N^{\rm hypothesis}$ reflects two realities:
(i) The total width of TeV-scale $N$ can be numerically large (though still perturbative), 
since $\Gamma_N \sim G_F m_N^3 \sum \vert V \vert^2$, and intrinsically broadens the true value of $\hat{M}_{T}$.
As we consider $m_N \geq150\GeV$, the mass window is never smaller than ${}^{+15\GeV}_{-20\GeV}$.
(ii) Experimental resolution and the presence of \met~also feed into the broadening of reconstructed $\hat{M}_{T}$.
The asymmetrical requirements reflects the asymmetric nature of multi-body cluster mass variables; see Fig.~\ref{fig:particleKin_dMass}.
The precise boundaries, $-15\%$ and $+10\%$, however, are somewhat arbitrary and can be tuned for optimization.
In practice,  application of the above selection cut is no different from the mass hypothesis-dependent diphoton or 4-charged lepton 
invariant mass requirements used in searches for the SM-like Higgs boson~\cite{Aad:2012tfa,Chatrchyan:2012xdj}.

\begin{table}[!t]
\begin{center}
\resizebox{\columnwidth}{!}{
 \begin{tabular}{ c }
\hline \hline
Shared Analysis Object Requirements at $\sqrt{s}=14$ TeV									\tabularnewline 
anti-$k_T(R=1)$ jets, \quad $I_{\rm Had.~Iso.}^{\rm \ell} < 0.1$, \quad  $\Delta R_{\ell_i \ell_j} > 0.3$,  	\tabularnewline
 $p_T^{e,~(\mu),~[\tau_h],~\{j\}} > 15~(15)~[30]~\{25\}\GeV,$                                           				\tabularnewline
 $\vert \eta^{\mu,\tau_h} \vert < 2.4,$ \quad  $\vert \eta^{ j} \vert <  4.5$, \quad 
 $\vert \eta^e \vert < 1.4 ~\text{or}~ 1.6 < \vert \eta^e \vert < 2.4$								\tabularnewline \hline
 Safe Jet Veto Analysis at $\sqrt{s}=14$ TeV												\tabularnewline
 $m_{\ell_i \ell_j} > 10\GeV,~\quad \vert m_{\ell_i\ell_j}-M_Z\vert > 15\GeV,~\quad \vert m_{3\ell}-M_Z\vert > 15\GeV$,  				\tabularnewline
 $\pTVeto = p_T^{\ell_1}$, \quad $S_T>125\GeV$, \quad 
 $  -0.15 \times m_N^{\rm hypothesis} < (\hat{M}_{MT} - m_N^{\rm hypothesis}) < 0.1 \times m_N^{\rm hypothesis}  $                                  \tabularnewline \hline
 Benchmark ``Standard'' Analysis at $\sqrt{s}=14$ TeV										\tabularnewline
 $m_{\ell_i \ell_j} > 10\GeV,~\quad \vert m_{\ell_i\ell_j}-M_Z\vert > 15\GeV,~\quad \vert m_{3\ell}-M_Z\vert > 15\GeV$,  	\tabularnewline
 $ p_T^{\rm b-Tagged}<25\GeV, ~p_T^{\ell_1} > 55\GeV, \quad p_T^{\ell_2} > 15\GeV, \quad m_{3\ell}>80\GeV$               \tabularnewline\hline\hline
\end{tabular}
} 
\caption{
Summary of $\sqrt{s}=14\TeV$ 
(top) analysis object requirements;
(middle) selection cuts for safe jet veto analysis (this study); and 
(bottom) selection cuts for benchmark analysis.
}
\label{tb:SelectionCuts14TeV}
\end{center}
\end{table}

We summarize the above selection cuts in the top two rows of Table~\ref{tb:SelectionCuts14TeV}.
The corresponding selection cut acceptance rates [ab] and acceptance efficiencies $[\%]$ in parentheses,
at NLO+PS, and with scale dependencies $[\%]$,  are tabulated in Table~\ref{tb:CutFlowWithScaleUnc}
for representative CC DY and $W\gamma$ fusion signal processes.
For concreteness, Table~\ref{tb:CutFlowWithScaleUnc} reflects the production of a heavy Majorana neutrino 
assuming the \texttt{EMU-II} flavor configuration of Table~\ref{tb:SignalRegions} with mixing normalization 
given by Eq.~(\ref{eq:mcMixingLFV}).
Other flavor categories reveal comparable rates.
Decays of $W$ bosons to $\tau$ leptons that can decay leptonically or hadronically are included;
for additional simulation inputs and modeling details, see Sec.~\ref{sec:Setup}.
In the third row are the generator-level cross sections and scale uncertainties
for inclusive $pp\to N\ell X$ production.
In the fourth row, cross sections, uncertainties, and efficiencies are listed after the application of
the \texttt{EMU-II} signal category criterial, momentum smearing, kinematic and fiducial cuts for 
analysis objects identification requirements (see top row of Table~\ref{tb:SelectionCuts14TeV}), and PID misidentification~(mis-PID).
A severe reduction of rate spanning $17-34\%$ is attributed to two major factors: branching fractions that contribute to the  \texttt{EMU-II}
signal category and selection efficiencies for light charged leptons.
For the CC DY process with $m_N=150\GeV$, the reduction is about $\varepsilon\sim50-52\%$ and $\varepsilon\sim34-40\%$, respectively.
The poor lepton efficiency is the compounded consequence of requiring three charged leptons, 
with an average acceptance rate of $70-75\%/$lepton, 
and is driven by charged leptons with $\vert \eta^\ell \vert >  2.5$ and  $p_T<15-30\GeV$.
For $m_N > 150\GeV$, this is a slightly weaker effect; for additional details, see text associated with Fig.~\ref{fig:partonKinElleta}.
The following three rows display the impact of lepton cuts, the dynamic jet veto, and mass-hypothesis cut.
For $m_N = 450-900\GeV$, we observe moderate-to-high acceptance rates for the lepton cuts,
universally high acceptance rate for the jet veto, and moderate rates for the mass hypothesis cut.
As discussed below Eq.~\ref{eq:DefMultibodyMT}, no attempt was made to optimize the efficiency of the multi-body transverse mass cut
but believe further investigation can yield fruitful results.
In summary, the overall acceptance rates, relative to signal categorization and object identification,
\begin{equation}
\mathcal{A}=\frac{\sigma^{\rm All~Cuts}}{\sigma^{\rm Cat.~Sm.+Kin.+Fid.}},
\end{equation}
span about $\mathcal{A} \approx 20\%$ for $m_N = 150$ and $\mathcal{A} \approx 45-55\%$ for $m_N = 450-900\GeV$.

\begin{table}[!t]
\begin{center}
\resizebox{\columnwidth}{!}{
 \begin{tabular}{ c | c | c | c | c | c }
\hline \hline
Selection Cut				&  \multicolumn{5}{c}{$\sigma^{\rm NLO+PS}$ [ab]~$(\varepsilon~[\%]) ~\qquad~ \sqrt{s}=14\TeV$} \tabularnewline	\hline
\multirow{2}{*}{Channel}		& \multicolumn{3}{c}{$q\overline{q}\to N\ell$}	& \multicolumn{2}{c}{$W\gamma\to N\ell$}	\tabularnewline
						& \multicolumn{1}{c}{$150\GeV$} & \multicolumn{1}{c}{$450\GeV$} &  \multicolumn{1}{c}{$900\GeV$} & \multicolumn{1}{c}{$450\GeV$} & \multicolumn{1}{c}{$900\GeV$} \tabularnewline\hline
Generator					& $125^{+0.5\%}_{-<0.5\%} $ 			& $1.74^{+2\%}_{-1\%}$ 			& $98.4^{+2\%}_{-2\%}\times10^{-3}$ 		& $408^{+5\%}_{-6\%}\times10^{-3}$ 	& $104^{+3\%}_{-3\%}\times10^{-3}$\tabularnewline\hline
Signal~Cat.~+~Smearing~ 							& 
\multirow{2}{*}{$21.0^{+<0.5\%}_{-<0.5\%}~(17\%)$} 		& 
\multirow{2}{*}{$0.481^{+1\%}_{-1\%}~(28\%)$} 			& 
\multirow{2}{*}{$33.0^{+2\%}_{-2\%}\times10^{-3}~(34\%)$} 	& 
\multirow{2}{*}{$104^{+5\%}_{-6\%}\times10^{-3}~(25\%)$} 	& 
\multirow{2}{*}{$29.7^{+4\%}_{-3\%}\times10^{-3}~(29\%)$}	\tabularnewline
mis-PID~+~Kin.~+~Fid		&	&	&	&	& 				\tabularnewline\hline
~+~$m_{\ell_i,\ell_j},~m_{3\ell},~S_T$ 			
& $6.02^{+<0.5\%}_{-0.5\%}~(29\%)$ 	& $0.429^{+2\%}_{-1\%}~(89\%)$ 	& $31.8^{+2\%}_{-2\%}\times10^{-3}~(96\%)$ 	& $81.9^{+4\%}_{-6\%}\times10^{-3}~(79\%)$ 	& $26.5^{+3\%}_{-3\%}\times10^{-3}~(89\%)$\tabularnewline\hline
+ $p_T^{\rm Veto} = p_T^{\ell_1}$ 			& $5.52^{+1\%}_{-<0.5\%}~(92\%)$ 		& $0.414^{+1\%}_{-0.8\%}~(97\%)$ 	& $31.2^{+2\%}_{-2\%}\times10^{-3}~(98\%)$ 	& $75.4^{+4\%}_{-6\%}\times10^{-3}~(92\%)$ 	& $25.9^{+3\%}_{-3\%}\times10^{-3}~(98\%)$\tabularnewline\hline
+ $\hat{M}_{MT}$			& $4.74^{+0.5\%}_{-<0.5\%}~(86\%)$ 	& $0.266^{+2\%}_{-0.5\%}~(64\%)$ 	& $18.0^{+2\%}_{-2\%}\times10^{-3}~(58\%)$ 	& $46.3^{+4\%}_{-6\%}\times10^{-3}~(61\%)$ 	& $14.0^{+5\%}_{-3\%}\times10^{-3}~(54\%)$\tabularnewline\hline
\hline
Acceptance 			&
\multirow{2}{*}{ $22.6\%$} & 
\multirow{2}{*}{ $55.3\%$} & 
\multirow{2}{*}{ $54.7\%$} & 
\multirow{2}{*}{ $44.5\%$} & 
\multirow{2}{*}{ $47.2\%$} \tabularnewline
$\mathcal{A}=\frac{\sigma^{\rm All~Cuts}}{\sigma^{\rm Cat.~Sm.+Kin.+Fid.}}$ 	& &   & & & \tabularnewline\hline
\hline
\end{tabular}
} 
\caption{Cut flow table, with efficiencies and scale uncertainties, 
corresponding to the dynamic jet veto analysis selection cuts in Table~\ref{tb:SelectionCuts14TeV},
for representative signal benchmarks.
}
\label{tb:CutFlowWithScaleUnc}
\end{center}
\end{table}

\subsubsection*{``Standard'' Trilepton Analysis at the LHC}
Searches for heavy neutrinos in multi-charged lepton final states have long been well-motivated since they compliment dilepton, one-lepton, and zero-lepton search channels.
Subsequently, strategies premised and centered on the existence of high-$p_T$ charged leptons, 
such as those as proposed in Refs.~\cite{Keung:1983uu,Han:2006ip,delAguila:2008cj,Atre:2009rg}, 
are now standardized; see, for example, the LHC searches of Ref.~\cite{Sirunyan:2018mtv}.
The analysis we propose qualitatively differs from these studies in several respects:
(i) The primary difference is our use of a flavor-independent, dynamic jet veto; aforementioned studies consider at most a veto on $b$-tagged jets.
(ii) Our requirements on individual charged leptons are relatively lax and uniform;
previous studies tend to employ different, but nonetheless stringent, kinematic criteria for leading and non-leading charged leptons.
(iii) We set stringent kinematic requirements on global leptonic activity. 
In a sense, our analysis takes advantage of and discriminates against the differences in local hadronic and global leptonic activities 
of signal and background processes in hard $pp$ collisions.
This analogy is made firmer by our use of $R=1$ jets.

It is therefore useful to quantify how our proposed analysis can improve (if at all) the anticipated LHC sensitivity.
To this extent, we also consider an alternative ``standard analysis'' based closely on the CMS collaboration's 
trilepton heavy neutrino search at Run II of the LHC~\cite{Sirunyan:2018mtv}.
We define this benchmark analysis assuming the same lepton flavor collider signature of Eq.~(\ref{cut:SigDef}).
After imposing the same leptonic invariant mass cuts of Eq.~(\ref{cut:Zpole}), we impose the kinematic criteria on an event's three charged leptons:
\begin{equation}
 p_T^{\ell_1} > 55\GeV, \quad p_T^{\ell_2} > 15\GeV, \quad m_{3\ell}>80\GeV.
\end{equation}
We refer readers to Ref.~\cite{Sirunyan:2018mtv} for the justification of these cuts.
Events with at least one analysis-quality, $b$-tagged jet are vetoed.
The ``standard analysis'' selection cuts are summarized in bottom row Table~\ref{tb:SelectionCuts14TeV}.
Remarkably, under the same active-sterile mixing hypothesis as Ref.~\cite{Sirunyan:2018mtv} 
and the same integrated luminosity at $\sqrt{s}=14$ TeV as reportedly used for 13 TeV $(\mathcal{L}\approx36\invfb)$,
\confirm{we find good agreement with their expected sensitivity}.
Our implementation is slightly less sensitive than Ref.~\cite{Sirunyan:2018mtv}. 
This serves as a highly nontrivial check of our signal and background modeling, including ``fake'' leptons, and our detector modeling.

\subsection{Results: Sensitivity at the LHC and Beyond}\label{sec:Results}

In this section, we report the main results of our study, namely, 
the anticipated sensitivity to heavy neutrinos via the trilepton signature at current and future hadron colliders.
Our findings are organized according to the following:
In Sec.~\ref{sec:ResultsDirac},
we present the sensitivity to heavy Dirac neutrinos at the 14 TeV LHC, under various active-sterile mixing hypotheses,
using our proposed dynamic jet veto analysis as well as the benchmark analysis.
In Sec.~\ref{sec:ResultsMajorana}, we repeat the exercise but for heavy Majorana neutrinos.
Finally, in Sec.~\ref{sec:ResultsBeyond}, we show the sensitivity to heavy Dirac neutrinos at the 
LHC's proposed 27 TeV upgrade and a hypothetical 100 TeV successor collider.

In all cases, we report the sensitivity assuming Gaussian statistics.
That is, for $N_{s~(b)}$ signal (background) events expected with an integrated luminosity of $\mathcal{L}$ 
and a cross section of  $\sigma^{\rm All~Cuts}_{s~(b)}$ after all selection cuts are applied,
the signal significance $(\mathcal{S})$ is quantified by
\begin{equation}
 \mathcal{S} = \frac{N_s}{\sqrt{N_s + N_b(1+\delta_b)}}, \quad N_{s~(b)} = \mathcal{L} \times \sigma^{\rm All~Cuts}_{s~(b)}.
 \label{eq:GaussStats}
\end{equation}
A background systematic factor of $\delta_b = 0.1$ is applied to account for mismodeling of SM background processes, 
e.g., missing higher order QCD corrections and subleading processes,
and detector effects, 
e.g., non-perfect electron and muon identification.
For a given $\mathcal{L}$ and $\sigma^{\rm All~Cuts}_{b}$, a particular mass and mixing hypothesis $(m_N,\vert V_{\ell N}\vert^2)$ 
can be falsified at the (approximately) 95\% confidence level (CL) if the corresponding signal rate $\sigma^{\rm All~Cuts}_{s}$ 
results in a significance of $\mathcal{S} > 2$.
Fixing instead $\mathcal{S}_{95}=2$, we can invert this inequality into an 95\% CL upper bound on $\sigma^{\rm All~Cuts}_{s}$ at a given $\mathcal{L}$,
which we label $\sigma_{95}$,
\begin{equation}
 \sigma_{95} = \frac{\mathcal{S}_{95}^2}{2\mathcal{L}}\left[1 + \sqrt{1 + \frac{4\mathcal{L} \times \sigma_b(1+\delta_b)}{\mathcal{S}_{95}^2}}\right].
\end{equation}
Using the relation between the bare cross section and mixing $S_{\ell\ell'}$ given in Eq.~\ref{eq:bareDecay},
the upper bound $\sigma_{95}$ can then be translated into an upper bound on $S_{\ell\ell'}$ at the 95\% CL:
\begin{equation}
 S_{\ell\ell'} = \frac{\vert V_{\ell N}\vert^2 \vert V_{\ell' N}\vert^2 }{\sum_{\ell_X} \vert V_{\ell_X N}\vert^2}
  ~<~ 
  S_{\ell\ell'}^{95} = \frac{\sigma_{95}}{\sigma_0}, \quad \sigma_0 = \frac{\sigma^{\rm All~Cuts}_s}{S_{\ell\ell'}^{\rm Hypo.}},
\end{equation}
where $\sigma_0$ is the bare cross section as derived from $\sigma^{\rm All~Cuts}_s$ and any of the mixing hypotheses $S_{\ell\ell'}^{\rm Hypo.}$
in Table~\ref{tb:SignalRegions}.
We report our results in terms of the 95\% sensitivity to $S_{\ell\ell'}$, 
or equivalently $\vert V_{\ell N}\vert^2$ given our mixing hypotheses, as a function of heavy neutrino mass $m_N$.

\subsubsection{Heavy Dirac Neutrinos at the $\sqrt{s}=14$ TeV LHC}\label{sec:ResultsDirac}

\begin{figure}[!t]
\begin{center}
\subfigure[]{\includegraphics[width=.48\textwidth]{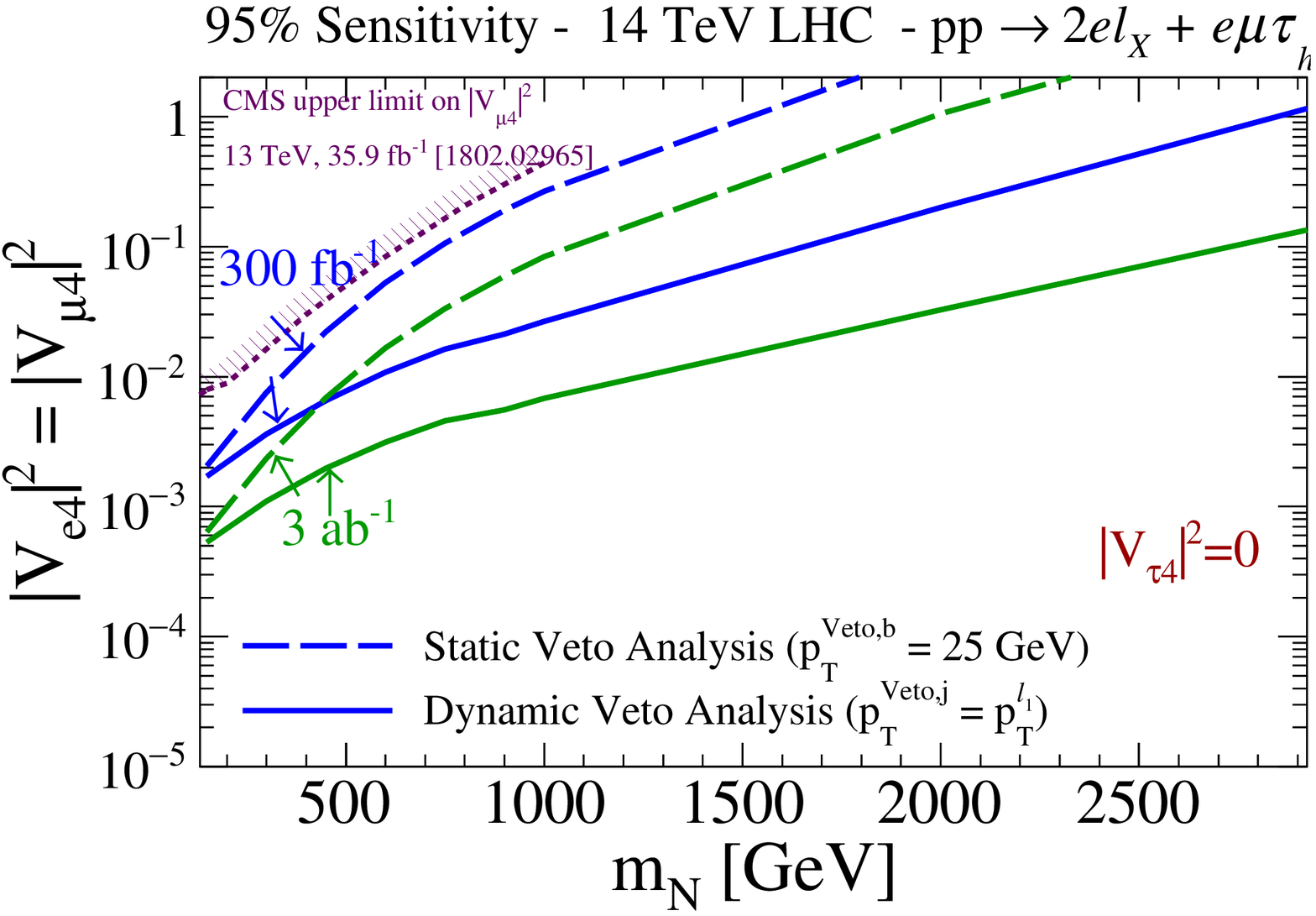}	}
\subfigure[]{\includegraphics[width=.48\textwidth]{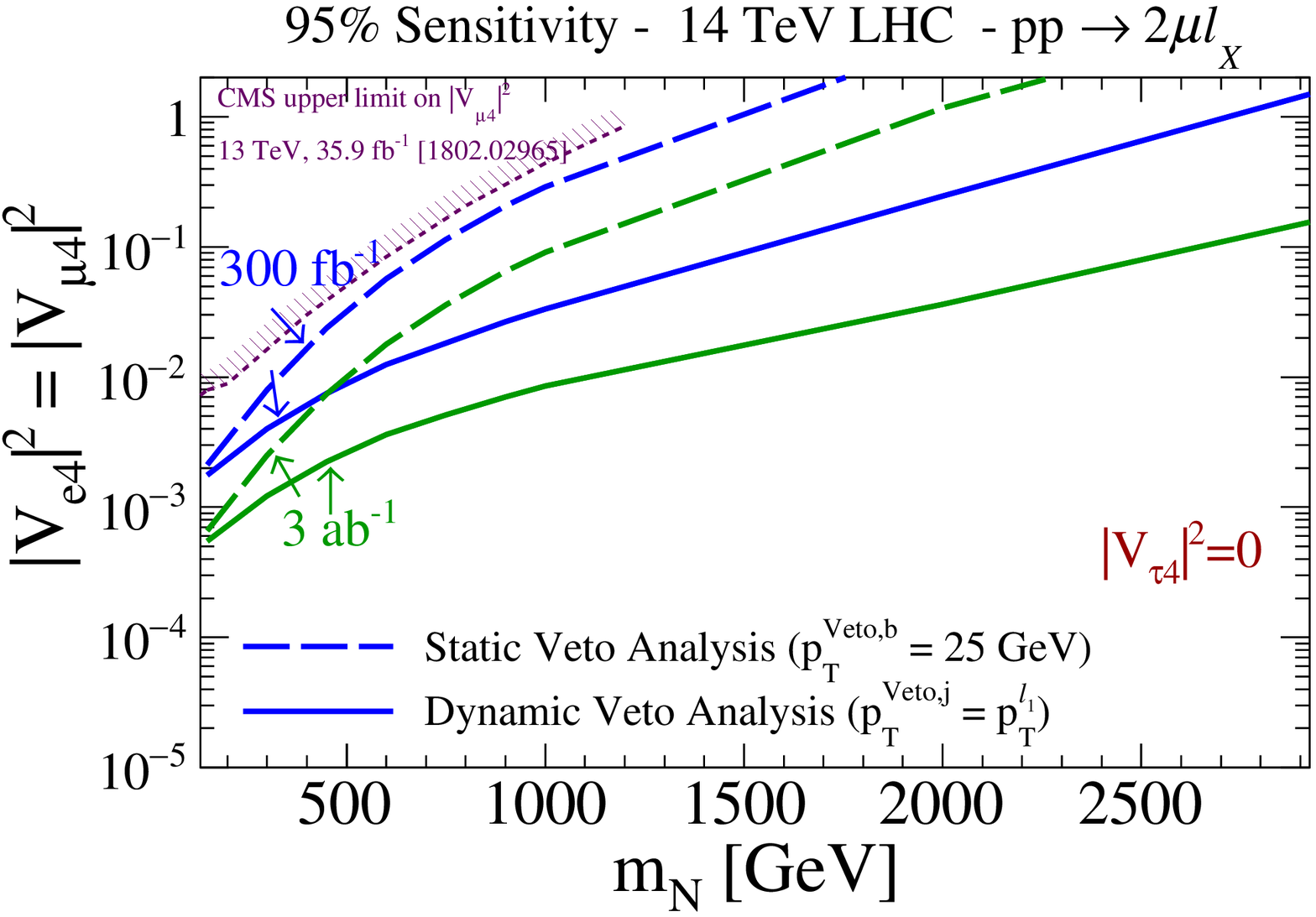}	}
\\
\subfigure[]{\includegraphics[width=.48\textwidth]{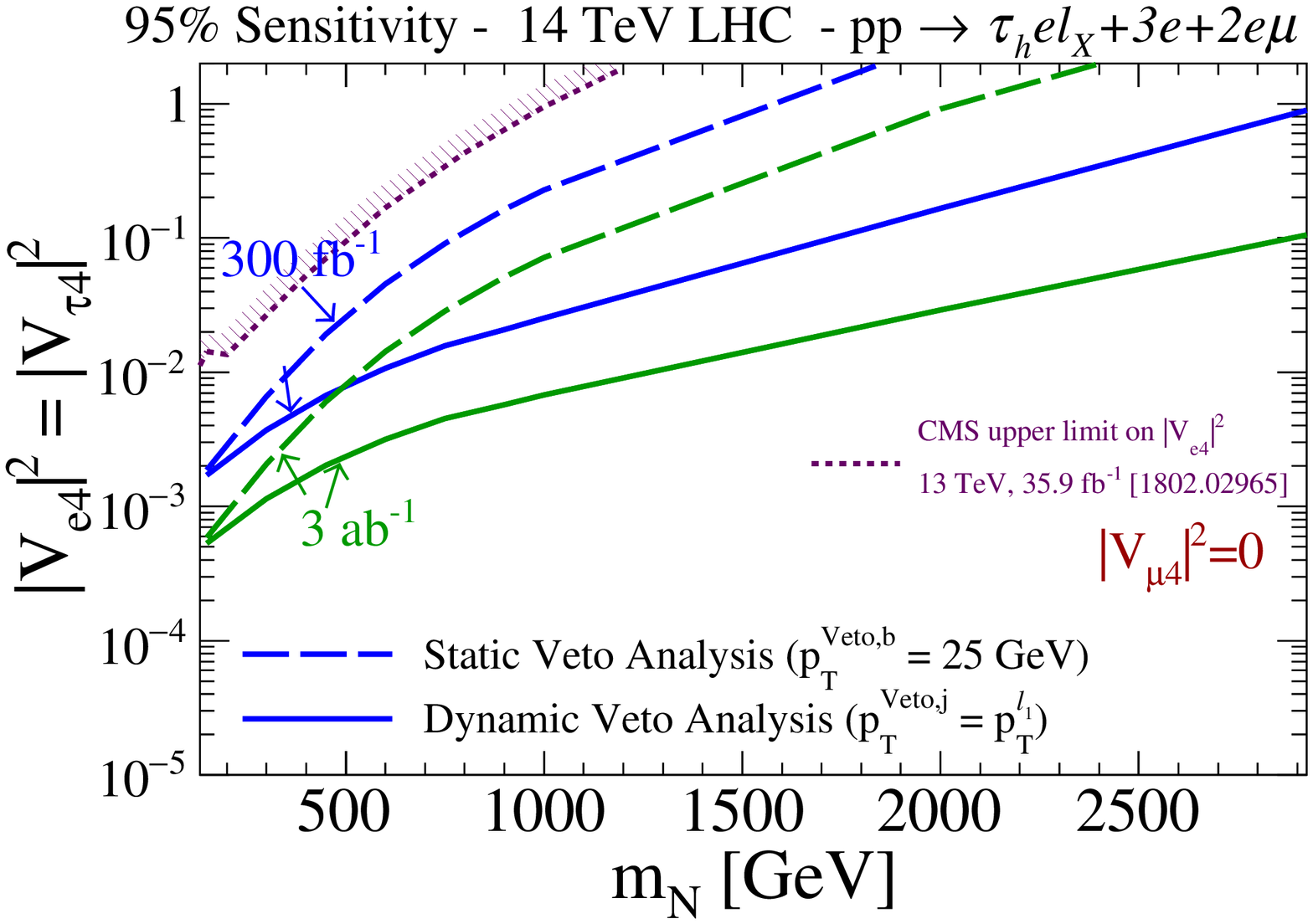}	}
\subfigure[]{\includegraphics[width=.48\textwidth]{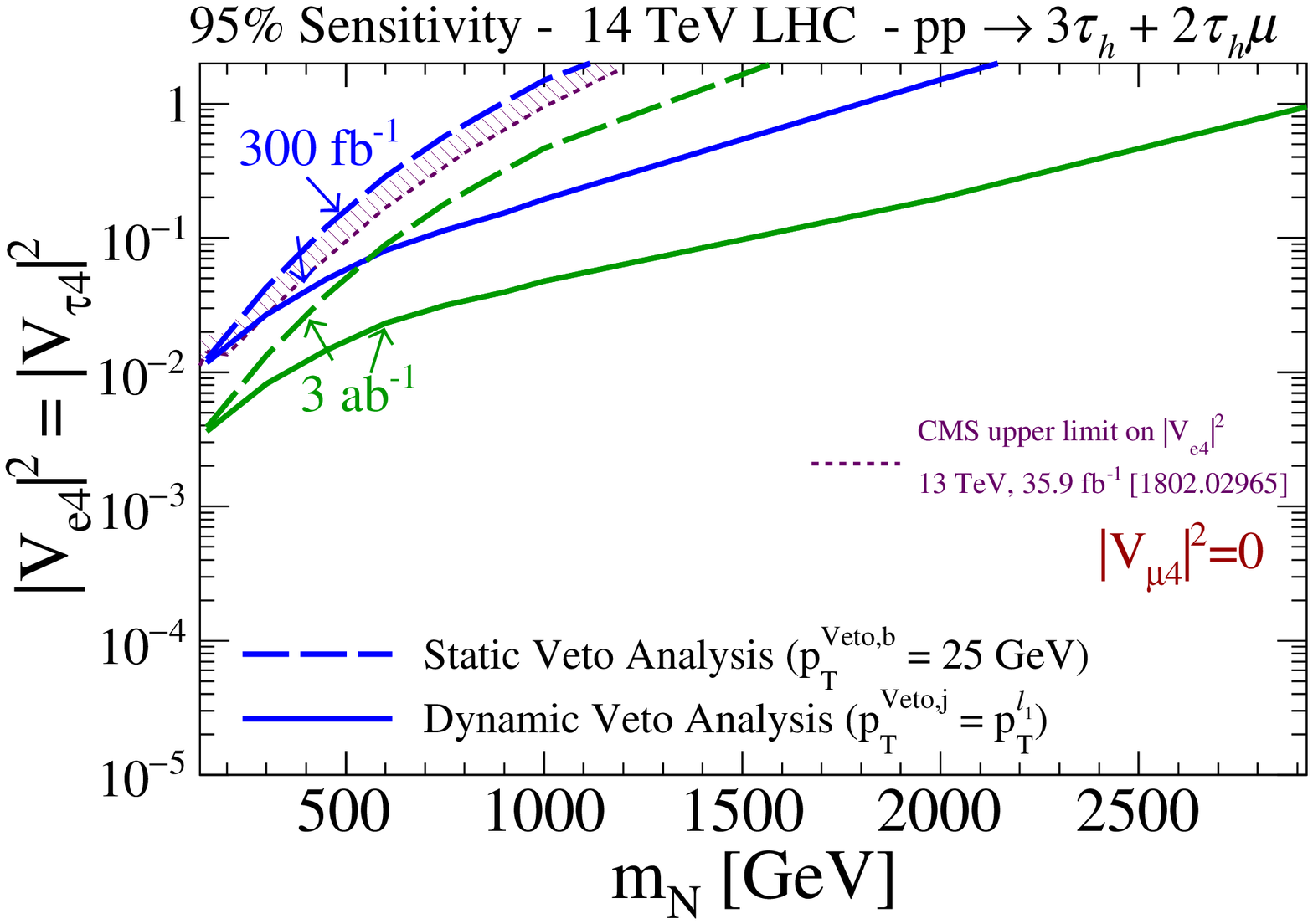}	}
\\
\subfigure[]{\includegraphics[width=.48\textwidth]{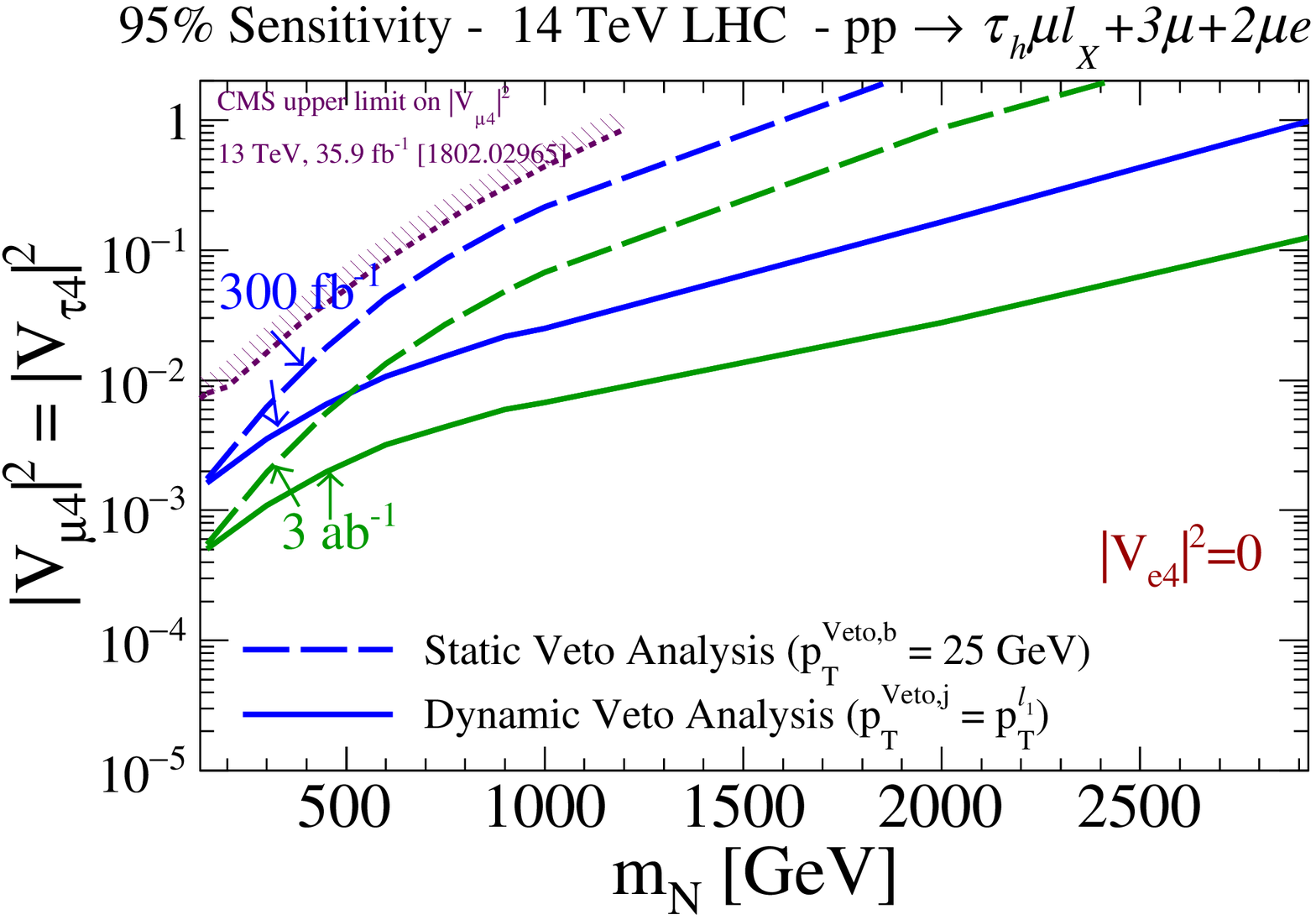}	}
\subfigure[]{\includegraphics[width=.48\textwidth]{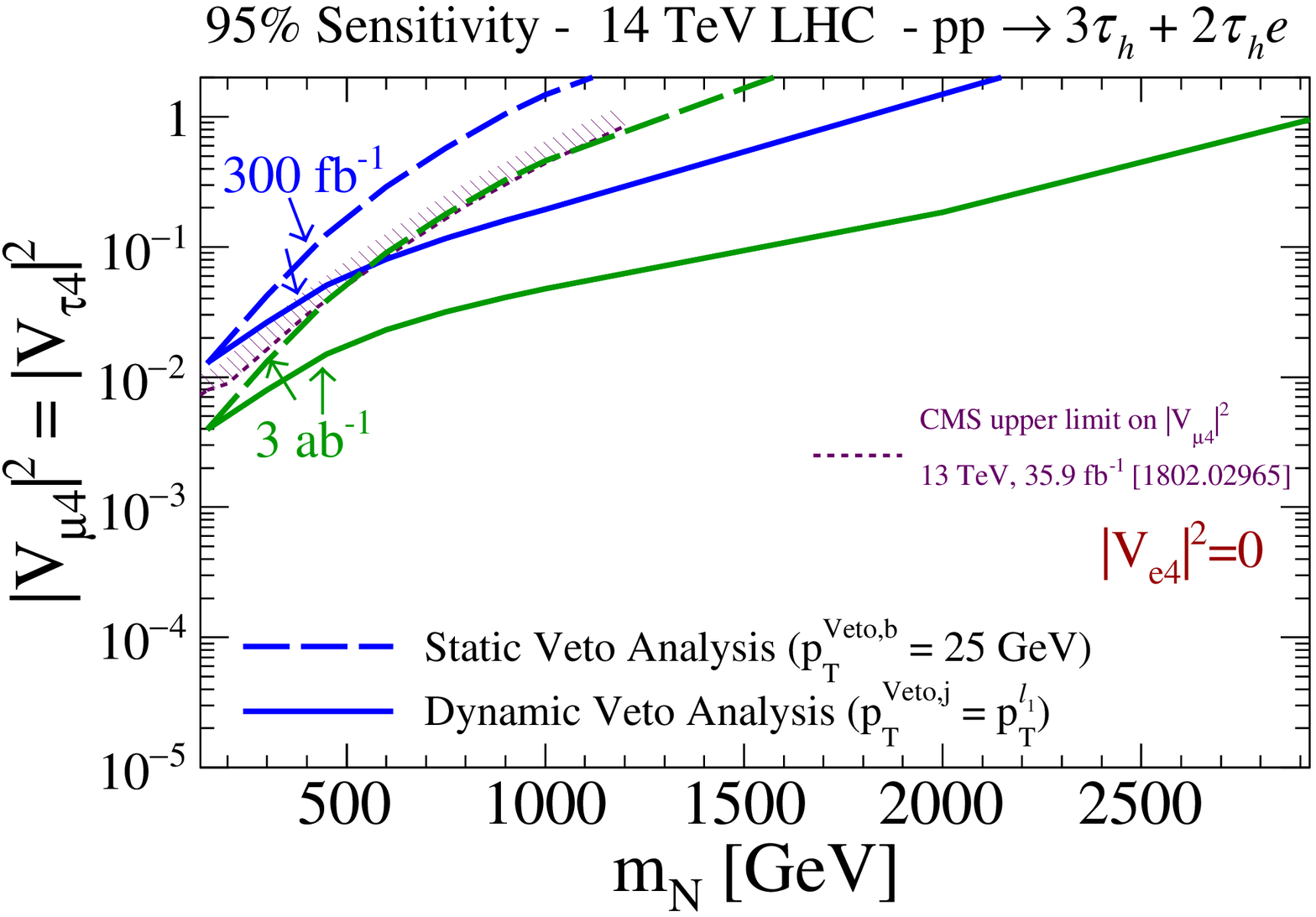}	}
\end{center}
\caption{
95\% CL sensitivity to active-sterile mixing $\vert V_{\ell 4}\vert$ as a function of heavy neutrino mass $N$ using 
the proposed dynamic jet veto trilepton analysis (solid) and the benchmark trilepton analysis (dash),
assuming $\mathcal{L} = 300\invfb$ (lower) and $3\invab$ (upper) at $\sqrt{s}=14\TeV$, 
for the charged lepton flavor violating signal categories
(a) \texttt{EMU-I}, (b) \texttt{EMU-II}, (c) \texttt{ETAU-I},  (d) \texttt{ETAU-II}, (e) \texttt{MUTAU-I},  (d) \texttt{MUTAU-II}, 
as defined in Table~\ref{tb:SignalRegions}. 
Also shown are limits from direct LHC searches~\cite{Sirunyan:2018mtv}.
}
\label{fig:results_LHC14_cLFV}
\end{figure}

\begin{figure}[!t]
\begin{center}
\subfigure[]{\includegraphics[width=.48\textwidth]{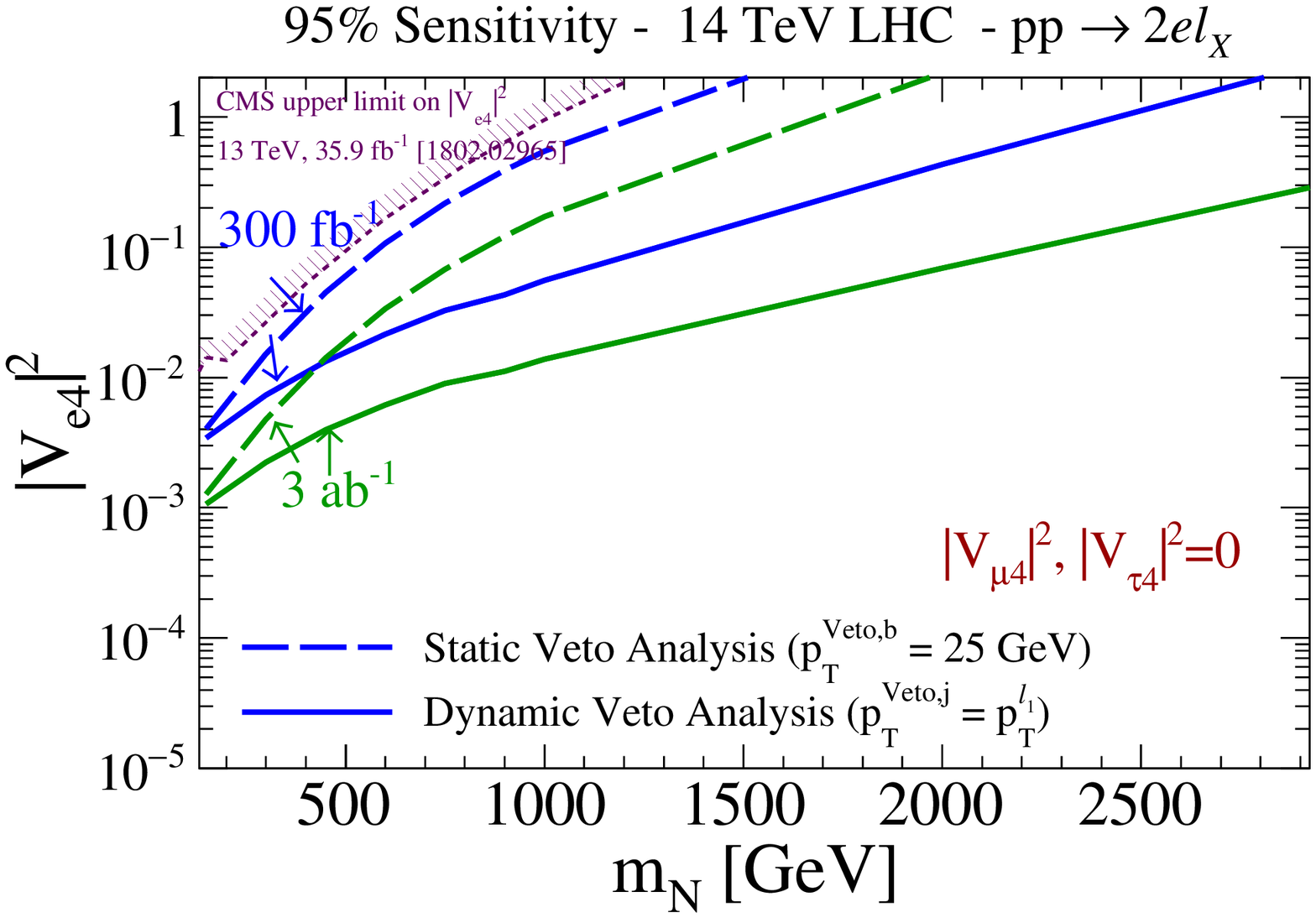}	}
\subfigure[]{\includegraphics[width=.48\textwidth]{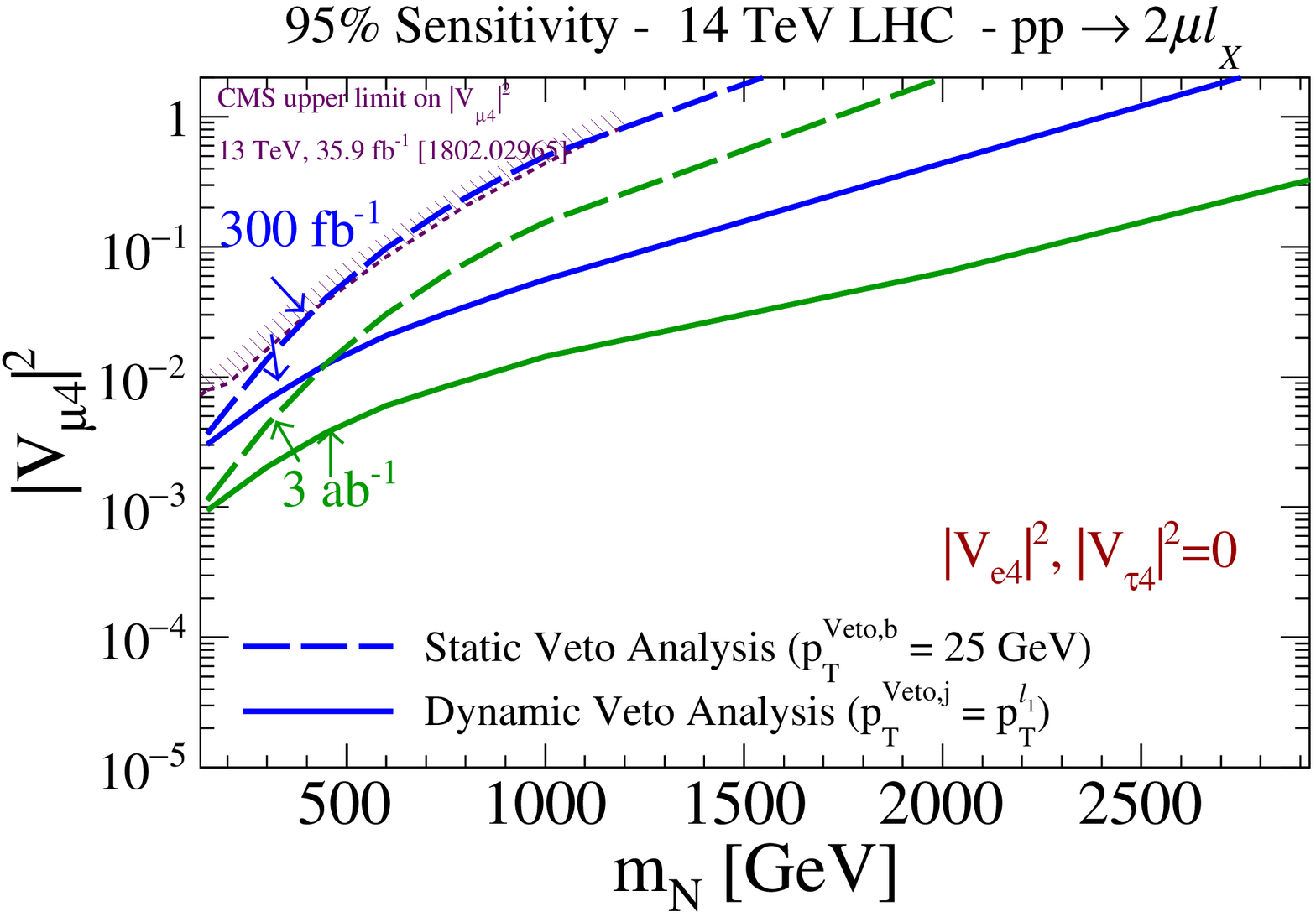}	}
\\
\subfigure[]{\includegraphics[width=.48\textwidth]{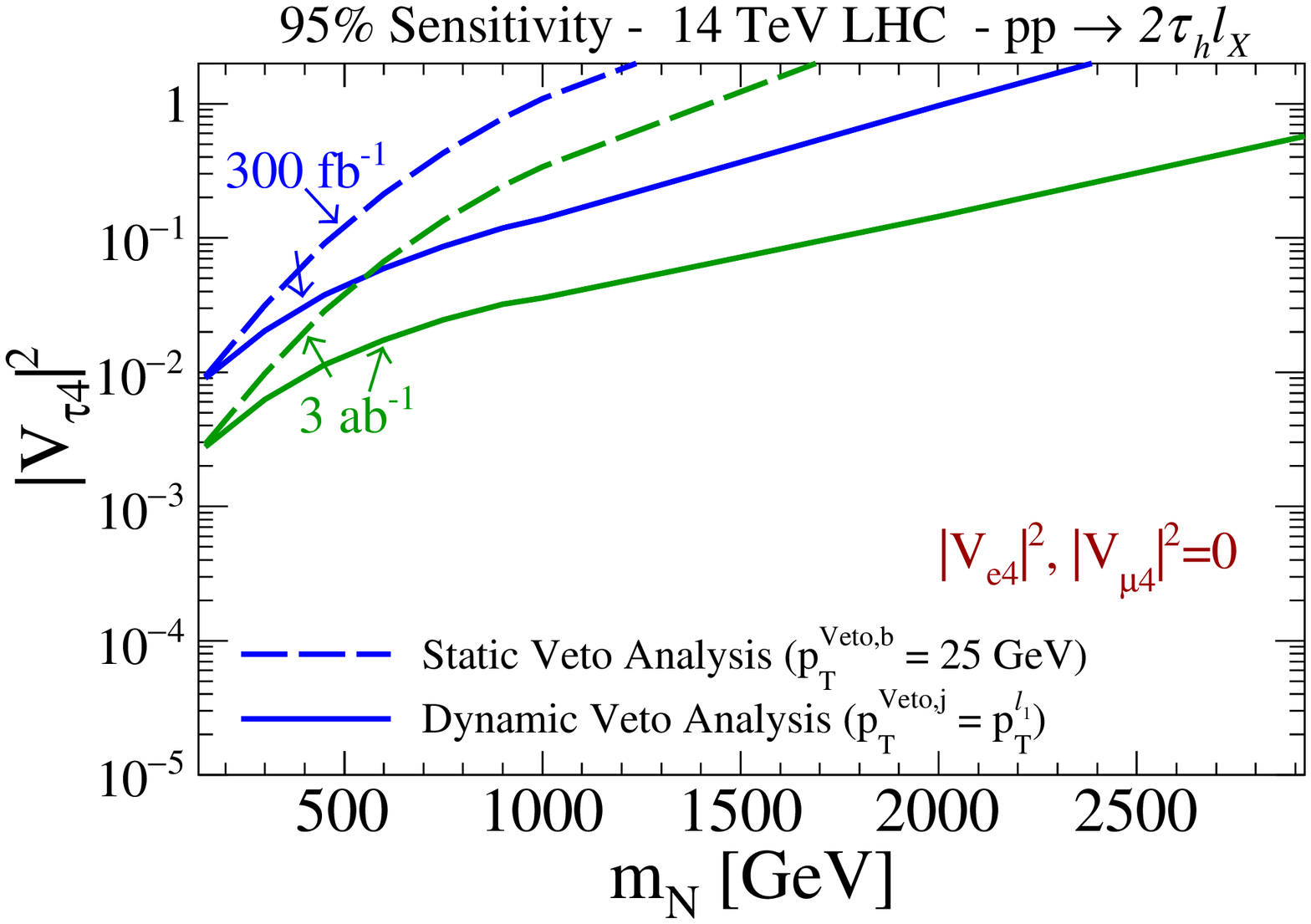}	}
\subfigure[]{\includegraphics[width=.48\textwidth]{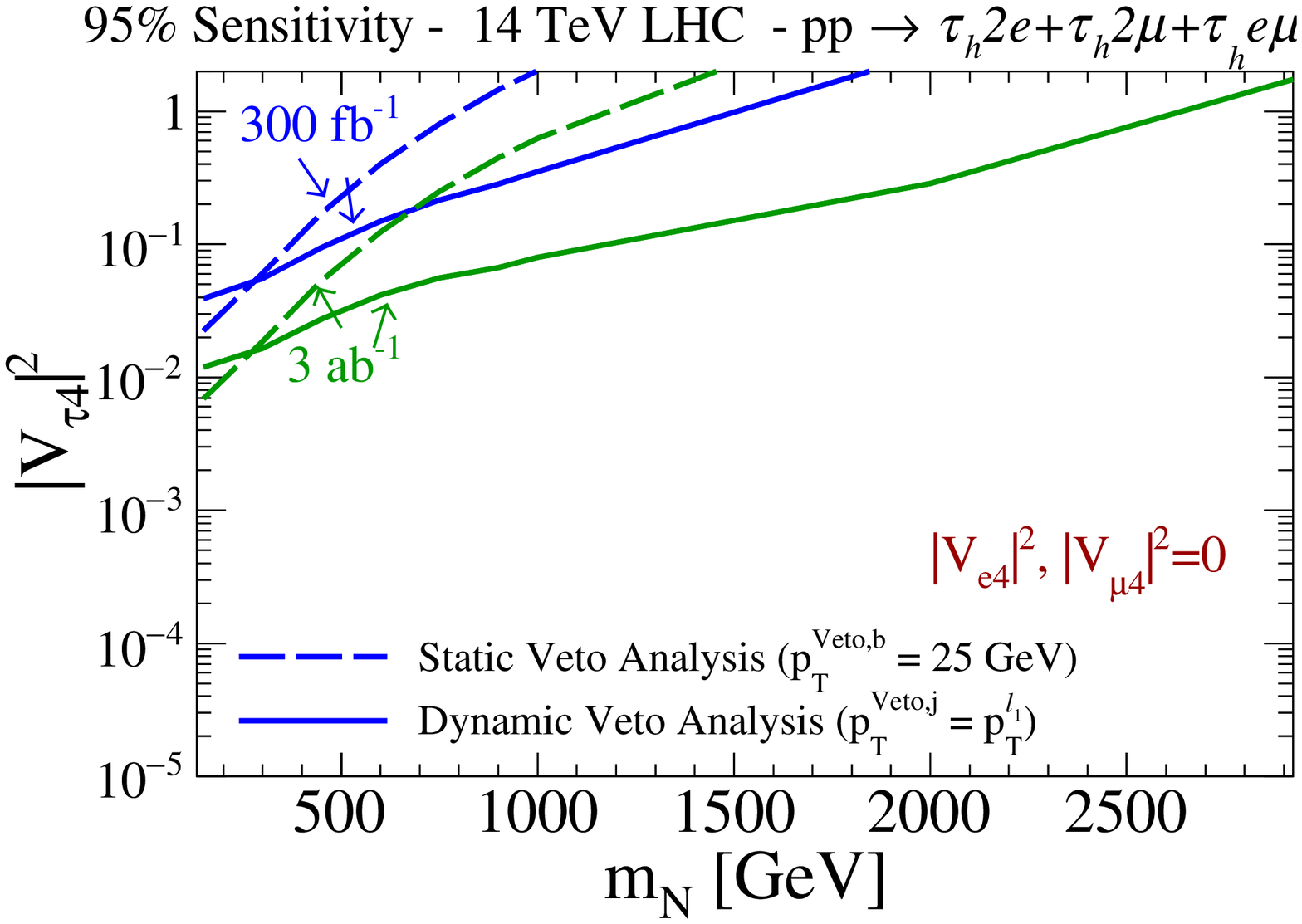}	}
\end{center}
\caption{
Same as Fig.~\ref{fig:results_LHC14_cLFV} but for the charged lepton flavor conserving signal categories
(a) \texttt{EE}, (b) \texttt{MUMU}, (c) \texttt{TAUTAU-I}, and (d) \texttt{TAUTAU-II} as defined in Table~\ref{tb:SignalRegions}. 
}
\label{fig:results_LHC14_cLFC}
\end{figure}

Throughout this section, we report the 95\% CL sensitivity to active-sterile mixing $\vert V_{\ell 4}\vert$ as a function of heavy neutrino mass $N$ using 
the proposed dynamic jet veto trilepton analysis (solid) and the benchmark trilepton analysis (dash) for the signal categories as defined in Table~\ref{tb:SignalRegions}.
We use the mixing hypotheses as given in Eq.~\ref{eq:mcMixingLFV}, which assumes $N$ couples to only two charged leptons with equal strength,
and Eq.~\ref{eq:mcMixingLFC}, which assumes $N$ couples to only one charged lepton.
We refer to the two mixing scenarios, respectively, as the charged lepton flavor violation (cLFV) scenario and the charged lepton flavor conservation (cLFC) scenario.
We assume the nominal LHC and HL-LHC luminosity benchmarks of $\mathcal{L} = 300\invfb$ (darker, upper curves) and $3\invab$ (lighter, lower curves) at $\sqrt{s}=14\TeV$.
Also shown for references are limits on $\vert V_{e4}\vert $ and $\vert V_{\mu 4}\vert$ from direct searches for the trilepton process by the CMS experiment at 
$\sqrt{s}=13\TeV$ using $\mathcal{L}\approx36\invfb$ of data~\cite{Sirunyan:2018mtv}.

\begin{figure}[!t]
\begin{center}
\subfigure[]{\includegraphics[width=.48\textwidth]{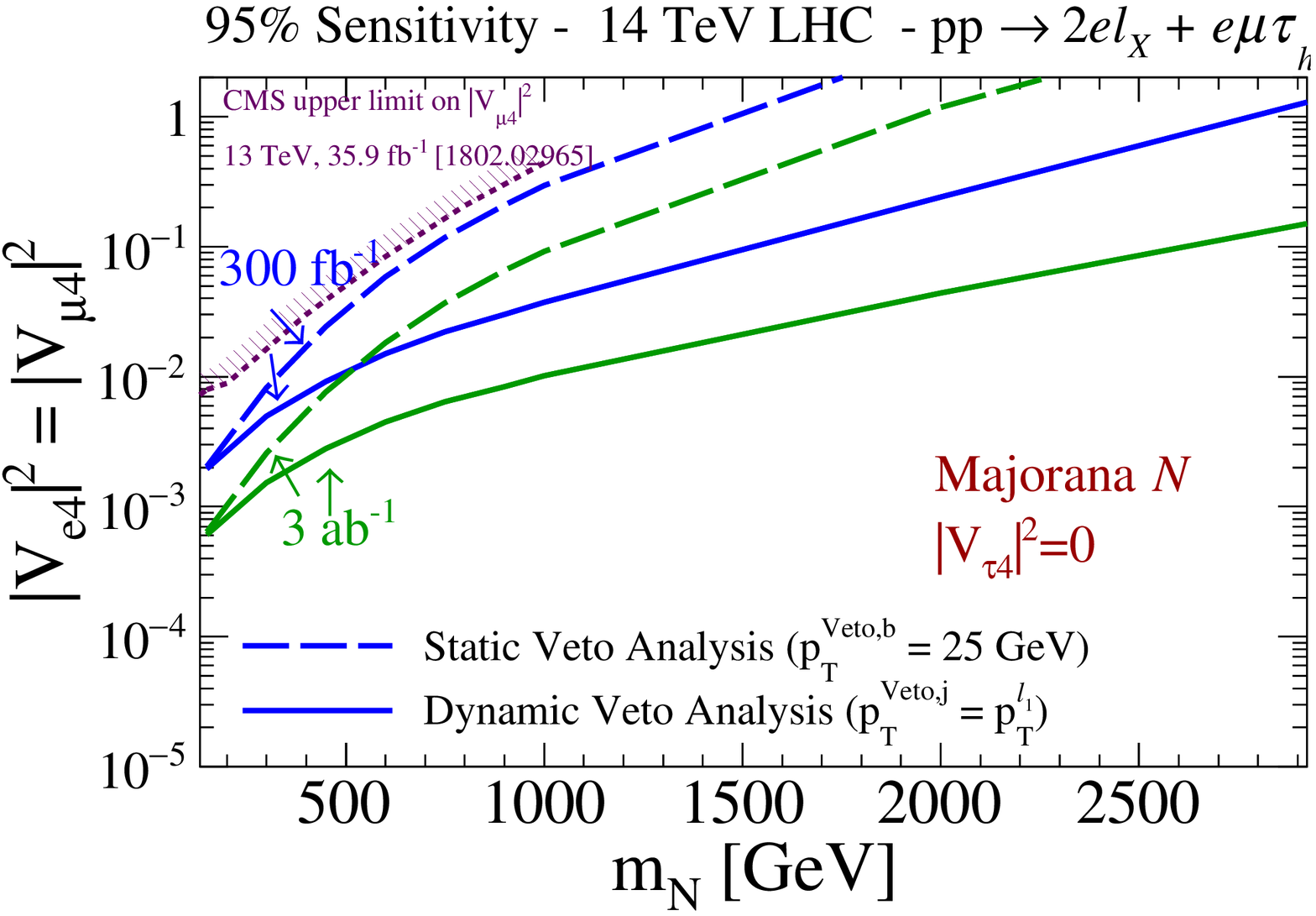}	}
\subfigure[]{\includegraphics[width=.48\textwidth]{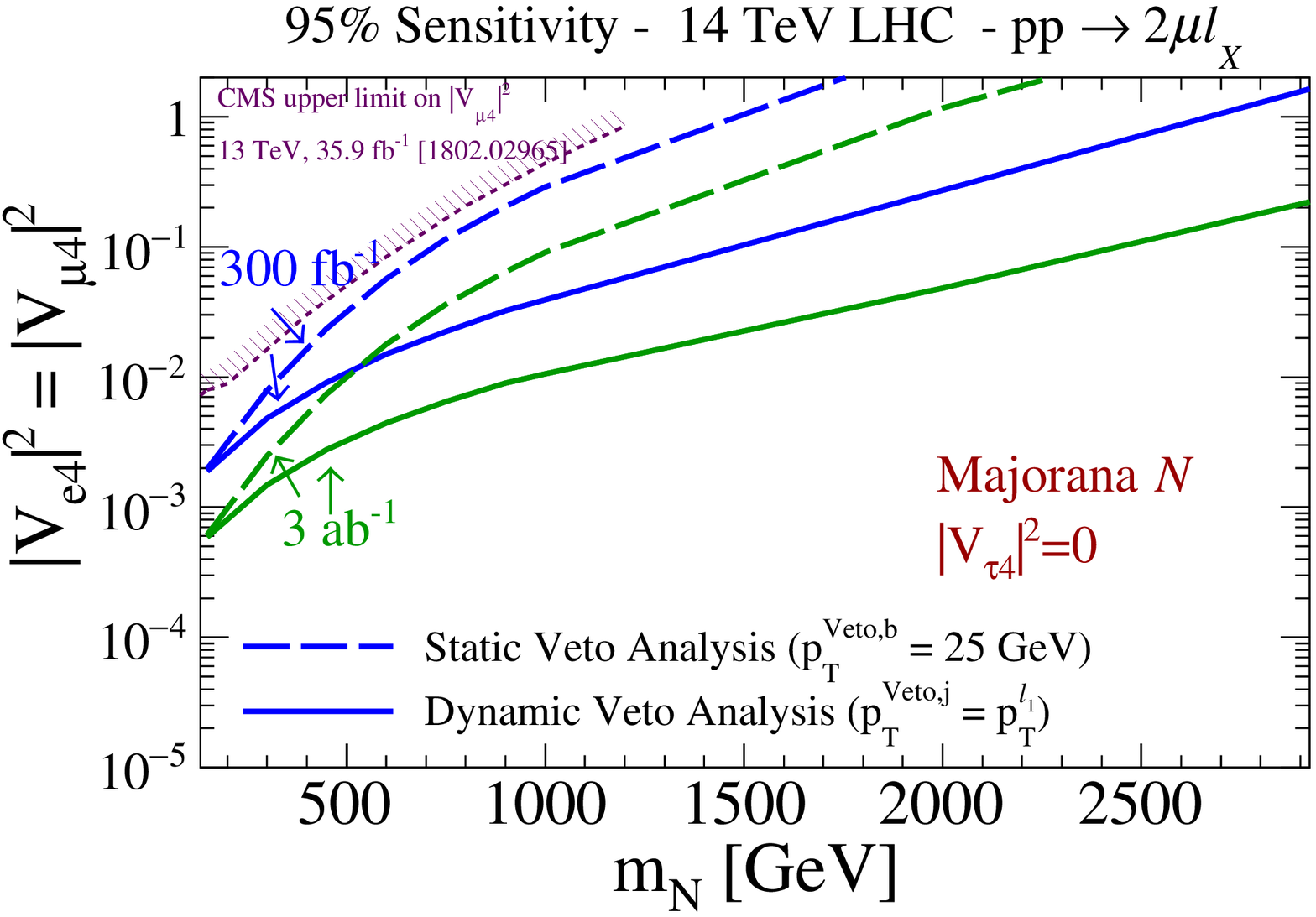}	}
\\
\subfigure[]{\includegraphics[width=.48\textwidth]{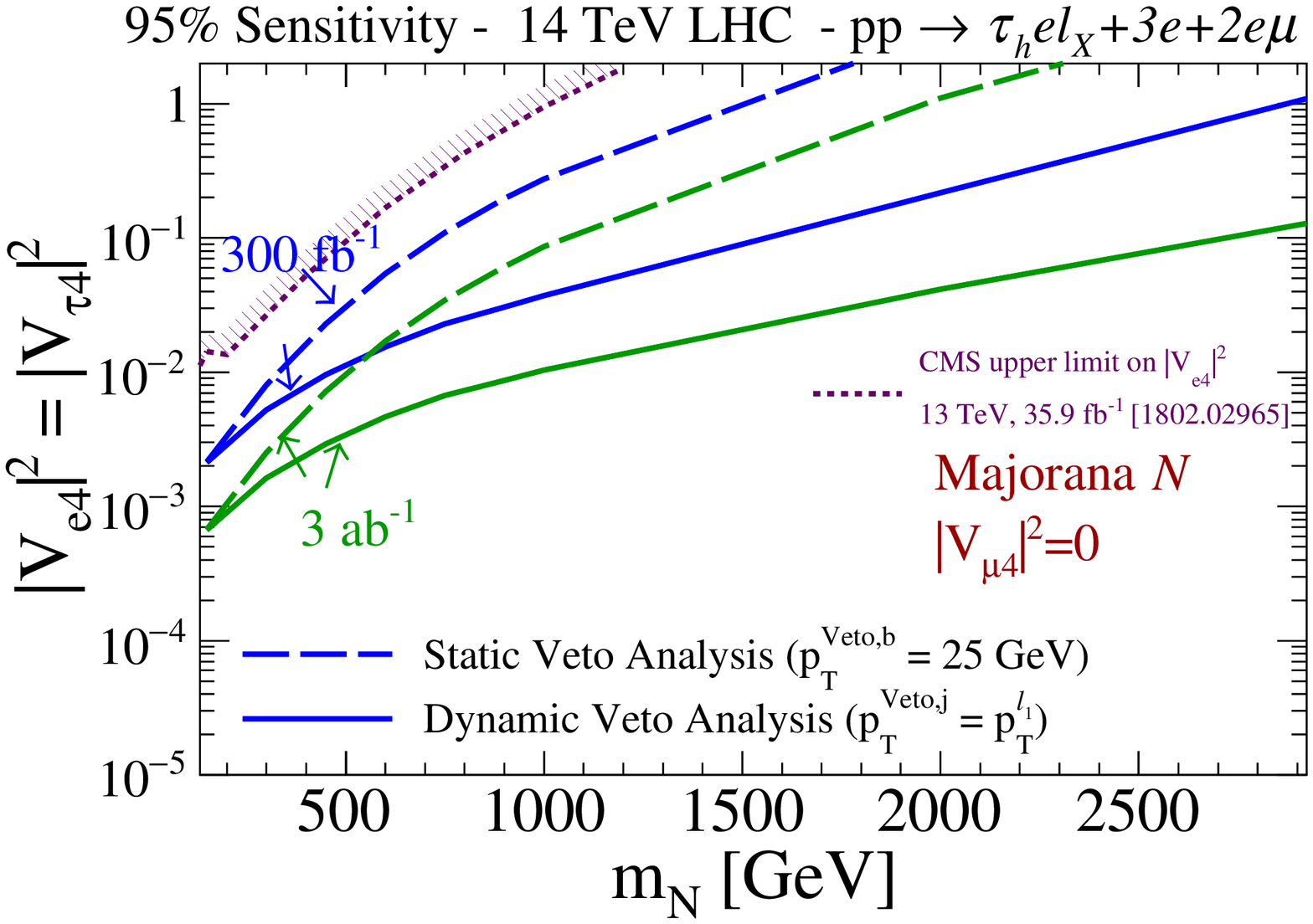}	}
\subfigure[]{\includegraphics[width=.48\textwidth]{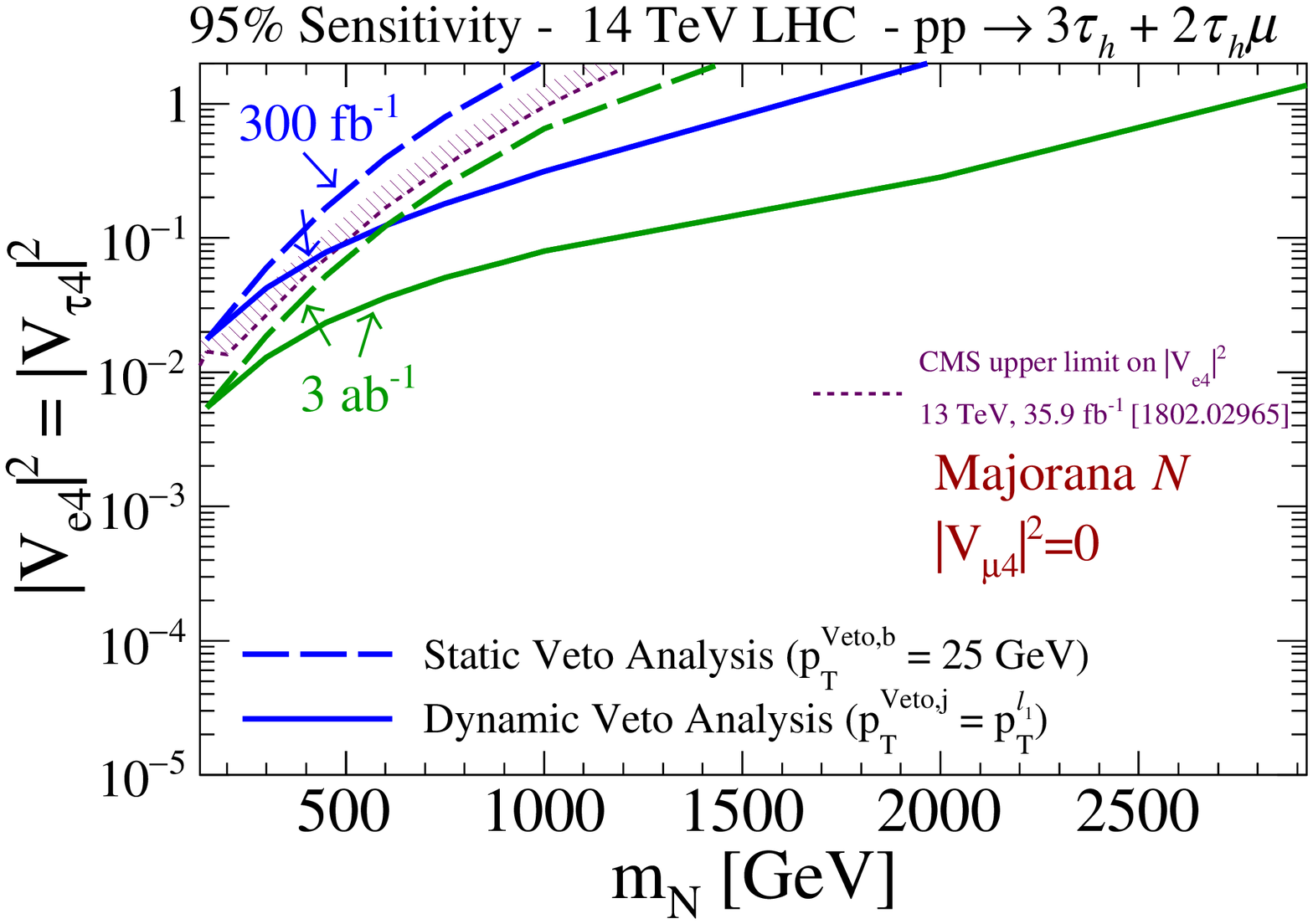}	}
\\
\subfigure[]{\includegraphics[width=.48\textwidth]{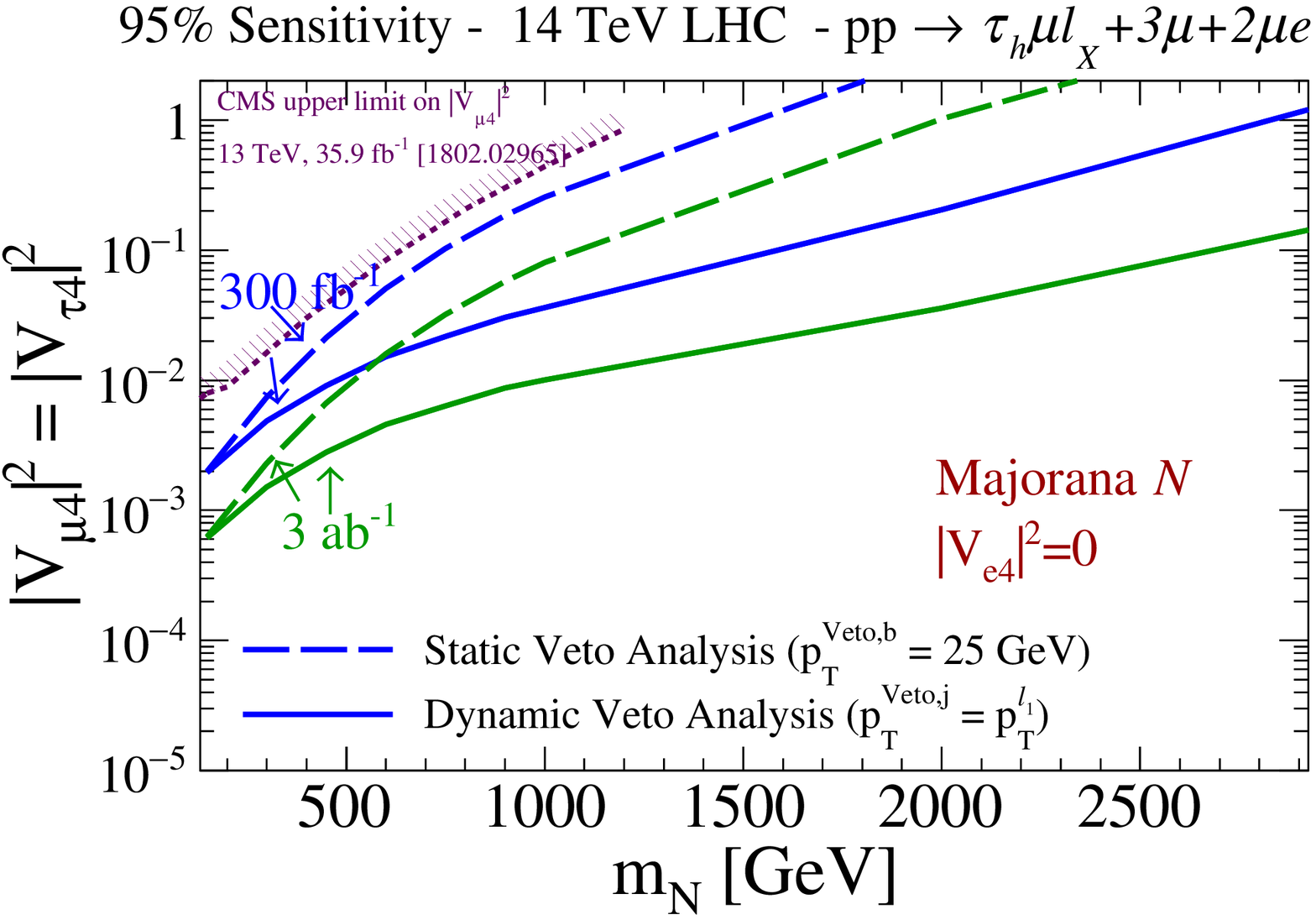}	}
\subfigure[]{\includegraphics[width=.48\textwidth]{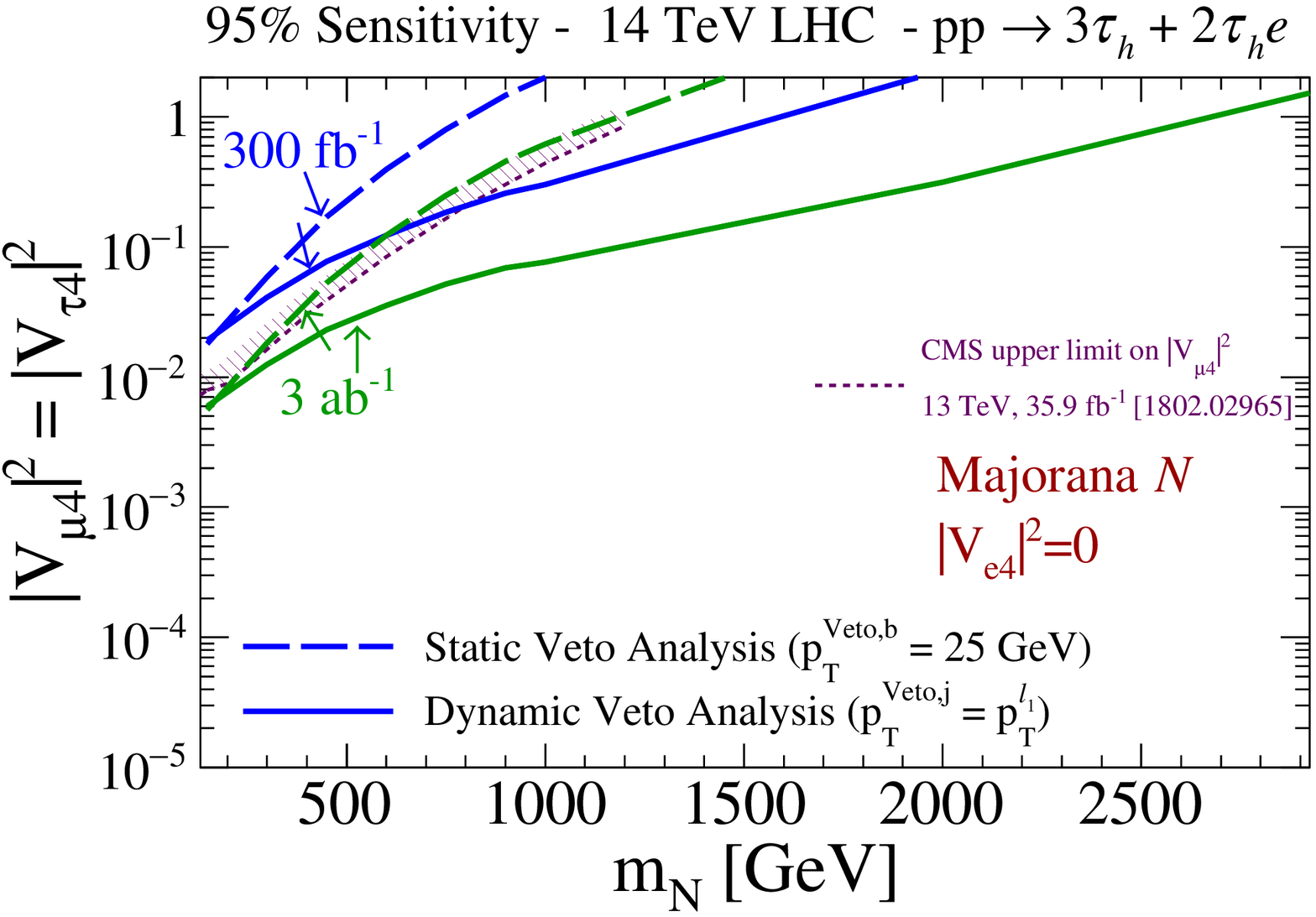}	}
\end{center}
\caption{Same as Fig.~\ref{fig:results_LHC14_cLFV} but for a single Majorana $N$ in a phenomenological Type I Seesaw.}
\label{fig:results_LHC14_cLFV_Major}
\end{figure}

In Fig.~\ref{fig:results_LHC14_cLFV}, we plot the anticipated sensitivity to the trilepton process for the cLFV scenarios:
(a) \texttt{EMU-I}, (b) \texttt{EMU-II}, (c) \texttt{ETAU-I},  (d) \texttt{ETAU-II}, (e) \texttt{MUTAU-I},  (f) \texttt{MUTAU-II}.
We observe several common features:
(i) For the lowest masses considered $(m_N = 150-300\GeV)$, we find that the dynamic jet veto analysis improves the sensitivity only marginally compared to present strategies.
This is attributed, in part, to the stringent $S_T$ cut in Eq.~\ref{cut:ST}, which greatly overlaps with the signal region for the lightest $N$,
and to our preferred choice of transverse mass variable mass (see discussion near Eq.~\ref{eq:DefMultibodyMT}).
(ii) For the highest masses $(m_N = 1-2\TeV)$, we find that the new analysis improves sensitivity by roughly $7-15\times$ at $\mathcal{L} = 300\invfb$
and well over $10\times$ at $\mathcal{L} = 3\invab$.
(iii) This increase in improvement is so large that for much of the mass range considered, 
the dynamic veto analysis with $\mathcal{L} = 300\invfb$ performs better than the standard analysis at $3\invab$.
(iv) For most cases, we see that mixing 
as low as $\vert V_{\ell 4}\vert^2 \sim 3-6\times10^{-4}$ can be probed for $m_N \lesssim 300\GeV$ and
as low as  $\sim 5\times10^{-3}$ for $m_N \sim 1\TeV$.
The exceptions are (d) \texttt{ETAU-II} and  (f) \texttt{MUTAU-II}, which include fewer sub-channels than other signatures and also suffers from double and triple $\tau_h$-tagging.
For these channels, the sensitivity is uniformly reduced by about one order of magnitude.
The similarity in reach across the various flavor scenarios follows from the near identical nature of the collider signatures themselves (see Table~\ref{tb:SignalRegions}),
meaning that differences are due to particle identification.
Furthermore, for $m_N\gtrsim300\GeV$, the improvement of the trilepton analysis now makes it competitive in sensitivity 
to hadron collider searches for LNV mediated by heavy Majorana neutrinos~\cite{delAguila:2008cj,Atre:2009rg,Deppisch:2015qwa,Antusch:2016ejd,Cai:2017mow}, 
which possess considerably fewer backgrounds than the trilepton signature.

\begin{figure}[!t]
\begin{center}
\subfigure[]{\includegraphics[width=.48\textwidth]{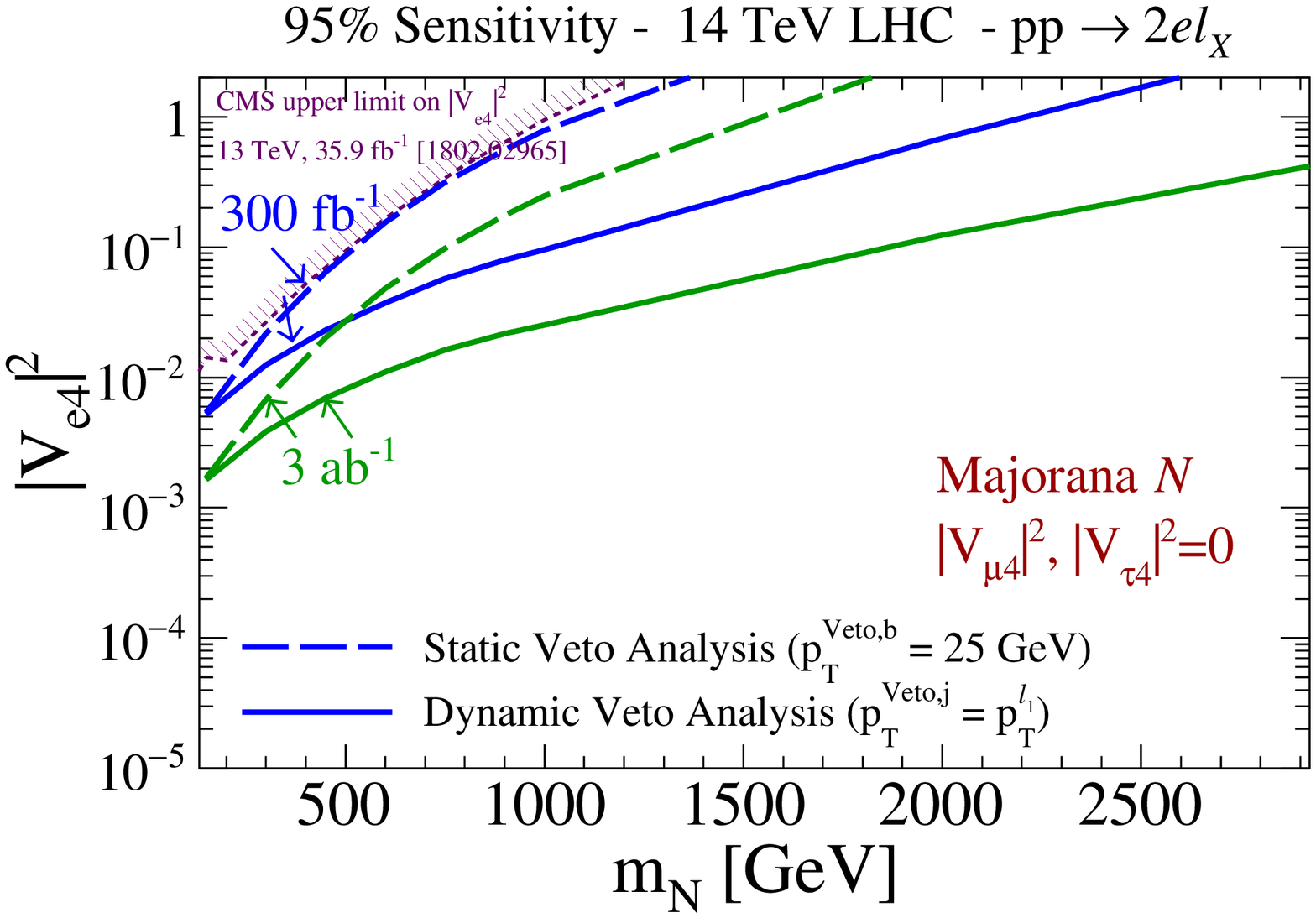}	}
\subfigure[]{\includegraphics[width=.48\textwidth]{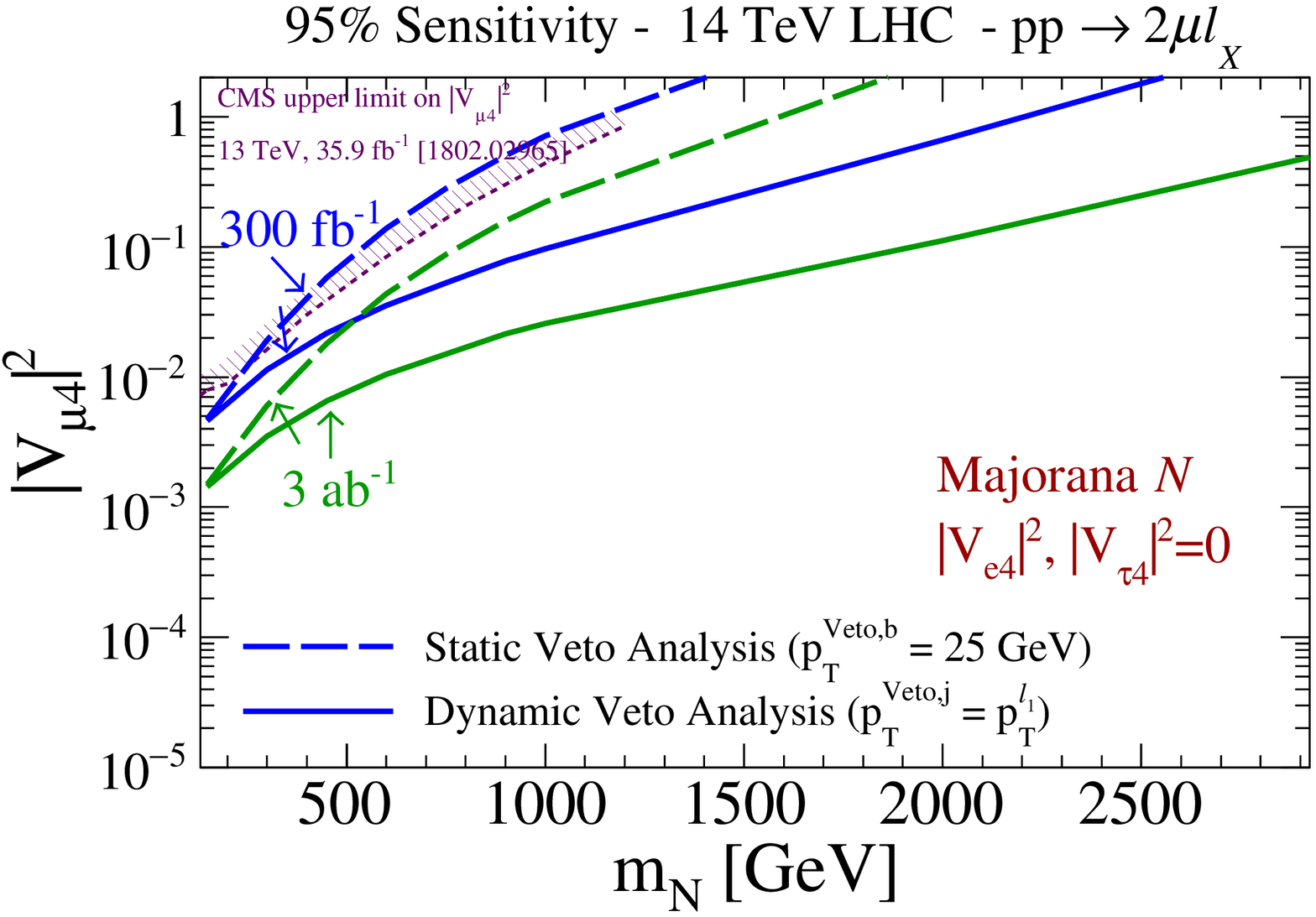}	}
\\
\subfigure[]{\includegraphics[width=.48\textwidth]{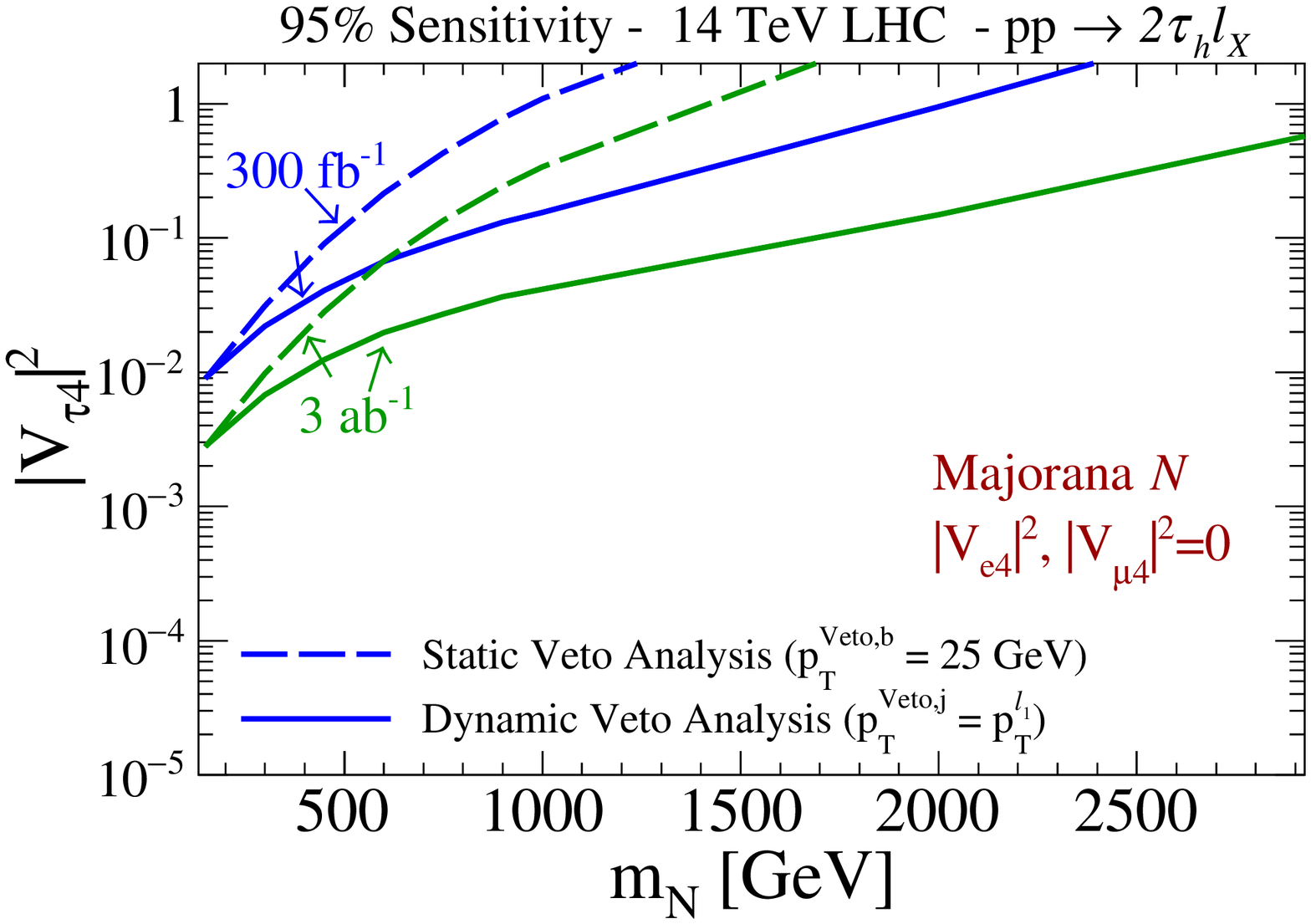}	}
\subfigure[]{\includegraphics[width=.48\textwidth]{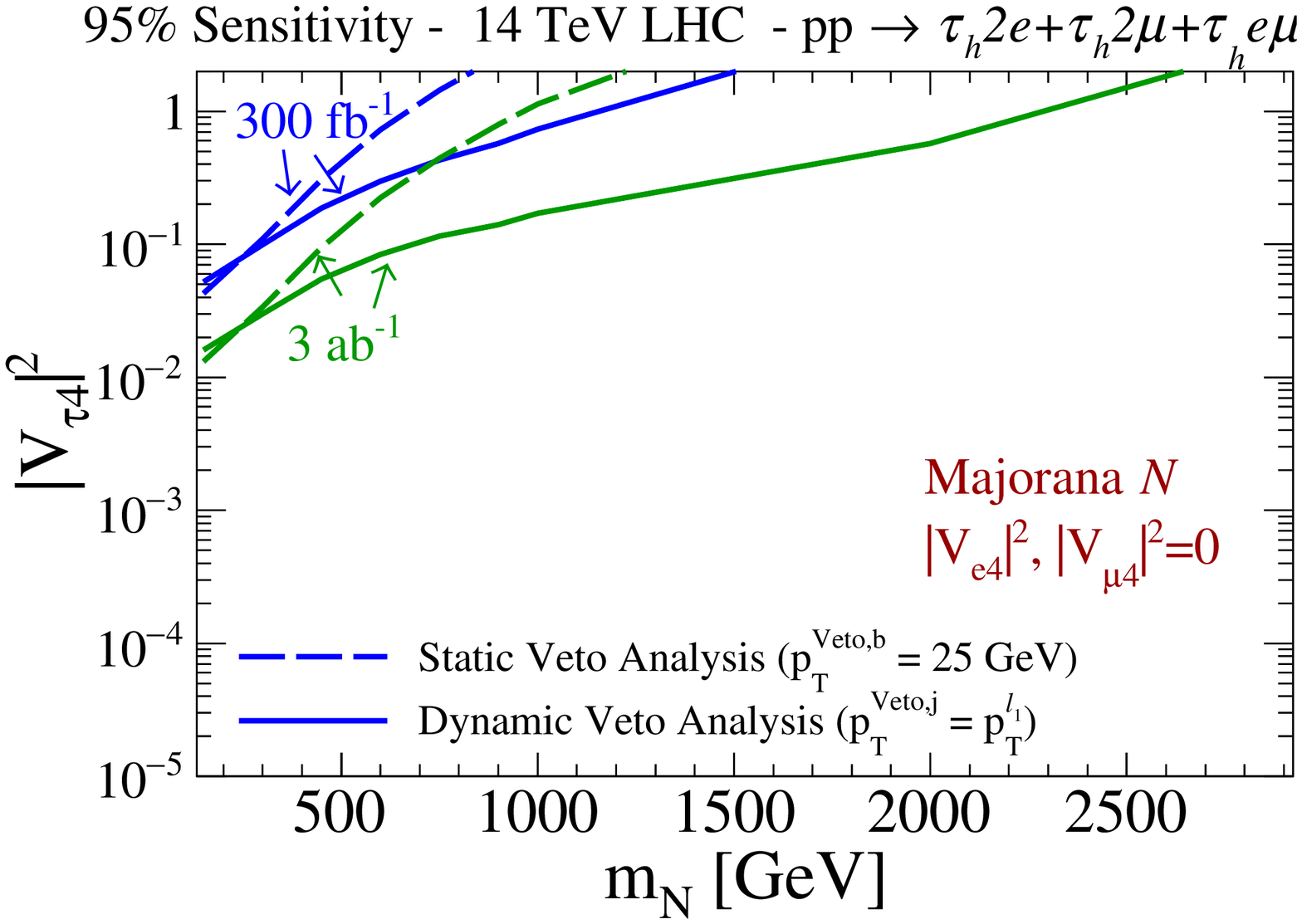}	}
\end{center}
\caption{Same as Fig.~\ref{fig:results_LHC14_cLFC} but for a single Majorana $N$ in a phenomenological Type I Seesaw.}
\label{fig:results_LHC14_cLFC_Major}
\end{figure}

In Fig.~\ref{fig:results_LHC14_cLFC}, we plot the anticipated sensitivity to the trilepton process for the cLFC scenarios:
(a) \texttt{EE}, (b) \texttt{MUMU}, (c) \texttt{TAUTAU-I}, and (d) \texttt{TAUTAU-II}.
Qualitatively, the relative improvement of the dynamic jet veto analysis over the standard analysis is largely the same.
An interesting exception to this is (d) the \texttt{TAUTAU-II} signature, where the performance of the standard analysis is noticeably better for $m_N\lesssim 300\GeV$.
Here, we believe that a dynamic veto of $\pTVeto=p_T^{\ell_1}$ actually does more harm than good since all three charged leptons carry significant 
less momentum than the scaling behavior summarized in Eq.~\ref{eq:pTlXSummary}. 
For $\tau \to \tau_h+\nu$ decays, one expects $\tau_h$ to carry only about half the momentum of the parent $\tau$ lepton;
likewise, for $\tau \to e/\mu+ 2\nu$, the $e/\mu$ carries only about a third of the original momentum.
This implies that the $\pTVeto$ threshold is much lower in reality, and reduces signal acceptance, as see in Fig.~\ref{fig:jetVetoEffdyxNLONNLL}.
Quantitatively,  we see a uniform reduction in sensitivity by nearly a factor of $2\times$ with respect to the cLFV cases.
This is due to a smaller $pp\to N\ell X$ production cross section for $N$ coupling to only one charged lepton,
the difference being  $\sigma_{\rm Tot.} = \sigma(N\ell_1X)$ as oppose to $\sigma_{\rm Tot.} = \sigma(N\ell_1X) + \sigma(N\ell_2X)$.
 
\subsubsection{Heavy Majorana Neutrinos at the $\sqrt{s}=14$ TeV LHC}\label{sec:ResultsMajorana}

For completeness, we now report the impact of our proposed dynamic jet veto analysis at $\sqrt{s}=14\TeV$ for a single Majorana $N$ in the context a phenomenological Type I Seesaw.
The analysis and signal categories are unchanged from the Dirac case, in Sec.~\ref{sec:ResultsDirac}.
The sole exception to this is the construction of the multi-body transverse mass variable $(M_{MT})$, in Eq.~\ref{eq:DefSigMultibodyMT}.
As described in Sec.~\ref{sec:ColliderParticleKin}, for Dirac neutrinos, only two permutations of the three-lepton system can possibly reconstruct to the true $M_{MT}$ of $N$
by charge conservation;
 for Majorana neutrinos, there are four.
Hence, we consider all four combinations in our  brute force guess-and-test determination of the $M_{MT}$ closest to our hypothesis $m_N$.

In Fig.~\ref{fig:results_LHC14_cLFV_Major}, we plot the anticipated sensitivity to the trilepton process mediated by a single Majorana neutrino for the cLFV scenarios:
(a) \texttt{EMU-I}, (b) \texttt{EMU-II}, (c) \texttt{ETAU-I},  (d) \texttt{ETAU-II}, (e) \texttt{MUTAU-I},  (f) \texttt{MUTAU-II}.
In all cases, we observe very comparable sensitivity for the Majorana neutrino scenario as we do for the Dirac neutrino scenario.
For the benchmark trilepton analysis, the two sets of results are nearly indistinguishable, which follows from the analysis being largely independent of the heavy neutrino's Majorana character.
For the dynamic veto analysis, the Majorana case features a slightly worse sensitivity, which we attribute to the increased likelihood of building the incorrect  $M_{MT}$. 

In Fig.~\ref{fig:results_LHC14_cLFC_Major} we plot the anticipated sensitivity to the trilepton process for the cLFC scenarios:
(a) \texttt{EE}, (b) \texttt{MUMU}, (c) \texttt{TAUTAU-I}, and (d) \texttt{TAUTAU-II}.
Again, little  difference between the Dirac and Majorana cases are observed and need not be discussed further.

\subsubsection{Heavy Dirac Neutrinos at $\sqrt{s}=27$ and $100$ TeV}\label{sec:ResultsBeyond}

\begin{table}[!t]
\begin{center}
 \begin{tabular}{ c }
 \hline\hline
 Analysis Object Requirements Changes at $\sqrt{s}=27$ TeV          \tabularnewline
 $p_T^{j} > 30\GeV$       \tabularnewline\hline
 \hline
 Analysis Object Requirements Changes at $\sqrt{s}=100$ TeV         \tabularnewline
 $p_T^{e,~(\mu),~[\tau_h],~\{j\}} > 20~(20)~[35]~\{35\}\GeV$       \tabularnewline\hline
 \hline
 Safe Jet Veto Analysis Changes at $\sqrt{s}=27~(100)$ TeV          \tabularnewline 
$S_T>150~(175)\GeV$                                      \tabularnewline\hline
\hline
\end{tabular}
\caption{Relative to Tb.~\ref{tb:SelectionCuts14TeV}, 
the changes to the analysis object requirements at (top) $\sqrt{s}=27\TeV$ and (middle) $\sqrt{s}=100\TeV$;
(bottom) changes to the safe jet veto analysis at  $\sqrt{s}=27~(100)\TeV$.}
\label{tb:SelectionCutsBeyondLHC}
\end{center}
\end{table}

\begin{figure}[!t]
\begin{center}
\subfigure[]{\includegraphics[width=.48\textwidth]{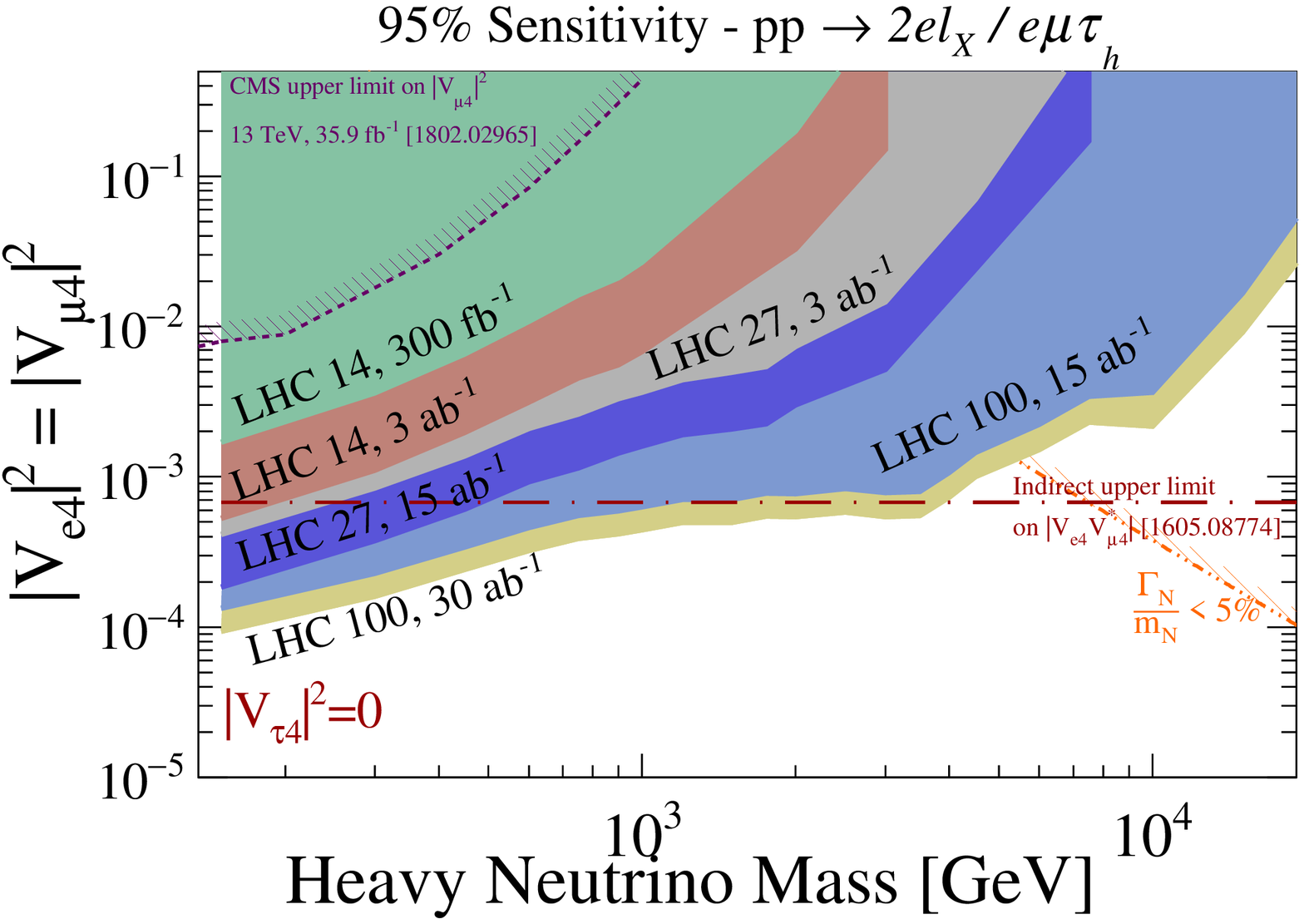}	}
\subfigure[]{\includegraphics[width=.48\textwidth]{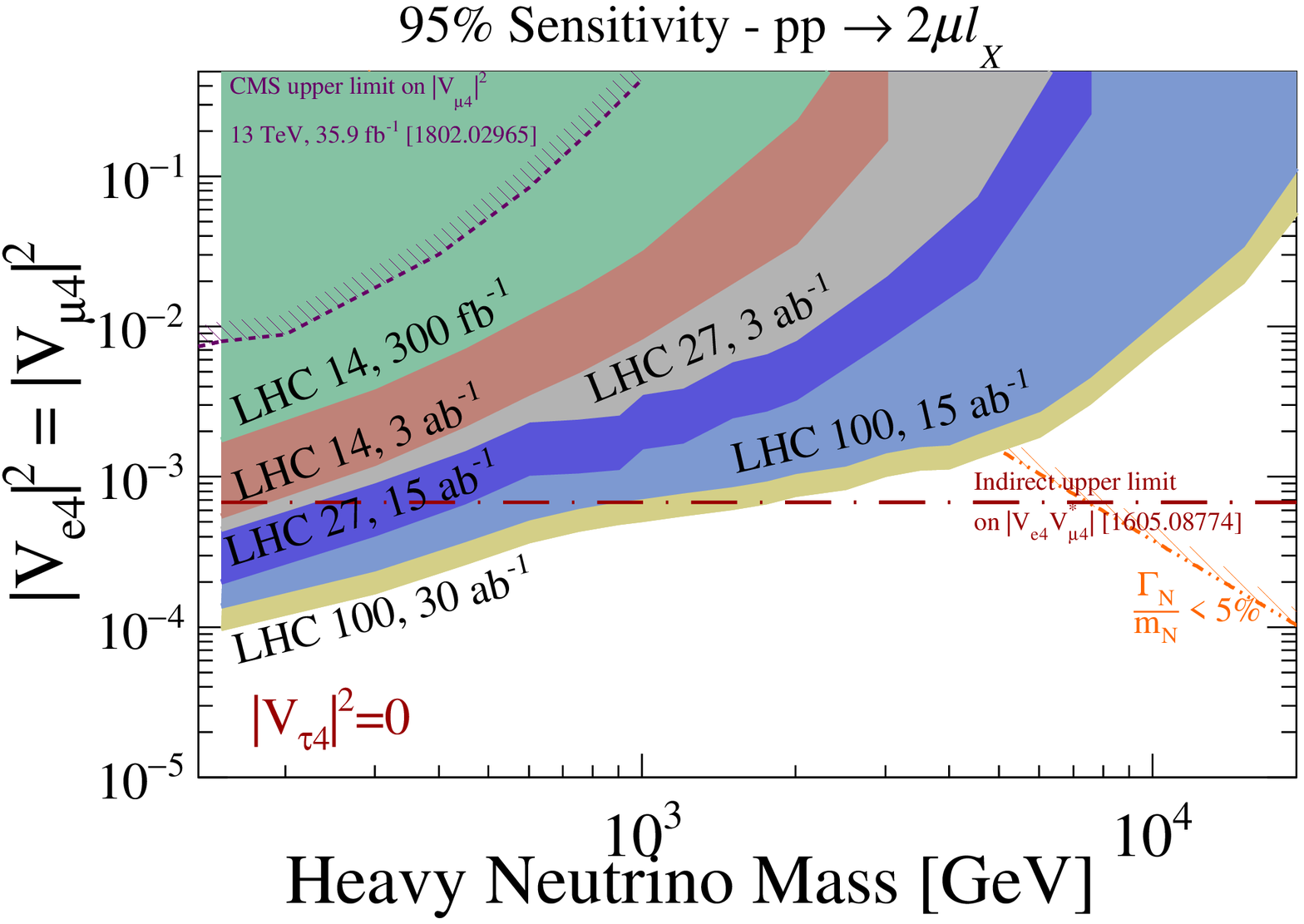}	}
\\
\subfigure[]{\includegraphics[width=.48\textwidth]{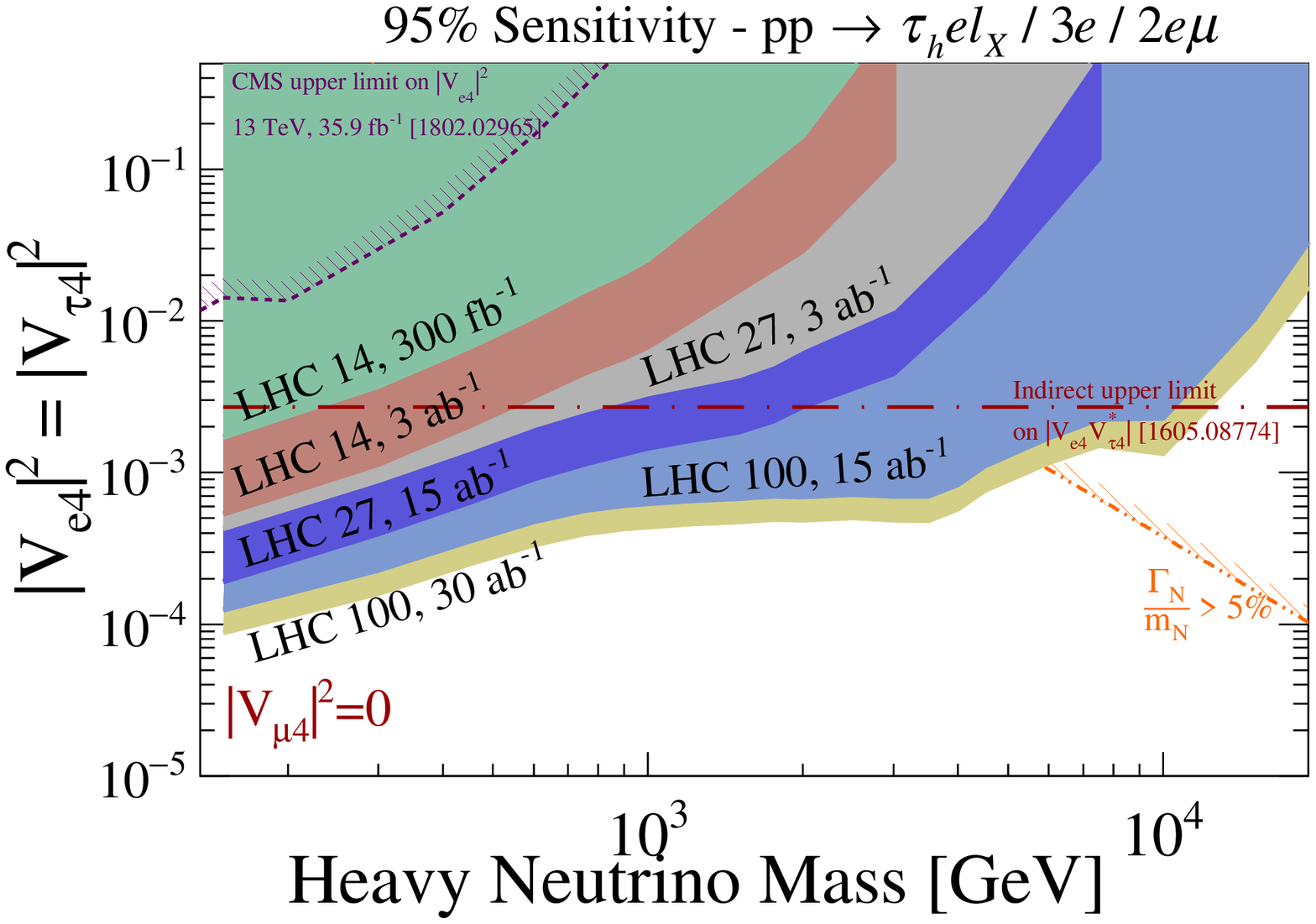}	}
\subfigure[]{\includegraphics[width=.48\textwidth]{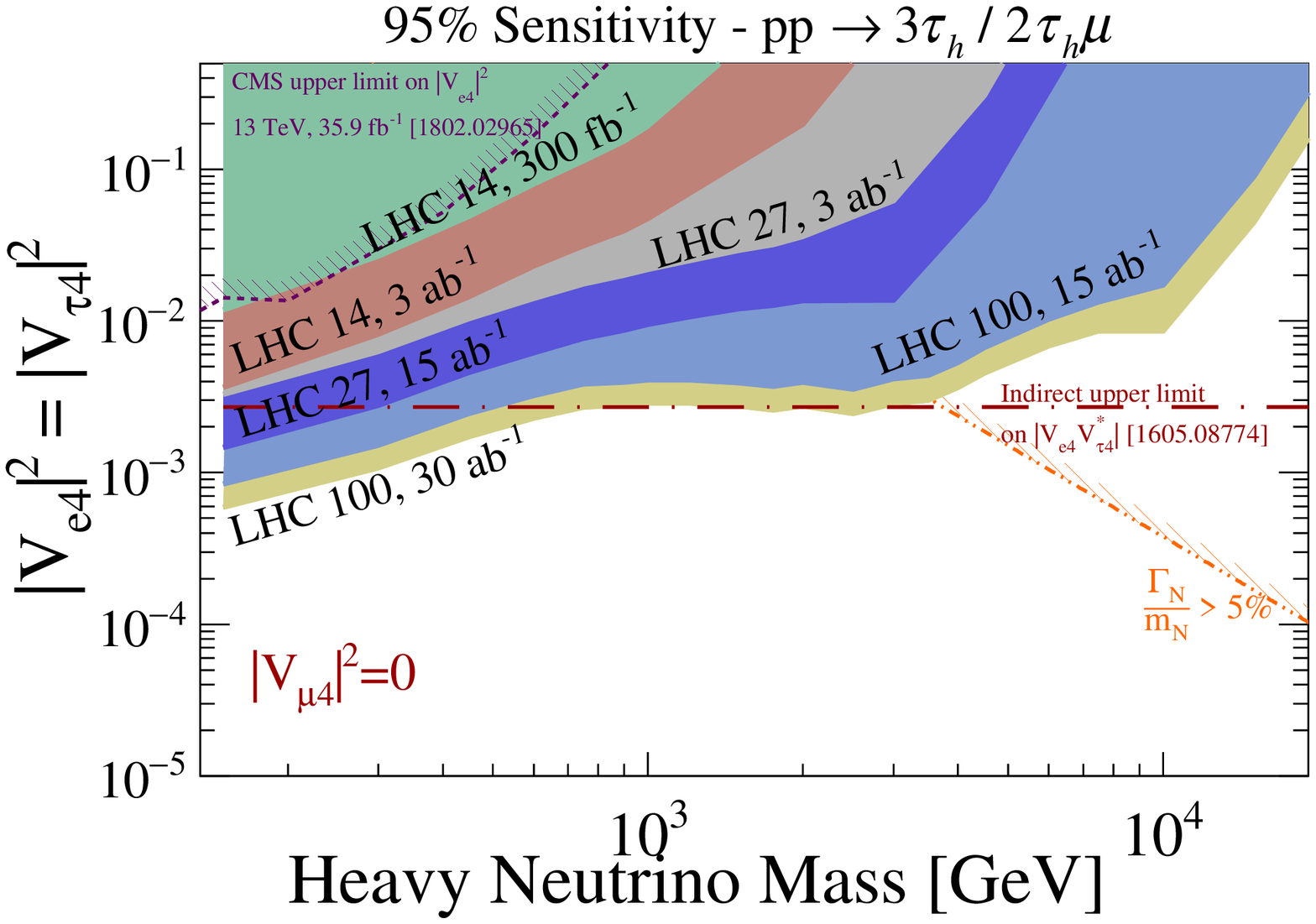}	}
\\
\subfigure[]{\includegraphics[width=.48\textwidth]{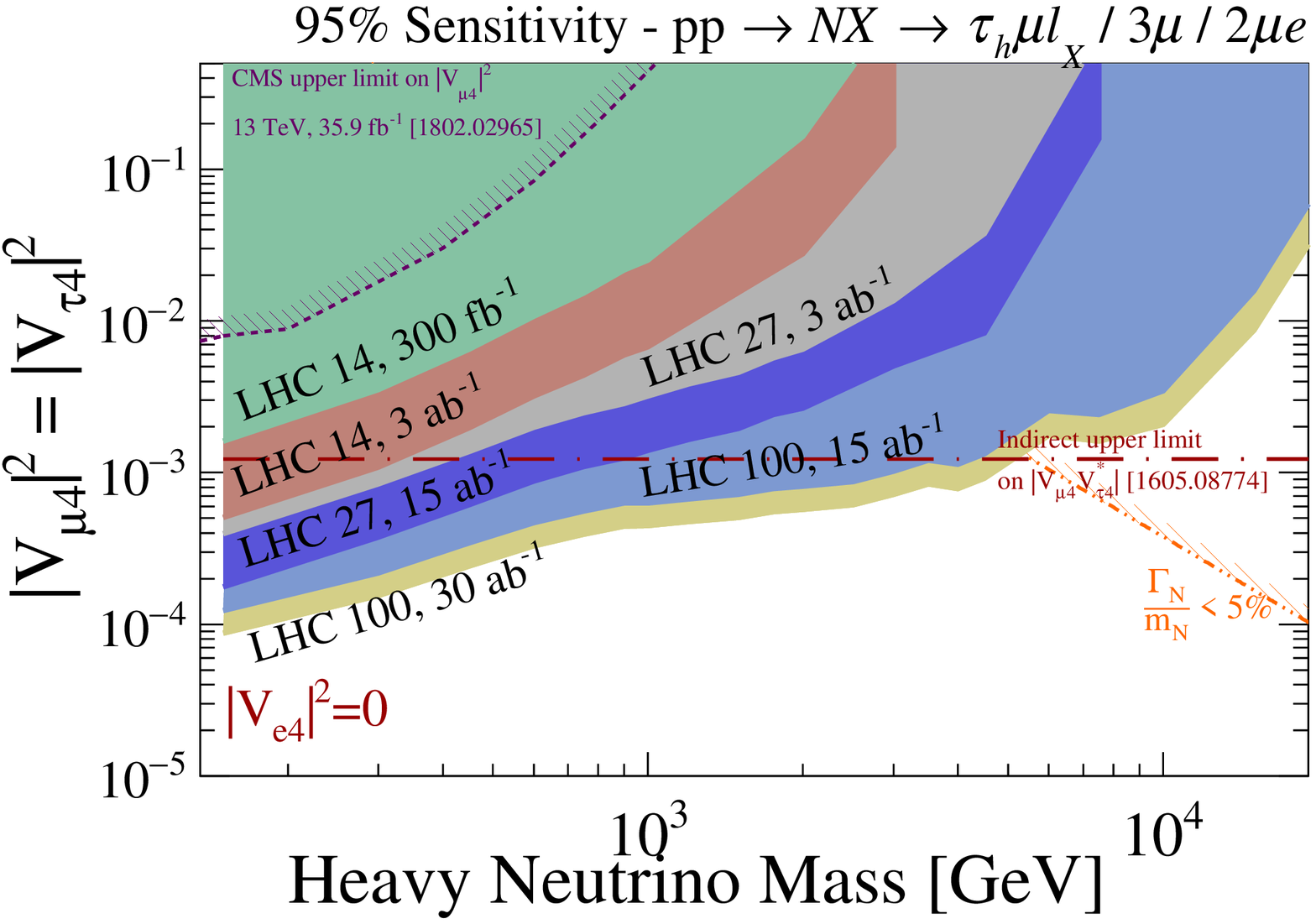}	}
\subfigure[]{\includegraphics[width=.48\textwidth]{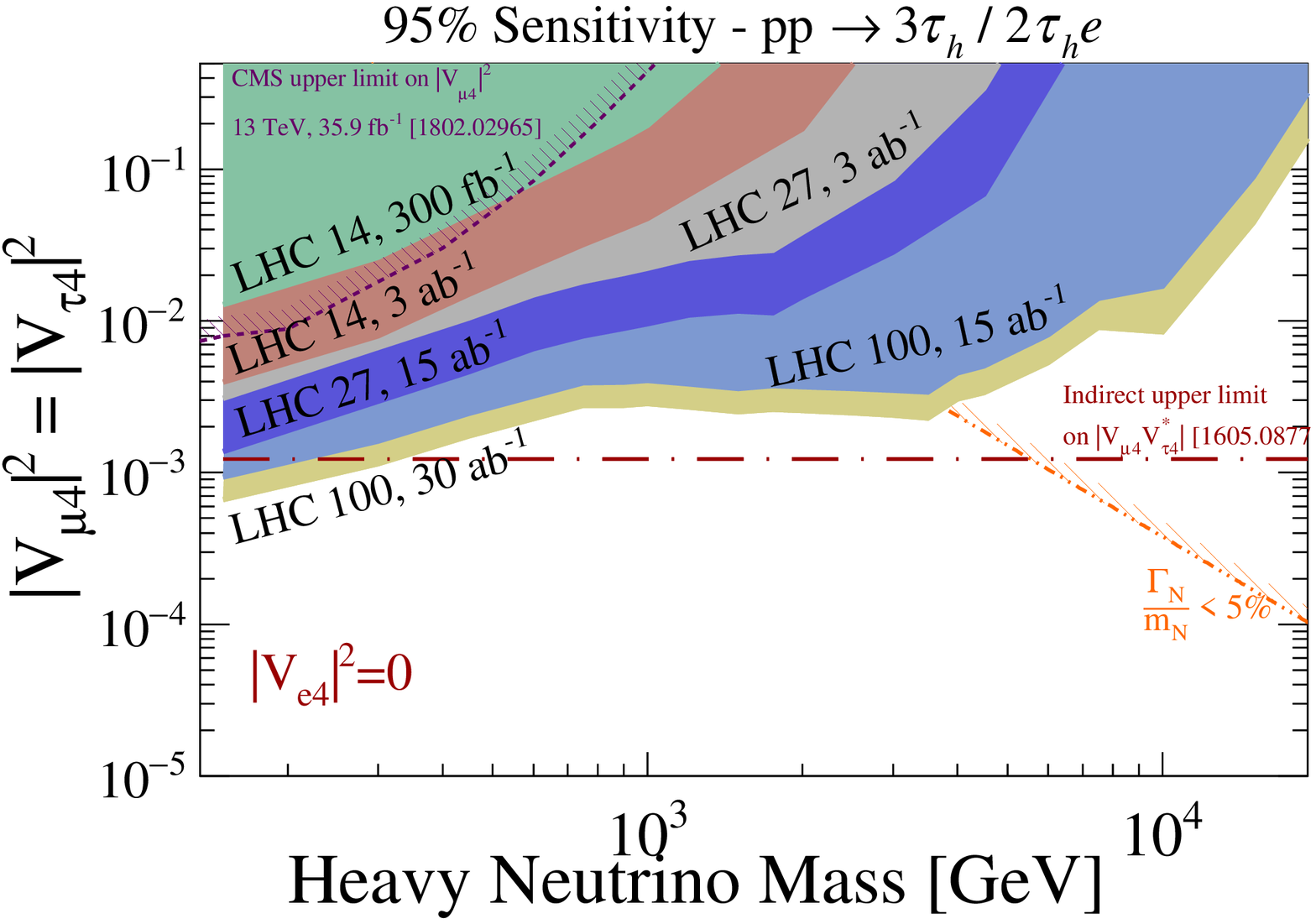}	}
\end{center}
\caption{
95\% CL sensitivity to active-sterile mixing $\vert V_{\ell 4}\vert$ as a function of heavy neutrino mass using 
the proposed dynamic jet veto trilepton analysis 
assuming benchmark integrated luminosities at $\sqrt{s}=14,~27$, and $100\TeV$,
for the charged lepton flavor violating signal categories
(a) \texttt{EMU-I}, (b) \texttt{EMU-II}, (c) \texttt{ETAU-I},  (d) \texttt{ETAU-II}, (e) \texttt{MUTAU-I},  (d) \texttt{MUTAU-II}, 
as defined in Table~\ref{tb:SignalRegions}. 
Also shown are limits from direct LHC searches~\cite{Sirunyan:2018mtv}, indirect global constraints~\cite{Fernandez-Martinez:2016lgt}, and perturbativity.
}
\label{fig:results_Global_cLFV}
\end{figure}

Encouraged by the improved sensitivity of trilepton searches to Dirac neutrinos at $\sqrt{s}=14\TeV$,
we consider the sensitivity one may have at a successor of the LHC~\cite{Avetisyan:2013onh,Arkani-Hamed:2015vfh,Mangano:2016jyj,Golling:2016gvc,CidVidal:2018eel}.
In particular, we consider the proposed HE-LHC at $\sqrt{s}=27\TeV$ with $\mathcal{L}=3$ and $15\invab$ of data, 
and a hypothetical $\sqrt{s}=100\TeV$ VLHC at $\mathcal{L}=15$ and $30\invab$.
We follow the Snowmass 2013 recommendations~\cite{Avetisyan:2013onh} and assume the LHC fiducial detector coverage and (mis)tagging efficiencies for all three collider scenarios.
As stressed in Figs.~\ref{fig:partonKinElleta} and \ref{fig:particleKinElleta}, restricting the $\eta$ coverage for charged leptons to only $\vert \eta \vert < 2.5$
has a detrimental impact on signal acceptance and background rejection efficiencies at higher collider energies. Hence, the results reported here are conservative.
All signatures remain unchanged from the definitions in Table~\ref{tb:SignalRegions}. 
Only minor changes to the definitions of analysis-quality objects given in Eq.~\ref{cut:AnaObjDef} for $\sqrt{s}=14\TeV$.
The only notable change in our analysis is a slight increase in the $S_T$.
These changes to our collider analysis are summarized in Table~\ref{tb:SelectionCutsBeyondLHC}.

In Fig.~\ref{fig:results_Global_cLFV}, we plot the 
95\% CL sensitivity to active-sterile mixing $\vert V_{\ell 4}\vert$ as a function of heavy neutrino mass using 
the proposed dynamic jet veto trilepton analysis 
assuming benchmark integrated luminosities at $\sqrt{s}=14,~27$, and $100\TeV$,
for the charged lepton flavor violating signal categories
(a) \texttt{EMU-I}, (b) \texttt{EMU-II}, (c) \texttt{ETAU-I},  (d) \texttt{ETAU-II}, (e) \texttt{MUTAU-I},  (d) \texttt{MUTAU-II}, 
as defined in Table~\ref{tb:SignalRegions}. 
We also show limits from direct LHC searches for the trilepton process~\cite{Sirunyan:2018mtv}, 
the indirect global constraints~\cite{Fernandez-Martinez:2016lgt} listed in Eq.~\ref{eq:EWPOconstraints}, 
and perturbativity requirements on the heavy neutrino's total width, as given in Eq.~\ref{eq:NuWidth}.

Globally, we see a promising picture:
In absolute numbers, with $\mathcal{L}=15~{\rm ab}^{-1}$ at $\sqrt{s}=27$ TeV, one  can probe mixing below $\vert V_{\ell N}\vert^2 = 10^{-2} ~(10^{-3})~[2\times10^{-4}]$ 
for $m_N \lesssim 3500~(700)~[200]$ GeV.
At 100 TeV with $\mathcal{L}=30~{\rm ab}^{-1}$, one can probe mixing 
as low as $9\times10^{-5}$ for $m_N \lesssim 200$ GeV,
below $10^{-3}$ for $m_N \lesssim 4$ TeV, and 
below $10^{-2}$ for $m_N \lesssim 15$ TeV.
In relative terms,
for the \texttt{EMU} cases, we see that the HL-LHC can surpass EWPD constraints for $m_N\lesssim 200\GeV$ over the program's lifetime.
This extends up to $m_N \lesssim 450~(4000)\GeV$ at that HE-LHC (VLHC) for its benchmark data cache.
For the \texttt{ETAU} scenario, remarkably, one can surpass sensitivity set by EWPD with as little as $\mathcal{L}=300\invfb$ at $\sqrt{s}=14\TeV$ for $m_N\lesssim200\GeV$.
For the HL-LHC, this extends to $m_N\lesssim 500\GeV$. At $\sqrt{s}=27~(100)\TeV$, this is extended further to $m_N\lesssim2~(12)\TeV$.
For the \texttt{MUTAU}, the case is similar in absolute reach but marginally weaker in relative reach due to stronger limits on $\mu-\tau$ mixing.

In Fig.~\ref{fig:results_Global_cLFC}, we show the results for the charged lepton flavor conserving signal categories
(a) \texttt{EE}, (b) \texttt{MUMU}, (c) \texttt{TAUTAU-I}, and (d) \texttt{TAUTAU-II}.
As seen for the $\sqrt{s}=14\TeV$ in Sec.~\ref{sec:ResultsDirac}, the reach is weaker for the cLFC channels due to the smaller production cross sections
but still of the same order of magnitude.

\begin{figure}[!t]
\begin{center}
\subfigure[]{\includegraphics[width=.48\textwidth]{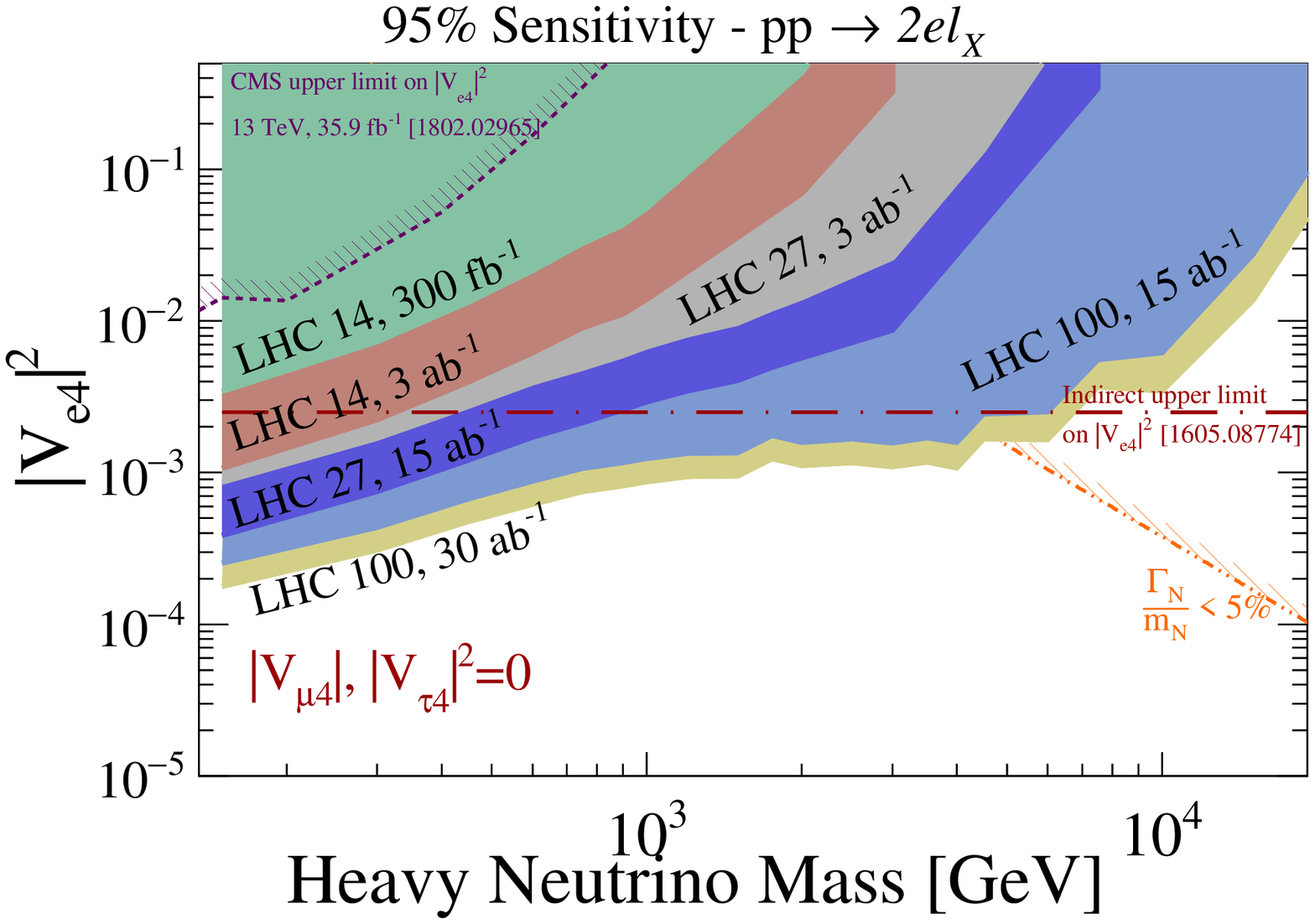}	}
\subfigure[]{\includegraphics[width=.48\textwidth]{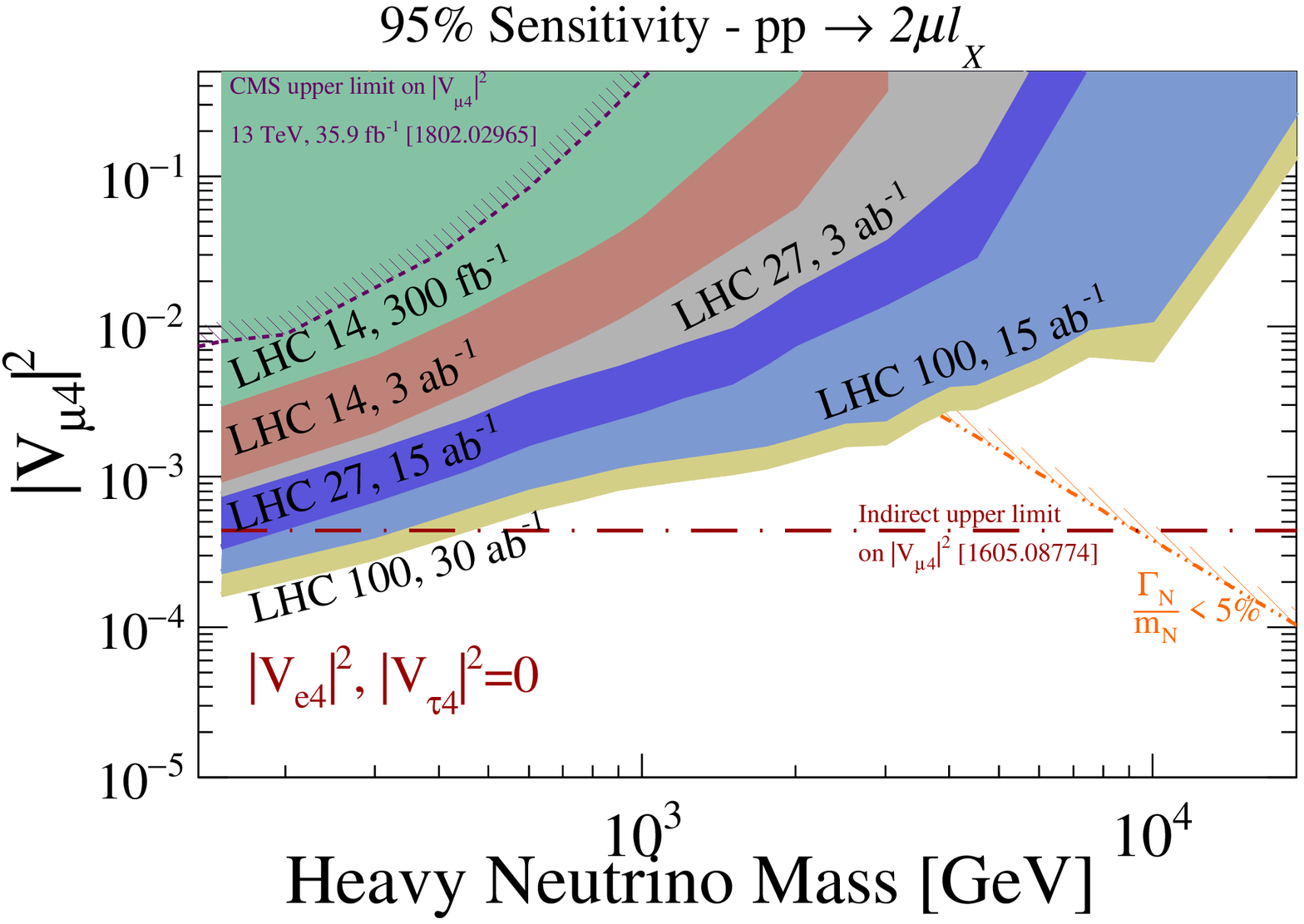}	}
\\
\subfigure[]{\includegraphics[width=.48\textwidth]{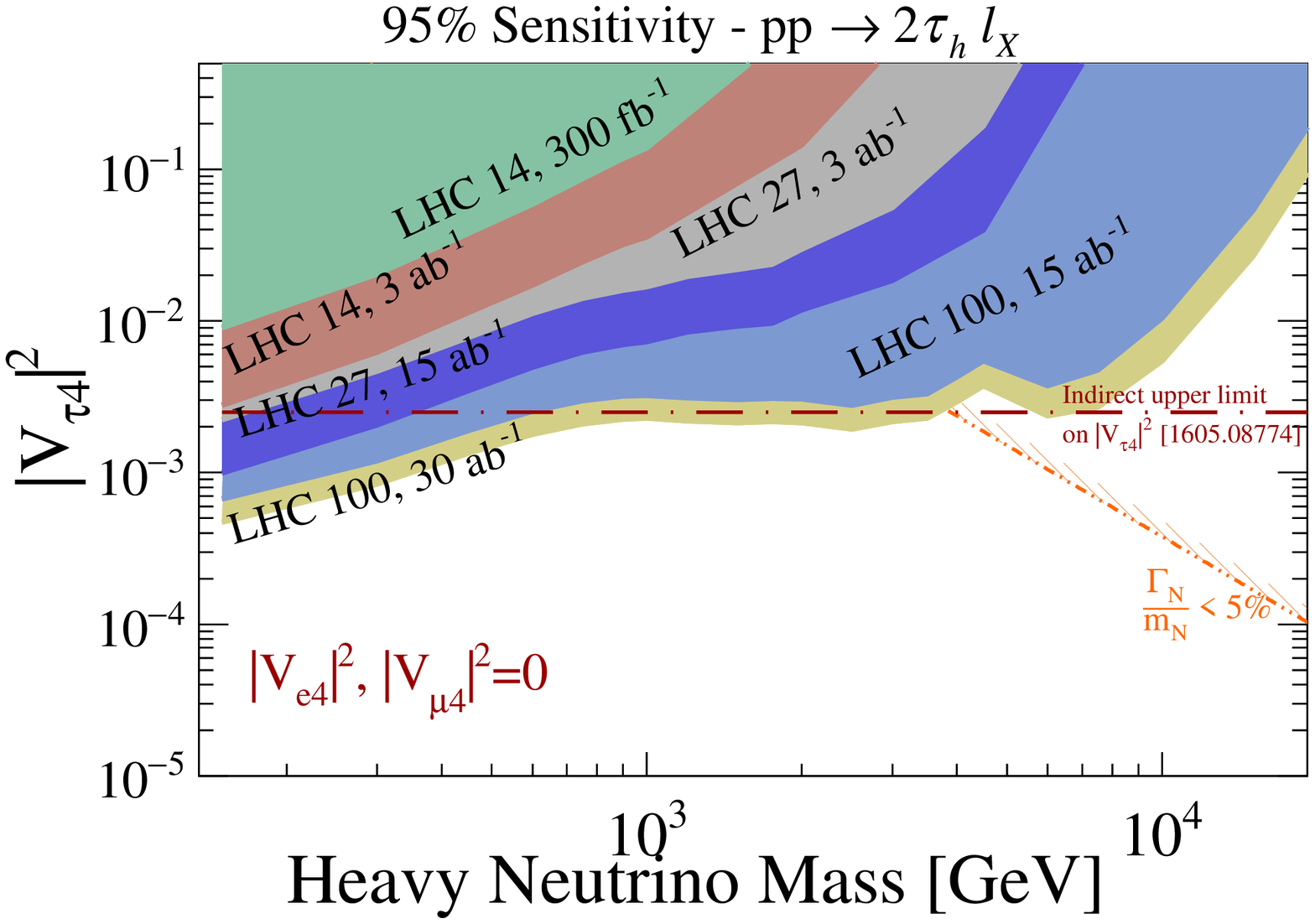}	}
\subfigure[]{\includegraphics[width=.48\textwidth]{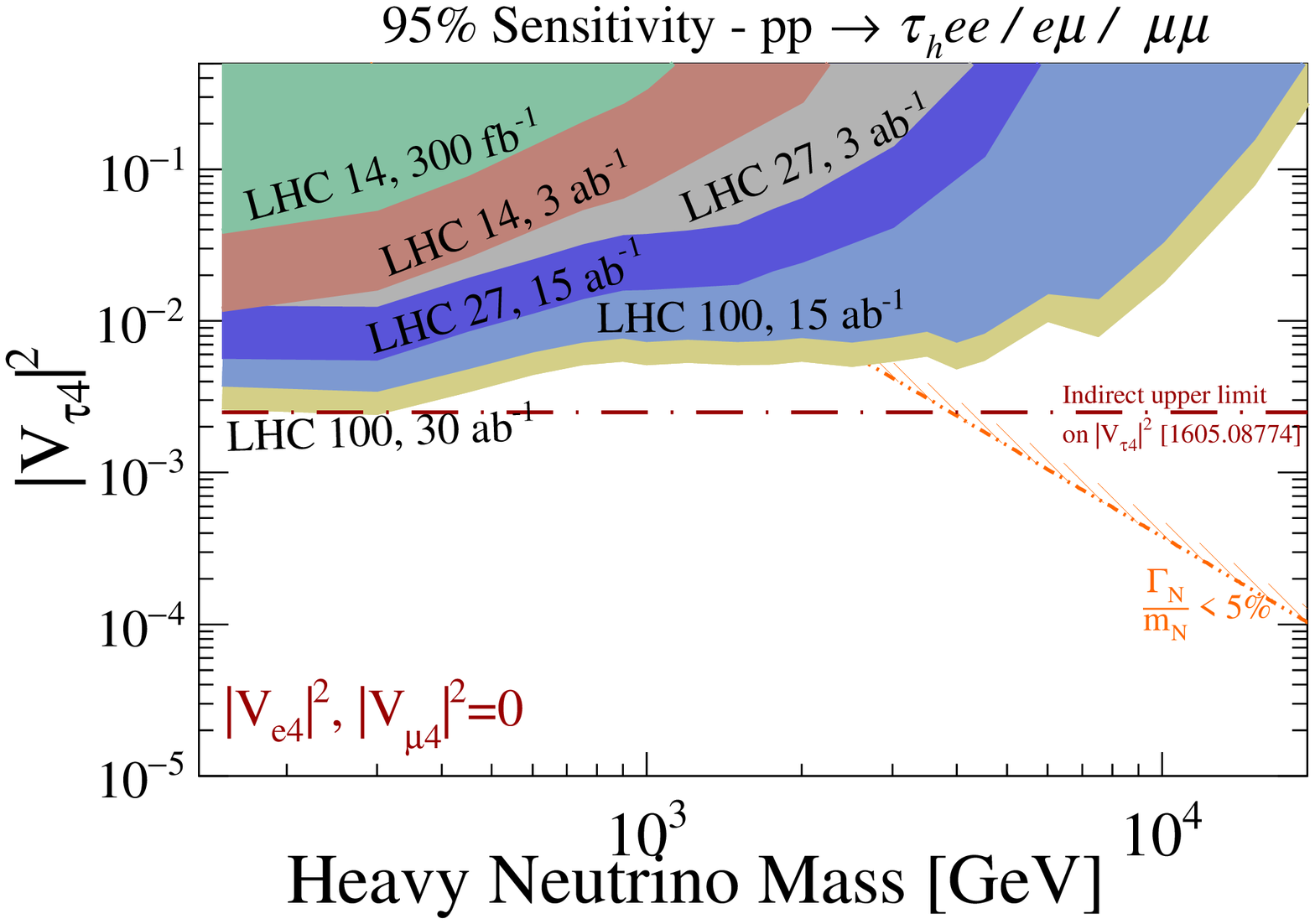}	}
\end{center}
\caption{
Same as Fig.~\ref{fig:results_Global_cLFV} but for the charged lepton flavor conserving signal categories
(a) \texttt{EE}, (b) \texttt{MUMU}, (c) \texttt{TAUTAU-I}, and (d) \texttt{TAUTAU-II} as defined in Table~\ref{tb:SignalRegions}. 
}
\label{fig:results_Global_cLFC}
\end{figure}

\section{Summary and Outlook}\label{sec:Summary}

Heavy neutrinos $(N)$ are one of the best-motivated (though far from only) solutions to  
how active neutrinos are so much lighter than fermions whose masses have 
been confirmed~\cite{Sirunyan:2018hoz,Aaboud:2018zhk,Aaboud:2018pen} to originate via the SM Higgs mechanism.
If they exist, however, their Majorana nature, the values of their masses, and the degree to which they couple to SM particles are far from clear.
Hence, one must take a broad approach in searching for heavy neutrinos,
particularly at colliders, which can directly probe $N$ with EW- and TeV-scale masses.

To this extent, we have systematically reassessed the discovery potential of heavy neutrinos with masses $m_N\geq150\GeV$,
 decaying to purely leptonic final states at the $\sqrt{s}=14$ TeV LHC, as given, for example, in Fig.~\ref{fig:feynman_DYX_Trilep}.
 The impetus for this examination is three-fold:
 (i) Present LHC searches for heavy neutrinos~\cite{Aad:2015xaa,Sirunyan:2018mtv,Sirunyan:2018xiv,Aaboud:2018spl} 
follow the seminal hadron collider analyses of Refs.~\cite{Keung:1983uu,Han:2006ip,delAguila:2008cj,Atre:2009rg},
which justifiably argue that search strategies should rely on the existence of high-$p_T$ charged leptons
when $N$ are produced solely through the charged current Drell-Yan (CC DY) mechanism.
It is now known, though, that alternative production mechanisms can compete or outright dominate over the CC DY mechanism.
(ii) New tools have been created that can more reliably model heavy neutrinos in hadron collisions.
Similarly, the understanding of jets has so significantly improved that jet observables can be used reliably as discriminates in measurements and searches.
(iii) Third, the community is presently assessing possible successors to the LHC 
program~\cite{Avetisyan:2013onh,Arkani-Hamed:2015vfh,Mangano:2016jyj,Golling:2016gvc,CidVidal:2018eel}.
Understandably, the analyses prescribed in Refs.~\cite{Keung:1983uu,delAguila:2008cj,Atre:2009rg}
were not designed to be robust against increasing collider energy.
It is unclear, therefore, if future projections based on current search strategies can reliably estimate the 
true discovery potential of a $\sqrt{s}=27$ HE-LHC or a $\sqrt{s}=100\TeV$ VLHC. 
We report investigating the matter and propose a new search strategy 
that both improves current sensitivity and remains robust at future colliders.
To reach this result, we have considered the following:

In Sec.~\ref{sec:Colliders} we have explored and compared the resonant production of heavy neutrinos $N$ through 
a variety of mechanisms for $\sqrt{s}=14-100\TeV$ and $m_N \geq 150\GeV$.
At 14 TeV, the CC DY and $W\gamma$ fusion (VBF) channels are dominant; 
at 27-50 TeV, gluon fusion (GF) emerges as an important and complementary channel for $m_N \approx 300-1000\GeV$;
and outright dominates for $m_N \lesssim 2\TeV$ at 100 TeV, beyond which VBF is the leading production vehicle.
We stress that few, if any, studies exist demonstrating the discovery potential of such heavy $N$ through the GF channel at hadron colliders.
Normalized distributions were then presented for a number of leptonic observables 
across $\sqrt{s}=14-100\TeV$ at LO and NLO+PS precision.
This revealed the existence of a class of observables, namely inclusive quantities built from the transverse momenta of an event's leading charged leptons,
 whose distribution shapes display only a weak sensitivity to changes in $\sqrt{s}$.
This suggests the ability to build a search analysis resilient across varying collider energies.

In Sec.~\ref{sec:jetVeto}, we turned to studying the behavior of hadronic observables over $\sqrt{s}=14-100\TeV$,
for both the signal and background processes, and at various fixed-order and resummed precision.
We focused exceptionally on hadronic activity in the context of a jet veto.
When the veto's transverse momentum threshold $(\pTVeto)$ is set to a static value of $\pTVeto=30-100\GeV$ 
for jet radii $R=0.1-1.0$ at $\sqrt{s}=14-100\TeV$, we find discouraging survival efficiencies for signal benchmarks.
Intriguingly, when using a dynamic jet veto, 
and in particular setting $\pTVeto$ on an event-by-event basis to the leading charged lepton $p_T$ in an event, we find qualitatively opposite behavior.
Survival efficiencies for signal benchmarks reach $\vareps(\pTVeto)\gtrsim 90-95\%$ and remain largely independent of $m_N$ 
(for $m_N\gtrsim 200\GeV$), $R$, and $\sqrt{s}$;
\confirm{QCD scale uncertainties reduce at NLO+PS(LL) and NLO+NNLL for signal and background processes};
and QCD background rejection capabilities in fact improve for increasing $\sqrt{s}$.
The impact of such a veto scheme on VBF-like topologies, 
events with final-state $\tau$ leptons that decay hadronically,
top quark and EW backgrounds, and, significantly, ``fake lepton'' backgrounds were also addressed.
A summary of our work for a broad audience was first reported in Ref.~\cite{Pascoli:2018rsg}.

Based on the findings reported in  Secs.~\ref{sec:Colliders} and Sec.~\ref{sec:jetVeto},
a new methodology is proposed in Sec.~\ref{sec:Observability} to search for heavy neutrinos $N$ in the trilepton final state,
i.e., $pp \to N\ell +X \to 3\ell + \met + X$.
The search analysis is based on employing a dynamic jet veto in conjunction with exclusive $S_T$.
The combination effectively discriminates according to the relative amount of leptonic and hadronic activities in an event.
For various active-sterile mixing/flavor hypotheses, the proposed analysis is compared with a state-of-the-art, 
benchmark analysis inspired by the analogous LHC Run II search~\cite{Sirunyan:2018mtv}.
At 14 TeV, we find that the proposed analysis can improve sensitivity to heavy Dirac neutrinos by an order of magnitude (in $(m_N,\vert V \vert^2)$-space) for $m_N\gtrsim150\GeV$.
This improvement is achievable at both \confirm{$\mathcal{L}=300\invfb$ and $3\invab$,} with the largest improvement occurring for the largest $m_N$, and is independent of flavor hypothesis.
In absolute terms, the $\sqrt{s}=14$ TeV LHC with $\mathcal{L}=3~{\rm ab}^{-1}$, 
can probe active-sterile mixing as small as $\vert V_{\ell N}\vert^2 = 10^{-2} ~(10^{-3})~[5\times10^{-4}]$  at 95\%~CL for heavy Dirac neutrinos masses $m_N \lesssim 1200~(300)~[200]$ GeV.
This is well  beyond the present $\vert V_{\ell N}\vert^2 \lesssim 10^{-3} -10^{-1}$ constraints for such heavy states set by indirect searches and precision measurements,
particularly for the $\tau$ channel where constraints are at the $\vert V_{\ell N}\vert^2 \lesssim 10^{-2} -10^{-1}$ level~\cite{Parke:2015goa,Fernandez-Martinez:2016lgt}.
For details, see Sec.~\ref{sec:ResultsDirac}.
For heavy Majorana neutrinos, \confirm{comparable improvements and reach are found}; see Sec.~\ref{sec:ResultsMajorana}.
For $\sqrt{s}=27$ and 100 TeV, and with only minor tunes to the analysis, the anticipated reach for Dirac neutrinos is extraordinary.
With $\mathcal{L}=15~{\rm ab}^{-1}$ at $\sqrt{s}=27$ TeV
one  can probe mixing below $\vert V_{\ell N}\vert^2 = 10^{-2} ~(10^{-3})~[2\times10^{-4}]$ for $m_N \lesssim 3500~(700)~[200]$ GeV.
At 100 TeV with $\mathcal{L}=30~{\rm ab}^{-1}$, one can probe mixing 
as low as $9\times10^{-5}$ for $m_N \lesssim 200$ GeV,
below $10^{-3}$ for $m_N \lesssim 4$ TeV, and 
below $10^{-2}$ for $m_N \lesssim 15$ TeV;
see Figs~\ref{fig:results_Global_cLFV} and~\ref{fig:results_Global_cLFC}.
In none of the above cases was the proposed analysis optimized using multi-variant or machine learning techniques;
we expect further improvement with such action.

Our overarching computational framework, which largely takes advantage of out-of-the-box, 
public Monte Carlo tools is described in great detail in Sec.~\ref{sec:Setup}.
This includes the publication of a Dirac neutrino-variant of the \texttt{FeynRules}-based 
\texttt{HeavyN} libraries~\cite{Degrande:2016aje}, and is available from the \texttt{FeynRules} 
database at \href{http://feynrules.irmp.ucl.ac.be/wiki/HeavyN}{feynrules.irmp.ucl.ac.be/wiki/HeavyN}.

The anomalous production of  charged leptons without hard, central hadronic activity in $pp$ collisions 
is a key prediction of low-scale Type I Seesaw models.
The prediction, however, is not unique:
The success of the dynamic jet veto relies on the color structure of the hard, signal process and the ability to measure (even by proxy) the mass scale on an event-by-event basis.
Hence, we believe much of our findings are immediately applicable to other searches for colorless particles that
can be produced and decayed through colorless force carriers.
For example:
the production of exotically charged scalars and fermions, respectively from the Types II and III Seesaws;
the production of heavy charged scalars that decay to long-lived stable, neutral particles, as commonly found in loop-level Seesaw and dark matter models;
pair production of slepton and electroweakino pairs in Supersymmetry~\cite{Fuks:2019iaj};
and pair production of $N$ through $Z'$ gauge bosons in $U(1)_{B-L}$ theories.

Moreover, we do not believe that setting the dynamic veto threshold $\pTVeto$ to the leading charged lepton in an event is always most optimal.
(Indeed, we found in some cases that the subleading charged lepton gave better performance.)
The exploration of other leptonic observables that are sensitive to the hard scale, e.g., $S_T$ and $\met$, is left to future studies.
Similarly, vetoing on alternative hadronic observables, just at $H_T$ or jet mass,
are also possibilities where further investigations are encouraged; see, e.g., Ref.~\cite{Fuks:2019iaj}.

\section{Conclusions}\label{sec:Conclusions}

Colliders offer a complementary probe of the origin of active neutrino masses, their mixing, as well as their potential Majorana nature.
If EW- or TeV-scale heavy neutrinos play a role in the generation of neutrino masses, and if such particles couple appreciably to the SM sector, 
then the LHC and its successors provide fertile ground for their discovery and properties determination.
In this study, we have systematically assessed the discovery potential of heavy neutrinos at $pp$ colliders with $\sqrt{s}=14-100\TeV$.
This has been motivated by the unequivocally improved understanding of neutrino and jet physics,
the sophistication of state-of-the-art Monte Carlo tools, 
and the mandates of the several HEP strategy updates ongoing at the time of this work.
Relative to ongoing search-strategies~~\cite{delAguila:2008cj,Sirunyan:2018mtv},
we find that the LHC's sensitivity can be improved by \confirm{an order of magnitude} in both the immediate term and over its lifetime.
We report that the LHC's anticipated sensitivity can exceed by several factors present constraints for such heavy states set by indirect searches and precision data.
The increase in sensitivity can be attributed to a newly proposed search strategy (see Sec.~\ref{sec:Observability}) 
for heavy neutrinos decaying into a purely leptonic final state that, intuitively, 
relies on the ability to discriminate an event's global leptonic activity from its hadronic activity.
Crucially, this employs an unusual jet veto scheme, 
one where the $p_T$ veto threshold is set on an event-by-event basis to the $p_T$ of the leading charged lepton in the event.
The functionality of this dynamic jet veto is described in technical detail in Sec.~\ref{sec:jetVeto};
a more broadly accessible summary is presented Ref.~\cite{Pascoli:2018rsg}.
In addition, the proposed methodology exhibits by design an unusual resilience against variable collider energy,
and therefore can serve as a baseline for future collider searches for multi-lepton searches of heavy neutrinos (see Sec.~\ref{sec:ResultsBeyond}) 
as well as other colorless exotica.

The results of this study are encouraging and we look forward to the prospect of at last unveiling the mystery underlying tiny neutrino masses.

\subsection*{Acknowledgments}
Karl Nordstr\"om and Martin Tamarit's isospin partner are thanked for valuable discussions.
The computing staff at Durham University's Institute for Particle Physics and Phenomenology and 
Universit\'e Catholique de Louvain's Centre for Cosmology, Particle Physics, and Phenomenology 
are thanked for their incredible technical support.
Use of Durham University's ADM and PAULI-CAT computing systems is acknowledged.
This work was funded in part by the UK STFC, and the European Union's Horizon 2020 research and innovation programme 
under the Marie Sklodowska-Curie grant agreements No 690575 (RISE InvisiblesPlus) and No 674896 (ITN ELUSIVE).
RR is supported under the UCLouvain MSCA co-fund ``MOVE-IN Louvain,''
the F.R.S.-FNRS ``Excellence of Science'' EOS be.h Project No 30820817,
and acknowledges the contribution of the COST Action CA16108.
SP and CW received financial support from the European Research Council under the European Unions Seventh Framework
Programme (FP/2007-2013)/ERC Grant NuMass Agreement No. 617143.
SP acknowledges partial support from the Wolfson Foundation and the Royal Society.
CW was supported in part by the U.S.~Department of Energy under contract DE-FG02-95ER40896 and in part by the PITT PACC.


\bibliography{issVetoRefs}

\end{document}